\documentclass[a4paper, 12pt, oneside]{Thesis}  % Use the "Thesis" style, based on the ECS Thesis style by Steve Gunn
\graphicspath{Figures/}  % Location of the graphics files (set up for graphics to be in PDF format)

\usepackage[square, numbers, comma, sort&compress]{natbib}  % Use the "Natbib" style for the references in the Bibliography
\usepackage{verbatim}  % Needed for the "comment" environment to make LaTeX comments
\usepackage{vector}  % Allows "\bvec{}" and "\buvec{}" for "blackboard" style bold vectors in maths
\hypersetup{urlcolor=blue, colorlinks=true}  % Colours hyperlinks in blue, but this can be distracting if there are many links.
\usepackage[utf8]{inputenc}
\usepackage{fancyhdr}
\usepackage{amsmath}
\usepackage{bm}
\usepackage{lmodern}
\usepackage{amsbsy}
\usepackage{amssymb}
\usepackage{amsthm}
\usepackage{amsfonts}
\usepackage{chngcntr}
\usepackage{tikz}
\usepackage{pgfplots}
\pgfplotsset{compat=newest}
\pgfplotsset{compat=1.6}
\usetikzlibrary{decorations.markings}
\usetikzlibrary{decorations.pathmorphing}
\usetikzlibrary{patterns,hobby}
\usetikzlibrary{calc}
\usetikzlibrary{arrows}
\usetikzlibrary{shapes}
\usetikzlibrary{positioning}
\usepackage{float}
\usepackage{parskip}
\usepackage{esvect}
\usepackage[normalem]{ulem}
\usepackage{mathtools}
\usepackage{bookmark}
\counterwithin{equation}{section}
\counterwithin{equation}{subsection}

\begin{document}
\frontmatter      % Begin Roman style (i, ii, iii, iv...) page numbering

% Set up the Title Page
\title  {A quantum probabilistic approach to the emergence of an inflationary Starobinsky Universe}
\authors  {\texorpdfstring
            {\href{your web site or email address}{Eftychios Kaimakkamis}}
            {}
            }
\addresses  {\groupname\\\deptname\\\univname}  % Do not change this here, instead these must be set in the "Thesis.cls" file, please look through it instead
\date       {January 2026}
\subject    {}

\maketitle
%% ----------------------------------------------------------------

%\setstretch{1.1}  % It is better to have smaller font and larger line spacing than the other way round

% Define the page headers using the FancyHdr package and set up for one-sided printing
\fancyhead{}  % Clears all page headers and footers
%\rhead{\thepage}  % Sets the right side header to show the page number
\lhead{}  % Clears the left side page header

\pagestyle{fancy}  % Finally, use the "fancy" page style to implement the FancyHdr headers

%% ----------------------------------------------------------------
% Declaration Page required for the Thesis, your institution may give you a different text to place here

%% ----------------------------------------------------------------
% The "Funny Quote Page"
\pagestyle{empty}  % No headers or footers for the following pages

\clearpage

\Validation{

\vspace{8pt}
\textbf{Doctoral Candidate:} Eftychios Kaimakkamis\\
\vspace{5pt}

    \textbf{Doctoral Thesis title:} A quantum probabilistic approach to the emergence of an inflationary Starobinsky Universe \\
\vspace{8pt}

\emph{The present Doctoral Dissertation was submitted in partial fulfillment of the requirements for the Degree of Doctor of Philosophy at the Department of Physics and was approved on January 20th 2026 by the members of the Examination Committee.}\\

\vspace{8pt}

\textbf{Examination Committee:}
%\vspace{2pt}
\begin{itemize}
    \item \underline{Research Supervisor}: Associate Professor Nicolaos Toumbas, University of Cyprus \\\\\_\_\_\_\_\_\_\_\_\_\_\_
    \item \underline{Head of the Committee}: Professor Haralambos Panagopoulos, University of Cyprus \\\\\_\_\_\_\_\_\_\_\_\_\_\_
    \item \underline{Committee Member}: Associate Professor Constantinos Moulopoulos, University of Cyprus \\\\\_\_\_\_\_\_\_\_\_\_\_\_
    \item \underline{Committee Member}: Professor Alexandros Kehagias, National Technical University of Athens \\\\\_\_\_\_\_\_\_\_\_\_\_\_
    \item \underline{Committee Member}: Assistant Professor Ioannis Florakis, University of Ioannina \\\\\_\_\_\_\_\_\_\_\_\_\_\_
\end{itemize}
}
\clearpage

\Declaration{

\addtocontents{toc}{\vspace{1em}}  % Add a gap in the Contents, for aesthetics
\vspace{20pt}
The present doctoral dissertation was submitted in partial fulfillment of the requirements for the degree of Doctor of Philosophy of the University of Cyprus. It is a product of original work of my own, unless otherwise mentioned through references, notes, or any other statements.
\vspace{30pt} 
 
Signed: Eftychios Kaimakkamis \\\\
\_\_\_\_\_\_\_\_\_\_\_\_
}
\clearpage

\Summary{

\addtocontents{toc}{\vspace{1em}}

Στην παρούσα διατριβή η πρότασή μας είναι ότι οι απαραίτητες συνθήκες για την ύπαρξη του πληθωρισμού θα μπορούσαν να εξηγηθούν σε πιθανολογική βάση. Ερμηνεύουμε τις φυσικές καταστάσεις $\psi$ που παίρνουμε από τον φορμαλισμό των συναρτησιακών ολοκληρωμάτων σαν πλάτη πυκνότητας πιθανότητας ώστε να μετρήσουμε το Σύμπαν σε κάποια κατάσταση με συγκεκριμένη γεωμετρία. Λύνοντας για αυτές τις κυματοσυναρτήσεις, μπορούμε να υπολογίσουμε σχετικές πιθανότητες για κάθε κοσμολογική περίοδο. Μπορούμε τότε να εξετάσουμε ποια από αυτές είναι πιθανότερη για διάφορες τιμές του παράγοντα κλίμακας $a$ που περιγράφει το φυσικό μέγεθος του Σύμπαντος. \par
Ξεκινάμε παραθέτοντας τα βασικά στοιχεία του μοντέλου της Θερμής Μεγάλης Έκρηξης και βρίσκοντας τις εξισώσεις $\text{Friedmann}$ που περιγράφουν την δυναμική του Σύμπαντος. Παρέχουμε μια ενδελεχή σύνοψη της θερμικής ιστορίας του Σύμπαντος και των διαφόρων κοσμολογικών εποχών, ενώ κάνουμε αναφορά στα ανοιχτά ερωτήματα που προκύπτουν, όπως το πρόβλημα του ορίζοντα, το πρόβλημα της επιπεδότητας, ή το πρόβλημα της τιμής της κοσμολογικής σταθεράς. \par
Στη συνέχεια προχωράμε στον πληθωρισμό, ξεκινώντας από μια εισαγωγή στον χώρο $\text{de Sitter}$. Παραθέτουμε τις απαραίτητες αρχικές συνθήκες, την μηχανική, και τις φυσικές συνέπειες του πληθωρισμού. Συζητάμε πώς ο πληθωρισμός μπορεί να παρέχει λύσεις στα προβλήματα της Θερμής Μεγάλης Έκρηξης, και σκιαγραφούμε τον μηχανισμό με τον οποίο δημιουργούνται αρχέγονες διαταραχές στην πυκνότητα ενέργειας από κβαντικές αναταράξεις από την πρώιμη πληθωριστική περίοδο. \par
Έπειτα, περιγράφουμε πληθωριστικά μοντέλα που περιλαμβάνουν ανώτερες τάξεις της καμπυλότητας στην βαρυτική δράση. Εστιάζουμε στην Σύμμορφη/$\text{Weyl}$ βαρύτητα, το μοντέλο $\text{Starobinsky}$, και την βαρύτητα $R^{2}$. Δείχνουμε πώς η θεωρία $R^{2}$ μπορεί να ιδωθεί σαν προσέγγιση του μοντέλου $\text{Starobinsky}$ κατά την πληθωριστική του φάση. Δίνεται μια κλασική ανάλυση των μοντέλων, αναπτύσσοντας επάνω στις τροποποιημένες βαρυτικές εξισώσεις και την μετατροπή της δράσης σε κανονική δράση $\text{Einstein-Hilbert}$ στην παρουσία βαθμωτού πεδίου ύλης $\phi$ με βαθμωτό δυναμικό $V(\phi)$. Αναλύουμε το φάσμα των διαταραχών της μετρικής γύρω από τις κλασικές λύσεις. Κλείνουμε το κεφάλαιο με μία ανάλυση για τους συνοριακούς όρους. \par
Στο πρώτο μας ερευνητικό έργο εξετάζουμε το πρόβλημα των ασαφειών της διάταξης τελεστών στην εξίσωση $\text{Wheeler-DeWitt}$. Αφιερώνουμε κάποιο χρόνο να εξηγήσουμε κάποια βασικά για την εξίσωση και τις εννοιολογικές βάσεις της κβαντικής κοσμολογίας. Αναπτύσσουμε έναν φορμαλισμό για την διαχείριση των ασαφειών στην κανονική κβάντωση ενός μονοδιάστατου μοντέλου $\text{minisuperspace}$ με το να ορίσουμε τις λεγόμενες συναρτήσεις ασαφειών στην εξίσωση $\text{WDW}$. Ορίζουμε ένα εσωτερικό γινόμενο χώρου $\text{Hilbert}$ για τον υπολογισμό μετρήσιμων φυσικών μεγεθών, και χρησιμοποιώντας το επιχείρημα της ερμιτιανότητας της Χαμιλτονιανής, παίρνουμε μία έκφραση για το μέτρο ολοκλήρωσης $\mu$ σε συνδυασμό με περιορισμούς στις συναρτήσεις ασάφειας. Με τον ορισμό μιας νέας κβαντικής κατάστασης $\Psi =\sqrt{\mu} \, \psi$ και κάνοντας χρήση του φορμαλισμού συναρτησιακών ολοκληρωμάτων, καταφέρνουμε να αφαιρέσουμε τις ασάφειες και να αποδείξουμε ότι η εξίσωση $\text{WDW}$ - και άρα η θεωρία - είναι καθολική. Γενικεύουμε τον φορμαλισμό για κάθε $M$-διάστατο σύστημα $\text{minisuperspace}$, για το οποίο μπορούμε να δείξουμε ότι η καθολικότητα ισχύει στην ημικλασική προσέγγιση για μια ευρεία τάξη μοντέλων. Επιπλέον, δείχνουμε ότι αυτή η καθολικότητα επεκτείνεται σε όλες τις τάξεις της κβαντικής θεωρίας έαν η πολλαπλότητα του $\text{target space}$ είναι επίπεδη. \par
Στην δεύτερη μας εργασία επιζητούμε να βρούμε πλάτη πιθανότητας για διάφορα κοσμολογικά σενάρια που αφορούν τον πληθωρισμό $\text{Starobinsky}$. Συνοψίζουμε την πρόταση $\text{no-boundary}$ των $\text{Hartle-Hawking}$ στην Γενική Θεωρία της Σχετικότητας και σχολιάζουμε τις ξεχωριστές συνέπειες από την επιλογή καμπύλης ολοκλήρωσης: την επιλογή προσήμου κατά $\text{Hartle-Hawking}$ ή την επιλογή των $\text{Linde-Vilenkin}$. Κάνουμε έναν ανάλογο υπολογισμό για την θεωρία $R^{2}$ και βρίσκουμε ότι για την επιλογή $\text{Linde-Vilenkin}$ είναι στατιστικά πιθανότερο το Σύμπαν να αναδυθεί στην πληθωριστική του φάση. Τέλος, υπολογίζουμε πλάτη μετάβασης όταν αποκλίνουμε από τις κλασικές εξισώσεις κίνησης, επιτρέποντας την κινητική ενέργεια του πεδίου $\phi$ να είναι μη μηδενική. Παραθέτουμε διάφορα σενάρια για το Σύμπαν, όπως μετάβαση από μία εξέλιξη τύπου Μεγάλης Έκρηξης-Μεγάλης Καταστροφής σε μία $\text{dS}$ τύπου επέκταση και το αντίστροφο.

}
\clearpage

% The Abstract Page
\addtotoc{Abstract}  % Add the "Abstract" page entry to the Contents
\abstract{In this thesis, our proposition is that the necessary conditions for the existence of inflation could possibly be explained through a quantum probabilistic approach. We interpret the physical states  $\psi$ obtained via the path integral formulation as probability density amplitudes for the Universe to be found in a certain state with a particular geometric configuration. By solving for these wave functions, we can calculate relative probabilities for each cosmological regime. We can then examine which regime is more likely for different values of the scale factor $a$ which describes the physical size of the Universe. \par
We start by reviewing basic properties of the Hot Big Bang cosmological model and deriving the Friedmann equations that describe the dynamics of the Universe. We provide a comprehensive summary of the thermal history of the Universe and the various cosmological epochs, and we elaborate on open questions and problems that arise, such as the horizon problem, the flatness problem, or the cosmological constant problem. \par
Next, we move on to the inflationary paradigm, starting from an introduction to de Sitter space. We showcase the necessary initial conditions, the basic mechanics, and physical implications of inflation. We discuss how inflation can provide solutions to the shortcomings of the standard Hot Big Bang cosmology, and we outline the mechanism of generating the primordial density perturbations from quantum fluctuations in the early inflationary era. \par
After that, we describe inflationary models that involve higher curvature terms in the gravitational action. We focus on Conformal/Weyl Gravity, the Starobinsky model, and scale invariant gravity $R^{2}$. We show that $R^{2}$ gravity can be seen as an approximation for the inflationary phase of the Starobinsky model. A classical analysis of the models is provided, by elaborating on the modified gravitational equations, and the transformation of the action into a conventional Einstein-Hilbert action in the presence of a scalar matter field $\phi$ with a scalar potential $V(\phi)$. We then analyze the spectrum of metric perturbations about the classical solutions. We close off the chapter with an analysis on boundary terms.\par
In our first piece of research, we examine the problem of ordering ambiguities in the Wheeler-DeWitt equation. We dedicate some time to explain some basics about the equation and the conceptual fundamentals of quantum cosmology. We develop a formalism to handle ambiguities in the canonical quantization of a 1d minisuperspace by defining so-called ambiguity functions within the WDW equation. We define a Hilbert space inner product for calculating physical observables, and using an argument from the hermiticity of the Hamiltonian, we obtain an expression for its measure $\mu$ in addition to constraints on the ambiguity functions. By defining a new quantum state $\Psi =\sqrt{\mu} \, \psi$ and using the path integral formalism, we are able to resolve ambiguities and prove the WDW equation - and thus the theory - is universal. We then generalize the formalism for any $M$-dimensional minisuperspace system, for which we can prove universality in the semi-classical approximation for a large class of models. We also show that this universality extends to all orders of quantum theory if the target space manifold of the theory is flat. \par
In our second work, we seek to calculate probability amplitudes for various cosmological scenarios involving Starobinsky inflation. We summarize the Hartle-Hawking no-boundary proposal for GR and we comment on the distinct possibilities of defining a contour of integration: the sign choice of Hartle-Hawking and the one of Linde-Vilenkin. We make an analogous calculation for $R^{2}$ gravity and find that for the Linde-Vilenkin choice it is statistically more likely for the Universe to emerge in its inflationary stage. Finally, we calculate transition amplitudes when we go off-shell with the kinetic energy of the inflaton field $\phi$ taking non-zero values. We formulate a number of possible scenarios for the Universe, such as transition from a Big Bang-Big Crunch evolution to a dS-like expansion and vice versa.}  

\clearpage  % Abstract ended, start a new page
%% ----------------------------------------------------------------

\setstretch{1.3}  % Reset the line-spacing to 1.3 for body text (if it has changed)

% The Acknowledgements page, for thanking everyone
\acknowledgements{
\addtocontents{toc}{\vspace{1em}}  % Add a gap in the Contents, for aesthetics

I would like to preemptively thank the assembled committee responsible for the thesis examination and other attendants during the presentation who showed interest in this work, and honour me with their presence and willingness to provide moral support. \par
I also appreciate the valuable feedback of fellow graduate students in preparation for the examination, as well as throughout the course of my studies and research leading up to it. Special thanks to Stavros Christodoulou who was consistently present whenever needed and was a constant source of assistance in all my years in Graduate School. \par
I am grateful for all the collaborators whose contributions to the current work were instrumental for gaining the required results. I especially want to thank Hervé Partouche, Karunava Sil, and Victor Franken for their meaningful input and fruitful discussions. \par
The warmest gratitude is directed towards my academic advisor Nicolaos Toumbas for the insightful feedback and assistance over the course of my PhD studies and the completion of this thesis. His guidance and expertise were a major reason for the novelty and scope of the research results we obtained.\par
Finally, I would like to thank all my friends and family for their persistent support  and encouragement in undertaking this task. It would have been impossible to complete this thesis without the help they provided during some very demanding periods of my PhD.

}
\clearpage  % End of the Acknowledgements
%% ----------------------------------------------------------------

\vspace*{\fill}
\begin{flushright}
    \Large\textgreek{Για την μητέρα μου.}
\end{flushright}
\vspace*{\fill}

\clearpage
%% ----------------------------------------------------------------

\pagestyle{fancy}  %The page style headers have been "empty" all this time, now use the "fancy" headers as defined before to bring them back

%% ----------------------------------------------------------------
\lhead{\emph{Contents}}  % Set the left side page header to "Contents"
\tableofcontents  % Write out the Table of Contents

%% ----------------------------------------------------------------
\lhead{\emph{List of Figures}}  % Set the left side page header to "List if Figures"
\listoffigures  % Write out the List of Figures

%% ----------------------------------------------------------------
\clearpage  % Start a new page

%% ----------------------------------------------------------------

\addtocontents{toc}{\vspace{2em}}  % Add a gap in the Contents, for aesthetics

%% ----------------------------------------------------------------
\mainmatter	  % Begin normal, numeric (1,2,3...) page numbering
\pagestyle{fancy}  % Return the page headers back to the "fancy" style

% Include the chapters of the thesis, as separate files
% Just uncomment the lines as you write the chapters
\fancyhead[LE,RO]{\thepage}

\fancyhead[RE,LO]{Introduction}

\chapter{Introduction}

\section{Cosmology}

Cosmology is the branch of Physics concerned with the evolution of the Universe and the physical description of its mechanisms. It is a relatively new branch, since for the majority of the history of science the predominant idea was that the Universe was static. It was presumed to have been created in its current state with no significant alterations. With the discovery of Einstein's General Theory of Relativity in 1915, however, we suddenly had field equations that could describe gravitational phenomena through dynamic spacetimes. Within a relatively short period of time, it was revealed to Einstein that his field equations predicted a Universe that is expanding. To amend this seeming issue from the perspective of someone believing in a static Universe, he introduced the cosmological constant $\Lambda$ in 1917\cite{Reference5,Reference39,Reference36}. \par
Things would change dramatically over the next decade. The first observational hints and theoretical framework for an expanding Universe were provided by Georges Lemaître in 1927, while Edwin Hubble made the first major astronomical observations for the cosmic red-shift in 1929. Since observed galaxies are moving away from us, then the Universe must be experiencing expansion. Through experimental data, the Hubble-Lemaître law (or just "Hubble's law") was discovered, prompting Einstein to drop the cosmological constant as redundant. In an ironic development, new observations in 1998 confirmed that the Universe is not only expanding, but expanding in an accelerated manner. This effect comes from the dominance of the so-called dark energy in the Universe, and the existence of a non-zero cosmological constant is the most likely explanation for it\cite{Reference34}. 

\begin{figure}[H]
\begin{center}
\includegraphics[scale=0.35]{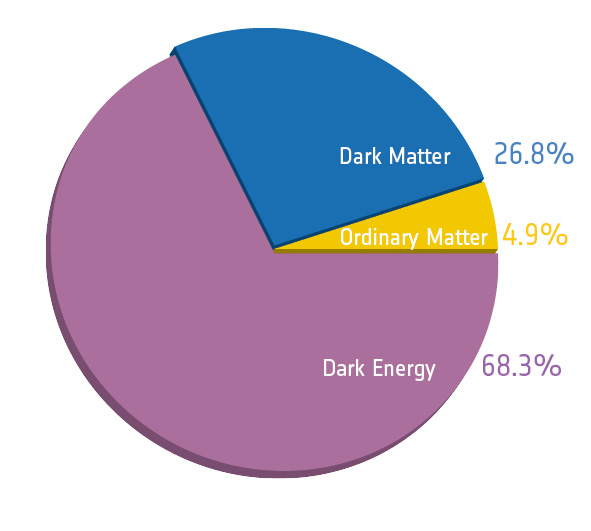}
\end{center}
\vspace{-0.6cm}
\caption{The cosmic pie of the observable Universe in its current cosmological era, with the latest measurements from the Planck telescope\cite{Reference35}. As we can see, dark energy dominates, followed by dark matter which serves to explain anomalous movements of astronomical bodies at the edge of galaxies. Ordinary matter is but a small fraction of the total matter and energy in the Universe.}
\label{cosmicpie}
\end{figure}

The knowledge that the Universe is expanding naturally led to discussions about its history. The simplest metric solution for a globally homogeneous and isotropic Universe - known as the Friedmann-Lemaître-Robertson Walker metric - predicts that the Universe had to begin at an initial singularity from which it expanded. This phenomenon was initially derided and pejoratively called a "Big Bang". However, as its theoretical merits started becoming obvious, the name was instead appropriated as a valid term to describe this phenomenon. Lemaître in 1931 theoretically predicted that the Big Bang could be traceable through radiation, while Ralph Alpher and Robert Herman theoretically predicted a Cosmic Microwave Background radiation (CMBR) in 1948. This was experimentally confirmed by Arno Penzias and Robert Woodrow Wilson in 1964. The subsequent modelling of the various cosmological eras the Universe went through came to be known as Hot Big Bang cosmology.

\section{Inflation}

Cosmic inflation was first introduced as a theoretical paradigm by Alan Guth in 1979, albeit similar ideas appeared around the same time, such as Aleksei Starobinsky's 1980 work on the concept of a "de Sitter phase" of the Universe\cite{Reference29}. These models relied on the idea that the Universe did not in fact begin with a Big Bang, but rather underwent an initial brief phase of exponential expansion. These prove viable extensions to conventional Hot Big Bang cosmology for a number of reasons. As we shall see later, inflation can help explain and solve a number of very important open questions that the Hot Big Bang is unable to do so.\par
As is suggested by the phrasing above, inflation is a paradigm, not a singular model. There are many different inflationary models that explain inflation through various mechanisms. Determing the ones which are correct has been a topic of major interest in recent years due to rapid developments and improvements of experimental techniques, as well as theoretical progress\cite{Reference2,Reference3,Reference8,Reference14}. This has helped to determine which of those models can be considered viable. One such model of interest which we will use as a convenient working basis is the Starobinsky model. The Starobinsky model is a model of modified gravity, where an additional higher curvature term is added to the gravitational action. We call it convenient because it is among the simplest possible modifications, as opposed to more complicated models which intertwine inflation with mechanisms of particle physics, such as that of the Higgs Boson. \par
The key question concerning inflation is how it could have potentially materialized. To be more precise, inflation generally requires rather strict initial conditions for the Universe that are sometimes regarded as "unphysical"\cite{Reference63}. It is therefore insufficient to provide a plausible mechanism for inflation. We also need to explain how and why its initial state emerged.

\section{Quantum Cosmology}

Our discussion thus far has been completely classical, as nearly all the inflationary models and their mechanisms can be handled as with classical field theory. It is ultimately necessary to eventually make quantum considerations in such models. We know that physical laws are dictated by quantum mechanical phenomena, therefore any inflationary model is incomplete without them. It is also necessary to eventually handle gravity more broadly in a quantum mechanical context. As it will be discussed later, there is a superficial contradiction between the study of cosmology and the study of quantum mechanics. This is addressed by the importance of quantum phenomena during an era when the Universe was microscopic and quantum field fluctuations were of significant magnitude by comparison. \par
The introduction of quantum mechanics to cosmology has had a remarkable effect. A particular success was its use to explain the creation and existence of large scale structures in the modern Universe through primordial fluctuations\cite{Reference14}\cite{Reference37,Reference30}. It also serves as the basis for inflation within the context of supersymmetric models and even superstring theory. It further serves as the basis for entirely new inflationary paradigms. A particularly interesting framework is that of chaotic inflation as proposed by Andrei Linde\cite{Reference14}. Rather than relying on a deterministic mechanism to achieve inflation, it is argued that it occurs due to quantum fluctuations of a scalar field that permeates space. Inflation in this case goes on forever, creating new inflationary patches ad infinitum.\par
At a more basic level, when canonically quantizing of gravity, we obtain a field equation called the Wheeler-DeWitt equation\cite{Reference22,Reference40}. The solution to that equation provides us with a "wave function of the Universe" which can be interpreted as the probability amplitude for a Universe to emerge out of nothing\cite{Reference1,Reference9,Reference17,Reference18, Reference19, Reference20, Reference21}. This interpretation is not without its conceptual or practical issues. A major conceptual issue is the way that such a wave function can be measured, given that an observer is by definition part of the Universe itself\cite{Reference14}. On the technical side, the equation is marred with operator ordering ambiguities that seem to suggest the existence of multiple quantum theories\cite{Reference14}\cite{Reference22}\cite{Reference31}. While we won't attempt to solve conceptual issues in this thesis, we have made remarkable progress in the latter. More specifically, we have managed to resolve the ambiguities for a large class of models, proving the quantum theory for a certain model is unique. \par
Regardless, it stands as a compelling tool to address the problem of initial conditions. That is, a comparison can be made between the quantum mechanical probability of the inflationary and non-inflationary phase of the Universe. In the case of the former being dominant, this would provide a statistical explanation for inflation. Here we encounter another open question regarding the method of calculation of the wave function. The two different choices of Hartle-Hawking and Linde-Vilenkin lead to different predictions: the former predicts a Universe without inflation, while the latter a Universe with it\cite{Reference17}\cite{Reference20}. We have made some progress in this aspect as well by showing both should be considered. \par
Lastly, our second goal was to calculate the wave function for the Starobinsky model as a plausible candidate for inflation. Based on that, we can then predict whether inflation or no inflation is the more likely state in which the Universe emerges. From the semi-classical results we were able to calculate, inflation appears qualitatively more probable. 

\section{Brief Summary}

In Chapter 2, we provide a comprehensive summary of Hot Big Bang cosmology. We introduce the FLRW metric. We then provide a variety of open questions that arise and which remain unresolved in this model. \par
In Chapter 3, we begin by providing the basics of de Sitter space and the mechanics of slow-roll inflation. The chapter goes over the subsequent reheating phase and its physical repercussions before we explain in detail how a variety of aforementioned open questions are resolved due to inflation. The chapter closes with the mechanism of structure formation due to energy density perturbations. \par
In Chapter 4, we introduce the Starobinsky model and $R^{2}$ theory as the former's approximation in the inflationary phase. The general Weyl theory is also explained. It is shown how these models can be transformed to a frame where they behave like GR in the presence of a scalar matter field and a potential term. The chapter continues with a study of metric perturbations around each classical solution, and finally we deal with boundary terms. \par
In Chapter 5, we present our work in dealing with the technical issues of the WDW equation. We give an introduction to the equation before explaining its conceptual and technical challenges. Starting from first principles, we analyze the special case of the 1d minisuperspace model to deal with operator ordering ambiguities and prove the uniqueness of the quantum theory. We proceed with the arbitrary-dimensional minisupercase with special emphasis for flat target space manifolds. In both cases we provide calculations for specific examples. \par
In Chapter 6, we move to our second work based on the calculation of the wave function. It begins with a short summary of Hartle-Hawking's no boundary proposal, and we mention the alternative Linde-Vilenkin contour choice, with an emphasis on their different cosmological predictions. We explain some technical details on how the calculation can be performed using instantons, and we apply the same methods to find a result for $R^{2}$ gravity. The chapter concludes with calculations for the excited states of the wave function. \par
In Chapter 7, the thesis ends with a review of our research findings and some remarks on possible future projects. Moreover, we provide two appendices: Appendix A contains a calculation on the hermiticity criteria for the WDW quantum Hamiltonian, while in Appendix B we calculate the Faddeev-Popov determinant which appears in the path integral.

\section{List of publications}

\begin{itemize}
    \item E. Kaimakkamis, H. Partouche, K. Sil, and N. Toumbas. The exact Wheeler-DeWitt equation for the scale-factor minisuperspace model. Class. Quant. Grav., 42, 2025. [arXiv:2412.18532 [hep-th]]
    \item E. Kaimakkamis and K. Sil. A formalism for the ambiguities of the Wheeler-DeWitt equation. PoS CORFU2023, 227, 2024. [arXiv:2407.03077 [gr-qc]]
    \item V. Franken, E. Kaimakkamis, H. Partouche, and N. Toumbas. Ordering-Independent Wheeler-DeWitt Equation for Flat Minisuperspace Models. 2025. [arXiv:2512.23656 [hep-th]]
    \item E. Kaimakkamis, H. Partouche, and N. Toumbas. Wave function of the universe in $R^{2}$ theory. TBD. Work in progress.
\end{itemize}

\fancyhead[RE,LO]{Hot Big Bang Cosmology}

\chapter{Hot Big Bang Cosmology}

The hot Big Bang model for cosmology has been a very successful cosmological model. It has gained experimental support with the discovery of the Cosmic Microwave Background Radiation (CMBR). It provides a convincing account for the evolution of the Universe that matches with observations and does not contradict other physical theories \cite{Reference15}. 
%Regardless of various possible details about the nature of our %Universe and further improvements eluding us, we can generally say 
The Hot Big Bang model is based on primarily three things: \par
\begin{enumerate}
    \item Our Universe has been expanding for roughly $1.38 \times 10^{10}$ years.
    \item At one point in the past the Universe collapses to a singularity (and thus the classical description breaks down).
    \item The Cosmological Principle holds. That is, on sufficiently large scales the Universe is homogeneous and isotropic.
\end{enumerate}

%These are by no means axioms in the sense that they are demanded a %priori - instead they occur naturally when one takes observational %data into account. However, they do not serve as definitive theorems %either - any single one of them may not hold true in a more general %cosmological model (that could also be quantum mechanical). In %particular, the Cosmological Principle and the existence of a starting %singularity point could be challenged by inflationary cosmology. %Despite these considerations, when describing the \textbf{observable} %Universe, the three aforementioned points are sufficient for Hot Big %Bang cosmology to hold; at least on this scale. \par

\section{Friedmann-Lemaître-Robertson-Walker metric}

The Friedmann-Lemaître-Robertson-Walker (FLRW) metric can be written generally as follows:

\begin{equation} \label{eq:1.1}
    ds^{2} = -dt^{2} + a^{2}(t)\left [\frac{dr^{2}}{1-kr^{2}} + r^{2}(d\theta ^{2} + \sin ^{2}{\theta} \, d\varphi ^{2})\right ]
\end{equation}

where $a(t)$ is the scale factor of the Universe, $k$ is a constant that describes the spatial curvature, $r$ is a dimensionless radial coordinate and $\theta , \varphi$ the polar and azimuthal spherical angles. 
%While we will be referring to the FLRW metric as a singular entity, it %is important to note that it is in fact an entire class of different %metrics. 
The scale factor $a(t)$ and $k$ differ according to the model. The form of the metric is dictated by the Cosmological principle. The spatial sections are homogeneous and isotropic. As a result, all points in the Universe look the same. The spatial metric is maximally symmetric, respecting the maximum amount of symmetries possible. 
%The set of coordinates used to describe this spacetime is called %"comoving", named as such because they correspond to the coordinate %system in which a comoving observer takes measurement. 
A comoving observer is inertial in these coordinates, and so his/her motion is entirely due to the expansion of the Universe. 
%while a moving observer will also tend to become inertial due to %expansion. 
The expansion is described by the scale factor, in terms of which we can compute distances between any two points in the universe. 
%If the scale factor were time-independent (just a constant), the %Universe would be completely static and unchanging in size. 
From the metric, it is fairly straightforward to calculate the scalar $4d$ curvature:

\begin{equation} \label{eq:1.2}
    R = 6\left (\frac{\dot{a}^{2}}{a^{2}} + \frac{\ddot{a}}{a} + \frac{k}{a^{2}} \right )
\end{equation}

where a dot indicates time derivative. 
%with respect to the comoving time appearing in the metric. 
%From this, the connection of $k$ with the curvature of a FLRW Universe %becomes explicit. 
The constant $k$ takes three values: -1, 0 and +1, which correspond to an open (negative curvature), flat (zero curvature) and closed (positive curvature) Universe. For $k=0$, the metric takes the form:

\begin{equation} \label{eq:1.3}
    ds^{2} = -dt^{2} + a^{2}(t)\left [dr^{2} + r^{2}(d\theta ^{2} + \sin ^{2}{\theta} \, d\varphi ^{2})\right ]
\end{equation}

If we define the conformal time $\eta$ as \cite{Reference15}:

\begin{equation} \label{eq:1.4}
    \eta (t) = \int _{0}^{t} \frac{dt'}{a(t')}
\end{equation}

the metric takes the conformally flat form:

\begin{equation} \label{eq:1.5}
     ds^{2} = a^{2}(t)\left [-d\eta ^{2} + dr^{2} + r^{2}(d\theta ^{2} + \sin ^{2}{\theta} \, d\varphi ^{2})\right ]
\end{equation}

%This is where term "conformal time" acquires its meaning; the new time %coordinate defines a conformal transformation of the original metric %to Minkowski:

%\begin{equation} \label{eq:1.6}
%    ds^{2} = -dt^{2} + \sum _{i=1}^{3} dx_{i}^{2}
%\end{equation}

%simply expressed in spherical coordinates. Conformal in this case %meaning that angles and thus shapes when transforming one into the %other are invariant, but the scales are different (scaled up if %expanded or scaled down if contracted). Indeed, the $k=0$ case being %conformal to Minkowski spacetime shows how this particular value leads %to a flat spacetime. 

If $k=1$, the $g_{rr}$ component of the metric, $1/1-r^{2}$, acquires a pole at $r = 1$ (since $0\leq r\leq 1$). This is not a physical singularity but a coordinate one. Indeed, the spatial metric is nothing but the metric on a compact 3-sphere with radius $R_{0}=a(t_0)$:

\begin{equation} \label{eq:1.7}
    ds^{2} = R_{0}^{2}\left [\frac{dr^{2}}{1-r^{2}} + r^2 d\Omega _{2}^{2} \right]
\end{equation}

where $d\Omega _{2}^{2} = (d\theta ^{2} + \sin ^{2}{\theta} \, d\varphi ^{2})$ is the metric on the unit 2-sphere. The space is homogeneous and positively curved. It is compact and finite, lacking any boundaries. The scale factor $a(t)$ is simply the physical radius of the Universe. 

Finally, for $k=-1$, the spatial metric reduces to that of a $3$-dimensional hyperboloid:

\begin{equation} \label{eq:1.8}
    ds^{2} = R_{0}^{2}\left [\frac{dr^{2}}{1+r^{2}} + r^2(d\theta ^{2} + \sinh ^{2}{\theta} \, d\varphi ^{2}) \right ]
\end{equation}

The space is now negatively curved, "open" and non-compact. 
%Here our initial $\theta $ coordinate has been analytically %continued to the semi-infinite complex plane such that the one %that appears in our metric now ranges from $0\leq \theta \leq %+\infty$ and thus our sine function becomes hyperbolic. 

To determine the evolution of the Universe, one has to derive the equations for $a(t)$. We can do so via the use of the Einstein equations \cite{Reference15}\cite{Reference5}:

\begin{equation} \label{eq:1.9}
    R_{\mu \nu } - \frac{1}{2} g_{\mu \nu }R + \Lambda g_{\mu \nu } = 8\pi GT_{\mu \nu }
\end{equation}

where $R_{\mu \nu }$ the Ricci curvature tensor (which is the trace of the Riemann curvature tensor $R_{\mu \nu  \rho \sigma }$), $g_{\mu \nu }$ the spacetime metric, $T_{\mu \nu }$ the stress-energy tensor, $\Lambda$ the cosmological constant and $G$ Newton's gravitational constant. The stress-energy tensor can be taken so as to have the properties of a perfect fluid \cite{Reference15}\cite{Reference14}:

\begin{equation} \label{eq:1.10}
T_{\mu \nu} =
\begin{pmatrix}    
                  -\rho & 0 & 0 & 0 \\
                    0   & P & 0 & 0 \\
                    0   & 0 & P & 0 \\
                    0   & 0 & 0 & P \\
\end{pmatrix}    
\end{equation}

where $\rho$ is the energy density and $P$ is the pressure. Both are strictly time-dependent to satisfy the Cosmological principle. The fact it is diagonal can also be explained by the isotropy and homogeneity of the Universe: these impose a strictly diagonal metric and curvature tensor, and as such the stress-energy tensor will be diagonal as well. 

Now we can derive the two Friedmann equations \cite{Reference15}\cite{Reference5}\cite{Reference14}:

\begin{equation} \label{eq:1.11}
    H^{2} + \frac{k}{a^{2}} = \frac{8\pi G\rho}{3} + \frac{\Lambda }{3}
\end{equation}

\begin{equation} \label{eq:1.12}
    \frac{\ddot{a} }{a} = -\frac{4\pi G}{3} (\rho + 3P) + \frac{\Lambda }{3}
\end{equation}

where $H=\dot{a}/{a}$ is the Hubble parameter, characterising the rate of expansion of the Universe. These equations can be combined to obtain an energy conservation law:

\begin{equation} \label{eq:1.13}
    \dot{\rho } + 3H(\rho + P) = 0
\end{equation}

%While it is not immediately obvious why this is an energy %conservation law, we invoke the fact 

Since the Universe is filled with a perfect fluid, the pressure is proportional to the energy density:

\begin{equation} \label{eq:1.14}
    P = w\rho
\end{equation}

where $w$ is taken to be constant (whose value is of great interest). Thus equation (\ref{eq:1.13}) can be written as:

\begin{equation} \label{eq:1.15}
\begin{split}
    &d(a^{3}\rho ) + P\, d(a^{3}) = 0 \\
    \Rightarrow &\, dE + P\, dV = 0
\end{split}
\end{equation}

This is simply the first law of thermodynamics for a system of constant entropy. From equation (\ref{eq:1.12}), we observe that in the absence of a cosmological constant (or for a small value of it), the acceleration of the Universe is negative, i.e. the rate of expansion decreases. For $k=1$, eventually this overtakes the positive value of the expansion rate and at some point the expansion stops and turns into contraction, eventually leading to a "Big Crunch" singularity. If $k=0$, the expansion never ceases, but the Universe does remain nearly static in the remote future. If instead $k=-1$, the deceleration never overtakes the expansion rate and the Universe keeps expanding forever. The following diagram illustrates the possibilities:

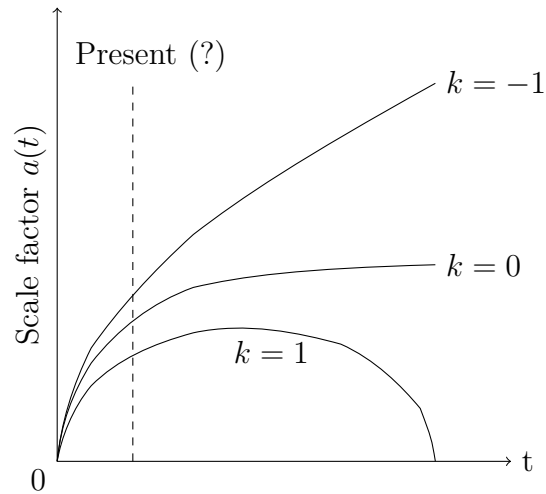
\begin{figure}[H]
\centering
\begin{tikzpicture} [scale=1]
\node at (-0.25,0) {} ;
\node at (0,-0.25) {} ;
\node at (-0.25,-0.25) {0} ;
\node at (1,-0.25) {} ;
\draw [->] (0,0) -- node[midway, above, sloped] {Scale factor $a(t)$} (0,6) ;
\draw [->] (0,0) -- (6,0) node[right] {t} ;
\draw [dashed] (1,0) -- (1,5) node[above] {\, \, \,Present (?)} ;
\draw (0,0) .. controls (0.05,0.5) and (0.2,1) .. (0.45,1.5) .. controls (0.8,2)  and (1.25,2.5) .. (1.8,3) .. controls (2.45,3.5) and (3.2,4) .. (5,5) node [right] {\(k = -1\)} ;
\draw (0,0) .. controls (0.05,0.5) and (0.2,0.9) .. (0.45,1.3) .. controls (0.8,1.75)  and (1.25,2.1) .. (1.8,2.3) .. controls (2.45,2.45) and (3.2,2.55) .. (5,2.6) node [right] {\(k = 0\)} ;
\draw (0,0) .. controls (0.05,0.3) and (0.2,0.7) .. (0.45,1) .. controls (0.8,1.35)  and (1.25,1.55) .. (1.8,1.7) .. controls (2.5,1.85) and (3.2,1.7) .. (3.75,1.55) node[midway, below] {\(k = 1\)} .. controls (4.2,1.35) and (4.55,1) .. (4.8,0.7) .. controls (4.95,0.3) and (4.95,0.3) .. (5,0) ;
\end{tikzpicture}
\caption{Scale factor evolution for the case of closed, open and flat Universe in the absence of a cosmological constant.}
\label{fig:1.1}
\end{figure}

Finally, we shall go over various interesting cases for the value of $w$ in equation (\ref{eq:1.14}):\par 
%There are three values of interest: 

(a) $\bm{w = 0}$ \par
This corresponds to a matter-dominated Universe. If cold, non-relativistic matter dominates the content of the Universe, the pressure is negligible. The dynamics is governed by the energy (or matter) density. From Friedmann equations, we find that $\rho \propto a^{-3}$. In other words, the matter density of is inversely proportional to its volume $V\sim a^{3}$ since the total mass is constant. \par

(b) $\bm{w = \frac{1}{3}}$ \par

This case corresponds to hot, relativistic matter, leading to a radiation--dominated Universe. Aside from massless gauge bosons (primarily photons), the thermal radiation may also include  high energy leptons, whose mass is effectively 0. The energy density scales as $\rho \propto a^{-4}$. From the Stefan-Boltzmann law, we obtain ($\rho \sim E \propto T^{4}$), leading to a very interesting relationship between the scale factor and the temperature: $T \propto a^{-1}$. \par

(c) $\bm{w = -1}$ \par

The energy density is now constant as it is countered completely by the pressure. The Universe is dominated by vacuum energy. In the inflationary stage of the Universe, the universe will be filled with a homogeneous scalar field $\phi$, with this equation of state. \par

\section{Universe age and Entropy}

Before we describe the stages of the cosmological evolution, it is useful to review some key concepts pertaining to it. Firstly we have the red-shifting effect, which can be explained as follows: if a photon of wavelength $\lambda$ is produced at some time $t_{1}$, observed at a later time $t_{0}$, the wavelength will have a larger value $\lambda _{0} >\lambda$. Thus the frequency of the photon decreases. We get a red-shifted spectrum. This can be understood in two ways: \par 

(a) The quantum mechanical explanation in which we use the the de Broglie relation $\lambda = \frac{h}{p}$ - where $h$ is Planck's constant and $p$ the photon momentum. The effects of the expansion of the Universe impact particle dynamics via the inverse relation of momentum and scale factor $p \propto a^{-1}$. Therefore $\lambda \propto a$ and indeed, the expansion of the Universe causes the photon wavelength to increase. \par

(b) The classical description in which we consider an infinitesimal time increment $\delta t$ between successive wave crests, during which the scale factor remains roughly constant, but since the quantity $\int _{t_{1}}^{t_{0}} \frac{dt'}{a(t')}$  is independent of the variable time $t$, it has to be the same for both the source of the emitted photon and the receiving observer. But since the observer is far away enough for the scale factor to have evolved to a value $a_{0} >a_{1}$, this means the time increment between wave crests will also increase, giving a lower frequency. \par 

We define the red-shift factor $z$ which via the relation \cite{Reference15}:

\begin{equation} \label{eq:1.16}
    1 + z = \frac{a_{0}}{a(t)} = \frac{\lambda_{0}}{\lambda(t)}
\end{equation}

As we mentioned above, the scale factor $a_{0}$ and wavelength $\lambda_{0}$ are measured at the time $t_{0}$ when we conduct our experiment. It is via the red-shift of light emitted from distant galaxies that astronomers have experimentally confirmed the fact the Universe is expanding. For small $z$, the red-shifting is roughly linear and matches to Hubble's law. 

Another useful quantity is the ratio of the energy density at some time $t$ to the critical density $\rho_{c}$ \cite{Reference15}\cite{Reference14}:

\begin{equation} \label{eq:1.17}
    \Omega (t) = \frac{8\pi G\rho(t)}{3H^{2}} \equiv \frac{\rho(t)}{\rho_{c}} 
\end{equation}

When $\Omega$ is one, the curvature term must be zero. Indeed, the Fridmann equation can be written as

\begin{equation} \label{eq:1.18}
    \Omega - 1 = \frac{k}{H^{2}a^{2}}
\end{equation}

If the energy density takes the critical value, equivalently $\Omega = 1$, $k=0$, and so we end up with a flat Universe. If $\rho$ is subcritical, the curvature must be negative. Finally if $\rho$ is supercritical, the curvature must be positive. 

We may also calculate the age of the Universe in terms of $\Omega(t_{0}) = \Omega_{0}$, $H_0$ and the red-shift factor $z$, using the relation

\begin{equation} \label{eq:1.19}
    t = \int _{0}^{a(t)} \frac{d[a(t')]}{\dot{a}(t')}
\end{equation}

where $t=0$ is chosen to be the time where the scale factor goes to 0 (initial singularity). 
%By making use of the red-shift quantity $z$ and value of the %ratio $\Omega(t_{0}) = \Omega_{0}$, 
Using the equation above, we can find expressions for the age of the Universe for both matter and radiation domination \cite{Reference15}:\par

\textbf{Matter domination:} 

\begin{equation} \label{eq:1.20}
    t = \frac{\Omega_{0}}{2H_{0}(\Omega_{0}-1)^{3/2}} \, \left [\cos ^{-1}{\left (\frac{\Omega_{0}z - \Omega_{0} +2}{\Omega_{0}z + \Omega_{0}}\right )} - \frac{2\sqrt{\Omega_{0}-1}\sqrt{\Omega_{0}+1}}{\Omega_{0}(1+z)} \right], \, \, \,\, \, (\Omega_{0} >1)
\end{equation}

\begin{equation} \label{eq:1.21}
     t = \frac{\Omega_{0}}{2H_{0}(1-\Omega_{0})^{3/2}} \, \left [-\cosh ^{-1}{\left (\frac{\Omega_{0}z - \Omega_{0} +2}{\Omega_{0}z + \Omega_{0}}\right )} + \frac{2\sqrt{1 -\Omega_{0}}\sqrt{\Omega_{0}+1}}{\Omega_{0}(1+z)} \right], \, \, \,\, \, (\Omega_{0}<1)
\end{equation}

\begin{equation} \label{eq:1.22}
    t = \frac{2}{3H_{0}} (1+z)^{-3/2}, \, \, \,\, \,  (\Omega_{0} = 1)
\end{equation}

\textbf{Radiation domination:}

\begin{equation} \label{eq:1.23}
    t = \frac{\sqrt{\Omega_{0}(1+z)^{2}} - \sqrt{\Omega_{0}(1+z)^{2} - \Omega_{0}+1}}{H_{0}(\Omega_{0}-1)(1+z)}
\end{equation}

For the matter case, at early enough times when the curvature term is negligible, we have $(1+z)\gg \Omega_{0}^{-1}$ and thus the above expressions for all three cases become:

\begin{equation} \label{eq:1.24}
    t \approx \frac{2}{3H_{0} \Omega_{0}^{1/2}} (1+z)^{-3/2}
\end{equation}

For the radiation case, we get:

\begin{equation} \label{eq:1.25}
    t \approx \frac{1}{2H_{0} \Omega_{0}^{1/2}} (1+z)^{-2}
\end{equation}

The Friedmann equations can also be used to derive the time-dependence of the scale factor. Namely, using

\begin{equation} \label{eq:1.26}
    \rho \propto a^{-3(1+w)}
\end{equation}

from energy conservation, we get:

\begin{equation} \label{eq:1.27}
    a \propto t^\frac{2}{3(1+w)}
\end{equation}

For a matter-dominated universe, $a\propto t^{2/3}$, and for the radiation-dominated case $a\propto t^{1/2}$. 
%(for the sake of completion) 
In the case of vacuum, we get $a\propto \exp{(H_{0}t)}$. This final relation leads to exponential expansion, and thus acceleration, a key feature not present in conventional Hot Big Bang cosmology. 

The final quantity we will discuss is the entropy per unit comoving volume, $S$ \cite{Reference15}: 

\begin{equation} \label{eq:1.28}
    S = \frac{a^{3}(\rho +P)}{T}
\end{equation}

where $T$ is the temperature. Using the Friedmann equations, we already saw that this quantity is conserved 
%(assuming that the chemical potential of the various bunches of particles can be taken to be roughly 0) 
and this in turn implies thermal equilibrium. Actually this is not accurate - the Universe was never in a state of perfect thermal equilibrium due to the fact that the temperature changes with time due to the expansion.
%the FLRW metric lacks a time-like Killing vector i.e. the time translation symmetry does not hold. 
However, throughout most of the cosmological history, conditions of near-thermal equilibrium held to a very good approximation, with all contents sharing the same temperature and creating a "cosmic soup" of particles. Once a certain species of particle decouples, its thermal evolution deviates from the rest, and ultimately this gives rise to the various important transitions that lead to the current structure of the Universe. \par

Since the entropy $S$ is constant, the scale factor scales inversely proportional to the temperature. The entropy at all cosmological epochs is dominated by relativistic particles, since these contribute more to the thermal state than cold non-relativistic particles. This implies $\rho \propto a^{-4}$ and $T\propto a^{-1}$. 
%Recall that this exact relation was predicted in the previous section by invoking the Stefan-Boltzmann law. 
%This can be generalized further if we consider the number of massless degrees %of freedom for our thermodynamic system when each of the species does not %necessarily share the exact same temperature \cite{Reference15}:

%\begin{equation} \label{eq:1.29}
%    g_{*} = \sum _{i=bosons} g_{i} \left(\frac{T_{i}}{T}\right )^{4} + %\frac{7}{8} \sum _{i=fermions} g_{i} \left(\frac{T_{i}}{T}\right )^{4}
%\end{equation}

Taking into account the fact that some species may not share the same temperature, we define the following quantity which counts the effective number of massless degrees of freedom:

\begin{equation} \label{eq:1.30}
    g_{*s} = \sum _{i=bosons} g_{i} \left(\frac{T_{i}}{T}\right )^{3} + \frac{7}{8} \sum _{i=fermions} g_{i} \left(\frac{T_{i}}{T}\right )^{3} 
\end{equation}

where $g_{i}$ stand for the number of degrees of freedom sharing the temperature $T_i$. 
%(notice how fermions acquire a special factor corresponding to their %arrangement limitations due to the Pauli principle which leads to the %well-known dichotomy between Bose-Einstein and Fermi-Dirac statistics). 
%For a state of (near-)thermal equilibrium, this is obviously just a constant %and therefore our proportionality relation is precise. But since the value of %$g_{*s}$ varies with time due to our Universe evolving, we expect the shared %equilibrium temperature to be dependent on it. 
Then the entropy density $s={S}/{V}$ is given by \cite{Reference15}:

\begin{equation} \label{eq:1.31}
    s = \frac{2\pi ^{2}}{45} g_{*s} T^{3}
\end{equation}

and $T\propto g_{*s}^{-1/3}a^{-1}$. 

So what does exactly happen as the Universe expands? The expansion causes the temperature to decrease. When major deviations from thermal equilibrium occur, e.g. when a particle species decouples, entropy conservation implies that the part of entropy previously carried to that species is now equally shared to the rest, and the effective number of degrees of freedom decreases. 
%Of course, this does not mean temperature for the mixture increases (as it is %inversely proportional to the degrees of freedom), but rather it continues to %fall off as before, just slightly faster. 
Decoupled massless species follow a separate evolution. They still behave as a thermal "soup" on their own with temperature equal to the temperature at decoupling, which scales as $T\propto a^{-1}$. Massive, non-relativistic particles cool-off more rapidly as $T\propto a^{-2}$. \par

\section{Thermal history of the Universe}

Now we can provide a brief thermal history of the Universe. We shall break it down to eras marked by important physical phase transitions at specific temperature scales \cite{Reference15}\cite{Reference14}\cite{Reference34}. A  departure from equilibrium can be determined by comparing the rate of a certain particle interaction $\Gamma$ with the expansion rate H. Thus $\Gamma > H$ for some interaction means the interacting particles remain in equilibrium, while $\Gamma < H$ signals departure from it. 
%That is because a very rapid interaction rate does not allow %that particular particle species to exist on its own for long %and isolate itself, quickly interacting with other coupled %particles species at any moment, while a slow rate would %indicate that the particle population has started interacting %much less with the cosmic soup that is in thermal equilibrium, %decoupling in the process. 
%This is not an exhaustive criterion though, as massless, non-interacting particles will tend to stay in thermal equilibrium and thus the rate of interaction is not a pertinent factor to determine its status. Rather, we should consider interactions that are significant in the sense that a drop in their rate does impact the thermal state of the cosmic soup and thus disturbs the equilibrium. For example, 
The breaking of symmetries and the acquiring of mass of force mediating particles via the Higgs mechanism constitute major turning points/cross-over transitions in the history of the Universe. \par

\textbf{1)} $\bm{t\lesssim 10^{-43} s \sim t_{P} \Rightarrow T \gtrsim 10^{19} \, GeV \sim m_{P}}$ \par
This period corresponds to the Planck era. At this point 
%we do not have a definitive description of the history of the %Universe, as 
we expect the classical theory of General Relativity to break down. There are candidate theories for a quantum description of gravity such as string theory and loop quantum gravity, which can help us understand the first moments of the Universe after the Big Bag. \par 
%We include this as an important milestone in the history of the %Universe however, as it can be seen the point where a possible %"theory of everything" which includes all four fundamental %forces first starts to break down. 

\textbf{2)} $\bm{t\sim 10^{-36} s\Rightarrow T\sim 10^{14}-10^{16} \, GeV}$ \par
The end of Grand Unification occurs, with the strong force separating from the electroweak interactions. The breaking of the grand unification symmetry group gives rise to the matter-antimatter asymmetry, the appearance of magnetic monopoles and the initiation of the inflation.  \par

\textbf{3)} $\bm{t\sim 10^{-11} s \Rightarrow T\sim 300 \, GeV}$ \par
The unification of the weak and electromagnetic interactions ends and the $W^{\pm}, \, Z^{0}$ gauge bosons acquire mass via the Higgs mechanism. \par 
%This means that particle species that interact exclusively or for the most part via these gauge bosons will now have a significantly reduced rate of interactions $\Gamma$ and as we explained in the preceding paragraph, they will decouple from the thermal equilibrium of the cosmic soup within a short period of time. More specifically, this happens when their temperature drops accordingly so that it matches the coupling strength of the interaction: $T\sim G_{X}^{-2/3} M_{P}^{-1/3}$ with $G_{X} \sim \frac{\alpha}{m_{X}^{2}}$ the coupling strength for the specific boson $X$, $\alpha$ a constant that determines the coupling strength of the gauge $g=\sqrt{4\pi \alpha}$ and $m_{X}$ the boson's mass. 

\textbf{4)} $\bm{t\sim 10^{-5} s\Rightarrow T\sim 100-300 \, MeV}$ \par
Chiral symmetry is broken and colour confinement takes place, forming baryons and mesons. The various quark species start combining in the most basic colour-singlet states (quark-antiquark and triplets of quarks of different colour). \par 

\textbf{5)} $\bm{t\sim 10^{-2}-10^{2} s \Rightarrow T\sim 10-0.1\, MeV}$ \par
Primordial nucleosynthesis takes place, as hydrogen and helium nuclei start to appear. This is a crucial step in the evolution of our Universe. The abundances of the light elements in the Universe can be predicted, serving as an important test of Hot Big Bang cosmology. 
%More specifically, there are two ways to quantify this: %abundance by mass and abundance by moles of said element. 
About $\sim 74\%$ of baryonic matter in the Universe consists of hydrogen, $\sim24\%$ consists of helium and only $\sim 2\%$ consists of heavier elements. \par 
%By moles the disparity is even more stark, as hydrogen %dominates at $\sim 86\%$, helium sits at $\sim 13\%$ and barely %just $\sim1\%$ for heavier elements. 

\textbf{6)} $\bm{t\sim 10^{11} s\Rightarrow T\sim 100 \, eV}$ \par
The energy density of non-relativistic matter becomes equal with that of radiation, initiating the matter-domination era of the Universe, which persists for roughly 9.8 billion years. Currently the Universe is dominated by dark energy. This era marks the beginning of structure formation with galaxies and clusters appearing for the first time.  \par 

\textbf{7)} $\bm{t\sim 10^{13} s\Rightarrow T\sim 10 \, eV}$ \par
As the temperature drops below $13.6 \, eV$, corresponding to the ground state energy of the electron in the hydrogen atom, electrons start to nuclei, and the first hydrogen and helium atoms form. 
%This is also why this epoch is also sometimes called %"recombination", as the previously decoupled quarks (now bound %inside particles making up the first nuclei) will reunite with %the newly-decoupled electrons. 
Matter and radiation completely decouple. 
%A massive increase in entropy ensues (at a much faster pace than before since adiabaticity no longer holds) which signifies a jump in the creation of matter contents of our Universe. 
It is important to note that the photons, which last scattered at this pivotal point, are none other than the ones detected by measuring the CMBR. \par

\section{Open problems with the Hot Big Bang model}

For all its success, the the Hot Big Bang model does not address some important questions \cite{Reference15}\cite{Reference14}.
\par
%They are not inconsistencies or contradictions by any means, %nor paradoxes in the same way UV catastrophe was, serving as a %prelude to quantum theory. 
%Rather they are questions that are impossible to explain within %the context of just this model, or natural questions that have %to do with the nature of the model itself. A common theme which %will become evident as we list these various problems is the %recurrence of "why" questions; the fundamentals of cosmology we %can agree on, but not everything has an adequate explanation. %In the following subsections we will go over some of the most %important issues, but obviously this does not constitute an %exhaustive list 

\subsection{The flatness problem} 

This particular problem can be equivalently seen in various ways. In a topological sense, it becomes a question regarding the global curvature of the Universe. Why is the Universe to such a great degree flat (zero curvature)? In a geometrical sense, 
%it is a question of the way distances and angles are measured; 
why the Universe can be approximated so well by Euclidean space? From a thermodynamical perspective, why is there such a vast amount of matter and energy (and by extension entropy) in the Universe? These questions can be linked by recalling the quantity $\Omega (t)$ defined in section 1.2: 

\begin{equation} \label{eq:1.32}
    \Omega (t) - 1 = \frac{\rho (t)}{\rho_{c}} - 1 = \frac{k}{H^{2}(t) a^{2}(t)}
\end{equation}

Flatness is achieved by simply setting the value of $k$ to 0, but also, via the definition of $\Omega$, via the energy density approaching the critical value. The way to see how the value of energy density can be extrapolated to the flatness problem is as follows: if $\rho$ at the beginning was above its critical value, this would lead to a closed Universe solution, and a Big Crunch singularity, 
%therefore would predict (as we saw before) a time where it %recollapses to a singularity 
long before it could stabilize into a flat solution and give rise to the Universe we see today. If on other hand it was subcritical, the amount of matter in the Universe would be much lower than it is observed. More specifically, the total entropy of our Universe contained within the Hubble volume today is about $10^{87}$ times higher than the entropy at early stages, a value that an open Universe could not replicate. 
%primarily because it would have become as cold as our Universe %is today at a much earlier time, not giving enough time for the %various crucial steps of its thermal evolution that we %described to take place. 
%That in turn would translate to much fewer deviations from %thermal equilibrium (therefore less production of entropy) and %less production of matter (therefore less "stuff" in the %Universe). 
And finally, the geometric question of Euclidicity can be trivially answered if we recall that a topologically flat surface is locally equivalent to the Euclidean plane. Just how flat/Euclidean/abundant in content is our Universe? It is expected that for the very early stages of the Universe  \cite{Reference14}:

\begin{equation} \label{eq:1.33}
    \lvert \Omega -1 \rvert \lesssim 10^{-59} 
\end{equation}

This is a very small value which strongly indicates that it is not accidental. It seems that there should be an underlying reason or explanation for why this value is so small and by extension why our Universe came to be this flat.  \par

\subsection{The issue of the Cosmological principle}

The Cosmological principle states that the Universe is globally homogeneous and isotropic. 
%is one of the fundamental tenants of the Hot Big Bang model and %stems from observation that confirms it. 
However, there is no real reason for why the Universe for \textbf{all} of its history should be homogeneous and isotropic. The fact it holds 
%true and appears to have done so for as long as our %conventional model can track back is yet another question that 
seems to have no convincing explanation.  \par
%We could go as far as to say that it is almost "unnatural" to %assume its validity since the very beginning; it would be %(literally) the biggest coincidence in the Universe that of all %the inhomogeneous and anisotropic initial conditions it could %possess, it happened to "opt" for none of them. 
%We could also be tempted to say that "the Cosmological %principle has been holding true since the beginning because the %consequences of this which we can measure validate it", but %that is not really an answer. It is no different than saying %"the Universe is the way it is because we observe that it is %so". Not a very convincing argument to say the least.

\subsection{The singularity problem}

A consequence of Friedmann equations is that going back in time %and reverse-engineering (as we have previously put it) the %evolution of the Universe 
leads to a certain point where we can no longer classically describe the state of the Universe. That point corresponds to the Universe being entirely concentrated in a single point ($a = 0$). This poses two major questions that the current cosmological model is ill-equipped to answer: (a) how can the existence of a singularity of infinite density and temperature be reconciled with fundamental ideas of what is both naturally and mathematically sound i.e. the demand that values of physical quantities must be finite, and (b) even if we do reconcile those ideas, how can we ever know what took place prior to that point of singularity? The second question is particularly concerning, as the basic premise of cosmology is to understand the origins and evolution of the Universe. The model is incapable of explaining what happened prior to the Big Bang. 
%its own designated beginning, we are essentially back to square %one. 
It is necessary to find extensions and new theories that have the capacity to at least speculate in a substantiated manner with respect to what preceded the Hot Big Bang singularity. \par

\subsection{The problem of unwanted relics}

A feature of early structure in the Universe is the variety of exotic, very massive particles that our fundamental theories predict and which are stable enough that we expect to find them in the Universe today. These can range from hypothetical ones such as gravitinos (featured in supergravity) to magnetic monopoles. These particles are both predicted to exist and have very slow annihilation rates. 
%making them hypothetically able to be measured today. 
The conditions for their formation in the thermal history of the Universe do not prohibit their considerable production either, meaning there is no known mechanism that would inhibit their existence in the first place. Given that they are so massive, that suggests that they will eventually contribute by an exceeding amount to the total energy density, and at some point they will even tend to dominate the Universe. 
%It is in fact expected that their energy density value would %roughly equal that of all baryons at the present. 
Had their energy density become dominant, this would lead to an early value of $\rho$ that is much higher than the critical value, which would have caused the Universe to close and to collapse at some time in the past. Without the Hot Big Bang model providing a mechanism or reasonable initial conditions that would regulate this inconsistency, it remains an unsolved issue in our understanding of particle physics and cosmology.

\subsection{The horizon problem}

This is yet another problem related to the observed global homogeneity of the Universe. We illustrate this, as follows: By observing the temperature of photons in the CMBR, we indeed find that virtually all of them share the same temperature $\sim 2.75\, K$ \cite{Reference15}\cite{Reference5}\cite{Reference34}. The fluctuations are small ($\mathcal{O}(10^{-4}) \, K$), which is consistent with the Cosmological principle. While locally this is very much expected, the global nature of this homogeneity is perplexing. The reason for this is that we expect photons to share the same temperature if they have previously been in thermal contact and established equilibrium. However, not every photon could have been in such thermal contact because some of them belong to regions of space that are not in causal contact with some other regions. Even if they are today, they weren't so at the time of last scattering. 
%To put it simply, it is impossible for all photons to be in agreement with each other due to a thermodynamic process because not enough time has elapsed for some group of photons to "observe" some other very remote groups. This is where the name "horizon problem" comes from; 
If we define the horizon distance as:

\begin{equation} \label{eq:1.34}
    d_{H}(t) = a(t)\int _{0}^{t} \frac{dt'}{a(t')} = a(t) \int _{0}^{a(t)} \frac{d[a(t')]}{\dot{a}(t') a(t')}
\end{equation}

then this delimits the distance which some observer will be able to "see" or come in causal contact with some region of space. %(notice the similarities with expression (\ref{eq:1.19}) ). 
The more time elapses, the more of the Universe an observer can see, since more time is given for the particles to have travelled. This the particle horizon, providing a limitation  for all photons of the CMBR to be in thermal contact. \par 
%The probability that this is merely a coincidence is extremely %small (roughly $10^{-90}$ \cite{Reference14}) which shows that %some other explanation must exist for this homogeneity to take %place. 

\subsection{The cosmological constant/vacuum energy problem}

Throughout most of our analysis in this chapter, we have set the cosmological constant $\Lambda$ in equation (\ref{eq:1.9}) to 0. But it is not exactly 0 today. Its value leads to an energy density that is today greater than the energy density of baryonic and dark matter in the Universe, due to the latter's dilution as a result of the long period of expansion since the Universe became matter-dominated. This dark energy amounts to 68.3\% of the total energy in the Universe according to most recent measurements, dwarfing ordinary matter's contribution at 4.9\% while the remaining 26.8\% is made up of dark matter \cite{Reference35}. Moreover, the nature of $\Lambda$ as a constant could be wrong. During past cosmological epochs the value of it was different. Among possible contributions, the term can be written as a potential of a scalar field that corresponds to the energy of the vacuum. Specifically \cite{Reference14}:

\begin{equation} \label{eq:1.35}
    \Lambda \approx 8\pi G V(\phi)
\end{equation}

where $V(\phi)$ is the potential and $\phi$ a scalar field. The problem that arises in our model has to do with the expected value of this "constant". Its present value has a great disparity with what we theoretically expect. The present value of $\Lambda$ is vanishingly small \cite{Reference35}: 

\begin{equation} \label{eq:1.36}
    \Lambda \approx 2.888 \times 10^{-122} \, l_{P}^{-2}
\end{equation}

expressed here in Planck units, $m_{P}$ being the Planck mass ($\sim 10^{19}$ GeV). By extension this leads to an extremely small value of the vacuum energy potential \cite{Reference14}:

\begin{equation} \label{eq:1.37}
    \lvert V(\phi _{0}) \rvert \lesssim 10^{-29} g/cm^{3}
\end{equation}

$\phi _{0}$ being the present value of the scalar field. The theoretical disparity stems from predictions in quantum field theory. If we treat this scalar field $\phi$ as a quantum entity, it should have quantum fluctuations which contribute to the value of the vacuum energy density. However, the associated calculations lead to a deviation from the present value by 60 orders of magnitude. The difference has been dubbed as "the largest discrepancy between theory and experiment in science" \cite{Reference36}. 
%We could potentially come up with a better fundamental physics theory in the context of QFT that will improve this value and make a better prediction, but then the problem of its evolution persists. That is, its present value being so small is still very much unexpected due to the massive decrease it must have experienced as the Universe evolved. For example, consider the $SU(5)$ theory where it is postulated that this greater symmetry preceded the known standard model symmetry of $SU(3)\times SU(2) \times SU(1)$. The value of $V(\phi)$ would have dropped by the massive amount of $10^{80} g/cm^{3}$ in the first phase transition (the breaking of the former symmetry into the latter). It would then drop by yet another incredible amount of $10^{25} g/cm^{3}$ in the second phase transition to $SU(3)\times SU(1)$, and then an additional $10^{14} g/cm^{3}$ in the time period that followed to reach its present value \cite{Reference14}. Yet despite these extremely steep falls that turned $\Lambda$ so small, it remains non-zero, dominating the current composition of the Universe. Such an effect is simply too unlikely to be an accident, and we expect that there is a nuanced physical explanation for why $\Lambda$ has acquired this particular value.

\pagebreak

 % Introduction

\fancyhead[RE,LO]{Inflation}

\chapter{Inflation}

Moving past Hot Big Bang cosmology, we shall present the current prevailing paradigm in the field. It is postulated that before the various stages in the thermal history of the Universe as we saw them in chapter 2, there was a brief period with extremely rapid, accelerating expansion of an exponential form. This is what we call inflation, and it has been the core idea behind many cosmological models that followed \cite{Reference15}\cite{Reference14}\cite{Reference34}. \par
This idea can be attributed to Alan Guth who was the first to propose that such a period in the evolution of the Universe took place, and described the mechanism by which it was undertaken \cite{Reference74}. The idea of an exponentially expanding spacetime was already entertained by various Soviet scientists beforehand \cite{Reference29}\cite{Reference30}.  Alexei Starobinsky - whose specific model and ideas we will go over in the next chapter - had also examined this possibility initially as a possible solution to the singularity problem, referring to this phenomenon as a "de Sitter phase" \cite{Reference28}\cite{Reference29}. As we shall see shortly, that term is equally apt, since during this exponential expansion, spacetime attains the geometry of de Sitter (dS) space. While Guth's original model shall not be examined in this text, it didn't last very long, as it became obvious that it had various problems of its own. Despite that, the novelties of the model and the arguments behind how it could explain many discrepancies and issues proved to be compelling. The general idea would remain, with new improvements added and the mechanics of inflation refined. This new category of new models are based on some fundamental principles which comprise the so-called "New Inflation", as opposed to Guth's "Old Inflation" \cite{Reference15}\cite{Reference14}.\par
In this chaper, we will go over the basics of New Inflation, elaborating on the conditions necessary for it to have taken place, and explain its physical repercussions. We will also mention some of its predictions and go over in due detail how all these lead to the partial or total solution of most of the issues we saw in the preceding chapter. Finally, we will mention some of its novel features, focusing on density perturbations. \par

\pagebreak

\section{Basics of de Sitter space}

While there are multiple ways in which one could choose to represent dS space, we choose the very general form of an $n$-dimensional hyperboloid embedded in the $(n+1)$-dimensional Minkowski plane \cite{Reference6}\cite{Reference11}:

\begin{equation} \label{eq:2.1}
    -(x^{0})^{2} + \sum _{i=1}^{n} (x^{i})^{2} = \ell ^{2}
\end{equation}

where $n$ corresponds to the number of spacelike dimensions and $\ell$ is a constant with dimensions of length. A visual representation of the hyperboloid is the following: \par

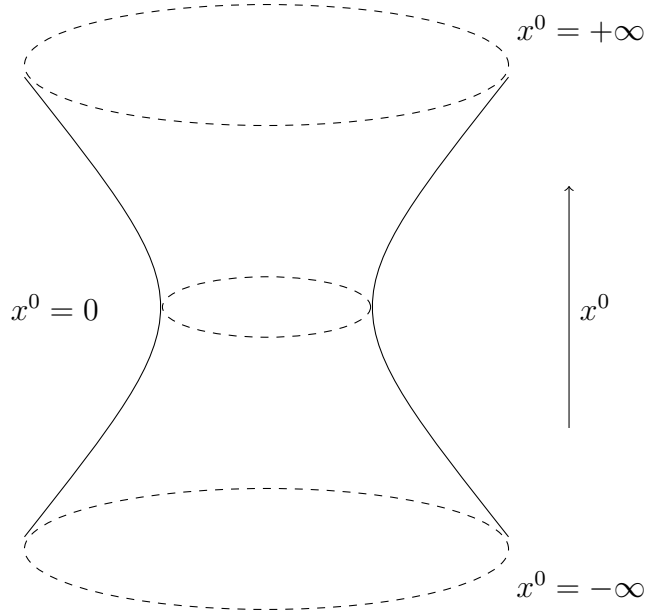
\begin{figure}[H]
\centering
\begin{tikzpicture} [scale = 1.6]
\node at (2.5,0) {} ;
\node at (-1.75,0) {\(x^0 = 0\)} ;
\node at (2.6,2.3) {\(x^0 = +\infty \)} ;
\node at (2.6,-2.3) {\(x^0 = -\infty \)} ;
\draw [dashed] (0,2) ellipse (2cm and 0.5cm) ;
\draw [dashed] (0,-2) ellipse (2cm and 0.5cm) ;
\draw [dashed] (0,0) ellipse (0.86cm and 0.25cm) ;
\draw (2,1.9) .. controls (0.5,0) and (0.5,0) ..  (2,-1.9) ;
\draw (-2,1.9) .. controls (-0.5,0) and (-0.5,0) ..  (-2,-1.9) ;
\draw [->] (2.5,-1) -- node[midway, right] {\(x^{0}\)} (2.5,1) ;
\end{tikzpicture}
\caption{Visual representation of the de Sitter hyperboloid. The dotted slice of constant $x^{0}=0$ serves as the smallest hypersurface of dS space after initial hyperbolic contraction, and preceding a symmetric hyperbolic expansion.}
\label{fig:2.1}
\end{figure}

As it is immediately obvious, dS space features an initial period of contraction from the infinite "past", reaches its smallest size at $x^{0}=0$ and then expands towards infinity in the "future". It is important to note that $x^{0}$ here is \textbf{not} a time coordinate, but does carry the negative sign that differentiates it from the other $n$ dimensions i.e. it is a \textbf{timelike} coordinate as opposed to the other $n$ spacelike ones. Restricting the analysis to the 5-dimensional case for the embedding space, in order to express the metric (\ref{eq:2.1}) in a form that is conformal to the FLRW metric, there are several coordinate systems which we can use. These coordinate systems describe specific parts and proportions of dS space, often "sliced" differently so that they cover different regions on the hyperboloid \cite{Reference37}. We shall focus on one that covers it in its entirety:

\begin{equation} \label{eq:2.2}
    \begin{split}
    &x^{0} = \ell\, \sinh{t} \\
    &x^{1} = \ell\, \cosh{t} \cos{\chi } \\
    &x^{2} = \ell\, \cosh{t} \sin{\chi } \cos{\theta } \\
    &x^{3} = \ell\, \cosh{t} \sin{\chi } \sin{\theta } \sin{\varphi } \\
    &x^{4} = \ell\, \cosh{t} \sin{\chi } \sin{\theta } \cos{\varphi }
\end{split}
\end{equation}

where $\chi, \, \theta$ are the two polar angles and $\varphi$ the azimuthal angle in the coordinate system of a 3-sphere. That gives us a form of the dS metric as such:

\begin{equation} \label{eq:2.3}
    ds^{2} = \ell ^{2}\left[-dt^{2} + \cosh ^{2}{\left(t \right)}d\Omega _{3}^{2}\right]
\end{equation}

with $d\Omega _{3}^{2} = d\chi ^{2} + \sin ^{2}{\chi } \,  d\Omega _{2}^{2}$ the metric of the aforementioned 3-sphere. The similarity with the FLRW metric is now rather obvious: the factor of $\ell ^{2}\cosh^{2}{(t)}$ will correspond to the scale factor $a^{2}(t)$. We will also assume that the time coordinate in the argument of the hyperbolic cosine is parametrized so that a dimensionless quantity appears:

\begin{equation} \label{eq:2.4}
    a(t) = \ell \, \cosh{\left (\frac{t}{\ell} \right )}
\end{equation}

This form of the scale factor as we shall see shortly corresponds to the $k=1$ version of the FLRW metric, and thus describes a closed dS space. For the sake of completion, in the other two cases we would have:

\begin{equation} \label{eq:2.5}
\begin{split}
    &a(t) = \ell \, e^{\frac{t}{\ell}} \, \, , \, \, \, (k=0) \\
    &a(t) = \ell \, \sinh{\left (\frac{t}{\ell} \right )} \, \, , \, \, \, (k=-1)
\end{split}    
\end{equation}

In the next subsection where we will discuss the dynamics of inflation, we will come to realize that this constant length quantity $\ell$ is none other than the horizon distance $H^{-1}$, or equivalently the physical radius of the closed Universe at that point in time, meaning that the expected form of the scale factor is going to be:

\begin{equation} \label{eq:2.6}
    a(t) = H^{-1} \, \cosh{(Ht)}
\end{equation}

In addition, if we calculate the scalar curvature of the dS hyperboloid, we will readily find that it is:

\begin{equation} \label{eq:2.7}
    R = \frac{12}{\ell ^{2}} = 12H^{2}
\end{equation}

The calculation can of course be done conventionally as with every other spacetime, using the trace of the Ricci curvature tensor, but in addition we could use the readily available form we found for the general FLRW metric (\ref{eq:1.2}). It is a matter of some brief calculations to verify how by inserting any of the three scale factor forms in (\ref{eq:2.4}) and (\ref{eq:2.5}) along with their appropriate $k$ value yields the exact same result as it appears above. Notice how, due to the length quantity being real and finite, $R$ is indeed strictly positive; the spacetime is positively curved just like the chosen coordinate system (being that of the $k=1$ case) predicts. \par
From this analysis, we can already draw some important conclusions about this spacetime. For starters, we see that the slices or "rings" of constant $x^{0}$ on the 4-dimensional hyperboloid (as dS space acts as a hypersurface in the 5-dimensional Minkowski space embedding) are spacelike 3-dimensional hypersurfaces. This justifies the previous observation that the contraction of dS space from the negative infinity and expansion towards the positive infinity do in fact coincide with the "past" and "future". We also notice that the spacetime has a $SO(1,4)$ Lorentz symmetry made manifest. It becomes immediately obvious that the appearance of the hyperbolic sine function in the definition of $x^{0}$ indeed provides an asymptotically exponential expansion/contraction $\sim e^{\pm \frac{t}{\ell} }$ in the remote future/past, justifying the assertion that the emergence of curved dS geometry and exponential expansion are equivalent statements. To make visually manifest how this spacetime is indeed topologically positively curved (i.e. closed) and to further earn an appreciation for its symmetries, we make a turn to Euclidean space by performing a Wick rotation, and performing a regular rotation by 90 degrees of the original time coordinate. To put it more succinctly:

\begin{equation} \label{eq:2.8}
    t \rightarrow i\tau = i\left (\frac{\pi}{2} - \omega \right )
\end{equation}

and the resulting metric will be:

\begin{equation} \label{eq:2.9}
    ds^{2} = \ell ^{2} \left [d\omega ^{2} + \sin^{2}{\omega }d\Omega _{3}^{2} \right ] = \ell ^{2} \, d\Omega _{4}^{2}
\end{equation}

which is none other than the metric of a 4-sphere. This compactification of dS space in Euclidean space will prove very useful in later discussions in chapters 5 and 6, but for now it serves us well in giving a more explicit visualization in how the space is compact, meaning it does not possess a boundary. Its infinities now lie on the two poles of the 4-sphere, while the hypersurfaces produced by slices of constant $t$ (or $\omega$ in this case) correspond to 3-spheres. By extension, the previous Lorentz symmetry transforms into a regular rotational symmetry in 5-dimensional space, $SO(5)$. The following figures elucidate these points further:

\begin{figure}[H]
\centering
\begin{tikzpicture} [scale=0.55]
\node at (0,4.25) {} ;
\node at (0,-4.25) {} ;
  \draw (0,0) circle (4cm);
  \draw (-4,0) arc (180:360:4 and 1.2) node[right] {\(t = 0 \Rightarrow \omega = \frac{\pi }{2} \)};
  \draw[dashed] (4,0) arc (0:180:4 and 1.2);
  \fill[fill=black] (0,0) circle (2pt);
  \draw[dashed] (0,0 ) -- node[above]{\(l\)} (4,0);
\end{tikzpicture}
\caption{Compactification of de Sitter space into a 4-sphere after a Wick rotation to Euclidean space. Notice how this particular metric also gives the length quantity $\ell$ further physical meaning, as it matches to the 4-sphere's radius in Euclidean space. This could also follow rather trivially from the scalar curvature, since the curvature of a sphere of radius $\ell$ follows $R \propto \frac{1}{\ell ^{2}}$. }
\label{fig:2.2}
\end{figure}
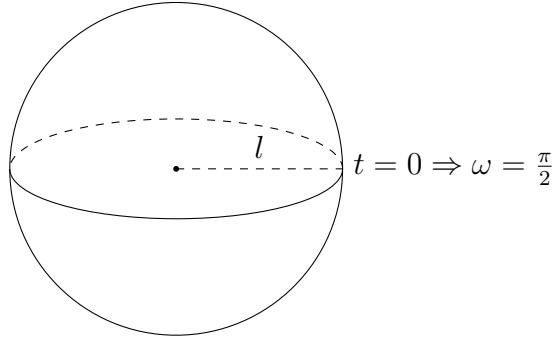

\begin{figure}[H]
\centering
\begin{tikzpicture} [scale=1.3]
\node at (2.5,0) {} ;
\node at (-1.75,0) {\(t = 0\)} ;
\node at (2.6,2.3) {\(t = +\infty \)} ;
\node at (2.6,-2.3) {\(t = -\infty \)} ;
\draw [dashed] (0,2) ellipse (2cm and 0.5cm) ;
\draw [dashed] (0,-2) ellipse (2cm and 0.5cm) ;
\draw [dashed] (0,0) ellipse (0.86cm and 0.25cm) ;
\draw (2,1.9) .. controls (0.5,0) and (0.5,0) ..  (2,-1.9) ;
\draw (-2,1.9) .. controls (-0.5,0) and (-0.5,0) ..  (-2,-1.9) ;
\draw [->] (2.5,-1) -- node[midway, right] {t} (2.5,1) ;
\draw (0,0) circle (0.86cm) ;
\end{tikzpicture}
\caption{de Sitter space fully embedded in Euclidean space juxtaposed with the hyperboloid of proper spacetime signature $(- \, + \, + \, + \, +)$. As we can see, the compactification did not affect the size and geometry of the spacelike hypersurface on the $t=0$ slice, since a Wick rotation leaves that unaffected. It did however "close off" the open spacelike slices of the hyperboloid extending towards $t=\pm \infty$.}
\label{fig:2.3}
\end{figure}
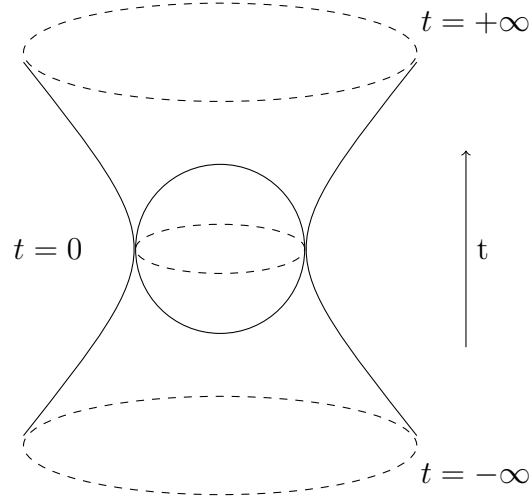

We close off our discussion on dS space by mentioning yet another one of its fascinating properties - its permanent causal fragmentation. In regular flat Minkowski spacetime, causality poses as nothing more than a temporary impediment to establishing a contact with its entirety. As we mentioned in chapter 2, there is the concept of the particle horizon which delimits the range up to which such a causal connection may be established and ultimately stems from the finite nature of the mediation of physical information. However, given enough time this limitation will slowly dissipate and we will be able to "see" all of Minkowski space. The case of dS space lends itself to a much more interesting phenomenon; the accelerating nature of its expansion will form a different type of horizon - an event horizon. Event horizons much like particle horizons also delimit the distance up to which causal interaction may take place and serve as boundaries for the region of spacetime currently not in such contact. The crucial difference is that ultimately event horizons owe this limitation they set to the geometry of space, and as such will never go away. This is what we refer to as "permanent causal fragmentation", as spacetime will be strictly divided into regions with varying degrees of causal contact between them. As for the reason for why this is fascinating, that's because we generally associate event horizons with the existence of singularities, black holes being the most prominent example. It is unusual for an otherwise homogeneous, isotropic and overall "well-behaved" spacetime to showcase this property. The way we can show this feature of dS space is by using a Penrose diagram. Similarly to turning the dS hyperboloid into a 4-sphere, a Penrose diagram helps to compactify a certain spacetime, turning its various infinities into finite points on the diagram. For the present case, we start off with the metric (\ref{eq:2.3}) and switch to a conformal time $\eta$ defined as:

\begin{equation} \label{eq:2.10}
\begin{split}
    &\cosh{t} = \frac{1}{\cos{\eta }} \\
    \Rightarrow \, &dt^{2} = d\eta ^{2}\frac{1}{\cos ^{2}{\eta } }
\end{split}
\end{equation}

and thus the metric will become:

\begin{equation} \label{eq:2.11}
    ds^{2} = \frac{1}{\cos ^{2}{\eta } } (-d\eta ^{2} + d\Omega ^{2}_{3})
\end{equation}

where we have set $\ell = 1$. We then perform a conformal transformation on the entirety of the metric:

\begin{equation} \label{eq:2.12}
    d\bar{s} ^{2} = \cos ^{2}{\eta } \, ds^{2} = -d\eta ^{2} + d\Omega ^{2}_{3}
\end{equation}

In order for the diagram to work as a 2D representation, we implicitly assume that in reality the only variable spatial coordinate is the second polar coordinate $\chi$ and therefore:

\begin{equation} \label{eq:2.13}
    d\bar{s} ^{2} = -d\eta ^{2} + d\chi ^{2}
\end{equation}

We notice that just as in the case of the 4-sphere after a Wick rotation and real rotation of 90 degrees, the infinities of dS space at $t=\pm \infty$ have also compactified into slices of constant $\eta$:

\begin{equation} \label{eq:2.14}
\begin{split}
    &i^{+} \rightarrow \eta = \frac{\pi }{2} \\
    &i^{-} \rightarrow \eta = -\frac{\pi }{2}
\end{split}
\end{equation}

Lastly, we define a new set of lightlike coordinates to serve as the diagonal 45-degree axes:

\begin{equation} \label{eq:2.15}
\begin{rcases}
    &U = \eta - \chi \\
    &V = \eta + \chi
\end{rcases}
\Rightarrow d\bar{s}^{2} = -dU\, dV
\end{equation}

The resulting Penrose diagram then is:

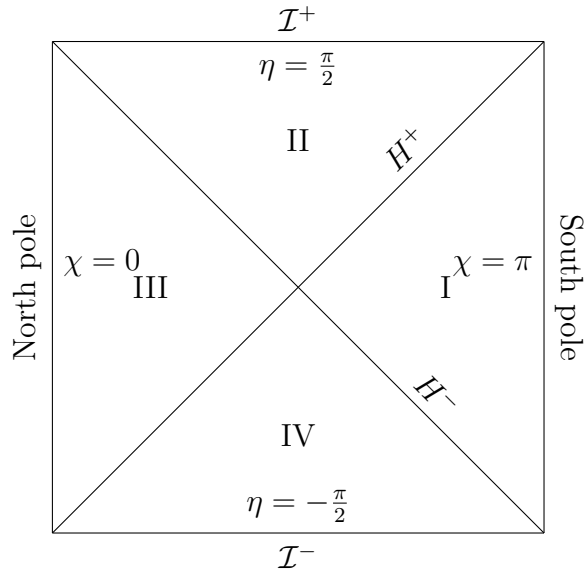
\begin{figure}[H]
\centering
\begin{tikzpicture}[scale=0.65]
\node (I)    at (3,0)   {I};
\node (III)   at (-3,0)   {III};
\node (II)  at (0,3) {II};
\node (IV)   at (0,-3) {IV};
\draw (-5,5) -- node[midway,above] {\(\mathcal{I} ^{+}\)} (5,5) ;
\draw (5,5) --  node[midway,above,sloped] {South pole} (5,-5) ;
\draw (5,-5) -- node[midway,below] {\(\mathcal{I} ^{-}\)} (-5,-5) ;
\draw (-5,-5) -- node[midway,above,sloped] {North pole} (-5,5) ;
\draw (0,0) -- node[midway,above,sloped] {\(H^{+}\)} (5,5) ;
\draw (0,0) -- node[midway,above,sloped] {\(H^{-}\)} (5,-5) ;
\draw (0,0) -- (-5,5) ;
\draw (0,0) -- (-5,-5) ;
\coordinate[label=below:{\(\eta = \frac{\pi }{2} \)}] (top) at (0,5) ;
\coordinate[label=above:{\(\eta = -\frac{\pi }{2} \)}] (bot) at (0,-5) ;
\coordinate[label=above left:{\(\chi = \pi \)}] (right) at (5,0) ;
\coordinate[label=above right:{\(\chi = 0 \)}] (left) at (-5,0) ;
\end{tikzpicture}
\caption{Penrose diagram of dS space. Horizontal slices of equal conformal time $\eta$ on the diagram will correspond to similar 3-spheres as those when slicing the Euclidean 4-sphere. The appropriate points for said value of $\eta$ which are located on the right and left vertical sides of the diagram form their south and north poles respectively.}
\label{fig:2.4}
\end{figure}

Here we can see the features which we went over earlier: dS space is fragmented into four regions, with a future and past event horizon $H^{+}$ and $H^{-}$ respectively forming as boundaries to those regions. An observer in region I may receive information from region IV, but cannot send any information themselves; the reverse is true for region II where - just like with a black hole - the observer may send information, but cannot receive any back. Any information sent from IV to I or from I to II will eventually seem to "freeze" close to the event horizons for the observer, as it asymptotically approaches the positive lightlike infinity $\mathcal{I} ^{+}$ and negative lightlike infinity $\mathcal{I} ^{-}$ respectively. Region III on the other hand is completely cut off from region I and no causal contact may ever be established. \par
With dS space as we have defined it expanding exponentially, the event horizons will not remain stationary, but will also increase in size. Despite all four regions growing similarly via this process, the total amount of space that is within any of the three regions outside the region in which an observer exists will appear to grow more compared to the latter. The resulting observation is rather grim: the observer will perceive their surroundings as an ever-increasing black hole that tends to encompass everything up to the spacetime point that they are initially located. Asymptotically in the future, the observer will have no causal contact with any other point in spacetime. Luckily for us, this exponential expansion during the inflationary stage of the Universe did not last forever, at least locally. Unfortunately though, as the vacuum energy begins to dominate space once again due to the dilution of other forms of energy density (like we explained previously), the spacetime once again begins to asymptotically tend to a dS state, marking the fate of the dS observer ultimately unavoidable. The Penrose diagram for the asymptotic dS space can be seen below.

\begin{figure}[H]
\centering
\begin{tikzpicture}[scale=0.4]
\node (I)    at (3,0) {} ;
\node (III)   at (-3,0) {} ;
\node (II)  at (0,3) {};
\node (IV)   at (0,-3) {};
\draw (-5,5) -- node[midway,above] {\(\mathcal{I} ^{+}\)} (5,5) ;
\draw (5,5) -- (5,-5) ;
\draw (5,-5) -- node[midway,below] {\(\mathcal{I} ^{-}\)} (-5,-5) ;
\draw (-5,-5) -- (-5,5) ;
\end{tikzpicture}
\caption{Penrose diagram for a Universe that is asymptotically dS in the remote past ($\mathcal{I} ^{-}$) and future ($\mathcal{I} ^{+}$). Horizons no longer exist, as the geometry in the region in between is not dS. An important feature is that the value of $H^{-1}$ differs greatly between the two points. The Universe starts off very small and highly curved (small $H^{-1}$) and ends up very large and almost completely flat (large $H^{-1}$).}
\label{fig:2.5}
\end{figure}
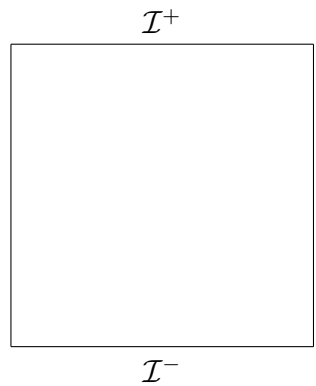

\section{Mechanics of inflation}

As we have explained in the second chapter, the dynamics of inflation can be explained by the existence of a scalar quantum field $\phi$ called the inflaton that is under the influence of a potential $V(\phi)$, itself corresponding to the vacuum energy of the Universe. It wouldn't be an unreasonable question to pose as to why we expect that $\phi$ is a scalar field. The reason has to do with the Cosmological principle; if the inflaton was a vector field or a tensor field like the spacetime metric, then it would have preferred directions which would lead to anisotropy. In addition, scalar fields are usually responsible for spontaneous symmetry breaking (such as the Higgs field), which makes them a more natural choice for fundamental models.\cite{Reference15}.\par
Just as we also explained, we expect a dS spacetime solution for inflation as a result of the domination of vacuum energy, which by using equation (\ref{eq:1.26}) and setting $w=-1$, we find that $\rho = constant$. Plugging that in the second Friedmann equation (\ref{eq:1.12}) and using that for $\rho = constant \Rightarrow \rho = -P$, we do in fact get that $a(t)$ is some hyperbolic function of $t$, as per (\ref{eq:2.4}) and (\ref{eq:2.5}). In the early stages where we predict inflation took place, the curvature term is negligible can be taken almost 0, giving us the standard exponential form:

\begin{equation} \label{eq:2.16}
    a(t) = H^{-1}e^{Ht}
\end{equation}

We also notice how the first Friedmann equation (\ref{eq:1.11}) predicts that during this phase, since $\rho$ is constant, so will $H$. This is not exactly accurate since the scale factor is time-dependent, but the change is expected to be extremely small and thus $H \approx constant$. The exact solution if we identify the vacuum energy density with the inflaton potential, $\rho = V(\phi)$:

\begin{equation} \label{eq:2.17}
    H \approx \sqrt{\frac{8\pi G \,V(\phi)}{3}} = \sqrt{\frac{8\pi V(\phi)}{3m_{P}^{2}}}
\end{equation}

where in natural units $G = m_{P}^{-2}$. We now consider the Lagrangian for the scalar field:

\begin{equation} \label{eq:2.18}
    \mathcal{L} = \frac{1}{2} \partial ^{\mu}\phi \, \partial _{\mu} \phi - V(\phi)
\end{equation}

When deriving the classical equations of motion for the inflaton, it is important not to forget that we are in curved spacetime and as such the free Euler-Lagrange equation does not hold. The equation of motion is rather:

\begin{equation} \label{eq:2.19}
    \tilde{\Box} \phi = \nabla ^{\mu} \nabla _{\mu} \phi = -\frac{dV}{d\phi}
\end{equation}

As we see, the D'Alembert operator here implies the square of the covariant derivative of $\phi$, where the covariant derivative for a 4-vector is defined as \cite{Reference5}\cite{Reference13}:

\begin{equation} \label{eq:2.20}
    \nabla _{\mu} A^{\nu} = \partial _{\mu} A^{\nu} + \Gamma ^{\nu}{}_{\mu \lambda} A^{\lambda}
\end{equation}

where $\Gamma ^{\nu}{}_{\mu \lambda}$ are the Christoffel symbols. A covariant derivative acts as a regular 4-gradient on a scalar field, but since we have the square of covariant derivatives, the second one does in fact produce connection terms. A useful identity for when the indices of the 4-vector and covariant derivative are the same is \cite{Reference13}:

\begin{equation} \label{eq:2.21}
    \nabla _{\mu} A^{\mu} = \frac{1}{\sqrt{g}} \partial _{\mu} (\sqrt{g} A^{\mu})
\end{equation}

where $g = \det{(-g_{\mu \nu})}$. By calculating the Christoffel symbols for the FLRW metric, the resulting equations of motion shall be \cite{Reference15}\cite{Reference14}:

\begin{equation} \label{eq:2.22}
    \ddot{\phi} - \tilde{\Delta} \phi + 3H\dot{\phi} = -V'(\phi)
\end{equation}

with $\tilde{\Delta}$ being the modified Laplacian operator, including the various spatial derivatives and the appropriate connection terms. It can be easily shown that those terms are in fact inversely proportional to the square of the scale factor, and therefore they are exponentially suppressed as the Universe expands, meaning we can ignore them in the equations of motion. From a physical standpoint such an approximation makes sense: as the Universe expands, the inflaton field becomes more and more spatially "smooth", taking almost perfect uniform properties. In addition, we take the liberty in including an additional term in the equations of motion \cite{Reference15}:

\begin{equation} \label{eq:2.23}
    \ddot{\phi} +\Gamma_{\phi} \dot{\phi} + 3H\dot{\phi} +V'(\phi) = 0
\end{equation}

This new term is none other than the decay of the inflaton, more specifically via the indirect decay by transitioning to other fields/particle species that are coupled to the former. $\Gamma_{\phi}$ in that case corresponds to the inflaton's decay width pertaining to that specific coupling e.g. for decay into light fermions $\Gamma_{\phi} = \frac{h^{2}m_{\phi}}{8\pi}$, where $h$ is the coupling strength of $\phi$ to said fermions and $m_{\phi}$ its mass (or effective mass, if it is massless). The equation of motion (\ref{eq:2.23}) has shaped up to be rather familiar; it is simply the equation of motion of a ball rolling down a hill (whose shape provides the potential term) with friction. Assuming that the ball does not slide, then the Lagrangian is:

\begin{equation} \label{eq:2.24}
    L = \frac{1}{2} mR^{2}\dot{\varphi}^{2} - V(\varphi)
\end{equation}

where $R$ the ball's radius and $\varphi$ the angle as it rotates while rolling. The friction will be provided by a dissipation function $F$ which for the sake of simplicity we can take in the form of the Rayleigh function:

\begin{equation} \label{eq:2.25}
    F = \frac{1}{2}\, k R^{2}\dot{\varphi}^{2}
\end{equation}

The Euler-Lagrange equations do in fact give a nearly identical equation:

\begin{equation} \label{eq:2.26}
    \ddot{\varphi} + \frac{k}{m}\dot{\varphi} + \frac{1}{m} V'(\varphi) = 0
\end{equation}

Excluding the inequivalence in terms of dimensionality, it becomes immediately apparent that the Hubble parameter and decay width act as friction terms as the inflaton tends to roll towards the minimum of its potential. The decay width does so because it directly reduces the amount of inflatons present, while the Hubble parameter does so because the bigger the expansion rate is and by extension the size of the Universe, the more prominent the red-shift becomes, reducing the kinetic energy of inflatons as they tend towards the aforementioned minimum. This motion can be further broken down into two phases: an initial "slow roll" phase where the value of $V(\phi)$ and the expansion rate $H$ barely change, and a "reheating" phase where rapid oscillations of the inflaton occur around its minimum. The first phase corresponds to the inflationary stage proper as we described it above and it can be approximated as:

\begin{equation} \label{eq:2.27}
    3H\dot{\phi} \approx -V'(\phi)
\end{equation}

The acceleration term $\ddot{\phi}$ is vanishingly small as a result of the slow roll, while the decay term is practically inactive. That is because the inflaton field is expected to be very weakly coupled to every other particle species, thus during the slow roll regime where the kinetic energy of the inflatons is very small, particle creation is extremely rare. This phase requires certain conditions in order for the motion to proceed as expected which have to do with the rate of change of the Hubble parameter. To be more precise, for this argument to hold we require that $|\dot{H}| \ll H^{2}$. Combined with the approximation in (\ref{eq:2.17}), we derive the so-called "slow roll conditions" \cite{Reference15}\cite{Reference14}:

\begin{equation} \label{eq:2.28}
\begin{split}
    |V''(\phi)| \ll 9H^{2} &\approx \frac{24\pi V(\phi)}{m_{P}^{2}} \\
    \left |\frac{V'(\phi)m_{P}}{V(\phi)} \right | &\ll \sqrt{48\pi}
\end{split}    
\end{equation}

These conditions ensure that the slope of the potential is initially very small and thereby the potential for that region is extremely flat, spanning most values of $\phi$ before reaching the minimum. This is a crucial criterion for two additional reasons: firstly, it ensures that enough time for the inflationary stage is granted so that the observable Universe and the various physical scales within it have grown enough to match present observations. While we will elaborate on this in the next section, the way one quantifies this duration is by e-folds - the number of factors of $e$ that the scale factor has grown \cite{Reference15}:

\begin{equation} \label{eq:2.29}
    \ln{\left [\frac{a(t_{2})}{a(t_{1})} \right ]} \equiv N(\phi _{1} \rightarrow \phi _{2}) \equiv \int _{t_{1}}^{t_{2}} H\, dt \approx -\frac{8\pi}{m_{P}^{2}} \int _{\phi _{1}}^{\phi _{2}} \frac{V(\phi)}{V'(\phi)} \, d\phi
\end{equation}

As it becomes immediately obvious, the validity of the slow roll conditions is equivalent with a sufficiently large amount of e-folds $N_{tot} = N(\phi_{i} \rightarrow \phi_{e})$ where $\phi_{e}$ the value of the inflaton field upon exiting the slow roll regime. The second reason for why the potential needs to be this flat are the quantum considerations. Thus far, the entire treatment has been classical and the inflaton dynamics are entirely consistent with classical field theory. In reality - even with a semi-classical treatment of gravity - the inflaton  is a quantum field, hence it is subject to quantum mechanical fluctuations. To analyze the system as we have done, we are using the semi-classical approximation. By defining:

\begin{equation} \label{eq:2.30}
    \phi = \phi _{cl} + \delta \phi
\end{equation}

we are assuming that the quantum fluctuations act as minuscule deviations ($\delta \phi \ll \phi _{cl}$) to the classical trajectory i.e. the expectation value of the inflaton field that obeys (\ref{eq:2.22}). This is tantamount to taking the semi-classical limit $\hbar \rightarrow 0$, and ignoring sub-leading terms. Given that massless scalar fields in dS space have a very specific type of spectrum of fluctuations (as we shall see in the following section), said spectrum has to possess an amplitude value that is small enough not to significantly alter the field's motion. That will prove to be connected to the slow roll conditions. \par
From a physical standpoint it's a rather straightforward idea, in that for an insufficiently flat potential during the inflationary stage, quantum fluctuations would be substantial in magnitude to "push" the inflaton field in either direction. When pushed towards the minimum, it causes the Universe to prematurely exit its period of inflation, while when pushed away from the minimum, it prolongs inflation indefinitely as there is no longer an energetically preferred direction for the inflaton to tend towards. An indicative version of a viable potential and the resulting motion are illustrated in the figure below.\par

\begin{figure}[H]
\centering
\begin{tikzpicture} [scale=0.6]
\node[draw, thick, circle, blue, inner sep=3pt] at (2.5,8) {} ;
\node[draw,dashed, thick, circle, blue,  inner sep=3pt] at (11.55,0.3) {} ;
\node at (11.5,-0.25) {$\sigma$} ;
\node at (2.5, -0.25) {$\phi _{i}$} ;
\node at (9,-0.25) {$\phi _{e}$} ;
\draw [dashed, thin]  (2.5,7.75) -- (2.5,0) ;
\draw [dashed ,thin] (9,6) -- (9,0) ;
\draw [->, thick, red] (2,8.1) arc (180:0:0.5) ;
\draw [<->, very thick, red] (12.05,0.45) arc (-10:-170:0.5) ;
\draw [->, dashed, red] (4,8.25) -- (9,6.7) ;
\draw [->] (-3,0) -- (-3,13) node[left] {$V(\phi)$} ;
\draw [->] (-3,0) -- (15,0) node[right] {$\phi$} ;
\draw [thick] (-3,8.13) .. controls (-2.5,8.12) and (-2,8.11) .. (-1.5,8.09) .. controls (-1,8.07) and (-0.5,8.04) .. (0,8) .. controls (1,7.95) and (2,7.85) .. (3,7.7) .. controls (4,7.55) and (5,7.35) .. (6,7.1) .. controls (7,6.8) and (8,6.55) .. (8.5, 6.3) .. controls (9,6) and (9.5,5.8) .. (10,4.2) .. controls (10.25,2.7) and (10.5,1.7) .. (10.75,1) .. controls (11,0.3) and (11.25,0.1) .. (11.5, 0) .. controls (11.75, 0.05) .. (12,0.3) .. controls (12.25,0.7) and (12.5,1.15) .. (12.75,1.85) .. controls (13.5, 4.5) and (14,7.3) .. (14.5,10.5) ;
\end{tikzpicture}
\caption{The classical motion of the inflaton as it rolls down its potential $V(\phi)$, juxtaposed with a ball rolling down a slope in the shape of the same potential. The field starts at a value $\phi _{i}$ and experiences a prolonged slow roll phase where the value of the potential (and by extension the expansion rate $H$) barely changes before it reaches the value $\phi _{e}$, upon which the slow roll ceases and it exits inflation. It then resumes by rolling quickly down its minimum located at $\sigma$ where $V(\sigma) = V'(\sigma) = 0$, whence rapid oscillations take place and the reheating phase begins.}
\label{fig:2.6}
\end{figure}
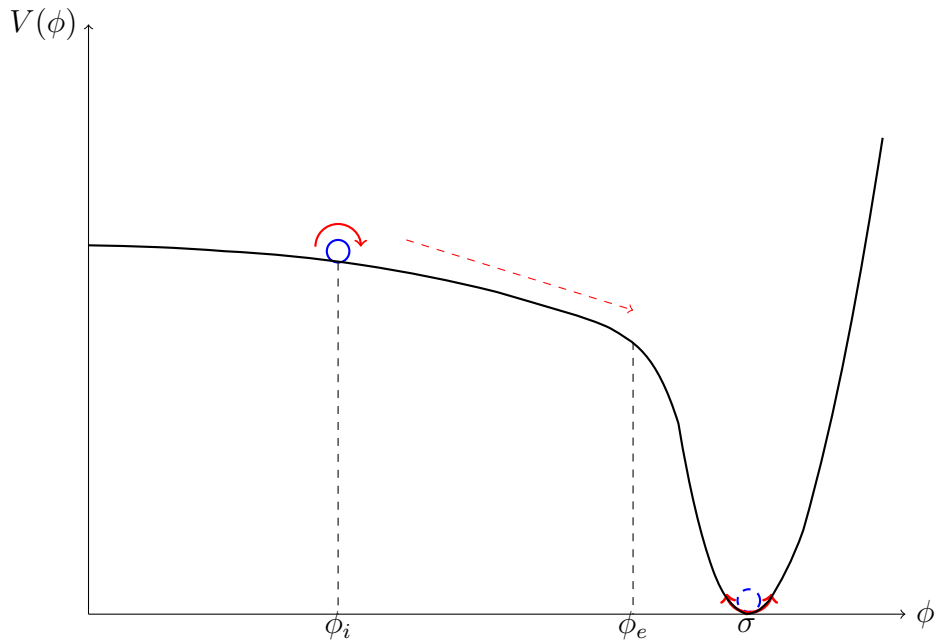

\section{Physical repercussions and reheating}

Just as one would expect, the exponential expansion of spacetime brings about some dramatic consequences. The Universe as a whole is expected to be many orders of magnitude larger than the observable Universe today, with numerous causally disconnected regions outside of it that form their own self-contained "mini-Universes". Furthermore, this massive expansion all but completely evaporates the contribution in energy density of just about anything else that existed in the Universe prior to inflation. While the energy density of the vacuum remains unchanged, every other form (regular matter, radiation etc) is exponentially suppressed due to their inverse dependence on the scale factor, and thus they are diluted in the vastness of space. This is a first solution to the problem of unwanted relics posed back in chapter 2; excluding superheavy species that might be produced as the inflaton decays (a pathology generally associated with supersymmetric models), any such primordial species will be diluted as well and their preexisting densities vanish, leading to present observations. \par
This also provides a reasonable explanation to another problem: baryon asymmetry. Even though under Hot Big Bang cosmology there wouldn't be any reason for antibaryons not to exist in the present epoch or in fact have complete symmetry with their baryon counterparts, there is an asymmetry heavily in the favour of the latter. In the context of inflation, this expected symmetry could have been (and most likely was) the state of the Universe preceding inflation, but both populations were diluted and withered away due to exponential expansion. It is up to the period of reheating to produce the necessary baryon asymmetry which we observe today. \par
The dilution of matter and radiation energy densities will also affect temperature, since temperature itself is inversely proportional to the scale factor. By extension, it will also drop off exponentially, effectively "supercooling" the Universe. Degrees of freedom stemming from primordial particle species disappear, and therefore the quantity $aT$ remains roughly constant. The reheating phase will once again be responsible for bringing things up to par. The particles which spawn due to the decay of the rapidly oscillating inflaton field will very quickly provide the Universe with a substantial amount of thermal contributions, the associated energy densities increase greatly and temperature rises. As a result entropy skyrockets, which serves as a fitting preamble to what the solution for the various cosmological problems will be, while also providing a smooth transition to the radiation-dominated cosmological era. In order to be more specific in this description of the reheating phase, it is advantageous to view the equations of motion for the inflaton in terms of its energy density. Starting from the stress-energy tensor:

\begin{equation} \label{eq:2.31}
    T^{\mu \nu} = \partial ^{\mu }\phi \partial ^{\nu }\phi - \mathcal{L}
\end{equation}

and ignoring the spatial derivatives of $\phi$ for the reasons explained before, it is easy to see how (\ref{eq:2.31}) takes the form of a perfect fluid once again, and as such the only non-zero components are the diagonal ones:

\begin{equation} \label{eq:2.32}
\begin{split}
    &\rho _{\phi } = \frac{\dot{\phi}^{2}}{2} + V(\phi) \\
    &P_{\phi} = \frac{\dot{\phi}^{2}}{2} - V(\phi)
\end{split}
\end{equation}

It is important to make the distinction between these values and those corresponding to the vacuum energy. While the inflaton field is inherently tied to the vacuum energy of space and drives inflation, vacuum energy and pressure as we explained will remain constant, while the quantities in (\ref{eq:2.32}) will not due to the non-static nature of the field. As such, the equation of motion (\ref{eq:2.23}) can be rewritten as:

\begin{equation} \label{eq:2.33}
    \dot{\rho}_{\phi} + 3H\dot{\phi}^{2} + \Gamma _{\phi }\dot{\phi}^{2} = 0
\end{equation}

However, due to the fact that the field is oscillating, we can take an average of the motion over a single cycle where $\langle \dot{\phi}^{2} \rangle = \rho _{\phi}$. This is none other than the virial theorem for a harmonic oscillator:

\begin{equation} \label{eq:2.34}
    \langle V(\phi) \rangle = \langle E_{K} \rangle = \left \langle \frac{\dot{\phi}^{2}}{2} \right \rangle
\begin{cases}
    \Rightarrow &\langle \rho_{\phi} \rangle = \langle \dot{\phi}^{2} \rangle \\
    \Rightarrow &\langle P_{\phi} \rangle = 0
\end{cases}
\end{equation}

The last relation is enlightening in regards to the nature of inflatons during reheating. Their coherent oscillations grant them an effective large mass $V''(\sigma)$ while their kinetic energy is comparatively not that large. Because of this they behave just like non-relativistic matter, as evidenced by their averaged 0 pressure. Eventually, the energy density equation becomes:

\begin{equation} \label{eq:2.35}
    \dot{\rho}_{\phi} + 3H\rho _{\phi} + \Gamma _{\phi }\rho_{\phi} = 0
\end{equation}

The solution to this equation is rather simple:

\begin{equation} \label{eq:2.36}
    \rho_{\phi} = M^{4} \left [\frac{a(t)}{a(t_{osc})} \right]^{-3} e^{-\Gamma _{\phi}(t-t_{osc})}
\end{equation}

where $t_{osc} \approx H^{-1} \sim \frac{m_{P}}{M^{2}}$ signifies the time where the period of rapid oscillations begins and $M^{4}$ is the value of the vacuum energy at that time. As expected, the growth of the scale factor will increase the energy density of the inflaton; not exponentially anymore since the slow roll is over, but as $t^{2/3}$ since inflatons behave like non-relativistic matter. As expected, their decay into coupled particle species will provide an exponential damping. Without assuming any specific model, we assume that the main products of the rapid oscillations will be light, highly relativistic particles. Following the conservation of energy, the equation for the evolution of energy density transitions to that for radiation \cite{Reference15}:

\begin{equation} \label{eq:2.37}
    \dot{\rho}_{R} + 4H\rho_{R} = \Gamma _{\phi} \rho _{\phi}
\end{equation}

Upon creation, the relativistic particles interact thermally between them, establishing an equilibrium, at which point $\rho_{R} = \frac{\pi^{2}}{30} g_{*} T^{4}$. Before they eventually manage to dominate, from the period between $t_{osc}$ to $t=\Gamma_{\phi}^{-1}$, the inflatons remain the predominant form of content in the Universe, causing it to be matter-dominated instead. Hence, the radiation energy density will approximately be \cite{Reference15}:

\begin{equation} \label{eq:2.38}
    \rho_{R} \approx \frac{m_{P}^{2}\Gamma_{\phi}}{10\pi t} \left [1-\left (\frac{t}{t_{osc}} \right )^{-\frac{5}{3}} \right] \approx \sqrt{\frac{6}{\pi}} \frac{m_{P}\Gamma_{\phi}M^{2}}{10} \left (\frac{a}{a_{osc}} \right )^{-\frac{3}{2}} \left [1- \left (\frac{a}{a_{osc}} \right )^{-\frac{5}{2}} \right ]
\end{equation}

where we have defined $a_{osc} = a(t_{osc})$.  What this means is that, while the radiation energy density increases substantially at the beginning of the reheating phase, it then proceeds to decrease as $\propto a^{-3/2}$. Thus in turn translates to the same effect for the temperature. On the other hand, just as we mentioned earlier, the entropy spikes and remains just as big, since $S \propto a^{3}\rho_{R}^{3/4} \sim a^{15/8}$. The final values of both temperature and entropy by the end of the entire process, and taking both regimes into account are going to be \cite{Reference14}: 

\begin{equation} \label{eq:2.39}
    T_{RH} \equiv T(t=\Gamma_{\phi}^{-1}) \approx 0.55 g_{*}^{-\frac{1}{4}} (m_{P}\Gamma_{\phi})^{\frac{1}{2}}
\end{equation}

\begin{equation} \label{eq:2.40}
    S_{patch} \approx e^{3N_{tot}} \frac{M^{4}}{T_{RH}^{4}} H^{-3}T_{RH}^{3} \approx e^{3N_{tot}} \frac{m_{P}^{3}}{M^{2}T_{RH}}
\end{equation}

$S_{patch}$ here implies that the entropy calculated refers to the initially small patch of spacetime in the early Universe that through the process of inflation grew to become the observable Universe. Entropy as we showed generally follows $aT$, which is why we expressed entropy as the initial value of $a(t_{i}) \approx H^{-1}$ times its factor of growth and the final value of the temperature as seen in (\ref{eq:2.39}), and all of this cubed. More specifically for the factor of growth, the exponential term comes directly from the total expansion during the slow roll phase, while the second term is due to reheating. That is because the energy density starts completely comprised of non-relativistic inflatons, remaining matter-dominated until its end. Therefore, if we recall (\ref{eq:1.26}), the ratio of scale factors is:

\begin{equation} \label{eq:2.41}
    \frac{a_{RH}}{a_{osc}} \propto \left [\frac{\rho (t=\Gamma _{\phi}^{-1})}{\rho (t_{osc})} \right ]^{-\frac{1}{3}} \approx \left (\frac{M^{4}}{T_{RH}^{4}} \right )^{\frac{1}{3}}
\end{equation}

From the form of (\ref{eq:2.40}), one can be deceived to believe that entropy increases due to the inflationary period as well, in contradiction to what we mentioned earlier about the slow roll regime being adiabatic. This is simply illusory though. Entropy does remain constant during that phase, but the immense size of the Universe by the end means that once temperature increases during reheating, the amount of entropy necessarily produced to assign that energy to such a vast region will also have to be huge, thereby the exponential growth indirectly influencing just how much entropy will be measured. \par

\section{Solutions to cosmological problems}

As we have seen, inflation has already provided us with a satisfying (if not incomplete) solution to the unwanted relics problem and the question of baryon asymmetry in the Universe via reheating. It also offers a rough outline on what could possibly a future solution to the cosmological constant problem. While it is still unclear why $\Lambda$ is at its present value, the dynamics of the inflaton and the evolution of its potential constitute a solid starting point in understanding why it had to be much larger in the past and eventually why it decreased to become nearly 0. Inflation does however answer the rest of the problems mentioned before much more convincingly or at the very least offers an explanation that works a lot better as a basis for further investigation. \par

\subsection{The horizon and flatness problems}

These are arguably the biggest success of inflationary cosmology. Not only does inflation resolve these problems, but also provides a very natural and intuitive physical explanation for this resolution with tangible predictions. The main tool in this investigation revolves around the duration of the slow roll phase in terms of e-folds $N$, as this will be able to determine whether the initial patch has really grown enough in order for current measurements to be consistent. In both cases, the measure by which we will examine the necessary amount of e-folds is going to be entropy. For the case of the flatness problem the connection is obvious; by examining the amount of entropy produced due to reheating within a volume defined by the curvature radius of the Universe, a sufficient production will provide roughly the same entropy as observed today due to the near-adiabatic growth that followed. To put it differently, if enough entropy was produced during inflation to match its current value, then the curvature radius that defines the volume which contains it will have also grown sufficiently. If what constitutes this sufficient amount truly is as massive as we expect, the curvature of this volume all but vanishes and it becomes extremely flat. More concretely, if we define the curvature radius as \cite{Reference15}:

\begin{equation} \label{eq:2.42}
    R_{curv}(t) \equiv a(t)|k|^{-\frac{1}{2}} = \frac{H^{-1}(t)}{|\Omega (t) - 1|^{\frac{1}{2}}}
\end{equation}

then the entropy at the end of reheating for the curvature volume will be:

\begin{equation} \label{eq:2.43}
    S_{curv} = R_{curv}^{3}(t=\Gamma ^{-1}{\phi}) T_{RH}^{3} \approx e^{3N_{tot}} \frac{m_{P}^{3}}{M^{2}T_{RH}} \frac{1}{|\Omega _{i} - 1|^{\frac{1}{2}}}
\end{equation}

where $\Omega _{i} = \Omega (t_{i})$ the initial value of the density parameter, since due to the slow roll conditions and the dynamics of the inflaton, it remains roughly constant for the duration. As we can see, the entropy contained in the curvature volume is very similar to the entropy contained in the initial patch (\ref{eq:2.40}), up to a factor involving the density parameter. As such, we demand that the amount of e-folds is larger or equal than a minimum value $N_{min}$ that leads to $S_{curv} \approx 10^{88}$. Solving in terms of the e-folds we get:

\begin{equation} \label{eq:2.44}
    N_{tot} \geq N_{min} = 53 + \frac{2}{3} \ln{\left (\frac{M}{10^{14} \, GeV} \right )} + \frac{1}{3} \ln{\left (\frac{T_{RH}}{10^{10} \, GeV} \right )} + \frac{1}{2} \ln{(|\Omega _{i} -1|)}
\end{equation}

For reasonable values of $M$ and $T_{RH}$ (somewhere between $1-10^{19}$ GeV) \cite{Reference15}\cite{Reference14}\cite{Reference34}, and for a value of $\Omega_{i}$ that is expectedly different enough from unity, $N_{min}$ ranges between approximately 24 to 68 e-folds. Given the slow roll conditions ensure that the potential is quite flat, this is a very modest requirement. We can also define a minimum amount of e-folds necessary so that the entropy $S_{patch}$ also gives the correct present value of entropy:

\begin{equation} \label{eq:2.45}
    N_{min} ' = 53 + \frac{2}{3} \ln{\left (\frac{M}{10^{14} \, GeV} \right )} + \frac{1}{3} \ln{\left (\frac{T_{RH}}{10^{10} \, GeV} \right )}
\end{equation}

which is effectively the same as $N_{min}$, bar the final term involving the density parameter. This minimum requirement will then correspond to the growth of the physical radius of the observable Universe, since that will just be the particle horizon at each cosmological era. If we then check the ratio of the present value of the curvature radius $(R_{curv})_{0}$ to the present value of the Hubble radius $H_{0}^{-1}$:

\begin{equation} \label{eq:2.46}
    \frac{(R_{curv})_{0}}{H_{0}^{-1}} = \frac{e^{(N_{tot} - N_{min} ')}}{|\Omega _{i} - 1|^{\frac{1}{2}}}
\end{equation}

we can then find the resulting value for the curvature of the Universe today:

\begin{equation} \label{eq:2.47}
    |\Omega _{0} - 1| = e^{2(N_{min} ' - N_{tot})} |\Omega _{i} -1|
\end{equation}

Indeed, no matter the value of $\Omega _{i}$, if inflation lasts long enough for $N_{tot} \gg N'_{min}$ to hold, then the flatness of the Universe follows suit. The density parameter is exponentially suppressed and eventually reaches near-unity. \par
The horizon problem solution is slightly less obvious, but the discussion thus far shows how it comes down to the same argument. That is, if we want the particle horizon to have evolved in a manner as to allow for all observed CMBR photons to have been in causal contact at some point in their past, then it is necessary for the initial patch we mentioned above to have grown from its very small size via inflation up to a point which with further adiabatic expansion following reheating, it eventually reaches its current size, thereby containing the same amount of entropy as we observe today. In other words, we want the physical radius to have started small enough for thermal contact to have been possible, and then to have grown as needed to get sufficiently big to match observational data. That is none other than the requirement $N_{tot} \geq N_{min} '$, which means that the flatness and horizon problems are solved by pretty much the exact same argument. This agreement between the two is in fact one of the most compelling features of inflationary models. \par
So, what exactly is the physical explanation behind these results? As we said, it is a rather satisfying intuitive picture. The flatness can be seen as the result of the extreme size of a curved spacetime, just like the surface of the earth seems quite flat to us due to our much smaller scale. Only in the case of inflation, the disparity between the total size of the Universe compared to the observable part we inhabit is expected to be immensely larger. That means the entire observable Universe seems more flat locally as an extract from this "superuniverse" than a tiny portion of the earth seems to a single human standing on its surface, making Euclidean geometry a much better approximation for astronomy than geography(!). \par
The horizon problem is slightly more subtle. Since the Universe grew so immensely, the initial patch must have been extremely small or equivalently, assuming a typically small Hubble radius $H^{-1}$ for the patch, it experienced a sufficient time portion of this immense growth. This allowed for the primordial photons of the early Universe to have come in causal contact and establish thermal equilibrium long before the era of last scattering. Therefore, despite photons in the CMBR being in causally disconnected regions, those regions would have been minuscule portions of the inflationary patch where it started.

\subsection{The Cosmological principle and singularity problems}

Even though at a first glance these two problems are completely different, they are in fact connected via their respective solutions, as both will make use of the cosmic no-hair theorem \cite{Reference15}. The no-hair theorem originates in the study of black holes \cite{Reference38}; because of the black hole's event horizon (as we explained in the first section), there is a limit up to which causal contact may be established and thus some physical properties of the black hole will forever remain obscure. The theorem states however that those unreachable properties don't matter, as going beyond the event horizon renders them impotent outside of it. The only quantities which matter for a black hole are external and able to be observed: mass, charge, and angular momentum. The existence of event horizons in dS space raises the interesting possibility that an analogous theorem might exist, rendering initial data (besides slow roll conditions) for a spacetime prior to inflation irrelevant. \par
Luckily such a theorem does exist, as proven by Wald \cite{Reference27}. The theorem manages to prove that almost all of the most general homogeneous, but anisotropic (or isotropic in some special cases) solutions to Einstein's field equations - the so-called the Bianchi models - do in fact tend to become asymptotically dS following exponential expansion, the details of which we shall delve into shortly. What is important to note though is that this theorem is not actually directly applicable to inflation, as it is about \textbf{permanently} exponentially expanding spacetimes. Given that inflation predicts a finite period of exponential expansion, the theorem applies insofar as those initial conditions and/or properties do not reappear some time later by "catching up" to the observable Universe.\par
For the singularity problem, this is as far as we can go. The cosmic no-hair theorem does not answer in any way what preceded the postulated singularity and what its conditions were, nor does it provide an explanation about whether that singularity really existed and how it can be made consistent. What it does say is that it is quite possible that the initial condition or state of the Universe is actually irrelevant in regards to the physical repercussions following inflation, since that information is gone due to the spacetime quickly becoming dS and losing its initial properties. The only hope for their reappearance and possible relevance is to rely on the chance of them "catching up" like we mentioned above. Hence to ponder about what came before the $t=0$ point might not even make sense physically. Perhaps, as Starobinsky mentions in his first paper describing inflation as a "dS phase" \cite{Reference29}, the singularity might have never existed at all and the entirety of the Universe's history can be initiated from this superhot, superdense initial dS state. In fact, in inflationary models where inflation does not take place simultaneously in the entirety of the volume of the Universe, the concept of the singularity is neither well-founded in terms of its uniqueness (as no sole singularity would exist), nor is the idea of singularities a necessary requirement to begin with \cite{Reference14}. \par
These are by no means satisfactory or even remotely complete answers, but the existence of inflation does entail this wider variety of options. In Hot Big Bang cosmology, the observed homogeneity and isotropy of the Universe practically forces us into a past that eventually leads to a singular point. Because those emerge from inflation locally in the observable Universe while they are not present globally, there is no need for such a postulate.\par
For the Cosmological principle, the application of the no-hair theorem is slightly more involved. The basic idea is quite obvious, in that the theorem renders initial conditions irrelevant for all homogeneous, anisotropic models. As we will see, those initial conditions may actually contain modes of inhomogeneity. Because of this, we are no longer bound by the need for the Universe to be homogeneous and isotropic by principle, but rather it occurs naturally due to inflation. \par
The way in which this is more involved has to do with the fact that the theorem only partially applies due to the finite duration of inflation. There are two additional conditions in order for the theorem to solve this problem: (a) the time for the spacetime to transpose into a dS state should not be longer or of the same duration as that of the inflaton rolling to its minimum, and (b) the possibly recurring modes of anisotropy and/or inhomogeneity should be suppressed enough as to not appear by the modern cosmological era. The reasons for both are as follows: (a) if the inflaton rolls to its minimum or close to it before we enter a dS state, inflation does not happen or lasts for too little, and (b) since we observe that the Universe today is homogeneous and isotropic on a large scale, there shouldn't be any physical modes that would cross the horizon and cause large-scale anisotropy or inhomogeneity. \par
To examine these, we assume that the anisotropy is expressed by the inequivalence of the scale factors $X_{i}$ in each of the three spatial dimensions \cite{Reference15}. Then we introduce a function $F(X_{1},X_{2},X_{3})$ which will contain said anisotropies of the metric in the Friedmann equation:

\begin{equation} \label{eq:2.48}
    H^{2} = \frac{1}{3} \left (\frac{\dot{\mathcal{V}}}{\mathcal{V}} \right )^{2} = \left (\frac{\dot{\bar{a}}}{\bar{a}} \right )^{2} = \frac{8\pi G \rho }{3} + F(X_{1}, X_{2}, X_{3})
\end{equation}

where $\mathcal{V} = X_{1}X_{2}X_{3}$ the volume scale factor and $\bar{a} \propto \mathcal{V} ^{1/3}$ the mean scale factor for each direction. Here the energy density $\rho$ is not restricted to that of the vacuum and could contain any form of matter. While $F$ can have a very general form and be complicated, its dependence on the mean scale factor can be deduced by the appropriate Bianchi model alone. It can be shown that for all models proportionality is at least as fast as $F \propto \bar{a} ^{-2}$, possibly even faster (e.g. for Bianchi I models it is actually $F \propto \bar{a} ^{-6}$) \cite{Reference15}\cite{Reference27}. This last term of (\ref{eq:2.48}) can be seen as a form of energy density of its own - the anisotropy energy density:

\begin{equation} \label{eq:2.49}
    F \equiv \frac{8\pi G \rho_{AN}}{3}
\end{equation}

This immediately helps clarify point (a): before inflation begins, the contribution of the anisotropy energy density increases the value of $H^{2}$. Due to the slow roll conditions (\ref{eq:2.28}), it is ensured that its contribution does not impede them as it only helps to make the rate of growth larger. That is, if $H^{2}$ is sufficiently large in the absence of the contribution of $\rho_{AN}$, then in its presence it is certain that it will remain as such. To actually enter the slow roll phase, the Universe has to become vacuum dominated, which is ultimately inevitable because the energy density of the vacuum remains constant, while that of anisotropy decreases. Depending on the specific time dependence of $\bar{a}$, it will take some time before that takes place and the Universe enters the slow roll regime. Then, as the Universe undergoes inflation, the contribution of $F$ falls off much  more rapidly and is exponentially suppressed, leading to a space that is to a very good approximation devoid of anisotropy. More specifically, for a mean scale factor following some kind of power law $\bar{a} \propto t^{n}$, it can be shown that the time needed for vacuum domination to occur and inflation to begin is equivalent to about one e-fold of inflation had it been isotropic to begin with. This is obviously not enough to alter the effects of inflation meaningfully. The physical interpretation of this result is that the larger $H$ as a consequence of the anisotropy existing effectively increases the "friction" parameter of the equation of motion. The $3H\dot{\phi}$ term becomes larger and the inflaton finds it more difficult to roll down its potential, limiting how far the latter decreases for as long as that "friction" remains high. \par
Next we need to check whether these anisotropic contributions reappear in the future, and if so to find the requirement for them to not have appeared by the current era. Assuming that the initial value of $\rho_{AN}$ was less or at most comparable to $V(\phi _{i})$ and taking the fastest growing (or rather slowest decaying) modes in the post-inflationary era $\rho_{AN} \propto \bar{a}^{-2}$, we can compare that its present value to that of the observed general energy density $\rho_{0}$ \cite{Reference15}:

\begin{equation} \label{eq:2.50}
    \frac{(\rho_{AN})_{0}}{\rho_{0}} \lesssim e^{-2N_{tot}} \left (\frac{M}{T_{RH}} \right )^{\frac{4}{3}} \left (\frac{T_{RH}}{10 \, eV} \right )^{2} \left (\frac{10 \, eV}{3\, K} \right )
\end{equation}

The first two terms, much like in the analysis for the horizon and flatness problems, come from the growth of the (mean) scale factor due to inflation and reheating respectively, while the third term corresponds to the adiabatic growth following reheating where $T \propto \bar{a} ^{-1}$. Everything else is a result of simplifications and shifting of constants. In order for anisotropy to be negligible today, we require that $\frac{(\rho_{AN})_{0}}{\rho_{0}} \lesssim 1$ which promptly provides us with a condition for the number of e-folds:

\begin{equation} \label{eq:2.51}
    N_{tot} \gtrsim 53 + \frac{2}{3} \ln{\left (\frac{M}{10^{14} \, GeV} \right )} + \frac{1}{3} \ln{\left (\frac{T_{RH}}{10^{10} \, GeV} \right )} = N_{min} '
\end{equation}

This is none other than the minimum number of e-folds to solve the horizon problem. The conclusion is that by solving the horizon problem with an adequate number of e-folds of inflation taking place, we ensure that any preexisting anisotropies of the Universe vanish and do not reappear by the current cosmological era. As for when they will actually reappear, we can estimate that the time $t_{AN}$ for which $\frac{(\rho_{AN})'}{(\rho_{0})'} \gtrsim 1$ (where the primes imply their contemporary values at that particular time):

\begin{equation} \label{eq:2.52}
    t_{AN} \gtrsim e^{3(N_{tot} - N_{min} ')} 10^{10} \,yr
\end{equation}

where we have further assumed matter-domination up until $t_{AN}$. This result confirms the previous suspicion that anisotropies are exponentially suppressed and - even though unavoidable - will not appear in any significant way for a very long time.\par
Lastly, we take a look at a more general case where the initial state of the Universe may also contain inhomogeneities. Here we introduce the concept of density perturbations. That is, physical modes $\lambda$ that exist within the early Universe start microscopic and have a quantum character, but since they are physical lengths they are also subject to inflation and as a result they exponentially increase. This allows them to reach macroscopic levels and therefore take the form of observable energy density perturbations in the future. The manner in which this occurs will be presented in more detail in the next section. In this case, it will suffice to establish how these modes when grown have to cross back into the horizon at some later point which is when the density perturbations are imprinted onto the spacetime. That is because, while they grow exponentially, the horizon distance $H^{-1}$ remains constant and will only start growing following the reheating phase. Physical modes that were smaller than $H^{-1}$  will surpass it and go back in later, whereas those that were larger will only perform the second crossing back into the Hubble volume that forms the observable Universe.\par
If we assume that inhomogeneities are carried by such physical modes $\lambda$, then we need to examine these two cases and how that would impact modern observations. For $a\lambda \gg H^{-1}$ i.e. way outside the horizon, they will remain so for the entire inflationary stage as well as the following eras up until the current one. More precisely, since $a\lambda$ grows by roughly $e^{N_{tot}}$, the entry into the horizon is exponentially delayed \cite{Reference15}. No physical mode corresponding to anything presently observable could have possibly been such a mode of inhomogeneity because then it would by definition mean that the latter have at some point crossed back into $H^{-1}$. Since they started out way outside it pre-inflation, they have a massive "head start" and it will take exponentially long time for the Hubble volume to catch up to them. If they had instead started out within $H^{-1}$ and therefore could have possibly crossed back in within a timespan less or comparable to the present age of the Universe, then they would still not have any significant impact on measurements. Since they grew within the Hubble volume that transposed into dS space during inflation, and they are at a scale today that is smaller or at most as large as $H_{0}^{-1}$, that means they either never crossed the horizon in order to imprint density perturbations or they did, but their amplitudes were set as metric perturbations on dS space. In either case, the physical result in terms of observed inhomogeneity is the same in that it is very small and only on a local level. The conclusion is that any inhomogeneities that might have existed, they were either snuffed out or exponentially suppressed and their appearance in the observable Universe postponed, just like in the case of anisotropy.

\section{Density perturbations}

Another novel feature of inflationary models is that they provide a compelling explanation for the existence of small-scale inhomogeneity in the Universe. While the Universe is homogeneous on a large-scale, at the local level we can find inhomogeneities with a physical scale that is much smaller than the size of the observable Universe. These are none other than the various structures we observe: clusters, galaxies, nebulae etc. Inflation provides an answer via the quantum fluctuations of $\phi$ we introduced in (\ref{eq:2.30}) \cite{Reference15}\cite{Reference14}\cite{Reference34}. Quantum fluctuations propagate as physical modes $\lambda$ like the ones we mentioned above, and they are subject to exponential growth themselves which causes them to become macroscopic and classical. \par
In Hot Big Bang cosmology, all physical modes of all scales cross inside the horizon once: when the horizon has expanded enough to encompass them. In inflationary cosmology though there are two cases which we have also mentioned in the previous section. The first one is when they are outside the horizon $H^{-1}$ to begin with, hence they will only cross the horizon once at some remote point in the future when the Hubble volume that matches the observable Universe today grows enough to let them in. The second one involves modes that start within $H^{-1}$ and grow to take the form of metric perturbations on the dS background. These modes cross outside the horizon first during inflation due to the fact they are growing in size while $H^{-1}$ remains roughly constant. When they are outside the horizon, they are "frozen" as these macroscopic perturbations, and then at some later time in the future they make another horizon crossing. Again, that is when $H^{-1}$ has increased sufficiently and thus the physical modes enter the Hubble volume for the second time. \par
Modes whose initial length was at a value that would eventually grow to be smaller or comparable to the modern Hubble radius ($\sim 3000$ Mpc) are the so-called "modes of interest". They have managed to reenter by the present cosmological epoch and by extension have an impact on modern observations. Other ones will not have had enough time to reenter (or enter for the first time like in the first case) and are therefore impossible to measure at this point in time. For the modes of interest, the entry of the "frozen" perturbations happens in the form of energy density perturbations which are now imprinted on the spacetime metric. Such density perturbations are precisely what is postulated to be the reason behind the small-scale inhomogeneities in the Universe and the formation of the various structures listed earlier. The mechanics of structure formation due to this phenomenon is obviously much more complicated, but lies outside the scope of this text. We shall proceed to elaborate on these density perturbations and their spectrum.\par
In order to set a limit to which modes are of interest, we consider the amount of e-folds before their first horizon crossing i.e. their exit. Since modes of interest should be smaller or equal to $3000$ Mpc today, they would have exited the horizon either at the same time or some e-folds later than the Hubble radius scale which exited at least $N_{min}'$ before inflation ended. If the scale $\lambda$ crossed the horizon at $N_{\lambda}$ e-folds before inflation ended, then the ratio of the e-folds is proportional to the ratio of their respective scales today:

\begin{equation} \label{eq:2.53}
    \frac{\lambda}{3000 \,Mpc} = e^{N_{\lambda} - N_{min}'}
\end{equation}

We can then readily find the expression for $N_{\lambda}$:

\begin{equation} \label{eq:2.54}
    N_{\lambda} = N_{min}' +\ln{\left (\frac{\lambda}{3000 \, Mpc} \right )} \approx 45 + \ln{\left (\frac{\lambda}{Mpc} \right )} + \frac{2}{3} \ln{\left (\frac{M}{10^{14} \, GeV} \right )} + \frac{1}{3} \ln{\left (\frac{T_{RH}}{10^{10} \, GeV} \right )}
\end{equation}

As expected, the amount of e-folds depends on the size of the physical mode; the smaller a scale is, the smaller the amount of e-folds before inflation ended it needed to cross the horizon. We need to find the spectrum of quantum mechanical fluctuations for $\phi$ in order to see how they later evolve macroscopically. Because the potential $V(\phi)$ is chosen to be extremely flat as part of the slow roll conditions, $V''$ which grants the inflaton its effective mass is much smaller than $H^{2}$ and therefore we can take $\phi$ to be massless during the slow roll phase. The spectrum of quantum fluctuations for a massless, minimally-coupled (i.e. has no coupling to the scalar curvature $R$) scalar field is standard \cite{Reference15}:

\begin{equation} \label{eq:2.55}
    (\Delta \phi )_{\vec{k}}^{2} = \bar{\mathcal{V}}^{-1} \frac{k^{3}}{2\pi ^{2}} |\delta \phi _{\vec{k}}|^{2} = \left (\frac{H}{2\pi} \right )^{2}
\end{equation}

where $\bar{\mathcal{V}}$ is the volume of a fundamental cube, $k$ the scale in k-space ($k \sim \lambda ^{-1}$) and $\delta \phi _{\vec{k}}$ are the quantum fluctuations which we have Fourier-decomposed to get the result of (\ref{eq:2.55}):

\begin{equation} \label{eq:2.56}
    \delta \phi _{\vec{k}} = \int d^{3}x \, e^{i\vec{k}\cdot \vec{x}} \, \delta \phi (\vec{x})
\end{equation}

The spectrum is none other than the square of the Gibbons-Hawking temperature $T_{GH} = \frac{H}{2\pi}$ that arises due to the existence of event horizons in dS space. A similar expression for the temperature is also produced in the presence of black holes, again on their event horizon. The mean square spectrum on the other hand is calculated by integrating over all possible modes in k-space for the expression (\ref{eq:2.55}). Since each mode is expected to give $T_{GH}^{2}$ regardless of its size, the mean square spectrum is going to be:

\begin{equation} \label{eq:2.57}
    (\Delta \phi )^{2} = N(t) \left (\frac{H}{2\pi} \right )^{2}
\end{equation}

where $N(t) \approx Ht$ the total number of e-folds elapsed by time $t$. That is because integrating over all modes $k$ is equivalent to summing all individual mode spectra which are identical as per (\ref{eq:2.55}). The resulting amount is therefore that quantity times all the e-folds that have passed in order for those scales to contribute. When these modes exit the initial Hubble volume set by $H^{-1}$, they become macroscopic and classical which means that the fluctuations will now obey the classical equation of motion:

\begin{equation} \label{eq:2.58}
    \ddot{\delta \phi }_{\vec{k}} + 3H\dot{\delta \phi }_{\vec{k}} + \frac{k^{2}}{a^{2}} \delta \phi _{\vec{k}} = 0
\end{equation}

When physical modes are already outside the horizon or have crossed it by quite some distance ($\lambda \gg (aH)^{-1} \Rightarrow k \ll aH$), then equation (\ref{eq:2.58}) simply gives $\delta \phi _{\vec{k}} = constant$, consistent with the idea of "frozen" macroscopic fluctuations that we mentioned earlier. The corresponding energy density perturbations will in turn be linked with these fluctuations by virtue of the vacuum energy:

\begin{equation} \label{eq:2.59}
    \delta \rho _{\phi} = \frac{\partial V}{\partial \phi} \delta \phi
\end{equation}

We now introduce the dimensionless gauge-invariant quantity $\zeta$ - sometimes called the Bardeen potential - that is linked to the perturbation spectrum \cite{Reference15}. For modes outside the horizon, $\zeta = constant$ as one would expect since the perturbations that are proportional to the fluctuations have "frozen". For modes approximately at the horizon crossing ($\lambda \approx (aH)^{-1}$), which is where the density perturbations shall be imprinted on the spacetime:

\begin{equation} \label{eq:2.60}
    \zeta = \frac{\delta \rho}{\rho +P}
\end{equation}

To find the spectrum as it is imprinted, we make use of the value of $\zeta$ at the first horizon crossing. There the Universe is still vacuum dominated and therefore $\rho +P \approx \rho_{\phi} + P_{\phi} = \dot{\phi} ^{2}$ and the only fluctuation is that of the inflaton energy density, in turn due to the fluctuation of $\phi$ (\ref{eq:2.55}): 

\begin{equation} \label{eq:2.61}
    \zeta _{N_{\lambda}} = \left (\frac{\delta \rho _{\phi}}{\dot{\phi}^{2}} \right )_{N_{\lambda}} = \left (\frac{\delta \phi V'}{\dot{\phi}^{2}} \right )_{N_{\lambda}} \approx \left (\frac{H^{2}}{\dot{\phi}} \right )_{N_{\lambda}}
\end{equation}

where the subscript $N_{\lambda}$ implies that we are calculating all these quantities at that amount of e-folds before the horizon crossing, setting the proper values which depend on the $\lambda$ at hand. Since beyond the horizon crossing $\zeta$ remains constant, (\ref{eq:2.61}) will match with the $\zeta$ at the second horizon crossing where the modes reenter the Hubble volume. At that point the Universe is either radiation or matter dominated and therefore:

\begin{equation} \label{eq:2.62}
    \zeta _{HOR} = \left (\frac{\delta \rho}{\rho +P} \right )_{HOR} =
\begin{cases}
&\left (\frac{\delta \rho}{\rho} \right )_{HOR} \, , \, \, (MD) \\
\frac{3}{4}&\left (\frac{\delta \rho}{\rho} \right )_{HOR} \, , \, \, (RD)
\end{cases}
\end{equation}

Considering that both $H$ and $\dot{\phi}$ change very slowly during the slow roll phase, the difference between the spectra of different modes $\lambda$ is actually minuscule. Furthermore, modes of interest are relatively "adjacent" to each other in the sense that they are spanning a small range of values that would leave even less room for variation in their spectrum amplitudes. This adjacency is quantified by the number of e-folds $N_{\lambda}$, which is only logarithmically dependent on $\lambda$. The resulting spectrum for the modes of interest is to a very good approximation scale-invariant i.e. more or less the same for all physical scales we can observe. That is called the Harrison-Zeldovich spectrum, and it is one of the definitive tools we can use as evidence for the existence of an inflationary stage. \par
A different way to view this is simply going back to the expression (\ref{eq:2.55}) and examining the correlation functions between fluctuations of different physical scales \cite{Reference34}:

\begin{equation} \label{eq:2.63}
   |\delta \phi _{\vec{k}}|^{2} \equiv \langle \delta \phi _{\vec{k}} \delta \phi _{\vec{k}'} \rangle = P_{\phi} (k) \, \delta ^{3}(\vec{k} + \vec{k}')  
\end{equation}

where $P_{\phi}(k)$ is taken to be a real positive function of the form:

\begin{equation} \label{eq:2.64}
    P_{\phi} = N_{\phi}^{2} k^{n-4}
\end{equation}

with $N_{\phi}$ being some constant. The delta function arises so that the correlation function $\langle \delta \phi (\vec{x}) \delta \phi (\vec{y}) \rangle$ depends only on $(\vec{x} - \vec{y})$. For $n=1$ which leads to $P_{\phi} \propto k^{-3}$, (\ref{eq:2.63}) gives a correlation function that is dependent on $k^{-3}$. When inserted in (\ref{eq:2.55}), we get the appropriate scale-invariant spectrum. In other words, the observed perturbation spectrum is the same for all physical scales. Constant curvature fluctuations however will indeed have a variant amplitude depending on the model at hand. This means that $n$ in reality takes a slightly different value to 1. In all cases, the desired amplitude  must range between $10^{-5}$-$10^{-4}$, matching the fluctuations observed in the CMBR. \par 
As a final note, we mention that the aforementioned density perturbation spectrum - depending on the cosmological era the mode has crossed the horizon for a second time - can also be a source of particle production that matches necessary requirements (massless/very light and minimally-coupled) \cite{Reference15}\cite{Reference14}. Examples of particular interest are gravitons and photons. The former provide us with an estimate on the value of the Hubble parameter and the vacuum energy at the time the inflation began, while the latter serve as an interesting possible explanation for the primeval magnetic fields we measure in the present.

\pagebreak % Background Theory 

\fancyhead[RE,LO]{Starobinsky model and Scale Invariant Gravity}

\chapter{Starobinsky model and Conformal Gravity}

Most inflationary models, while fulfilling the necessary criteria for successful inflation, have the issue of either leading to inconsistencies in some of their predictions or have an issue regarding how they can get the proper initial conditions to kickstart it. The Starobinsky model is one of the few that (at least thus far) remains fully consistent with observation and has the distinct advantage in that it can be applied to various different types of theories of inflation. Because of that, the Starobinsky model constitutes one of the strongest contenders for explaining the mechanics of inflation. The model achieves is rather modest in that it makes no assumptions about a grand unified theory (GUT) in which the cosmological evolution is tied with particle physics. Rather, it introduces a modification to the classical description of gravity. Classical gravity is described by the Einstein-Hilbert action \cite{Reference5}:

\begin{equation} \label{eq:3.1}
    S_{EH} = \kappa ^{2} \int d^{4}x \, \sqrt{g} \, (R - 2\Lambda ) 
\end{equation}

where $\kappa ^{2} = \frac{1}{16\pi G} = \frac{m_{P}^{2}}{16\pi}$, while we have also assumed the existence of a cosmological constant $\Lambda$. If we consider a matter component in the Lagrangian of the theory:

\begin{equation} \label{eq:3.2}
    S_{M} = \int d^{4}x \,\sqrt{g} \, \mathcal{L} _{M}
\end{equation}

so that the total action is $S = S_{EH} + S_{M}$. Applying the principle of least action, we get the Einstein field equations (\ref{eq:1.9}). The sign choice for the gravitational action is arbitrary, since in the absence of matter both a positive and a negative action provide the same field equations. Once we consider matter, this freedom remains, but the sign of the kinetic terms will have to change accordingly in order to be opposite. In this text we take the gravitational actions to be positive, hence any added kinetic matter term should have a negative sign. \par
In the Starobinsky model, the correction in the Einstein-Hilbert action is a $R^{2}$ term. For simplicity's sake, we shall assume an empty space (devoid of matter) in the absence of a cosmological constant; both of these assumptions will turn out to be rather apt. The former is to a great extent correct, as all forms of matter become extremely diluted during inflation, like we explained in chapter 2. The latter will also turn out correct, as the correction term will be the driving force behind inflation instead. Then the gravitational action can be written as \cite{Reference28}\cite{Reference29}:

\begin{equation} \label{eq:3.3}
    S_{S} = \int d^{4}x\, \sqrt{g} \, (\lambda R^{2} + \kappa ^{2}R)
\end{equation}

where $\lambda$ - not to be confused with the physical scales of the last chapter - is a dimensionless coupling constant. This new term is by no means conspicuous, since the effects on the field equations are significant. In order for this correction to work, we need to make sure that - at least in the present - the additive term does not contribute to the action significantly and therefore the gravitational action becomes effectively just the Einstein-Hilbert one, leading to an effective GR theory. This will be the subject of the next sections. By contrast, it is the $R$ term that becomes negligible in the early stages of the Universe, as to allow the $R^{2}$ to dominate and drive inflation. Hence we can approximate the Starobinsky model for the cosmological era of interest (i.e. the inflationary epoch) as a theory of quadratic gravity:

\begin{equation} \label{eq:3.4}
    S_{G} = \lambda \int d^{4}x\, \sqrt{g} \, R^{2}
\end{equation}

also known as scale invariant gravity, a feature of which we shall examine later in the chapter. This is a theory of great interest in its own right, and owes this to more than just the fact it is scale invariant. The fact its coupling constant is dimensionless allows for this term to be renormalizable (in the conventional dimensional regularization scheme), unlike regular GR. Moreover, it is a theory that - as we shall see later - is devoid of ghost fields. Ghost fields or just ghosts are field modes from the solutions of the theory that have an "incorrect" sign for the kinetic energy term. If those modes propagate this is a major pathology, therefore $R^{2}$ gravity not containing any is a great asset. Treating the correction as quantum mechanical, a spontaneous symmetry breaking (SSB) may produce the conventional Einstein gravity, and the mechanism that initiates inflation in the early stages where it dominates. In addition, the model can be embedded in supergravity and string theory. All these make both the Starobinsky model and $R^{2}$ very promising and exciting prospects \cite{Reference2}\cite{Reference3}\cite{Reference8}. 

\section{General Weyl theory and $R^{2}$ gravity}

The pure $R^{2}$ theory sits at the crossroads of two different types of generalized theories of gravity. It is one of the simplest of the $f(R)$ theories which involve any kind of function of the scalar curvature in the definition of the gravitational action, and it is also the simplest form of quadratic gravity that may also involves quadratic terms of tensors relating to the curvature of the manifold. The former's definition can be expanded to a tensorial form so that it includes the latter, but generally they are treated distinctly. Starting from the latter, we note that the most general form for quadratic gravity has the form \cite{Reference2}:

\begin{equation} \label{eq:3.5}
     S_{W} = \int d^{4}x\, \sqrt{g} \, (c_{1}C_{\mu \nu \rho \sigma }C^{\mu \nu \rho \sigma } + c_{2}R^{2} + c_{3}\Hat{R} _{\mu \nu }\Hat{R}^{\mu \nu})
\end{equation}

where we have defined the Einstein tensor $\Hat{R} _{\mu \nu } = R_{\mu \nu } - \frac{1}{4} g_{\mu \nu }R$, $C_{\mu \nu \rho \sigma }$ is the Weyl tensor and $c_{1}$, $c_{2}$ and $c_{3}$ are constants. The Weyl tensor is constructed from the Riemann curvature tensor in such a way as to be traceless \cite{Reference5}\cite{Reference13}:

\begin{equation} \label{eq:3.6}
     C_{\mu \nu \rho \sigma } = R_{\mu \nu \rho \sigma } + \frac{1}{2} (g_{\mu \sigma }R_{\rho \nu } + g_{\nu \rho }R_{\sigma \mu } - g_{\mu \rho }R_{\sigma \nu } - g_{\nu \sigma }R_{\rho \mu }) + \frac{1}{6} (g_{\mu \rho }g_{\sigma \nu } - g_{\mu \sigma }g_{\rho \nu })R 
\end{equation}

Theories that involve this tensor are called Weyl theories, which is where the "general Weyl theory" name comes from for this action. They are also "Conformal Gravity" theories which at the simplest form may include a quadratic term of the Weyl tensor. Both of these names are in reference to their conformal or Weyl invariance. While we have talked about conformal transformations before, we have not defined them concretely or explained their significance. A conformal or Weyl transformation of the metric is defined as:

\begin{equation} \label{eq:3.7}
    g_{\mu \nu} \rightarrow g'_{\mu \nu} = e^{2\Omega (x)} g_{\mu \nu}
\end{equation}

where $\Omega (x)$ is some arbitrary function of the spacetime coordinates. Such a transformation has the effect of changing the scale of the metric, but retaining local angles and orientation. For example, when conformally transforming the dS space in chapter 3 in order to map it onto a Penrose diagram, we have obviously compromised the correct size, but angles were preserved as to correctly showcase causal relationships within it. It is also a more powerful symmetry than the global scaling symmetry which pure $R^{2}$ gravity possesses. The reason is that the conformal group $SO(2,4)$ which characterizes general conformal/Weyl transformations includes not just the scaling (also known as dilatation) transformation, but also Poincaré transformations (Lorentz group $SO(1,3)$ plus spacetime translations) and special conformal transformations. This group is extremely inclusive, which in turn makes the constructing of theories that follow this symmetry much easier due to the stringent constraints. A global scaling transformation of the metric is very similar to the Weyl one:

\begin{equation} \label{eq:3.8}
    g_{\mu \nu} \rightarrow g'_{\mu \nu} = e^{2\Omega } g_{\mu \nu}
\end{equation}

with the crucial difference that $\Omega$ is now a constant, not a local function. Equivalently, we could keep the metric intact, but rescale the coordinates: 

\begin{equation} \label{eq:3.9}
    x^{\mu} \rightarrow x'^{\mu} = \epsilon \, x^{\mu}
\end{equation}

where again $\epsilon$ is some constant. Such a transformation increases the size of the metric, but it only preserve angles and orientation if it is uniform as it is defined in (\ref{eq:3.8}) and (\ref{eq:3.9}). A simple example of uniform scaling is enlarging the image of a photograph with zoom, while non-uniform would be the case of enlarging it without maintaining the original aspect ratio i.e. an anisotropic rescaling. Such a change would distort the image and not preserve local angles. We would then not be able to construct a conformal map between the original image and the result. \par
This shows how Weyl transformations are a more powerful symmetry to possess, but scale invariance is also potent when discussing inflation. Since inflation quickly leads to a dS state with global isotropy and homogeneity, scaling and Weyl transformations of the metric behave very similarly for any initial inflationary patch of size $\mathcal{O}(H^{-1})$. The only exceptions are the metric's dependence on time via the scale factor, and considerations of global inhomogeneity and anisotropy. The latter is not very concerning, since those inflationary patches will end up losing causal contact, therefore gravitational effects in each of those regions will not be distorted. The former is also not a big problem, since inflation for each such region lasts for a very short period of time. \par
Coming back to action (\ref{eq:3.5}), we may rewrite the first term differently so that the regular curvature tensors appear explicitly:

\begin{equation} \label{eq:3.10}
    C_{\mu \nu \rho \sigma }C^{\mu \nu \rho \sigma } = R_{\mu \nu \rho \sigma }R^{\mu \nu \rho \sigma } - 2 \Hat{R} _{\mu \nu }\Hat{R} ^{\mu \nu } - \frac{1}{6} R^{2} = GB + 2\Hat{R} _{\mu \nu }\Hat{R} ^{\mu \nu } - \frac{1}{6} R^{2}
\end{equation}

where GB is the topological Gauss-Bonnet term. This term is provided by the generalized Gauss-Bonnet theorem, with its expression in four dimensions being \cite{Reference2}:

\begin{equation} \label{eq:3.11}
    GB = R_{\mu \nu \rho \sigma }R^{\mu \nu \rho \sigma } - 4 R_{\mu \nu }R^{\mu \nu } + R^{2}
\end{equation}

This term is topological in the sense that it is strictly independent of the manifold's size, curvature, shape etc. This is generally expressed by:

\begin{equation} \label{eq:3.12}
    \int d^{4}x \, \sqrt{g} \, GB = 32\pi ^{2} \chi (M)
\end{equation}

$\chi (M)$ is the Euler characteristic of the manifold $M$, in turn related to the genus $g$ - the number of "holes" on the manifold. In addition, this term is described as topological in the sense that it does not ultimately contribute to the field equations. This means that this term is somewhat irrelevant in the classical consideration of gravity. In a semi-classical context though, it may provide a non-trivial contribution to the path integral.  We can thereby restate the action as:

\begin{equation} \label{eq:3.13}
    S_{W} = \int d^{4}x\, \sqrt{g} \, \left [a \left (R_{\mu \nu }R^{\mu \nu } - \frac{1}{3} R^{2} \right ) + bR^{2} + c\, GB \right ]
\end{equation}

with $a = 2c_{1} + c_{3}$, $b = c_{2} + \frac{c_{3}}{12}$ and $c = c_{1}$. The field equations for this action can be expressed via the definition of some new tensorial quantities \cite{Reference2}:

\begin{equation} \label{eq:3.14}
    W_{\mu \nu} \equiv bJ_{\mu \nu} + aB_{\mu \nu} = 0
\end{equation}

where $J_{\mu \nu} =  RR_{\mu \nu } + g_{\mu \nu }\nabla ^{2}R - \nabla _{\mu }\nabla _{\nu }R - \frac{1}{4} g_{\mu \nu }R^{2}$ and $B_{\mu \nu}$ is the Bach tensor:

\begin{equation} \label{eq:3.15}
    B_{\mu \nu} = \left (\nabla ^{\rho} \nabla ^{\sigma} +\frac{1}{2} R^{\rho \sigma} \right )\, C_{\mu \nu \rho \sigma}
\end{equation}

We remind that the operators $\nabla _{\mu }$ and $\nabla ^{2} = \nabla _{\mu } \nabla ^{\mu}$ refer to the covariant derivative. \par
Conversely, the solutions for the $f(R)$ family of theories are somewhat simpler to derive. From the gravitational action $S_{f(R)} = \int d^{4}x \, \sqrt{g} \, f(R)$, the resulting field equations in this case are \cite{Reference8}\cite{Reference4}:

\begin{equation} \label{eq:3.16}
     f'(R)R_{\mu \nu } + g_{\mu \nu }\nabla ^{2}f'(R) - \nabla _{\mu }\nabla _{\nu }f'(R) - \frac{1}{2} g_{\mu \nu }f(R) = 0
\end{equation}

The primed quantities imply derivatives with respect to $R$. While we could derive the analogous expression for $R^{2}$ gravity from (\ref{eq:3.4}) alone, we can use the general result above by setting $f(R) = R^{2}$ to get the equations:

\begin{equation} \label{eq:3.17}
    RR_{\mu \nu } + g_{\mu \nu }\nabla ^{2}R - \nabla _{\mu }\nabla _{\nu }R - \frac{1}{4} g_{\mu \nu }R^{2} = 0
\end{equation}

Notice how (\ref{eq:3.17}) is none other than the tensor quantity $J_{\mu \nu}$ of (\ref{eq:3.14}). Indeed, if we had instead used the general solutions for the quadratic gravity theory and set $a = 0$ so that the only (classically) relevant term of the action was $bR^{2}$, then the solution $W_{\mu \nu} = J_{\mu \nu} = 0$ of the previous gravitational action would be fully consistent with the result above. In both equations (\ref{eq:3.14}) and (\ref{eq:3.17}), there are two types of solutions that we are interested in:\par
\textbf{1)} $\bm{R_{\mu \nu} =0}$ \\
These are the Ricci flat solutions. This trivially includes solutions of completely flat spacetimes where the curvature tensor is also 0.  Examples of such spaces are the Minkowski space, the Schwarzschild solution that describes black holes, and gravitational waves. \par
\textbf{2)} $\bm{R_{\mu \nu} = \beta \, g_{\mu \nu}}$ \\
These are called Einstein spaces and for some value of the constant $\beta$, they in turn give a constant scalar curvature $R = 4\beta$. These solutions correspond to dS and AdS (anti-de Sitter) spacetime solutions, depending on the sign of $\beta$: dS for $\beta >0$ which is positively curved, and AdS for $\beta <0$ which is negatively curved. The dS solution will obviously correspond to the inflationary stage we desire to get, which is partially why $R^{2}$ is said to drive inflation. The fact the curvature in either case is constant is a manifestation of the scale invariance of the theory: no matter how much the metric grows, the curvature is globally maintained. \par

\section{Transformation to Einstein frame}

The discussion thus far has been rather general. Now we shall shed some light on why the Starobinsky model, $R^{2}$ gravity or general higher curvature gravitational actions are viable inflationary models. In order to do this, we will transition to the Einstein frame. We shall transform the gravitational action to produce regular the Einstein theory plus some extra terms that will be responsible for inflation. First, let us consider pure $R^{2}$ gravity. We start by introducing a parameter $\Phi$ \cite{Reference2}\cite{Reference3}:

\begin{equation} \label{eq:3.18}
    S_{G} = \int d^{4}x \, \sqrt{g} \, \left (R\Phi - \frac{1}{4\lambda } \Phi ^{2} \right )
\end{equation}

From this action we get the equation of motion for $\Phi$:

\begin{equation} \label{eq:3.19}
    \Phi = 2R\lambda
\end{equation}

where $R$ implies any of the metric solutions of the theory. When inserted back into (\ref{eq:3.18}) and minimizing the action, we get the field equations we derived earlier. We are allowed to do that because the equation of motion for this field does not contain time derivatives, thus $\Phi$ does not propagate. It is not a physical mode solution, but instead acts as a constraint on the field equations, which means it's a Lagrange multiplier. If we further assume that $\lambda$ and $R$ are positive (both of which comply with expanding dS solutions), we may redefine $\Phi$:

\begin{equation} \label{eq:3.20}
    \Phi = \frac{M_{P}^{2}}{2} e^{\alpha \phi } \Rightarrow \, \phi = \frac{1}{\alpha } \ln{(4M_{P}^{-2} R\lambda )}
\end{equation}

where $\alpha$ is some constant, $M_{P} = \frac{m_{P}}{\sqrt{8\pi}}$ the reduced Planck mass which provides $\Phi$ with the correct dimensions, and $\phi$ the redefined scalar field. It is clear that $\Phi$ has units of mass squared, as it has the same dimensions with the scalar curvature. The dimensions of $\alpha$ on the other hand depend on the definition of $\phi$. For example, in the case of a scalar matter field where the $\phi$ has dimensions of mass, $\alpha$ should have dimensions of length. We then perform the conformal transformation \cite{Reference2}\cite{Reference3}:

\begin{equation} \label{eq:3.21}
    g_{\mu \nu } \rightarrow \tilde g_{\mu\nu}= e^{\alpha \phi }g_{\mu \nu } \Rightarrow
    \begin{cases}
      &\sqrt{\tilde{g} } = e^{2\alpha \phi }\sqrt{g} \\
    &\widetilde{R} = e^{-\alpha \phi }(R - 3\alpha \square \phi - g^{\mu \nu }\partial _{\mu }\phi \partial _{\nu }\phi )
    \end{cases}
\end{equation}

In order to see how this action will shape up, we recall a useful expression following a conformal transformation of the metric \cite{Reference3}:

\begin{equation} \label{eq:3.22}
    \int d^{d}x \, \sqrt{g} \, Y \,R = \int d^{d}x\, \sqrt{\tilde{g} } \, \left [\widetilde{R} - \frac{(d-1)}{(d-2)} \frac{\partial _{\mu }Y\partial ^{\mu }Y}{Y^{\frac{d}{d-2} }} \right ]
\end{equation}

where $Y$ is a scaling function and $d$ the number of dimensions. Here we have $Y = e^{\alpha \phi}$. Choosing $\alpha = \sqrt{\frac{2}{3}} M_{P}^{-1}$ for the sake of convenience, we get the final expression for the $R^{2}$ action in the Einstein frame:

\begin{equation} \label{eq:3.23}
    S_{G} = \int d^{4}x \, \sqrt{\tilde{g} } \, \frac{M_{P}^{2}}{2} \left (\widetilde{R} - \frac{M_{P}^{2}}{8\lambda } - M_{P}^{-2} \, \tilde{g} ^{\mu \nu }\partial  _{\mu }\phi \partial  _{\nu } \phi \right )
\end{equation}

The resulting action is fully equivalent with the Einstein-Hilbert action (\ref{eq:3.1}) where the coupling constant of the theory plays the role of the cosmological constant ($\Lambda = \frac{M_{P}^{2}}{16\lambda}$). The action also includes the kinetic term of the $\phi$ field and can therefore be seen as the matter component of the theory that only includes a massless scalar field. Note that the positive sign of $\lambda$ as we have assumed above ensures that the cosmological constant term has the correct sign. The field equations are derived as:

\begin{equation} \label{eq:3.24}
    \widetilde{R} _{\mu \nu } - \frac{1}{2} \tilde{g} _{\mu \nu }\widetilde{R} + \frac{M_{P}^{2}}{16\lambda } \tilde{g} _{\mu \nu } = 8\pi G \left (\partial  _{\mu }\phi \partial  _{\nu }\phi - \frac{1}{2} \tilde{g} _{\mu \nu }\partial  ^{\sigma }\phi \partial  _{\sigma }\phi \right )
\end{equation}

Before we comment on the possible solutions, it is interesting to see what becomes of the scale invariance of the theory. Understandably, this switch to the Einstein frame must not eliminate the symmetry, but regular Einstein gravity cannot possibly be scale invariant. Therefore it is expected that the symmetry transforms into a different kind. Indeed, considering the way the new metric has been defined by (\ref{eq:3.21}), a scaling transformation in the original frame would be simply equivalent to shifting the scalar field in the new one:

\begin{equation} \label{eq:3.25}
    \alpha \phi \rightarrow \alpha \bar{\phi} = \alpha \phi + 2\Omega \equiv \alpha \phi + \sigma
\end{equation}

This new field shifting symmetry imposes the constraint that there can be no potential dependent on $\phi$, as that would mean that shifting the field by the arbitrary quantity $\sigma$ would not leave the total action invariant. These are consistent with the mechanics of inflation we have already discussed. More specifically, the slow roll conditions impose a very flat, barely changing potential for the duration of inflation. The inflaton field will roll on the long plateau, and its flatness leads to $\phi$ being effectively massless in that region. Now we shall examine the four different possible cases for the solutions of (\ref{eq:3.24}): \par
\textbf{1)} $\bm{R>0 \, , \, \lambda>0}$ \par
This is the solution we have assumed from the beginning with positive $\Phi$. The signature of the metric remains unchanged, we get a positive cosmological constant, and a constant positive curvature. In other words, a dS solution. The kinetic term of the scalar field is also positive, allowing for a ghost-free solution. As such, the theory is not pathological, and gravity functions as an attractive force.\par
\textbf{2)} $\bm{R<0 \, , \, \lambda<0}$ \par
With both signs flipped, $\Phi$ is still positive. Again no ghosts appear due to the kinetic term also being positive. However, the negative curvature gives rise to an AdS space solution, and we end up with a negative cosmological constant. Even though gravity still emerges as an attractive force, the solution does not match with the inflationary. \par
\textbf{3)} $\bm{R<0 \, , \, \lambda>0}$\par
We once again arrive at an AdS solution, but the opposite signs of $R$ and $\lambda$ turn $\Phi$ negative, forcing us to define it differently in order for the conformal transformation (\ref{eq:3.21}) to work and leave the theory intact. As a result, the eventual gravitational term and the kinetic term of $\phi$ are both negative. The theory contains ghosts and gravity becomes a repulsive force instead.\par
\textbf{4)} $\bm{R>0 \, , \, \lambda<0}$\par
The exact same result as 3) happens, with the only difference being that the background solution now corresponds to a dS space.\par
In order to showcase how the entirety of inflation is mapped in this frame for this model and to make sure the mechanics are consistent outside of the slow roll approximation, we shall do the same for the complete Starobinsky model action  (\ref{eq:3.3}). We once again begin by removing the quadratic terms via the introduction of a Lagrange multiplier \cite{Reference3}:

\begin{equation} \label{eq:3.26}
    S_{S} = \int d^{4}x\, \sqrt{g } \, \left (\kappa ^{2}R + R\Phi - \frac{1}{4\lambda } \Phi ^{2} \right )
\end{equation}

where it once again obeys the equation of motion (\ref{eq:3.19}). Because of the additional linear $R$ term, the redefinition of $\Phi$ changes slightly:

\begin{equation} \label{eq:3.27}
    \Phi = \frac{M_{P}^{2}}{2} e^{\alpha \phi} - \kappa ^{2}  \Rightarrow \, e^{\alpha \phi} =  1 + \kappa ^{-2} \Phi
\end{equation}

In the absence of the extra term and knowing that $\kappa ^{2} = \frac{M_{P}^{2}}{2}$, we immediately get (\ref{eq:3.20}) like before. By performing the exact same conformal transformation (\ref{eq:3.21}), redefining the volume element and scalar curvature as (\ref{eq:3.22}), and carefully calculating the action, we end up with:

\begin{equation} \label{eq:3.28}
    S_{G} = \int d^{4}x\, \sqrt{\tilde{g} } \, \kappa ^{2} \left [\widetilde{R} - \frac{3\alpha ^{2}}{4} \partial _{\mu }\phi \partial ^{\mu }\phi - \frac{M_{P}^{2}}{8\lambda } (1 - e^{-\alpha \phi })^{2} \right ]
\end{equation}

We have once again produced regular Einstein gravity in the presence of a scalar field $\phi$, but the crucial difference is that the final term is no longer simply a constant that can be matched to the $\Lambda$ of General Relativity. Rather, it is a function of the scalar field that acts as a positive-definite potential:

\begin{equation} \label{eq:3.29}
    V(\phi) = \frac{M_{P}^{2} \, \kappa ^{2}}{8\lambda } (1 - e^{-\alpha \phi })^{2} = \frac{M_{P}^{4}}{16\lambda } (1 - e^{-\alpha \phi })^{2}
\end{equation}

This instead corresponds to the vacuum energy potential that dominates the Universe early on, and does in fact possess the appropriate properties. Namely, for large values of $\phi$ it is extremely close to the value $\frac{M_{P}^{4}}{16\lambda}$ that (when accounting for the factor of $M_{P}^{2}$ which is factored out in the Einstein-Hilbert action) acts as a cosmological constant in $R^{2}$ gravity. It remains flat for a large range of values until it slowly starts to decrease, approaching its minimum at $\phi = 0$ where the potential vanishes. About that point we expect to have the rapid oscillations and reheating before an exponential increase. We then transition to the later eras governed by Hot Big Bang cosmology. The only difference from the example of a potential we showed in \ref{fig:2.6} is that the inflaton field decreases rather than increases as inflation unfolds and the minimum is in fact located at $\phi = 0$ rather than some later value $\sigma$. Neither of these are a cause for concern though; as long as the initial value of $\phi$ is remote enough from the minimum and the slow roll conditions are met, inflation will follow just as we presented it. If we check analytically:

\begin{equation} \label{eq:3.30}
\begin{split}
    &V'(\phi) = e^{-\alpha \phi} \, \frac{\alpha M_{P}^{4}}{8\lambda } (1 - e^{-\alpha \phi }) \\
    &V''(\phi) = -e^{-\alpha \phi} \, \frac{\alpha ^{2} M_{P}^{4}}{8\lambda } (1 - 2e^{-\alpha \phi })
\end{split}
\end{equation}

Both end up being exponentially small for large positive values of $\phi$, comfortably fulfilling the slow roll conditions. Then it is only a matter of finding which initial value $\phi_{i}$ is correct to get the necessary amount of e-folds to solve various cosmological problems and match observational data. The exact form of the effective potential in the Starobinsky model can be seen below.

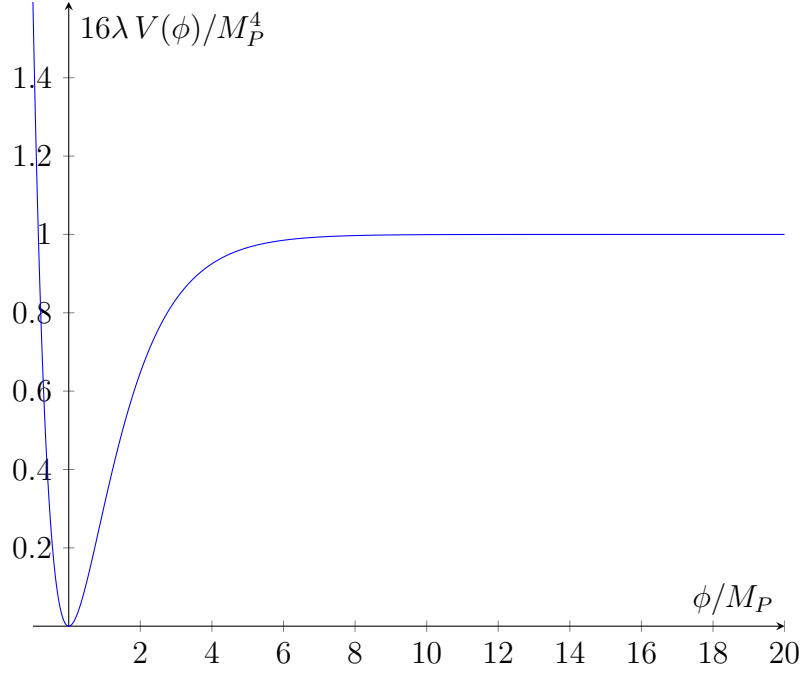
\begin{figure}[H]
\centering
\begin{tikzpicture}
\begin{axis}[
    xlabel = {$\phi / M_{P}$},
    ylabel = {$16\lambda \, V(\phi) / M_{P}^{4}$},
    axis lines = middle,
    scale = 1.45
    ]
\addplot [
    domain=-1:20, 
    samples=1000, 
    color=blue,
    ]
    {(1-e^((-sqrt(2/3))*x))^2};
\end{axis}
\end{tikzpicture}
\caption{The Starobinsky effective potential for $\alpha = \sqrt{\frac{2}{3}} M_{P}^{-1}$. For nearly all positive values of $\phi$ it is extremely flat, forming a plateau about the value $\sim M_{P}^{4}$. The field decreases during slow roll and at some point exits inflation. After that, the roll becomes faster and the inflatons eventually fall to the minimum $V(0) = 0$ where the reheating phase ensues. The steepness of the potential thereafter effectively prevents the inflatons from ever escaping the vicinity of the minimum classically. This ensures the cosmological constant today is vanishingly small (albeit not 0 as the potential predicts).}
\label{fig:3.1}
\end{figure}

\section{Metric perturbations around spacetime solutions}

In order to study the propagating physical modes of the theories presented, we must examine the spectrum produced by perturbing their metric solutions. We shall consider the two cases of interest to us: a Minkowski solution, and a dS solution. The latter matches the background during inflation, while the former for all cosmological periods thereafter. We will then perturb each metric by a small tensorial quantity $h_{\mu \nu}$. In this analysis, we shall only retain terms up to second order of this perturbation. For the sake of simplicity, we restrict the discussion to the case of the Starobinsky model, but we shall also see how the results apply for pure $R^{2}$ gravity when $\kappa ^{2} \rightarrow 0$. \par

\subsection{Flat space solutions}

Firstly, we introduce the perturbation around the Minkowski metric:

\begin{equation} \label{eq:3.31}
    g_{\mu \nu } = \eta _{\mu \nu } + h_{\mu \nu }
\end{equation}

It is then straightforward to calculate the induced shift on the rest of the quantities of the Starobinsky action, for which get \cite{Reference2}:

\begin{equation} \label{eq:3.32}
    S_{0} \approx \int d^{4}x \, \sqrt{g} \,  \left [\frac{\kappa ^{2}}{2} \left (\frac{1}{2} h^{\mu \nu }\square h_{\mu \nu } - \frac{1}{2} \varphi \square \varphi - \nabla_{\mu }h^{\mu \nu }\nabla_{\nu }\varphi + \nabla_{\mu }h^{\mu \nu }\nabla^{\sigma }h_{\sigma \nu } \right ) + \lambda (\partial ^{\rho }\partial ^{\mu }h_{\rho \mu } - \square \varphi )^{2} \right ]
\end{equation}

where we have defined $\varphi = h^{\mu}{}_{\mu}$. Here it is implied that the covariant derivative involves just the unperturbed part of the metric for terms where the inclusion of the perturbation would go beyond the second order limit we set. It is important to mention, though, that for the action we may keep up to third order instead, as one power of $h_{\mu \nu}$ is lost by minimizing the action to get the field equations. Next we apply a tensor decomposition on the metric perturbation using the fact it is symmetric by definition \cite{Reference2}:

\begin{equation} \label{eq:3.33}
    h_{\mu \nu } = h^{\perp }_{\mu \nu } + \partial _{\mu }\alpha ^{\perp }_{\nu } + \partial _{\nu }\alpha ^{\perp }_{\mu } + \left (\partial _{\mu }\partial _{\nu } - \frac{1}{4} \eta _{\mu \nu } \square \right )\alpha + \frac{1}{4} \eta _{\mu \nu } \varphi  
\end{equation}

with the newly introduced components satisfying the following conditions:

\begin{equation} \label{eq:3.34}
    \partial ^{\mu }h^{\perp }_{\mu \nu } = 0, \quad \quad \eta ^{\mu \nu }h^{\perp }_{\mu \nu } = 0, \quad \quad \partial ^{\mu }\alpha ^{\perp }_{\mu } = 0
\end{equation}

These quantities - as the way they are written suggests - are transverse i.e. perpendicular to the propagation of the physical fields that will appear, while $h^{\perp }_{\mu \nu }$ is also traceless. Next, we impose a set of infinitesimal diffeomorphisms:

\begin{equation} \label{eq:3.35}
    x^{\mu } \rightarrow x^{\mu } + \xi ^{\mu }(x)
\end{equation}

This corresponds to just a local change of spacetime coordinates. The resulting change in the metric perturbation is:

\begin{equation} \label{eq:3.36}
     h_{\mu \nu } \rightarrow h_{\mu \nu } + \partial _{\mu }\xi _{\nu } + \partial _{\nu }\xi _{\mu }
\end{equation}

The diffeomorphisms can also be decomposed into a transverse and a longitudinal  part:

\begin{equation} \label{eq:3.37}
    \xi _{\mu } = \xi ^{\perp }_{\mu } + \partial _{\mu }\xi \Rightarrow 
    \begin{cases}
     &\partial ^{\mu }\xi ^{\perp }_{\mu } = 0 \\
    &\partial ^{\mu }\xi _{\mu } = \square \xi
    \end{cases}
\end{equation}

and thus the quantities that emerged from the tensor decomposition (\ref{eq:3.34}) take the form:

\begin{equation} \label{eq:3.38}
    h^{\perp }_{\mu \nu } \rightarrow h^{\perp }_{\mu \nu }, \quad \quad \alpha ^{\perp }_{\mu } \rightarrow \alpha ^{\perp }_{\mu } + \xi ^{\perp }_{\mu }, \quad \quad \alpha \rightarrow \alpha + 2\xi , \quad \quad \varphi \rightarrow \varphi + 2\square \xi 
\end{equation}

As expected, the transverse component of the metric perturbation remains unchanged, since the diffeomorphism for that component is a gauge symmetry transformation. The vector $\alpha ^{\perp }_{\mu }$ does shift, but since the additive term is just another transverse element, it does not lead to any physical repercussions. The latter two of the changes are significant, therefore we wish to rearrange them in a way in order to make the gauge invariance manifest. We do so by introducing the scalar field:

\begin{equation} \label{eq:3.39}
    \Phi = \varphi - \square \alpha
\end{equation}

not to be confused with the Lagrange multiplier we saw in the previous section of the chapter. It's easy to see that this new scalar field is indeed invariant under the diffeomorphisms of (\ref{eq:3.36}). By extension, this gives rise to the field equation of motion \cite{Reference2}:

\begin{equation} \label{eq:3.40}
    \partial ^{\mu} \partial ^{\nu} h_{\mu \nu} = -\frac{3}{4} \square \Phi
\end{equation}

There physical meaning of the field $\Phi$ is now clear: it is the physical mode associated with the longitudinal component of the perturbation. The action eventually is expressed as \cite{Reference2}:

\begin{equation} \label{eq:3.41}
    S_{0} = \int d^{4}x\, \sqrt{g} \,  \left [\frac{\kappa ^{2}}{4} (h^{\perp }_{\mu \nu }\square h^{\perp \mu \nu } + \partial _{\mu }\Phi \partial ^{\mu }\Phi ) + \frac{9\lambda}{16} (\square \Phi )^{2} \right ]
\end{equation}

Since this action simply consists of the kinetic terms for tensors and fields, we may Fourier-decompose both and quantize them to get their QFT propagators. We will not show how to perform this in detail, but the results in momentum space are \cite{Reference2}:

\begin{equation} \label{eq:3.42}
    \Delta ^{(2)}_{\mu \nu ,\rho \sigma } = -\frac{2}{\kappa ^{2}} \frac{1}{q^{2}} P^{(2)}_{\mu \nu ,\rho \sigma }
\end{equation}

\begin{equation} \label{eq:3.43}
    \Delta ^{(0)}_{\mu \nu ,\rho \sigma } = \frac{1}{6\lambda}\frac{1}{q^{2}} \frac{1}{q^{2} + \frac{\kappa ^{2}}{6\lambda} } P^{(0)}_{\mu \nu ,\rho \sigma }
\end{equation}

where $P^{(2)}_{\mu \nu ,\rho \sigma }$ and $P^{(0)}_{\mu \nu ,\rho \sigma }$ are the projection operators:

\begin{equation} \label{eq:3.44}
    P^{(2)}_{\mu \nu ,\rho \sigma } = \frac{1}{2} (\theta _{\mu \rho }\theta _{\nu \sigma } + \theta _{\nu \rho }\theta _{\mu \sigma }) - \frac{1}{3} \theta _{\mu \nu }\theta _{\rho \sigma }
\end{equation}

\begin{equation} \label{eq:3.45}
    P^{(0)}_{\mu \nu ,\rho \sigma } = \frac{1}{3} \theta _{\mu \nu }\theta _{\rho \sigma }
\end{equation}

\begin{equation} \label{eq:3.46}
    \theta _{\mu \nu } = \eta _{\mu \nu } - \frac{q_{\mu }q_{\nu }}{q^{2}}
\end{equation}

Notice that the denominator signs are flipped from the usual QFT convention. That is because we follow the metric $(-\, +\, +\, +)$ convention to be consistent with the same choice for gravity, not the $(+\, -\, -\, -)$ convention used in most QFTs. The superscripts (0) and (2) correspond to the spin of the physical modes. The transverse tensor mode $h^{\perp }_{\mu \nu }$ provides the spin-2 mode, whereas the scalar field $\Phi$ the spin-0 mode. The pole of each of these propagators provides us with the mass of the corresponding particles for the produced physical modes. For the spin-2 propagator this is trivial, since it is just the massless spin-2 graviton that appears in GR, too. Its dependence on $\kappa ^{2}$ is telling, as this mode would not have appeared in pure $R^{2}$ gravity. The spin-0 propagator requires a more careful treatment by applying the $i\epsilon$ prescription and separating the two poles:

\begin{equation} \label{eq:3.47}
    \Delta ^{(0)}_{\mu \nu ,\rho \sigma } = -\frac{1}{\kappa ^{2}} \left (\frac{1}{q^{2} + \frac{\kappa ^{2}}{6\lambda} - i\epsilon} - \frac{1}{q^{2} + i\epsilon} \right ) P^{(0)}_{\mu \nu ,\rho \sigma }
\end{equation}

The result is much clearer. The first of the produced modes is a massive spin-0 particle with $m^{2} = \frac{\kappa ^{2}}{6\lambda}$. This is not a gravitational mediation particle, but a scalar field vacuum excitation. The second mode is a massless ghost field, as indicated by its overall opposite sign. This is why we choose an opposite sign $i\epsilon$ term, as a same-sign choice with the other mode would violate the optical theorem. Before we comment on this ostensible pathology, we want to check what happens with these modes in pure $R^{2}$ gravity. Because $\kappa ^{2}$ appears in (\ref{eq:3.47}) in the denominator, we go back to (\ref{eq:3.43}) and set $\kappa^{2} = 0$:

\begin{equation} \label{eq:3.48}
     \Delta ^{(0)}_{\mu \nu ,\rho \sigma } = \frac{1}{6\lambda}\frac{1}{q^{4}} P^{(0)}_{\mu \nu ,\rho \sigma }
\end{equation}

This propagator lacks any pathologies as it produces a pair of massless scalar particles, the significance of which we will analyze soon. What this tells us is that the ghost which appears in the Starobinsky model is actually not a novelty, but a consequence of the GR part of the theory, and it can be shown that it is in fact harmless. There are two ways to illustrate this. \par
The first way is an argument of locality. From (\ref{eq:3.41}) we can see that the longitudinal component of the perturbation is given by the kinetic term of $\Phi$. Hence, if we want to solve for $\Phi$, we have to calculate the inverse of the D'Alembertian $\square ^{-1}$ for the perturbation term on the lhs. By Taylor expanding, we will have an infinite sum of derivative terms, and as a result $\Phi$ will be non-local. We conclude that it does not propagate and the pathology is gone. The second way in which we can showcase this is by coupling the spin-0 component of the action to matter \cite{Reference2}:

\begin{equation} \label{eq:3.49}
\begin{split}
    S^{(0)}_{0} &= \int d^{4}x\, \sqrt{g} \, \left [\frac{3\kappa ^{2}}{32} \partial _{\mu }\Phi \partial ^{\mu }\Phi + \frac{9\lambda}{16} (\square \Phi )^{2} + \Phi T \right ]\\
    &= \int d^{4}x\, \sqrt{g} \, \left (\frac{3\kappa ^{2}}{32} \partial _{\mu }\Phi \partial ^{\mu }\Phi + \Psi \square \Phi - \frac{4\lambda}{9} \Psi ^{2} + \Phi T \right )
\end{split}
\end{equation}

where $T$ is the trace of the stress-energy tensor for the newly introduced matter field $\Psi$ ($T^{\mu }{}_{\mu } = 8T$). Minimizing (\ref{eq:3.49}), we get the equations of motion:

\begin{equation} \label{eq:3.50}
    \left (\square -\frac{\kappa ^{2}}{6\lambda} \right )\Psi = -T
\end{equation}

\begin{equation} \label{eq:3.51}
    \square \Phi = \frac{8}{9\lambda} \Psi
\end{equation}

$\Psi$ as expected is a well-behaved, propagating physical mode with mass $m^{2} = \frac{\kappa ^{2}}{6\lambda}$. In fact, it is the first of the two spin-0 modes we found from the propagator (\ref{eq:3.47}). $\Phi$ - since it's dependent on another field - comes out non-local and not propagating. As for $R^{2}$ gravity, we may remove the first term in the action (\ref{eq:3.49}) and once again set $\kappa ^{2} = 0$, which is fully equivalent to setting this limit directly in the equation of motion (\ref{eq:3.50}):

\begin{equation} \label{eq:3.52}
    \square \Psi = -T
\end{equation}

$\Psi$ becomes a massless mode as we expected, whereas $\Phi$ can be dealt with in the same way we have described above. Alternatively, we can remove $\Phi$ with a gauge transformation. On the other hand, the spin-2 mode does not appear at all in this theory because it is entirely dependent on the existence of the transverse tensorial perturbation which would not appear had we not included the $\kappa^{2}R$ term in the original action. Since the only remaining physical propagating mode is the massless scalar excitation of the Minkowski vacuum, there is no particle that mediates gravity. In other words, $R^{2}$ does not produce any gravitational force on flat geometries, explaining why for the extremely flat post-inflationary Universe the $R^{2}$ corrective term does not alter classical gravity. \par
As a final note, we briefly mention the more general case where we include a Weyl term as per the action (\ref{eq:3.13}). We may expand this action by an extra $R$ term in order to draw a direct comparison with the Starobinsky model: 

\begin{equation} \label{eq:3.53}
     S_{0,W} = \int d^{4}x\, \sqrt{g} \, \left [a \left (R_{\mu \nu }R^{\mu \nu } - \frac{1}{3} R^{2} \right ) + bR^{2} + \kappa^{2} R \right ]
\end{equation}

where we have also ignored the classically irrelevant GB term. The spin-0 propagator will not change from its form in (\ref{eq:3.43}), as no new scalar term contributes to that component of the action. The only difference is that the total factor of $R^{2}$ this time is $b - \frac{a}{3}$ which we can arbitrarily match to $\lambda$ from the preceding analysis. However, the first term in this case is the trace of tensors, contributing to the tensor component of the action (\ref{eq:3.41}). This affects the spin-2 propagator:

\begin{equation} \label{eq:3.54}
    \Delta ^{(2)}_{\mu \nu, \rho \sigma } = \frac{2}{q^{2}} \frac{1}{a\left (q^{2} - \frac{\kappa ^{2}}{a} \right )} P^{(2)}_{\mu \nu, \rho \sigma }
\end{equation}

By carefully treating this propagator just like the spin-0 case we get:

\begin{equation} \label{eq:3.55}
    \Delta ^{(2)}_{\mu \nu, \rho \sigma } = -\frac{2}{\kappa ^{2}} \left (\frac{1}{q^{2} - i\epsilon} - \frac{1}{q^{2} - \frac{\kappa ^{2}}{a} +i\epsilon} \right ) P^{(2)}_{\mu \nu, \rho \sigma }
\end{equation}

In addition to the regular massless spin-2 graviton which we expected to get due to the GR component of the action, we also get a massive spin-2 ghost particle. Unlike the harmless ghost of the spin-0 propagator that can be gauged away, this ghost is not salvageable. The mode possesses a negative squared mass $m^{2} = -\frac{\kappa ^{2}}{a}$ and therefore its kinetic term will always be negative. The conclusion is that the more general case of the Weyl theory is pathological and cannot work as a viable alternative theory of gravitation on the perturbative level. This changes in the context of string theory, but it lies outside of the scope of this thesis.

\subsection{Curved space solutions}

Now we move on to the Einstein space solutions. We consider the perturbative contributions only for the $R^{2}$ component of the metric, since this is the theory for which such curved space solutions exist, plus we expect that only $R^{2}$ gravity will be relevant in this background. We introduce the same kind of metric perturbation:

\begin{equation} \label{eq:3.56}
     g_{\mu \nu } = \bar{g} _{\mu \nu } + h_{\mu \nu }
\end{equation}

where $\bar{g} _{\mu \nu }$ is the dS/AdS space metric for which $\Bar{R} _{\mu \nu } = \beta \, \bar{g} _{\mu \nu }$ and $\beta = \frac{\Bar{R}}{4}$. The following discussion will be restricted to the dS solutions ($\beta > 0$) that concern inflation, but it could easily be extended to AdS solutions as well. Calculating the various terms of the action up to third order of the perturbation (thus second order for the field equations) we have \cite{Reference2}:

\begin{equation} \label{eq:3.57}
\begin{split}
    S_{2} = \lambda \int d^{4}x&\, \Bigg [\left (\nabla _{\mu }\nabla _{\nu }h^{\mu \nu } - \square h - \frac{1}{4} \Bar{R} \varphi \right )^{2} - \Bar{R} \bigg (-\frac{1}{2} h_{\mu \nu } \square h^{\mu \nu } + \frac{1}{2} \varphi \square \varphi \\ 
    & + \nabla _{\mu }\phi \nabla _{\nu }h^{\mu \nu } - \nabla ^{\mu }h_{\mu \rho }\nabla _{\nu }h^{\nu \rho } \bigg ) + \frac{\Bar{R} ^{2}}{6} \left (h_{\mu \nu }h^{\mu \nu } + \frac{1}{4} \varphi ^{2} \right ) \Bigg ]
\end{split}
\end{equation}

where $\varphi = h^{\mu}{}_{\mu}$ like in the flat space solution. We proceed by tensor decomposing the perturbation \cite{Reference2}:

\begin{equation} \label{eq:3.58}
     h_{\mu \nu } = h^{\perp }_{\mu \nu } + \nabla _{\mu }\alpha ^{\perp }_{\nu } + \nabla _{\nu }\alpha ^{\perp }_{\mu } + \left (\nabla _{\mu }\nabla _{\nu } - \frac{1}{4} \eta _{\mu \nu }\square \right )\alpha + \frac{1}{2} \Bar{g} _{\mu \nu }\varphi 
\end{equation}

The newly introduced tensor and 4-vector satisfy the following conditions:

\begin{equation} \label{eq:3.59}
    \nabla ^{\mu }h^{\perp }_{\mu \nu } = 0, \quad \quad \eta ^{\mu \nu }h^{\perp }_{\mu \nu } = 0, \quad \quad \nabla ^{\mu }a^{\perp }_{\mu } = 0
\end{equation}

Similarly to the flat space case, these tensors are transverse and as such vanish by the action of the covariant derivative, while $h^{\perp }_{\mu \nu }$ is traceless again. The following steps where we impose a set of diffeomorphisms $\xi_{\mu} (x)$ and introduce a new gauge-invariant scalar field $\Phi$ are identical to the ones of the previous subsection. For the sake of brevity we will skip them, with the final expression for the action being:

\begin{equation} \label{eq:3.60}
    S_{2} = \lambda \int d^{4}x\, \left [\frac{\Bar{R} }{2} h^{\perp }_{\mu \nu }\left (\square - \frac{\Bar{R} }{6} \right )h^{\perp \mu \nu } + \frac{9}{16} \Phi \left (\square ^{2} + \frac{\Bar{R} }{3} \square \right )\Phi \right ]
\end{equation}

The form of this action can be derived if we equivalently consider \cite{Reference2}:

\begin{equation} \label{eq:3.61}
\begin{split}
    S_{2} &= \lambda \int d^{4}x \, \sqrt{g} \, \left [(\delta _{1}R)^{2} \sqrt{ g} + 2\Bar{R} (\delta _{1}R)(\delta _{1}\sqrt{g} ) + \Bar{R} ^{2}(\delta _{2}\sqrt{g} ) + 2\Bar{R} \sqrt{g} (\delta _{2} R) \right ]\\
   &=\lambda\int d^{4}x\, \sqrt{g} \, \left [(\delta _{1}R)^{2}\sqrt{ g} + 2\Bar{R} \delta _{2}(R\sqrt{g} ) - \Bar{R} ^{2}(\delta _{2} \sqrt{g} ) \right ]
\end{split}    
\end{equation}

where $\delta _{1}$ and $\delta _{2}$ are the first order and second order variations due to the perturbation respectively. The first term of the action in the second line will give us the kinetic term of $\Phi$, matching the last term of (\ref{eq:3.60}). The last two terms of (\ref{eq:3.61}) on the other hand eventually take the form of the rest of (\ref{eq:3.60}). These terms are none other than the second order variation of the Einstein-Hilbert action in the presence of a cosmological constant $\Lambda = \frac{M_{P}^{2}}{16\lambda}$. As expected from the discussion in the transformation to the Einstein frame, pure $R^{2}$ gravity in a dS background is fully consistent with regular Einstein gravity in the presence of a massless scalar field. Therefore, the only propagating physical modes are the standard spin-2 graviton and a spin-0 massless excitation that is $\phi$. The theory is yet again ghost-free with no pathologies detected. The fact it behaves identically to GR is an exceptional result, because it means that it is capable of describing (classical) gravity during inflation. Moreover, it means that the impotence of the $R$ term of the action in the early stages of the Universe is fully compensated by the effective form of the $R^{2}$ term. 

\section{Boundary terms}

When mentioning the Einstein-Hilbert action and the produced field equations, we took for granted that we are dealing with compact spacetimes. As we saw back in chapter 2, compact spaces lack boundaries e.g. the surface of a sphere of any dimensions. This is a natural assumption cosmologically, since the Universe is compact. In this context, the principle of least action works by having the extremal terms of the calculation go to 0. This is also reasonable, as the spacetime points of the integral go to infinity where fields and their derivatives vanish. By introducing a boundary to the spacetime, this calculation becomes non-trivial. To counteract boundary terms in order maintain physical consistency in the field equations, we insert a new action term \cite{Reference4}:

\begin{equation} \label{eq:3.62}
     S_{GYH} = M_{P}^{2} \oint _{\partial \mathcal V} d^{3}y \, \epsilon \,  \sqrt{|h|} K
\end{equation}

where $\partial \mathcal{V}$ is the 3-dimensional hypersurface that serves as the boundary for the hypervolume $\mathcal{V}$ of the spacetime, $y^{i}$ are the coordinates on the boundary, $h=\det{(h_{ij})}$ the determinant of the induced metric $h_{ij}$ that maps it (not to be confused with the metric perturbation of the last section), $\epsilon$ a constant that takes the value $\pm 1$ when the boundary is timelike/spacelike and $K$ is the scalar extrinsic curvature:

\begin{equation} \label{eq:3.63}
    K = K_{\mu \nu }h^{\mu \nu }
\end{equation}

$K_{\mu \nu}$ being the appropriate extrinsic curvature tensor (also called the second fundamental form) which depends on the orientation given by the unit normal 4-vector $n^{\mu}$ \cite{Reference13}:

\begin{equation} \label{eq:3.64}
    K_{\mu \nu} = h_{\mu}{}^{\rho} h_{\nu}{}^{\sigma} \nabla _{\rho}n_{\sigma}
\end{equation}

The physical interpretation of this quantity has to do with the perceived shape of the hypersurface. As we embed new structures in the original manifold described by the metric $g_{\mu \nu}$, we are defining submanifolds within it that are described by their own induced metrics $h_{ij}$. These submanifolds form the aforementioned hypersurfaces of the original space. We can always calculate their scalar curvature with the Riemann tensor, and that constitutes their intrinsic curvature. It is intrinsic because it is independent of the choice of embedding for the submanifold. Thus it could be measured by a hypothetical inhabitant of that hypersurface without any consideration of the encompassing higher-dimensional "Universe". However, there is a second perceived curvature that is extrinsic, as it is formed by the particular embedding chosen. Depending on how they are embedded i.e. what kind of constant slices we take, the perceived shape changes, following the orientation of the original manifold. The rate at which the orientation changes determines this shape, therefore the extrinsic curvature acts effectively as a "shape tensor". In turn, the induced metric is related to the spacetime metric via the general coordinate transformation for the particular embedding:

\begin{equation} \label{eq:3.65}
    h_{ij} = g_{\mu \nu} \, \partial _{i}x^{\mu} \, \partial _{j} x^{\nu}
\end{equation}

where $\partial_{i} = \frac{\partial}{\partial y^{i}}$. The indices $i, \, j$ may take any three values out of the total four of the original spacetime, depending on the type of boundary that exists. A projection of this metric on the submanifold as to include all possible indices is given by \cite{Reference13}:

\begin{equation} \label{eq:3.66}
    h_{\mu \nu} = g_{\mu \nu} - \epsilon \, n_{\mu } n_{\nu} = h_{ij} \, \partial _{\mu} x^{i} \, \partial _{\nu } x^{j} 
\end{equation}

From this projector of the induced metric it is easy to see that $h_{\mu}{}^{\rho} = \delta _{\mu}{}^{\rho} - \epsilon \, n_{\mu} n^{\rho}$ which greatly simplifies (\ref{eq:3.64}). It is also easy to show that $h_{\mu}{}^{\rho} n_{\rho} = 0$. This is why this form is sometimes called the transverse metric, since it is perpendicular to the unit normal of the hypersurface. The factor of $\epsilon$ mentioned above is also directly related to the orientation of the manifold \cite{Reference4}:

\begin{equation} \label{eq:3.67}
    \epsilon = n^{\mu} n_{\mu}
\end{equation}

Even though the value $+1$ for timelike and $-1$ for spacelike hypersurfaces is used here, it can be reversed if the metric convention used is $(+ \, - \, - \, -)$ instead of $(- \, + \, + \, +)$. The action (\ref{eq:3.62}) is called the Gibbons-York-Hawking boundary term, and one can verify that minimizing the total action $S_{EH}+S_{GYH}$ yields the same Einstein field equations as before for any kind of boundary. Since the action provides first-order differential equations with respect to the metric $g_{\mu \nu}$, the Dirichlet boundary conditions need just one constraint:

\begin{equation} \label{eq:3.68}
    \delta g_{\mu \nu} \big |_{\partial \mathcal{V}} = 0
\end{equation}

If we want to extend the GYH term to $f(R)$ theories, we have to be careful as to impose an additional Dirichlet boundary condition accordingly. Examining the field equations (\ref{eq:3.16}) for these theories, we realize that they are all second-order differential equations with respect to the metric, therefore we need one more boundary condition. A natural choice is to set $R$ constant on the boundary:

\begin{equation} \label{eq:3.69}
    \delta R \big |_{\partial \mathcal{V}} = 0
\end{equation}

since that is effectively a boundary condition on the derivatives of $g_{\mu \nu}$. We shall not delve into the mathematical derivation for the boundary term, so we directly present the most general result \cite{Reference4}:

\begin{equation} \label{eq:3.70}
     \big (S_{GYH} \big )_{f(R)} = 2\oint _{\partial \mathcal{V} }d^{3}y \, \epsilon \, \sqrt{|h|} \, f'(R) \, K
\end{equation}

It is then easy to find the special case for scale invariant gravity:

\begin{equation} \label{eq:3.71}
    \big (S_{GYH} \big )_{R^{2}} = 4\lambda \oint _{\partial \mathcal{V} }d^{3}y \, \epsilon \, \sqrt{|h|} \, R \,K
\end{equation}

Given that the two metric solutions of interest for the latter provide constant scalar curvature ($0$ or $4\beta$), (\ref{eq:3.71}) is essentially just the regular GYH boundary term multiplied by a constant. \par
As a final note, we examine the example of simple spacelike hypersurfaces for a FLRW spacetime (\ref{eq:1.1}). The orientation of the hypersurface is simply given by the unit normal parallel to the direction of the comoving time, perpendicular on a constant time slice ($t-t_{0}$):

\begin{equation} \label{eq:3.72}
    f = t- t_{0} \Rightarrow \, n_{\mu} = \partial _{\mu}f = \delta ^{\mu}{}_{t}
\end{equation}

The only non-zero components of the induced metric will be:

\begin{equation} \label{eq:3.73}
    h_{rr} = \frac{a^{2}}{1-kr^{2}}, \qquad h_{\theta \theta } = a^{2}r^{2}, \qquad h_{\varphi \varphi } = a^{2}r^{2}\sin ^{2}\theta
\end{equation}

It is then straightforward to calculate the extrinsic curvature quantities:

\begin{equation} \label{eq:3.74}
\begin{split}
     &K_{ij} = \nabla _{i} n_{j} = \frac{1}{2} \partial _{t}h_{ij} \\
     \Rightarrow \, &K = \frac{1}{2} h^{ij}\partial _{t} h_{ij} = \frac{3\dot{a} }{a} = 3H
\end{split}
\end{equation}

As we can see, the result is related to the radius of the spacelike 3-spheres that these hypersurfaces form.

\pagebreak % Experimental Setup

\fancyhead[RE,LO]{Quantum Cosmology and the Wheeler-DeWitt equation}

\chapter{Quantum Cosmology and the Wheeler-DeWitt equation}

At a first glance the concept of quantum cosmology seems like an oxymoron; the study of quantum mechanics pertains to the physics at the smallest scales, whereas cosmology is meant to study the Universe, which is by definition the most macroscopic thing there is. This naive dichotomy is dispelled by considering the very early stages in which the Universe was microscopic in scale. Moreover, whereas the Hot Big Bang model postulates a state of the Universe prior to expansion with an abundance of initial particles, inflation does not and is instead guided by the mechanics (be it classical or quantum) of the inflaton field. \par
The first significant attempt within this domain was the introduction of the Wheeler-DeWitt equation \cite{Reference22}. It was an attempt at formally reconciling the Schrödinger equation with gravity in the form of GR. The wave functions don't predict the position of particles, but the Universe itself being in a certain state. The equation reads \cite{Reference22}\cite{Reference1}:
\begin{equation} \label{eq:4.1}
    \left [-G_{ijkl}\frac{\delta ^{2}}{\delta h_{ij}\delta h_{kl}} - {}^{3}R(h)\sqrt{h} + 2\Lambda \sqrt{h} \right ]\psi (h_{ij}) = 0
\end{equation}
where $h_{ij}$ the induced metric of a hypersurface like we mentioned in the previous chapter, ${}^{3}R(h)$ the associated scalar curvature and $G_{ijkl}$ the metric of what we call "superspace". Superspace refers to the (mathematical) space of all 3-dimensional metrics $h_{ij}$. The wave function $\psi$ is no longer a function of spatial coordinates, but the geometry of spacetime instead. In other words, its degrees of freedom are the spacetime metrics. Those metrics are purely spacelike and are hypersurfaces of 4-dimensional metrics $g_{\mu \nu}$ for constant time slices. This choice is deliberate; another way to express the Wheeler-DeWitt equation more generally is \cite{Reference14}\cite{Reference22}:
\begin{equation} \label{eq:4.2}
    \hat{\mathcal{H}} | \psi \rangle = 0 \Rightarrow \frac{\partial \psi}{\partial t} = 0
\end{equation}
This Hamiltonian constraint shows that the wave function is independent of time, and causes the system in question to have 0 energy. We shall delve into the particulars of canonical quantization and details in its derivation in the next sections.\par
Firstly, we notice a conceptual problem. Since the wave function is meant to describe the Universe - which is far from static - time-independence seems at odds with the conventional understanding of cosmology. This discrepancy stems from a fundamental question called "the problem of time" \cite{Reference14}\cite{Reference59}. In quantum mechanics time is absolute and universal when quantifying the wave function's evolution. In QFT where special relativity is accounted for, time is relative with respect to reference frames and seen as a dimension of a 4-dimensional spacetime.  Time in GR is not just relative, but an emergent characteristic of spacetime, where the metrics which describe it act as their own degrees of freedom. As a result, while in practical terms the wave function is not going to be static in terms of the evolution of the Universe, there is no absolute flow of time. After all, since the Universe includes everything - including the concept of space and time - there can be no external "clock" by which we could measure this wave function's evolution. \par
A second conceptual problem has to do with what the wave function is supposed to describe \cite{Reference14}\cite{Reference1}\cite{Reference18}\cite{Reference20}\cite{Reference21}. Namely, the wave function is generally meant to give a probability distribution for the Universe. However, since the Universe includes everything that there has ever been and ever will be, there is no external observer to make this measurement. We as observers affect the system because we are part of the Universe i.e. we are part of the system we are trying to analyze. It is not trivial to formulate a physical argument as to how measurements could be relevant as a means to observing this probability distribution. DeWitt made the argument that upon making a measurement as observers we are implicitly "sundering" the Universe in two: one macroscopic part that is the observer, and another macroscopic part that is everything else \cite{Reference14}\cite{Reference22}. By extension the observation of everything else as we perceive it is entirely dependent on the measurement, and the Universe is there only because we were there to measure it. That of course doesn't mean the Universe does not exist in general before any observation happens. It simply means that what is observable to us is measured as such because the measurement happened in the first place, leading the wave function to collapse in a certain state. This is entirely consistent with the Copenhagen interpretation, where the act of observation is enough to cause the wave function to collapse in one of its possible states. This view of the wave function of the Universe faces the same issues as conventional quantum mechanics. \par
Another conceptual difficulty concerns the applicability of solutions in measurement theory. The goal of calculating a wave function and deriving the probability distribution is to formulate expectation values of observables to be tested and measured multiple times, validating or invalidating the probabilities predicted mathematically. For one, the solutions aren't normalizable and require special care to give a physically sound probability current. But beyond that, the Universe lacks the ability for repetition and reproducability of results. It is not a system which we can replicate or perform multiple iterations. This might have a possible solution, though. The existence of inflation gives rise to an enormous amount of causally disconnected regions like our own observable Universe, and in some models like chaotic inflation, not simultaneously \cite{Reference14}\cite{Reference12}. In this context, a quantum mechanical probability would not have to correspond to the entirety of the Universe, but rather individual regions of initial size $\mathcal{O}(H^{-1})$ that each have their own wave function. Each will evolve accordingly, leading to different probability distributions. The end result is a system of exponentially many mini-universes in a potentially eternal process, serving as multiple iterations of the same types of measurement. \par 
Here we will also clarify the meaning of "observer" in the cosmological sense. It is anything able to interact with the system, and as a result whose existence allows for the conceptualization of "rulers" and "clocks" which lead to measurements. The idea is that a measurement is any form of interaction that quantifies observables, which effectively measures the associated physical quantities. Thus a quantum cosmological observer is essentially just an extension of the regular quantum observer in the Copenhagen interpretation. Once again, the conceptual problems and paradoxes of the latter are inherited by quantum cosmology. As such, a more concrete understanding of quantum cosmology is incumbent on a more concrete understanding of quantum mechanics as a whole. \par
It is clear that quantum cosmology is a rich domain with great possibilities to apply a variety of models, with a variety of fundamental questions not settled. Nonetheless, there is still a considerable amount of physical knowledge that we may extract from it. In the following sections, we shall examine the WDW equation more closely. We shall tackle the emergent ambiguities, prove its universality in specific cases, and derive solutions for some meaningful examples such as minisuperspace JT gravity and the Starobinsky problem. \par

\pagebreak

\section{1d minisuperspace model}

The WDW equation is generally very difficult to solve for an arbitrary system. However, we are fortunate in that cosmology allows us to limit our scope to just the minisuperspace approximation which is easier to handle. The minisuperspace is a subset of the original superspace that only includes closed, homogeneous, and isotropic metrics. That is a reasonable assumption to make given our knowledge for the Universe. In technical terms, it implies that we simply consider $h_{ij}$ to be FLRW metrics. With this simplification in mind, we shall proceed to outline the process of quantization. To start off, we assume a simple Lagrangian for the early Universe \cite{Reference14}:
\begin{equation} \label{eq:4.3}
    \mathcal{L} (g_{\mu \nu} ,\phi ) = \frac{1}{2} \left (R - 2\Lambda \right )
\end{equation}
The gravitational interactions are entirely described by GR in the presence of a cosmological constant $\Lambda$, and we have used the convention $\kappa ^{2} = \frac{1}{2}$. The FLRW metric which will be used is going to be slightly generalized from its form in (\ref{eq:1.1}):
\begin{equation} \label{eq:4.4}
    ds^{2} = -N^{2}(t) \, dt^{2} + a^{2}(t) \, d\Omega _{3}^{2}
\end{equation}
The function $N(t)$ is called the lapse function which tells us how we foliate spacetime. Foliation is the process of breaking the 4-dimensional spacetime into a series of 3-dimensional hypersurfaces of constant time. This allows for the timelike unit normal on these hypersurfaces to parametrize time and form a time flow in that direction. The lapse function will then simply be the flow of proper time $\tau $ of a moving observer along the unit normal with respect to the original coordinate time $t$. This is not always doable, as there are spacetimes which cannot be foliated in this way e.g. many which containing singularities. In our case we have the freedom to do so. Another way to look at this is to view the choice of $N(t)$ as a form of gauge fixing \cite{Reference31}\cite{Reference32}. This is why in the previous chapters we freely took $N=1$. \par 
We may reduce the degrees of freedom of the Lagrangian by integrating over the spatial coordinates, thus leaving only the scale factor as a parameter. Since the spatial component is simply a 3-sphere, the 3-dimensional integral yields $2\pi ^{2} a^{3}$. For brevity, we shall symbolize $v_{3} = 2\pi ^{2}$. The scalar curvature is taken to be (\ref{eq:1.2}) for $k=1$, but slightly tweaked in order to account for the existence of the lapse function. If we make use of the lack of boundary of the metric and take the spatial derivatives of $\phi$ to be integrated to zero, we finally arrive at the proper Lagrangian \cite{Reference42}:
\begin{equation} \label{eq:4.5}
    L(a,N,\phi) = -3v_{3} \left (\frac{\dot{a}^{2} a}{N} - Na \right ) - v_{3} a^{3} N \Lambda 
\end{equation}
where we have also integrated by parts to eliminate any $\ddot{a}$ terms. Making use of the general definition of the canonical conjugate momentum of a quantity $X$, $\pi _{X} = \frac{\partial L}{\partial \dot{X}}$, we define the conjugate momenta for the Lagrangian above:
\begin{equation} \label{eq:4.6}
    \pi _{a} = -\frac{6v_{3}\dot{a} a}{N} \, , \quad \pi _{N} = 0
\end{equation}
We can construct the Hamiltonian:
\begin{equation} \label{eq:4.7}
    \mathcal{H} = \pi _{a} \dot{a} + \pi _{N} \dot{N} - L = -\frac{N}{a} \left (\frac{\pi _{a} ^{2}}{12 v_{3} + 3v_{3} a^{2}} \right ) + N\left (v_{3}a^{3} \Lambda \right )
\end{equation}
Just as we expected, the Hamiltonian due to being purely gravitational - which now corresponds to the degree of freedom of the scale factor - is negative. Due to the conjugate momentum of the lapse function being 0, it follows trivially that its time variation will also be 0, hence $\frac{\partial \mathcal{H}}{\partial N} = 0$. But since the Hamiltonian is only linearly dependent on $N$, we have:
\begin{equation} \label{eq:4.8}
    \frac{\partial \mathcal{H}}{\partial N} = \frac{\mathcal{H}}{N} = -\frac{1}{a} \left (\frac{\pi _{a} ^{2}}{12 v_{3}} + 3v_{3} a^{2} \right ) + v_{3}a^{3} \Lambda = 0
\end{equation}
This is precisely why the Hamiltonian constraint (\ref{eq:4.2}) holds and why any Hamiltonian involving gravity for a spacetime that can be foliated as such is ultimately 0. A different way to understand this property is as follows. Let us assume a general coordinate transformation $x^{\mu} \rightarrow x'^{\mu}$. Any gravitational theory remains invariant under such a transformation due to the action itself being invariant. This kind of transformation includes time reparametrization under which the time coordinate may be redefined as:
\begin{equation} \label{eq:4.9}
    t'(t) = t + f(t)
\end{equation}
for arbitrary local function $f(t)$. This redefinition will satisfy:
\begin{equation} \label{eq:4.10}
    N'(t')dt' = N(t)dt
\end{equation}
which is equivalent to saying that any choice of a time variable leaves the theory unchanged. Any degree of freedom $q(t)$ will also have to satisfy $q'(t') = q(t)$. For a general theory involving gravitation described by a Lagrangian $\bar{L} [q(t), \dot{q}(t), N(t)]$, the action variation under time reparametrization will be:
\begin{equation} \label{eq:4.12}
    \delta S = \int dt \left \{ N \delta q \left [\frac{\partial \bar{L} }{\partial q} - \frac{1}{N} \partial _{t} \left (\frac{\partial \bar{L} }{\partial (\dot{q} N^{-1})} \right ) \right ] + \delta N \left [\bar{L} - \frac{\dot{q} }{N} \frac{\partial \bar{L} }{\partial (\dot{q} N^{-1})} \right ] \right \} = 0
\end{equation}
In order for this to hold, both of the expressions in parentheses should vanish. The first one is none other than the Euler-Lagrange equation of the theory. The second is a Legendre transformation of the Langrangian where we have the conjugate momentum:
\begin{equation} \label{eq:4.13}
    \pi _{q} = \frac{\partial \bar{L} }{\partial (\dot{q} N^{-1})}
\end{equation}
Since this transforms it into the Hamiltonian of the system $\bar{\mathcal{H}} = \pi _{q} \frac{\dot{q}}{N} - \bar{L}$, we inevitably get a system of 0 energy. This is why we say that the Legendre transformation acts as a constraint on the theory. The redefinition (\ref{eq:4.9}) is therefore a gauge transformation, and its effect on the lapse function (\ref{eq:4.10}) can be seen as gauge fixing, like we mentioned before. We consider a generic redefinition of the scale factor:
\begin{equation} \label{scalefactor}
    a = \mathcal{A}(q)
\end{equation}
where $\mathcal{A}(q)$ is an arbitrary positive, invertible function of the new field $q$. The Lagrangian can be rewritten as such:
\begin{equation} \label{lagrangian}
    L = 3v_{3}N \left (-\frac{\mathcal{A} \mathcal{A}'^{2}}{N^{2}} \dot{q}^{2} + \mathcal{A} - \ell ^{2} \mathcal{A}^{3} \right )
\end{equation}
where we have defined $\ell = \sqrt{\frac{\Lambda}{3}}$, and with primes we imply derivatives with respect to $q$. It is useful to bring this to a non-linear $\sigma$-model form, identifying the following target-space metric:
\begin{equation} \label{metric}
    \gamma _{qq} = -6v_{3}\mathcal{A}\mathcal{A}'^{2}
\end{equation}
This metric describes a one-dimensional manifold $\cal T$ parametrized by the values of the field  $q$. The fact it is negative has to do with the fact that the gravitational degree of freedom in a quantum model is non-propagating ghost, just as we have shown in the previous chapter. For the sake of clarity, we henceforth symbolize the corresponding coordinate in $\cal T$ as $\textrm{q}$. The Lagrangian we get is:
\begin{equation} \label{newlagrangian}
    L(N,q,\dot q) = N \left({1\over 2}\, \gamma _{\textrm{q} \textrm{q}}(q)\, \frac{\dot{q}^{2}}{N^{2}} - V(q) \right)
\end{equation}
where we additionally defined $V(q) = 3v_{3}\left(-\mathcal{A}+\ell ^{2}\mathcal{A}^3\right)$. Following from these, the Hamiltonian becomes:
\begin{equation} \label{hamiltonian}
    \frac{\mathcal{H}}{N} =  \frac{1}{2}\, \gamma ^{\textrm{qq}} \pi _{q}^{2} + V   = 0
\end{equation}

\subsection{Canonical quantization and ordering ambiguities}

We proceed with the canonical quantization of the model. We promote $q$ and its conjugate momentum $\pi _{q}$ to operators which should satisfy the canonical commutation relation: 
\begin{equation} \label{eq:4.14}
    [q, \, \pi _{q}] = i\hbar 
\end{equation}
which in the $q$-picture of the associated Hilbert space this leads to:
\begin{equation} \label{operator}
    q \rightarrow \textrm{q} \, , \qquad \pi _{q} \rightarrow -i\hbar \frac{d}{d\textrm{q}}
\end{equation}
This however leads to the well-known problem of the ordering ambiguities of the WDW equation \cite{Reference22}\cite{Reference47}. Since $\gamma _{\textrm{qq}}$ is a generic function of $q$ in the classical picture and $\textrm{q}$ in the quantum one after the promotion to operators, it is non-trivially affected by the differential operator of (\ref{operator}). It therefore makes a difference on the quantum level whether the metric will be written before or after the conjugate momentum operator. The most generic way to show this is by introducing two complex "ambiguity functions" $v^{\textrm{q}}$ and $s$\cite{Reference31}\cite{Reference42}:
\begin{equation} \label{eq:4.15}
    \gamma ^{\textrm{qq}} \pi _{q} \pi _{q} = \frac{ \gamma ^{\textrm{qq}} }{v^\textrm{q} \, s}\, \pi_{q} \, v^\textrm{q} \, \pi _{q} \, s
\end{equation}
Classically this is a tautology, but on the quantum level it covers every possibility of reordering. These ambiguity functions are generic functions of $\textrm{q}$, but not of $\pi_{q}$, since we assume that the WDW should remain a differential equation of second order. Assuming the existence of vanishing states $\psi$ under the effect of the quantum Hamiltonian operator, the WDW equation is:
\begin{equation}\label{wdw}
 \frac{\widehat{\mathcal{H}}}{N}\,\psi\equiv -\frac{\hbar ^{2}}{2} \frac{\gamma^{\textrm{qq}}}{v^\textrm{q}\, s}\frac{d}{d \textrm{q}}\bigg(v^\textrm{q}\,\frac{d(s\psi)}{d \textrm{q}}\bigg)+V\psi =0\,,
\end{equation}
Something we notice from our notation is that $v^{\textrm{q}}$ is written as a vector, while $s$ as a scalar. This choice isn't coincidental. The WDW equation and the solutions $\psi$ should be independent of the coordinate system chosen on $\mathcal{T}$. In other words, the theory should not be dependent on the reference frame chosen for the fields. A change of coordinates such that $\textrm{q} = \textrm{q}(\check{\textrm{q}})$ - where $\check{\textrm{q}}$ are the new coordinates - should leave the WDW equation the same, while $\psi$ transforms as a scalar. This leads to the transformation rules:
\begin{equation} \label{eq:4.16}
    s\rightarrow s\, , \qquad v^\textrm{q}\rightarrow v^\textrm{q}\, \frac{d \check{\textrm{q}}}{d \textrm{q}}\equiv v^{\check{\textrm{q}}}
\end{equation}
It becomes apparent that under such a transformation $s$ transforms as a scalar, while $v^{\textrm{q}}$ transforms as a vector. For the sake of convenience, we define two new ambiguity functions:
\begin{equation} \label{ambiguities}
    \rho^\textrm{q} = v^\textrm{q} s^2 \,,\qquad  \omega = {\gamma^{\textrm{qq}}\over v^\textrm{q} \,s}{d\over d \textrm{q}}\!\left(v^\textrm{q} {d s\over d \textrm{q}}\right)\equiv {\nabla^\textrm{q}(v^\textrm{q} \,\nabla_{\!\textrm{q}} s)\over v^\textrm{q} \,s}
\end{equation}
bringing the WDW to the form:
\begin{equation} \label{wdw2}
\begin{split}
    \frac{\widehat{\mathcal{H}}}{N}\,\psi&=  -\frac{\hbar ^{2}}{2} \!\left[{\gamma^{\textrm{q}\textrm{q}}\over \rho^\textrm{q}} \frac{d}{d \textrm{q}}\! \left(\rho^\textrm{q}\, \frac{d\psi}{d \textrm{q}}\right) \!+ \omega \psi \right]\!+V\psi\\
    & =  -\frac{\hbar ^{2}}{2}\!\left[\gamma^{\textrm{q}\textrm{q}}\psi^{\prime\prime}+\gamma^{\textrm{q}\textrm{q}}\,{\rho^{\textrm{q}\prime}\over\rho^\textrm{q}}\,\psi'+\omega\psi\right]\!+V\psi\\
    & =  -\frac{\hbar ^{2}}{2}\!\left[\nabla^2\psi+{\nabla^\textrm{q}\rho^\textrm{q}\over\rho^\textrm{q}}\nabla_{\!\textrm{q}}\psi+\omega\psi\right]\!+V\psi = 0
\end{split}
\end{equation}
where primes indicate the derivatives $\frac{d}{d\textrm{q}}$.

\subsection{Ambiguities in path integral quantization}

Given that the canonical quantization gives rise to such ambiguities, a natural question would be whether a path integral quantization somehow resolves them. As we shall see though, ambiguities do in fact appear in this scheme as well. We start by a generic time parametrization within the interval $\left [x^{0}_{i}, \, x^{0}_{f} \right ]$ for which the field $q$ takes the values:
\begin{equation} \label{interval}
    q\left (x^{0}_{i} \right) = \textrm{q}_{i} \, , \qquad q\left (x^{0}_{f} \right ) = \textrm{q}
\end{equation}
The path integral for a state $\psi$ integrates over two fields - $q$ and the lapse function $N$\cite{Reference31}\cite{Reference42}:
\begin{equation} \label{pathintegral}
    \psi(\textrm{q})=\int {\mathcal{D} N \over \text{Vol}(\text{Diff}[-N^2])} \int_{\textrm{q}_i}^{\textrm{q}}\mathcal{D} q \; e^{{i\over \hbar}\int_{x^0_i}^{x^0_f}d x^0\, L(N,q,\dot q)}
\end{equation}
This is to be expected, since the resulting equation (\ref{wdw}) is of second order and linear, thus the Hilbert space is 2-dimensional. It also means that the lower boundary condition $\textrm{q}_{i}$ is sufficient to fully define a solution. \par
It is important to note that the path integral yields physically distinct solutions only when the choice of the spacetime metric results in different geometries, and it is over those geometries which the path integral sums. However, as we have shown earlier, the choice of $N$ for a certain configuration constitutes a gauge choice that doesn't physically alter the system. This is why we divide by the volume of the diffeomorphism group $\text{Vol}(\text{Diff}[-N^2])$. \par
Another way to look at this is that the $g_{00}$ component of the original spacetime metric describes a line segment of proper length:
\begin{equation} \label{line}
    \ell = \int_{x^0_i}^{x^0_f} d x^0 \,\sqrt{-g_{00}}
\end{equation}
not to be confused with the $\ell$ defined earlier for the potential of the system. Distinct metrics are therefore those which have different proper lengths $\ell$, while those with the same proper length belong to the same equivalence class $g_{00}[\ell]$. We may then make the gauge choice $N=\ell$ for the interval $\left [x^{0}_{i}, \, x^{0}_{f} \right ] = [0, \,1]$, which means the metric of the line is\cite{Reference31}\cite{Reference42}:
\begin{equation} \label{g00}
    g_{00}[\ell] = -\ell ^{2}
\end{equation}
The diffeomorphism group of $[-N^{2}]$ (or equivalently that of $g_{00}$) therefore contains all metrics belonging to the same equivalence class. Since the physically distinct states we sum over are those of different proper length, we may implement the Faddeev-Popov method to get rid of the redundant degrees of freedom from the diffeomorphisms, using the gauge choice mentioned in the previous paragraph. The path integral may then be expressed as\cite{Reference31}\cite{Reference42}:
\begin{equation} \label{pathintegral2}
    \psi(\textrm{q})=\int_0^{+\infty}\!\!d\ell \; \Delta_{\rm FP}(\ell) \int_{q(0)=\textrm{q}_i}^{q(1)=\textrm{q}}\mathcal{D} q   \; e^{{i\over \hbar}\int_{0}^{1}d x^0\, L(\ell,q,\dot q)}
\end{equation}
where $\Delta_{\rm{FP}}$ is the resulting Faddeev-Popov determinant, which for a line segment comes out to be equal to 1. This we shall address in Appendix B. To illustrate the source of ambiguities in this prescription, consider an arbitrary field redefinition $q=Q(\check{q})$, where $Q$ is some invertible function. The Lagrangian of the theory will change form:

\begin{equation}
    \check L(\ell,\check q,\dot{\check q}) = \ell \left({1\over 2}\, \gamma _{\check {\rm{q}} \check {\rm{q}}}(\check q)\, \frac{\dot{\check q}^{2}}{\ell^2} - \check V(\check q) \right)
\end{equation}
where $V(\check{q})=V\left(Q^{-1}(\check{q})\right)$. This, however, leaves the action of the theory invariant, since the Lagrangian is a scalar under field redefinitions. Hence the exponential in the integrand is not affected by our transformation. In turn, the boundaries of integration will transform accordingly:

\begin{equation}
    \check{q}(1)=\check{\rm{q}}\, , \qquad  \check{q}(0)=\check{\rm{q}}_{\rm{i}}
\end{equation}
where:
\begin{equation}
\begin{split}
    \check {\rm{q}}&=Q^{-1}(q(1))=Q^{-1}(\rm{q})\:\equiv \check {\rm{q}}(\rm{q})\, ,  \\
    \check {\rm{q}}_{\rm{i}}&= \check{\rm{q}}(\rm{q}_{\rm{i}})
\end{split}
\end{equation}

By contrast, the path integral measure does change due to the existence of a Jacobian determinant:

\begin{equation}
    \mathcal{D} q = \mathcal D \check q \;\mathcal J
\end{equation}
where we the Jacobian is defined as:
\begin{equation}
    \mathcal J=\prod_{x^0\in[0,1]}\left|\frac{dq(x^0)}{d \check{q}(x^0)}\right|=\prod_{x^0\in[0,1]} \big|Q'(\check{q}(x^0))\big|
\end{equation}

As a result, the original wave function in $\rm{q}$ coordinates after this transformation will become:

\begin{equation} \label{newpsi}
    \psi (\textrm{q}(\check{\textrm{q}})) =\int_0^{+\infty} d\ell \int_{\check{\rm{q}}_{\rm{i}}}^{\check{\rm{q}}} \mathcal{D} \check{q} \;\mathcal{J}\, e^{\frac{i}{\hbar}\int_0^1 dx^0\, \check{L} (\ell,\check{q},\dot{\check{q}})}
\end{equation}

Compare that to the wave function had we defined it in this frame to begin with:

\begin{equation}
    \check{\psi} (\check{\textrm{q}}) =\int_0^{+\infty} d\ell \int_{\check{\rm{q}}_{\rm{i}}}^{\check{\rm{q}}} \mathcal{D} \check{q} \, e^{\frac{i}{\hbar}\int_0^1 dx^0\, \check{L} (\ell,\check{q},\dot{\check{q}})}
\end{equation}

This expression lacks the Jacobian in the integrand, which makes it generally different to (\ref{newpsi}). A quantum theory based on the states above will therefore differ from one based on the definition (\ref{pathintegral2}). Given that both definitions are valid, this seems to be leading to multiple quantum theories, just as in the case of different ambiguity functions $\rho ^{\rm{q}}$ and $\omega$ in the canonical quantization prescription. We could then say that there is an ambiguity in the way we define the wave function in the path integral formulation, which has to do with the way we choose the measure of integration. \par

\subsection{Hilbert space inner product and probability amplitudes}

As we have explained above, the choice of path integral measure $\mathcal{D} q$ changes the wave function solutions for a particular system. As such, we have to be careful when defining the Hilbert space inner product for our theory. For one such specific choice, we define it as\cite{Reference42}:

\begin{equation} \label{amp}
\begin{split}
    \langle \psi_1,  \psi_2 \rangle_{\mathcal D q} &= \int \limits_{\mathcal T} d \textrm q \, \sqrt{-\gamma} \, \mu \, \psi_1^* \psi_2\\
    &= \text{sign}(\mathcal A ')\int^{\mathcal A^{-1}(+\infty)}_{\mathcal A^{-1}(0)} d \textrm q \, \sqrt{-\gamma} \, \mu \, \psi_1^* \psi_2\,.
    \end{split}
\end{equation}
where $\mu$ is a real positive function that acts as a measure, and the sign of the function $\mathcal{A}'$ is introduced to ensure that the inner product is positive-definite. The latter function, in turn, has a fixed sign due to being an invertible function (as per (\ref{scalefactor})) and therefore monotonic. The boundaries of integration correspond to the upper and lower bounds of the scale factor $a$, which are $+\infty$ and 0 respectively. Notice that we write down the volume element for the manifold $\mathcal{T}$ explicitly. We could have included it in the definition of $\mu$, but this form ensures that $\mu$ behaves as a scalar under diffeomorphisms of $\mathcal{T}$.\par
The value of the measure can be determined by imposing Hamiltonian hermiticity. To be more precise, let us assume two completely arbitrary complex functions $f_{1}$ and $f_{2}$ chosen such that they don't vanish when the Hamiltonian acts on them. As a result, their inner product after an integration by parts reads\cite{Reference42}:

\begin{equation}\label{hermiticity1}
\begin{split}
    \left \langle f_1 ,  {\widehat{\mathcal{H}} \over N} \,f_2 \right\rangle = & \left\langle {\widehat{\mathcal{H}}^\dagger\over N}\, f_1 ,  f_2 \right \rangle  -\text{sign}(\mathcal{A}')\,{\hbar^2\over2}\,\times\\
    &~~~~~~~~~~~~\,\left[\rho^\textrm q\left({\gamma^{\textrm q \textrm q}\over \rho^\textrm q}\, \sqrt{-\gamma}\,\mu\, f_1^*\, {d f_2\over d\textrm q}-{d\over d\textrm q}\!\left({\gamma^{\textrm q \textrm q}\over \rho^\textrm q}\, \sqrt{-\gamma}\,\mu\, f_1^*\right)f_2\right)\right]_{\mathcal A^{-1}(0)}^{\mathcal A^{-1}(+\infty)}
\end{split}
\end{equation}
where the Hermitian conjugate of the Hamiltonian is:
\begin{equation}\label{wdw+}
 \frac{\widehat{\mathcal{H}}^ \dagger}{N}\,f= -\frac{\hbar ^{2}}{2} {1\over \sqrt{-\gamma}\,\mu}{d\over d\textrm q}\!\left(\rho^{\textrm q *} {d\over d\textrm q}\Big({\gamma^{\textrm q \textrm q}\over \rho^{\textrm q *}}\, \sqrt{-\gamma}\,\mu\, f \Big)\right)  + \left(V-{\hbar^2\over 2}\,\omega^*\right)f
\end{equation}

There are two criteria to impose hermiticity. The first criterion we impose is that the Hamiltonian and its Hermitian conjugate must be algebraically the same:

\begin{equation}\label{HH+}
    {\widehat{\mathcal{H}} \over N}\, f={\widehat{\mathcal{H}} ^\dagger \over N}\, f
\end{equation}

The second criterion is that the boundary term in (\ref{hermiticity1}) should be 0. If our generic functions $f_{1},\, f_{2}$ are replaced by states $\psi _{1}, \, \psi_{2}$ which vanish under the influence of the Hamiltonian, then this criterion is trivially satisfied. There is, however, a more formal proof of how it vanishes which we include in Appendix A. Back to the first criterion, it is convenient for us to first analyze the ambiguity function $\rho ^{\rm q}$ as a complex number:

\begin{equation}\label{polar}
    \rho^\textrm q = r^\textrm q \, e^{i\theta}\, , \qquad \text{where} \qquad r^\textrm q \in \mathbb{R}\,, ~~\theta\in\left(-{\pi\over 2},{\pi\over 2}\right]
\end{equation}

As evidenced by the notation, the "length" of the complex number $r^{\rm q}$ transforms as a vector, while the argument $\theta$ as a scalar. That is a consequence of the diffeomorphisms of $\mathcal{T}$ - and consequently field redefinitions - being strictly real. Therefore the complex phase $e^{i\theta}$ shouldn't be affected by such a transformation. Moreover, it's important to notice that $r^{\textrm{q}}$ is not strictly positive definite, and $\theta$ doesn't span the usual $(-\pi ,\, \pi ]$ range for complex numbers\cite{Reference42}. Our modified choice is specifically made so that the invertible mapping from one reference frame to the other gives a constant negative sign for $\frac{d\textrm{q}}{d\check{\textrm{q}}}$. We shall see why that is necessary shortly. \par
Applying the above definition to (\ref{HH+}), we get the condition:

\begin{equation} \label{condition}
     2\gamma^{\textrm{qq}}\!\left[{r^{\textrm q \prime}\over r^\textrm q}-{(\gamma^{\textrm{qq}}\sqrt{-\gamma}\,\mu)'\over \gamma^{\textrm{qq}}\sqrt{-\gamma}\,\mu}\right]\!f'=\!\left[-2i \, \text{Im}\omega+\gamma^{\textrm{qq}}{\displaystyle \Big({\gamma^{\textrm{qq}}\over \rho^{\textrm q *}}\sqrt{-\gamma}\, \mu\Big)^{\!\prime\prime}+{\rho^{\textrm q *\prime}\over \rho^{\textrm q *}}\Big({\gamma^{\textrm{qq}}\over \rho^{\textrm q *}}\sqrt{-\gamma}\, \mu\Big)^{\!\prime}\over \displaystyle {\gamma^{\textrm{qq}}\over \rho^{\textrm q *}}\sqrt{-\gamma}\, \mu}\right]\!f
\end{equation}

Since $f$ is an arbitrary function, the brackets both on the lhs and the rhs should vanish so that the equation holds. If not, then it becomes a differential equation which, when solved, gives a particular form for $f$. Setting the bracket on the lhs equal to 0 gives the form of the measure:

\begin{equation} \label{mu}
    \mu={|r^{\textrm q}|\over -\gamma^{\textrm{qq}}\sqrt{-\gamma}} =\sqrt{-\gamma}\, |\rho^{\textrm q}|
\end{equation}

It is now made clear why our previous choice was necessary. The domains of definition chosen ensure that from (\ref{condition}) we get a strictly positive-definite $\mu$ as seen in (\ref{mu}). The result is that the measure is linearly dependent on the real part of the ambiguity $\rho ^{\rm q}$. Naively this seems to suggest a vectorial quality to $\mu$, but here it is understood that the value of $\left |\rho ^{\textrm{q}} \right |$ is a scalar number. However, it is immediately obvious that $\mu$ is not universal, since it depends on the choice of ambiguity function. Or if we express it in the language of the path integrals, a different choice of path integral measure $\mathcal{D}q$ generally corresponds to a different choice of Hilbert space inner product measure $\mu$\cite{Reference42}.\par
If we further demand that the rhs bracket vanishes, we additionally obtain a constraint on the second ambiguity function $\omega$:

\begin{equation} \label{omega}
    \text{Im} \omega ={\gamma^{\textrm{qq}}\over 2}\!\left(\theta^{\prime\prime}+{r^{\textrm q \prime}\over r^\textrm q}\,\theta'\right) ={1\over 2}\!\left(\nabla^2\theta+{\nabla^\textrm q r^\textrm q\over r^\textrm q}\, \nabla_{\!\textrm q}\theta\right)
\end{equation}

Notice how the expression found combines to create explicitly scalar quantities. That is consistent with what we know about $\omega$, i.e. that is it a scalar itself. In addition, we observe that the hermiticity criterion has constrained one unknown quantity from the ambiguities, which we have found to be the imaginary part of $\omega$. \par
Given that the states $\psi$ and the measure $\mu$ are what we could call ambiguity-dependent, we shall introduce a new quantity\cite{Reference42}:

\begin{equation} \label{newpsi2}
    \Psi = \sqrt{\mu} \, \psi
\end{equation}

Since the volume element of the integral is an ambiguity-free quantity, proving that these new states $\Psi$ are also ambiguity-free would mean that the theory itself is universal, and any choice of path integral measure leads to the same measurable quantities. If we plug (\ref{newpsi2}) into the WDW equation, we find:

\begin{equation}
    \sqrt{\mu}\,{\widehat{\mathcal{H}}\over N}\,\psi\equiv-{\hbar^2\over 2}\big[\nabla^2\Psi+Y^\textrm q \nabla_{\!\textrm q}\Psi+W\Psi\big]\!+V\Psi = 0
\end{equation}
where the introduced new quantities read\cite{Reference42}:

\begin{equation}
    Y^\textrm q =\gamma^{\textrm{qq}}\,{\rho^{\textrm q \prime}\over \rho^{\textrm q}}-{(\gamma^{\textrm{qq}}\sqrt{-\gamma\mu})'\over \sqrt{-\gamma\mu}}-\gamma^{\textrm{qq}}\, {\nabla_{\!\textrm q}\sqrt{\mu}\over \sqrt{\mu}}
\end{equation}
\begin{equation}
    W=\omega-{\nabla^2\!\sqrt{\mu}\over \sqrt{\mu}}-Y^\textrm q \,{\nabla_{\!\textrm q}\sqrt\mu\over \sqrt{\mu}}   
\end{equation}

Applying the result for $\mu$ from (\ref{mu}) greatly simplifies the first expression:
\begin{equation}
    Y^\textrm q =i\nabla^\textrm q \theta
\end{equation}
while applying the constraint (\ref{condition}) simplifies the second one:
\begin{equation}
    W=\text{Re} \omega-{\nabla^2\!\sqrt{\mu}\over \sqrt{\mu}}+{i\over 2}\, \nabla^{2}\theta
\end{equation}

As things currently stand, $\Psi$ is itself still an ambiguity-dependent quantity. A differently chosen $\rho ^{\textrm{q}}$ gives different values for $\mu$ and $\theta$, and combined with a potentially differently chosen $\omega$, these lead to different $Y^{\textrm{q}}$ and $W$. If we can prove that for an arbitrary change in the path integral measure we have:
\begin{equation}
    \check\theta(\check{\textrm{q}})=\theta(\textrm{q}(\check{\textrm{q}}))+\text{constant}\, , \qquad\check W (\check{\textrm{q}})= W(\textrm{q}(\check{\textrm{q}}))
\end{equation}
then the states $\Psi$ for the two frames will also be the same:
\begin{equation}
    \check\Psi(\check{\textrm{q}})=\Psi(\textrm{q}(\check{\textrm{q}}))
\end{equation}
and therefore the Hilbert space inner product will be invariant and the theory universal:
\begin{equation}
\begin{split}
    \langle \psi_1,  \psi_2 \rangle_{\mathcal D q}&=\int \limits_{\mathcal{T}} d\check{\textrm{q}}\, \sqrt{-\gamma_{\check{\textrm{q}}\check{\textrm{q}}}} \, \Psi_1^* (\textrm{q}(\check{\textrm{q}}))\Psi_2(\textrm{q}(\check{\textrm{q}}))\\
    &=\int \limits_{\mathcal{T}} d\check{\textrm{q}}\, \sqrt{-\gamma_{\check{\textrm{q}}\check{\textrm{q}}}} \, \check\Psi_1 ^* (\check{\textrm{q}}) \check \Psi_2(\check{\textrm{q}})\\
    &=\langle \check\psi_1,  \check\psi_2 \rangle_{\mathcal D \check q}
\end{split}
\end{equation}
It will be our goal in the next subsection to try and prove that this is the case, thus showing that despite ordering ambiguities, the quantum theory for a given system is unique. \par

\subsection{Resolution of ambiguities}

A way to determine the ambiguity functions and examine the uniqueness of the quantum theory is to try and derive the equivalent of the Schrödinger equation from the path integral. To that end, we need to have a frame where the Lagrangian of theory becomes that of a non-relativistic particle in 1-dimensional space under the influence of a potential\cite{Reference42}. \par
We begin by considering the generic transition amplitude from an initial state at the point $\textrm{q}_{i}$ to a final state at $\textrm{q}$:

\begin{equation}
\begin{split}
    U(\textrm{q}_{i},\textrm{q},\ell)&=\int_{\textrm{q}_i}^{\textrm{q}}\mathcal{D} q   \; e^{{i\over \hbar}\int_{0}^{1}d x^0\, L(\ell,q,\dot q)}\\
    &=\int_{\textrm{q}_i}^{\textrm{q}}\mathcal{D} q   \; e^{{i\over \hbar}\int_{0}^
    \ell d t\,  \scriptsize \big(\gamma_{\textrm{qq}}(q)({d q\over d t})^2-V(q) \scriptsize \big)} 
\end{split}
\end{equation}
where we apply the gauge $N = \ell \, , \, \ell \in [0,1]$ as we have discussed before. Now let us perform a field redefinition from $q$ to $q_{0}$ with the scale factor mapping being $\mathcal{A}_{0}(q_{0})$ such that the metric $\gamma _{\textrm{qq}}$ becomes the flat Minkowski metric. We demand then:
\begin{equation}
    6v_{3}\mathcal{A}_0(q_0)\mathcal{A}_0'^2(q_0) = 1
\end{equation}
which leads to\cite{Reference31}\cite{Reference42}:
\begin{equation}
    \mathcal{A}_0(q_0)=\left({3\over 2}\, {1\over \sqrt{6v_3}}\right)^{2\over3}|q_0-\textrm{q}_{0*}|^{2\over3}
\end{equation}
where $\textrm{q}_{0*}$ an arbitrary integration constant. $\mathcal{A}_{0}$ should be invertible, so to avoid singularities we define the domain definition of $q_{0} \in [\textrm{q}_{0*} , \, +\infty )$. The inverse mapping is therefore:
\begin{equation}
    q =\mathcal{A}^{-1}(a)=\mathcal{A}^{-1}(\mathcal{A}_0(q_0)) \equiv Q_0(q_0)
\end{equation}
and the points for the initial and final state in this frame become:
\begin{equation}
    \textrm{q}_0=Q_0^{-1}(\textrm{q})\, , \qquad \textrm{q}_{0i}=Q_0^{-1}(\textrm{q}_i)
\end{equation}

The transition amplitude now reads:
\begin{equation} \label{transamp}
\begin{split}
    U(\textrm{q}_{i},\textrm{q},\ell)&=\int_{\textrm{q}_{0i}}^{\textrm{q}_0}\mathcal{D} q_0 \;\mathcal{J}_0  \; e^{-{i\over \hbar}\int_{0}^
    \ell d t\, \left({1\over 2}({d q_0\over d t})^2+V_0(q_0)\!\right)}\\
    &\equiv U_0(\textrm{q}_{0i},\textrm{q}_0,\ell)
\end{split}
\end{equation}
where $V_{0}(q_{0}) = V(Q_{0}(q_{0}))$ and we are careful to include a Jacobian determinant:
\begin{equation}
    \mathcal{J}_0=\prod_{t\in[0,\ell]}J\big(q_0(t)\big) 
\end{equation}
with $J\big(q_0(t)\big)= \big|Q_0'\big(q_0(t)\big)\big|$. This transition amplitude can be used to derive the Schrödinger equation. We follow the procedure outlined here\cite{Reference42}\cite{Reference44}; we discretize the interval $\ell \in [0,1]$ in $n$ slices of duration $\varepsilon = \frac{\ell}{n}$. The field $q_0$ instead of a continuum becomes a collection of values $q_{0k}$ at each discretized point in time $t = k\varepsilon$, where $k \in \{0, ... ,n\}$ the index keeping track of the temporal sequence. The first and last of these points remain fixed, so we integrate over all possible values of each $q_{0k} \in [\textrm{q}_{0*}, \, +\infty)$ for the range $k \in \{1,...,n-1\}$. The discretized transition amplitude can be written in the following manner:

\begin{equation}
\begin{split}
    U_0^{(n)}(\textrm{q}_{0i},\textrm{q}_0,\ell)={1\over \mathcal{N}(\varepsilon,\textrm{q}_{0i},\textrm{q}_0)} &\prod_{k'=1}^{n-1} \left(\int_{\textrm{q}_{0*}}^{+\infty}{d q_{0k'}\over \mathcal{N}(\varepsilon,\textrm{q}_{0i},\textrm{q}_0)} \, J(q_{0k'})\right)\\
    &\!\times \exp{\Bigg \{{-{i\over \hbar}\sum_{k=0}^{n-1}\varepsilon \Big [{1\over 2}\displaystyle \left ({q_{0(k+1)}-q_{0k}\over \varepsilon}\right )^2+V_0\textstyle({q_{0k}+q_{0(k+1)}\over 2})\Big ]} \Bigg \}}
\end{split}
\end{equation}
where $q_{00} = \textrm{q}_{0i}$ and $q_{0n} = \textrm{q}_{0}$ and we have introduced a mid-point discretization for $V_{0}$. We have also used the primed variable $k'$ in the first product to distinguish from the one appearing in the sum in the exponent. The quantity $\mathcal{N}$ introduced in each slice corresponds to a normalization factor which as we can see only depends on the duration of the slices, as well as the boundary points. Allowing it to depend on any of the intermediate points would allow it to enter the integral, which would modify the weights of each slice. We should be diligent in handling the discretization of the path integral measure as to maintain time reparametrization invariance. We shall prove later on in the chapter that the form above gives back the old measure in the continuum limit. \par
If we isolate the integration over the last slice, the transition amplitude can be expressed as:

\begin{equation} \label{amplitude}
\begin{split}
    U_0^{(n)}(\textrm{q}_{0i},\textrm{q}_0,\ell)= \int_{\textrm{q}_{0*}}^{+\infty}{d q_{0(n-1)}\over \mathcal{N}(\varepsilon,\textrm{q}_{0i},\textrm{q}_0)} \, J(q_{0(n-1)})\, e^{-{i\over 2\hbar\varepsilon}(\textrm{q}_{0}-q_{0(n-1)})^2}  \, &e^{-{i\over \hbar}\varepsilon V_0{\textstyle(}{q_{0(n-1)}+\textrm{q}_{0}\over 2}{\textstyle)}}\\ 
    &\!\!\!\!\!\!\!\!\!\!\!\!\!\!\! \times\,   U_{0}^{(n-1)}(\textrm{q}_{0i},q_{0(n-1)},\ell-\varepsilon)
\end{split}
\end{equation}
where we have the transition amplitude from the initial state to the penultimate point of the time interval, before the omitted slice:
\begin{equation}
\begin{split}
    U_0^{(n-1)}(\textrm{q}_{0i},q_{0(n-1)},\ell-\varepsilon)={1\over \mathcal{N}(\varepsilon,\textrm{q}_{0i},\textrm{q}_0)} &\prod_{k'=1}^{n-2} \left(\int_{\textrm{q}_{0*}}^{+\infty}{d q_{0k'}\over \mathcal{N}(\varepsilon,\textrm{q}_{0i},\textrm{q}_0)} \, J(q_{0k'})\right)\\
    &\!\!\times e^{-{i\over \hbar}\sum_{k=0}^{n-2}\varepsilon \textstyle[{1\over 2}\textstyle({q_{0(k+1)}-q_{0k}\over \varepsilon}\textstyle)^2+V_0\textstyle({q_{0k}+q_{0(k+1)}\over 2})\textstyle]} 
\end{split}
\end{equation}
Notice that the transition amplitude $U^{(n-1)}$ contains the rest of the inserted normalization constants $\mathcal{N}$, leaving just one in the expression (\ref{amplitude}). Now we can expand the rhs of that expression for small $\varepsilon$ and Taylor-expanding the various functions of $q_{0(n-1)}$ around the point $\textrm{q}_{0}$. We introduce a regularization parameter $i\kappa \, , \, \kappa >0$ in the exponential to make it converge, and in combination with the following Gaussian integral identities\cite{Reference42}:
\begin{equation}
    \int_{-\infty}^{+\infty}d q\, e^{-\alpha q^2}=\sqrt{\pi\over \alpha}\, , \qquad \int_{-\infty}^{+\infty}d q\, q^2\, e^{-\alpha q^2}={1\over 2\alpha}\sqrt{\pi\over \alpha}
\end{equation}
where in our case $\alpha = \frac{i}{2\hbar \varepsilon}(1-i\kappa)$, we find that the expression up to order $\varepsilon$ reads:
\begin{equation} \label{diff}
    \begin{split}
U_0^{(n)}(\textrm{q}_{0i},\textrm{q}_0,\ell)={1\over \mathcal{N}(\varepsilon,\textrm{q}_{0i},\textrm{q}_0)}\,&\sqrt{2\pi\hbar\varepsilon\over i}\,\Big(1-i\,{\varepsilon\over \hbar}\, V_0(\textrm{q}_0)+{\hbar\varepsilon\over 2i}\,{\partial^2\over \partial q_{0(n-1)}^2}+{\mathcal{O}}(\varepsilon^2)\Big)\\
&~~~~~~\times\big[J(q_{0(n-1)})\, U_0^{(n-1)}(\textrm{q}_{0i},q_{0(n-1)},\ell-\varepsilon)\big]\Big|_{q_{0(n-1)}=\textrm{q}_0}
\end{split}
\end{equation}

In the limit $n\rightarrow +\infty$ (and by extension $\varepsilon \rightarrow 0$) the expression becomes:
\begin{equation}
    U_0(\textrm{q}_{0i},\textrm{q}_0,\ell)\underset{\varepsilon\to 0}\sim{1\over \mathcal{N}(\varepsilon,\textrm{q}_{0i},\textrm{q}_0)}\, \sqrt{2\pi\hbar\varepsilon\over i}\, J(\textrm{q}_0)\, U_0(\textrm{q}_{0i},\textrm{q}_0,\ell)
\end{equation}
so in order to maintain consistency, the normalization constant has to be:
\begin{equation}
    \mathcal{N}(\varepsilon,\textrm{q}_{0i},\textrm{q}_0)=\sqrt{2\pi\hbar\varepsilon\over i}\, J(\textrm{q}_0)
\end{equation}

Going back to the expression (\ref{diff}), moving the term of order $\mathcal{O}(\varepsilon ^{0})$ to the lhs and take the same limit as above, we end up with the differential equation:
\begin{equation}
    \begin{split}
&i\hbar\,{\partial \over \partial\ell}\, U_0(\textrm{q}_{0i},\textrm{q}_0,\ell)=\\
&\,~~~~~~~~~~~~\left\{{\hbar^2\over 2}\left[{\partial^2\over \partial\textrm{q}_0^2}+{2\over J(\textrm{q}_0)}\, {d J\over d\textrm{q}_0}(\textrm{q}_0)\, {\partial\over \partial\textrm{q}_0}+{1\over J(\textrm{q}_0)}\, {d^2 J\over d\textrm{q}_0^2}(\textrm{q}_0)\right]+V_0(\textrm{q}_0)\right\} U_0(\textrm{q}_{0i},\textrm{q}_0,\ell)
\end{split}
\end{equation}

By using the known form of the Jacobian, we can relate it to the metric coordinate transformation as such:
\begin{equation}
    \gamma_{\textrm{qq}}(\textrm{q})=\gamma_{\textrm{q}_0 \textrm{q}_0}(\textrm{q}_0)\left({d\textrm{q}_0\over d\textrm{q}}\right)^2=-{1\over 2J(\textrm{q}_0)^2}
\end{equation}
which leads us to the equation:
\begin{equation} \label{expression}
    \begin{split}
i\hbar\,{\partial \over \partial\ell}\, U(\textrm{q}_{i},\textrm{q},\ell)&= \left\{ -\frac{\hbar ^{2}}{2}\!\left[\gamma^{\textrm{qq}}{\partial^2\over \partial\textrm{q}^2}+{\gamma^{\textrm{qq}}\over P^\textrm{q}}\,{d P^{\textrm{q}}\over d\textrm{q}}\,{\partial\over \partial\textrm{q}}+\Omega \right]\!+V\right\}U(\textrm{q}_{i},\textrm{q},\ell) \\
&\equiv \widehat{\mathcal{H}} \,U(\textrm{q}_{i},\textrm{q},\ell)
\end{split}
\end{equation}
where we have dropped the 0 in the indices for simplicity, and we have the defined the quantities:
\begin{equation} \label{newamb}
    \begin{split}
P^\textrm{q}&= (-\gamma_{\textrm{qq}})^{-{3\over 2}}\, ,  \\
\Omega &=  - {1\over 2\,(\gamma_{\textrm{qq}})^2}\,{d^2\gamma_{\textrm{qq}}\over d\textrm{q}^2} + {1\over (\gamma_{\textrm{qq}})^3}\left({d\gamma_{\textrm{qq}}\over d\textrm{q}}\right)^2\
\end{split}
\end{equation}
which (as we will rigorously demonstrate shortly) are none other than the ambiguity functions for this frame, with $P^{\textrm{q}}$ being the vector-like function and $\Omega$ being the scalar. Notice how they depend on the Jacobian (and by extension the metric) and its derivatives. Had the transformation been different, these functions would have had a different form. This is another illustration of our earlier point that ambiguities in the path integral picture come from the inequivalence of states after field redefinitions. The operator $\widehat{\mathcal{H}}$ is none other than the quantum Hamiltonian of the system for our gauge choice of $N$. The expression (\ref{expression}) even though it leads to the Hamiltonian of the system, it is still not quite the WDW equation, since the transition amplitudes are not the vanishing states $\psi$ themselves. The latter are obtained via\cite{Reference42}:
\begin{equation}
    \psi(\textrm{q})=\int_0^{+\infty}d\ell \;U(\textrm{q}_{i},\textrm{q},\ell)
\end{equation}
so the differential equation becomes:
\begin{equation} \label{bound}
    i\hbar\, \big[U(\textrm{q}_{i},\textrm{q},\ell)\big]^{\ell=+\infty}_{\ell=0}=  \widehat{\mathcal{H}}\,\psi\equiv -\frac{\hbar ^{2}}{2}\!\left[\gamma^{\textrm{qq}}\psi^{\prime\prime}+\gamma^{\textrm{qq}}\,{P^{\textrm{q}\prime}\over P^\textrm{q}}\,\psi'+\Omega\psi\right]\!+V\psi
\end{equation}

The lhs boundary terms involve terms of order $\ell$ and $\ell ^{-1}$ in the complex exponential due to the full transition amplitude having the form:
\begin{equation}
    U(\textrm{q}_{i},\textrm{q},\ell)=\int_{\textrm{q}_i}^{\textrm{q}}\mathcal{D} q   \; e^{-{i\over \hbar}\int_{0}^
1 d x^0\, \scriptsize \big({1\over 2}|\gamma_{\textrm{qq}}|{\dot q^2\over \ell}+\ell V(q)\scriptsize \big)}
\end{equation}
We need to regularize this in order to determine the values of the boundary terms. We perform a complex deformation of $\ell$ ($i\epsilon$ prescription) for a small value $\epsilon \rightarrow 0^{+}$ in the appropriate direction of the imaginary axis depending on the sign of the potential at each point $q(x^{0})$:
\begin{equation}
\begin{split}
    U(\textrm{q}_{i},\textrm{q},\ell)&=\prod_{x^0\in(0,1)}\int_{-\infty}^{+\infty}d q(x^0)   \; e^{-{i\over \hbar} \scriptsize \big({1\over 2}|\gamma_{\textrm{qq}}|{\dot q^2\over \ell}+\ell\scriptsize \big[1-i\epsilon \, \text{sign}(V(q))\scriptsize \big] V(q)\scriptsize \big)}\\
    &=\prod_{x^0\in(0,1)}\int_{-\infty}^{+\infty}d q(x^0)   \; e^{-{i\over \hbar}{\scriptsize\big(}{1\over 2}|\gamma_{\textrm{qq}}|{\dot q^2\over \ell}+\ell V(q)\scriptsize \big)}e^{-{\epsilon\over \hbar}\ell |V(q)|}
\end{split}
\end{equation}

When $\ell \rightarrow +\infty$, the first exponent in the second line is ill-defined, but finite. The second exponential vanishes if we evaluate it before taking the limit $\epsilon \rightarrow 0^{+}$. If we consider the $\ell =0$ limit, we make the deformation strictly along the positive imaginary axis, therefore:
\begin{equation}
    U(\textrm{q}_{i},\textrm{q},0)=\prod_{x^0\in(0,1)}\int_{-\infty}^{+\infty}d q(x^0)   \; e^{-{i\over \hbar} \scriptsize \big({1\over 2}|\gamma_{\textrm{qq}}|{\dot q^2\over i\epsilon}+i\epsilon V(q)\scriptsize \big)}
\end{equation}
where we have implicitly ignored terms of order $\mathcal{O}(\epsilon ^{2})$ and above. For the limit $\epsilon \rightarrow 0^{+}$ this expression also vanishes. By extension the transition amplitude vanishes at the boundaries of the lhs in equation (\ref{bound}). The states $\psi$ vanish when $\widehat{\mathcal{H}}$ acts on them as expected, so the rhs of (\ref{bound}) is the WDW equation. This validates earlier assertions that $P^{\textrm{q}}$ corresponds to $\rho ^{\textrm{q}}$ and $\Omega$ corresponds to $\omega$. In covariant form, the derived equation becomes:
\begin{equation}
    -\frac{\hbar ^{2}}{2}\!\left[\nabla^2\psi+{\nabla^\textrm{q}\rho^\textrm{q}\over\rho^\textrm{q}}\nabla_{\!\textrm{q}}\psi+\omega\psi\right]\!+V\psi=0
\end{equation}
What we immediately notice is that the ambiguity functions as seen in (\ref{newamb}) are strictly real. The argument $\theta$ is therefore 0, making the vector quantity $Y^{\textrm{q}}$ in the equation for $\Psi = \sqrt{\mu} \, \psi$ also exactly 0. It also simplifies the value of $W$ to:
\begin{equation} \label{neww}
    W=\omega-{\nabla^2\!\sqrt{\mu}\over \sqrt{\mu}}
\end{equation}
The measure $\mu$ itself is found to be:
\begin{equation}
    \mu = \left (-\gamma ^{\textrm{qq}} \right )^{-1}
\end{equation}
which if we plug into (\ref{neww}) we get $W=0$. The result is that the WDW equation for the states $\Psi$ is now a completely ambiguity-free Klein-Gordon type equation in covariant form:
\begin{equation} \label{finaleq}
    -{\hbar^2\over 2} \nabla^2\Psi+V\Psi=0
\end{equation}
The calculated states from this equation are universal to all orders in the quantum theory as a result, and the quantum theory described for this system is unique. The various measurable quantities from the Hilbert space inner products provide only a single answer. \par
To ratify this result, we need to check whether the path integral measure in the early steps of the preceding analysis is time-reparametrization invariant in the continuum limit. Consider two frames for which the path integral measures are $\mathcal{D}q$ and $\mathcal{D}\tilde{q}$ respectively. Let the ambiguity function in the second frame be the value of $P^{\textrm{q}}$ as seen in (\ref{newamb}):
\begin{equation}
    \tilde \rho^{\tilde{\textrm{q}}}=(-\gamma_{\tilde{\textrm{q}}\tilde{\textrm{q}}})^{-{3\over 2}}
\end{equation}
which is connected to the function in the first frame via the transformation:
\begin{equation}
    \tilde \rho^\textrm{q}=\tilde \rho^{\tilde{\textrm{q}}}\, {d \textrm{q}\over d\tilde{\textrm{q}}}
\end{equation}
The metrics in the two frames are connected by the coordinate transformation:
\begin{equation}
    -\gamma_{\tilde{\textrm{q}}\tilde{\textrm{q}}}=(-\gamma_{\textrm{qq}})^\nu =-\gamma_{\textrm{qq}}\, \Big({d \textrm{q}\over d\tilde{\textrm{q}}}\Big)^2
\end{equation}
where $\nu$ is some real number. From this, the pullback becomes ${d\tilde{\textrm{q}}\over d\textrm{q}}=\pm (-\gamma_{\textrm{qq}})^{1-\nu\over 2}$ while the ambiguity function in the first frame is calculated as:
\begin{equation}
    \tilde \rho^\textrm{q}=\pm (-\gamma_{\textrm{qq}})^{-\nu-{1\over 2}}
\end{equation}
For $\nu = \frac{1}{4}$, the ambiguity function above is (up to a multiplicative constant) the same as the one derived via the path integral in the literature\cite{Reference31}:
\begin{equation} \label{pathamb}
    \rho^\textrm{q}_{*} =\mathcal{A}^{-{3\over 4}}\,  |\mathcal{A}'|^{-{3\over 2}} =\left(\!{-{\gamma_{\textrm{qq}}\over  6v_3}}\right)^{-{3\over 4}}
\end{equation}
This means that the path integral measure $\mathcal{D}q$ is the same as that which corresponds to the calculation in the cited paper, which we shall denote here as $\mathcal{D}q_{*}$. This proves that we may always find a frame of coordinates $q$ such that the transformation from the path integral result of our present work leads to $\mathcal{D}\tilde{q} = \mathcal{D}q_{*}$. Since $\mathcal{D}q_{*}$ is explicitly invariant under time-reparametrizations, the same can be said of the one we have used in our analysis.\par

\subsection{Explicit calculation}

As an addendum, it would be useful to provide an explicit calculation. We know there is an exact solution for the full equation in any frame, but it is not clear whether there is a way to retrieve it analytically. Therefore it will suffice to approximate it using the WKB method:
\begin{equation} \label{wkb}
    \psi = \exp{\left [ \frac{i}{\hbar} \left (\mathcal{F}_{0} + \frac{\hbar}{i} \mathcal{F}_{1} + \mathcal{O}(\hbar^{2} ) \right ) \right ]}
\end{equation}
where $\mathcal{F}_{0} , \, \mathcal{F}_{1}$ are functions of $\textrm{q}$. Anything of order $\mathcal{O}(\hbar^{2})$ is ignored for reasons that will be made clear shortly. Assuming a generic coordinate frame, we take that the ambiguities are generic. Plugging (\ref{wkb}) in equation (\ref{wdw2}):
\begin{equation} \label{wkb2}
    \left [-\frac{1}{\hbar^{2}} (\mathcal{F}_{0}^{'})^{2} + \frac{i}{\hbar} \left ( \mathcal{F}_{0}^{''} + 2\mathcal{F}_{0}^{'} \, \mathcal{F}_{1}^{'} + \frac{(\rho ^{\textrm{q}})'}{\rho ^{\textrm{q}}} \mathcal{F}_{0}^{'} \right ) + \mathcal{O}(\hbar ^{0}) \right ] \psi = -\frac{2 \gamma _{\textrm{qq}} V}{\hbar ^{2}} \psi 
\end{equation}
where primes indicate derivatives with respect to $\textrm{q}$ and we have also divided both sides by $\frac{\hbar^{2}}{2} \gamma ^{\textrm{qq}}$. It is evident why we ignored terms beyond $\mathcal{O}(\hbar^{2})$ in the WKB approximation: anything they contribute is of order $\mathcal{O}(\hbar^{0})$ and above in the WDW equation, corresponding to higher order corrections. We can derive a solution up to an arbitrary order in the quantum theory, but here we will adopt the semiclassical approximation $\hbar \rightarrow 0$, so we keep the terms up to $\mathcal{O}(\hbar ^{-1})$. From the equation above we get two differential equations for the respective orders in $\hbar$, giving two types of solutions based on the value of $-2 \gamma _{\textrm{qq}} V$. If it lies in the regime $-2 \gamma _{\textrm{qq}} V >0$, then the equations yield\cite{Reference31}:
\begin{equation} \label{1dresult}
    \mathcal{F}_{0} = \pm i \frac{2v_{3}}{\tilde{\ell}^{2}} \left [1 - (\tilde{\ell} \mathcal{A})^{2} \right ]^{\frac{3}{2}} + \text{constant}
\end{equation}
\begin{equation}
    \mathcal{F}_{1} = -\frac{1}{4} \log{\left (-2 \gamma _{\textrm{qq}} V \, \left (\rho ^{\textrm{q}} \right )^{2} \right )} + \text{constant}
\end{equation}
where we have written $\tilde{\ell} = \sqrt{\frac{\Lambda}{3}}$ to distinguish it more clearly from the length  of the line segment in the cosmological time gauge. This regime essentially corresponds to the classically allowed region $0 <\tilde{\ell}\mathcal{A} <1$. For the classically forbidden region $ \tilde{\ell}\mathcal{A} >1$ (or equivalently $-2 \gamma _{\textrm{qq}} V <0$) the solutions become\cite{Reference31}:
\begin{equation}
    \mathcal{F}_{0} = \pm \frac{2v_{3}}{\tilde{\ell}^{2}} \left [(\tilde{\ell} \textrm{q})^{2} -1 \right ]^{\frac{3}{2}} + \text{constant}
\end{equation}
\begin{equation}
    \mathcal{F}_{1} = -\frac{1}{4} \log{\left (2 \gamma _{\textrm{qq}} V \, \left (\rho ^{\textrm{q}} \right )^{2} \right )} + \text{constant}
\end{equation}

In both cases we notice that we get a linear combination of two independent solutions:
\begin{equation} \label{refsol1}
    \psi _{0 < \tilde{\ell}\mathcal{A} <1} = \sum _{s=\pm 1} \mathcal{N}_{s} \frac{\exp{\left \{ s\displaystyle\frac{2v_{3}}{\hbar \tilde{\ell}^{2}} \left [1 - (\tilde{\ell} \mathcal{A})^{2} \right ]^{\frac{3}{2}} \right \} }}{\left |\rho ^{\textrm{q}} \right |^{\frac{1}{2}} \mathcal{A}^{\frac{1}{2}} \left | \mathcal{A}' \right |^{\frac{1}{2}} \left [1 - (\tilde{\ell} \mathcal{A})^{2} \right ]^{\frac{1}{4}}} \, \left (1+ \mathcal{O}(\hbar) \right )
\end{equation}
\begin{equation} \label{refsol2}
    \psi _{\tilde{\ell}\mathcal{A} >1} = \sum _{s=\pm 1} \mathcal{N}_{s} \frac{\exp{\left \{ is\displaystyle\frac{2v_{3}}{\hbar \tilde{\ell}^{2}} \left [(\tilde{\ell} \mathcal{A})^{2} -1 \right ]^{\frac{3}{2}} \right \} }}{\left |\rho ^{\textrm{q}} \right |^{\frac{1}{2}} \mathcal{A}^{\frac{1}{2}} \left | \mathcal{A}' \right |^{\frac{1}{2}} \left [(\tilde{\ell} \mathcal{A})^{2} -1 \right ]^{\frac{1}{4}}} \, \left (1+ \mathcal{O}(\hbar) \right )
\end{equation}
where $\mathcal{N}_{s}$ the corresponding constant factors for each term (not to be confused with the normalization constants defined earlier in the discretization scheme). The main difference between the two regimes is that the solutions are real exponentials for the classically allowed region, but oscillatory for the forbidden one. What we notice further is that the ambiguity $\omega$ doesn't appear in these solutions. This was expected, given that it is part of the higher order $\mathcal{O}(\hbar^{0})$ terms in (\ref{wkb2}). The ambiguity $\rho ^{\textrm{q}}$ in turn appears only in the subleading term coming from the $\mathcal{O}(\hbar^{-1})$ differential equation. If we calculate the corresponding states $\Psi$ for either case, we see that $\rho ^{\textrm{q}}$ goes away as expected, leaving universal solutions:
\begin{equation}
    \Psi _{0 < \tilde{\ell}\mathcal{A} <1} = \sqrt{\mu}  \, \psi _{0 < \tilde{\ell}\mathcal{A} <1} = \sum _{s=\pm 1} \mathcal{N}_{s} \frac{\exp{\left \{ s\displaystyle\frac{2v_{3}}{\hbar \tilde{\ell}^{2}} \left [1 - (\tilde{\ell} \mathcal{A})^{2} \right ]^{\frac{3}{2}} \right \} }}{\mathcal{A}^{\frac{1}{4}} \left [1 - (\tilde{\ell} \mathcal{A})^{2} \right ]^{\frac{1}{4}}} \, \left (1+ \mathcal{O}(\hbar) \right )
\end{equation}
\begin{equation}
    \Psi _{\tilde{\ell}\mathcal{A} >1} =\sqrt{\mu} \,  \psi _{\tilde{\ell}\mathcal{A} >1} = \sum _{s=\pm 1} \mathcal{N}_{s} \frac{\exp{\left \{ is\displaystyle\frac{2v_{3}}{\hbar \tilde{\ell}^{2}} \left [(\tilde{\ell} \mathcal{A})^{2} -1 \right ]^{\frac{3}{2}} \right \} }}{\mathcal{A}^{\frac{1}{4}} \left [(\tilde{\ell} \mathcal{A})^{2} -1 \right ]^{\frac{1}{4}}} \, \left (1+ \mathcal{O}(\hbar) \right )
\end{equation}
where we used the derived expression (\ref{mu}) and absorbed constant terms in $\mathcal{N}_{s}$. We are mostly interested in the solutions for the classically allowed region for reasons that will be made clear in chapter 6. It is important to point out that the value of $s$ is crucial for determining whether a solution favours an early Universe undergoing inflation or not. This will also be explored in depth in the next chapter. \par

\pagebreak

\section{Generalization to $M$-dimensional minisuperspace}

\noindent We move on to generalize the formalism of the previous section for the most general minisuperspace model possible. We consider a gravitational action in $d+1$ spacetime dimensions, including homogeneous scalar matter fields that are minimally coupled to gravity\cite{Reference43}\cite{Reference64}:
\begin{equation} \label{eq:0}
    S = \frac{1}{2} \int \limits_{\mathcal{M}} d^{d +1}x \, \sqrt{-g} \, \left [R-2\Lambda - g^{00} \sum _{\alpha ,\beta} C_{\alpha \beta} \dot{\phi}^{\alpha} \dot{\phi}^{\beta} - 2V \left (\left \{\phi ^{\alpha} \right \} \right ) \right ].
\end{equation}
In the ADM formalism, the spacetime metric can be written as \cite{Reference22}\cite{Reference39}\cite{Reference47}:
\begin{equation} \label{eq:2}
    ds^{2} = -\left [N^{2}(t) - N_{i}(t)N^{i}(t) \right] dt^{2} +2N_{i}(t)dt \, dx^{i} + h_{ij}(t) dx^{i} dx^{j} 
\end{equation}
where $N$ is the lapse function, $N_{i}$ are shift factors and $h_{ij}$ stand for the spatial components of the metric. The scalar fields interact between themselves both through the potential, as well as the mixed kinetic terms with factors $C_{\alpha \beta} = C_{\beta \alpha}$. \par
While the Universe is closed and homogeneous, this time we allow for spatial anisotropies. The spatial metric $h_{ij}$ has $D={d(d+1)}/{2}$ independent degrees of freedom as a result. Combining these dof accordingly, we can find a quantity corresponding to the dilatation of the spatial manifold, with other combinations providing other features of the manifold\cite{Reference22}\cite{Reference61}\cite{Reference62}. These metric degrees of freedom are defined as:
\begin{equation} \label{eq:23}
\begin{split}
    &a(t) = f_{0} \left ( \{ h_{ij} \} \right ) \\
    &\eta^{k}(t) = f^{k} \left ( \{ h_{ij} \} \right ) \quad , \, k \in \{1,\dots, D-1\}.
\end{split}
\end{equation}
$a$ has a negative kinetic term in the Lagrangian, while the remaining $D-1$ degrees of freedom have positive kinetic terms\cite{Reference61}. If we then include the $m$ scalar matter fields $\phi ^{\alpha}$, we end up with $m+D-1\equiv M-1$ fields with positive kinetic terms. We arrange all the fields of the model:
\begin{equation} \label{eq:3}
    q^{A} =
\begin{cases}
    &a \, , \quad A = 0 \\
    &\eta^{k} \, , \quad A = k \in \left \{1,\dots, D-1 \right \} \\
    &\phi^{\alpha} \, , \quad A = \alpha + D-1 \in \left \{D,\dots, M-1 \right\}.
\end{cases}
\end{equation}
Notice that the lapse function $N$ and shift factors $N_{i}$ do not appear, since they are not dynamic variables on account of having kinetic terms equal to 0. This matches what we encountered in the 1-dimensional case i.e. these functions being a matter of gauge choice. In our case, we are free to use same gauge as before where $N = \ell \, , \, N_{i} = 0$ which corresponds to (\ref{line}) and (\ref{g00}). We shall examine how the quantum states are defined in due time. \par
We now derive the full potential of the theory:
\begin{equation} \label{eq:V}
    \mathcal{V} \left ( \left \{ q^{A} \right \} \right ) = \frac{1}{2} \int \limits _{\partial \mathcal{M}} d^{d}x \sqrt{-g} \, \left [ 2V \left (\left \{\phi ^{\alpha} \right \} \right ) + 2\Lambda - {}^{d}R\right ],
\end{equation}
where ${}^{d}R$ is the curvature on a constant time slice $\partial \mathcal{M}$ of metric $h_{ij}$. Just like in the 1-dimensional model, we may perform a generic field redefinition of all fields $q^{A}$:
\begin{equation} 
    q^{A} = Q^{A} \left ( \left \{ \tilde{q}^{\textrm{I}} \right \} \right )
\end{equation}
We retrieve a non-linear $\sigma$-model Lagrangian: 
\begin{equation} \label{eq:1}
    L = N\left[\frac{1}{2}\gamma _{\textrm{IJ}} \left(\left \{ \tilde{q}^{\textrm{I}} \right \} \right )\,  \dot{\tilde{q}}^{\textrm{I}} \dot{\tilde{q}}^{\textrm{J}} - \widetilde{V} \left (\left \{ \tilde{q}^{\textrm{I}} \right \} \right ) \right ]
\end{equation}
where $\gamma _{\textrm{IJ}}$ is the metric on the target space manifold $\mathcal{T}$ which is of Lorentzian signature, and $\widetilde{V}$ is the potential in the new frame. From the Lagrangian (\ref{eq:1}) we derive the conjugate momenta $\pi_{\textrm{I}}$ and the classical Hamiltonian:
\begin{equation} \label{eq:4}
    \frac{\mathcal{H}}{N} = \frac{1}{2} \gamma ^{\textrm{IJ}} \pi _{\textrm{I}} \pi _{\textrm{J}} + \widetilde{V} = 0
\end{equation}
Moving on to canonical quantization, we promote $\tilde{q}^I$ and $\pi_I$ to operators:
\begin{equation} \label{eq:6}
    \tilde{q}^{\textrm{I}} \rightarrow \textrm{q}^{\textrm{I}} \, , \qquad \pi _{\textrm{I}} \rightarrow -i\hbar \frac{\partial}{\partial \textrm{q}^{\textrm{I}}}
\end{equation}
and we impose the canonical commutation relations:
\begin{equation} \label{eq:5}
\begin{split}
    \begin{split}
    [\tilde{q}^{\textrm{I}},\, \pi _{\textrm{J}}] &= i\hbar \, \delta ^{\textrm{I}}{} _{\textrm{J}} \\
     [\tilde{q}^{\textrm{I}}, \tilde{q}^{\textrm{J}}] &= 0 \\
    [\pi _{\textrm{I}} , \pi _{\textrm{J}}] &= 0
\end{split}
\end{split}
\end{equation}
Since the metric $\gamma^{\textrm{IJ}}$ is a general function of all fields, it admits ordering ambiguities like before. The difference in the multidimensional case is that we could potentially introduce ambiguity functions of a higher tensorial class which would allow a variety of cross-terms. For example, consider the following\cite{Reference64}:
\begin{equation}
    \gamma ^{\textrm{IJ}} \pi _{\textrm{I}} \pi_{\textrm{J}} \rightarrow \gamma ^{\textrm{IJ}} \,T^{\textrm{CD}} \, S^{\textrm{KL}} \,\pi _{\textrm{C}} \, T_{\textrm{DI}} \,  \left (\pi_{\textrm{K}} \, S_{\textrm{LJ}} \right )
\end{equation}
where for the classical equivalency to hold we assume:
\begin{equation}
    T^{\textrm{CD}} T_{\textrm{DI}} = \delta ^{\textrm{C}}{}_{\textrm{I}} \quad , \quad S^{\textrm{KL}} S_{\textrm{LJ}} = \delta ^{\textrm{K}}{}_{\textrm{J}}
\end{equation}
However, this creates more terms than it is necessary. $\gamma ^{\textrm{IJ}}$ has $M^{2}$ terms, while each conjugate momentum has $M$ components. That means there are $M^{4}$ combinations (ignoring what's classically allowed due to contractions). To capture all operator reorderings in the quantum theory, we only need to introduce ambiguity functions for those $M^{4}$ terms. We choose:
\begin{equation} 
    \gamma ^{\textrm{IJ}} \pi _{\textrm{I}} \pi _{\textrm{J}} = \gamma^{\textrm{IJ}} \, s^{-1} \, v_{\textrm{J}} \, \Xi ^{\textrm{K}}{}_{\textrm{I}} \, \pi_{\textrm{K}} (v^{\textrm{L}} \, \pi _{\textrm{L}} \, s)
\end{equation}
where we impose the condition: 
\begin{equation}
     \gamma ^{\textrm{IJ}} \, v_{\textrm{J}} \, \Xi ^{\textrm{K}}{}_{\textrm{I}} \, v^{\textrm{L}} = v^{\textrm{I}} \, \Xi ^{\textrm{K}}{}_{\textrm{I}} \, v^{\textrm{L}} = \gamma ^{\textrm{KL}}
\end{equation}
Since the metric for a given system is known, we have $M^{2}$ equations for $M^{2} +M$ unknowns; $M^{2}$ from the quantity $\Xi ^{\textrm{K}}{}_{\textrm{I}}$ and another $M$ from $v^{\textrm{L}}$. Therefore only $M$ terms remain undetermined which we can arbitrarily choose. In total we have $M+1$ ambiguity terms due to the existence of the scalar $s$. Let us rewrite this more concisely using the following:
\begin{equation} \label{eq:8}
\begin{split}
    &\rho  ^{\textrm{I}} = v _{\alpha}^{\textrm{I}} s^{2} \quad , \quad \tilde{\rho} _{\textrm{J}} = \tilde{v}_{\textrm{J}} s^{-2} \, \, ,\quad \text{where} \quad \gamma ^{\textrm{IJ}}\tilde{\rho}_{\textrm{J}} \, \Xi ^{\textrm{K}}{}_{\textrm{I}} \, \rho^{\textrm{L}} = \gamma ^{\textrm{KL}} \\
    &\omega  = \gamma ^{\textrm{IJ}} s^{-1} \, v_{\textrm{J}} \, \Xi ^{\textrm{K}}{}_{\textrm{I}}\frac{\partial}{\partial \textrm{q}^{\textrm{K}}} \left (v^{\textrm{L}} \frac{\partial s}{\partial \textrm{q}^{\textrm{L}}} \right ) = s^{-1} \,v^\textrm{I} \, \Xi ^{\textrm{K}}{}_{\textrm{I}}\nabla_\textrm{K}(v^\textrm{L} \,\nabla_{\textrm{L}} s)
\end{split}
\end{equation}
Under field redefinitions these transform as:
\begin{equation}
    \check{\rho}^{\check{\textrm{I}}} = \rho ^{\textrm{I}} \, \frac{\partial \check{\tilde{q}} ^{\check{\textrm{I}}}}{\partial \tilde{q} ^{\textrm{I}}} \, \, , \qquad \check{\tilde{\rho}} _{\check{\textrm{I}}} = \tilde{\rho}_{\textrm{I}} \, \frac{\partial \tilde{q} ^{\textrm{I}}}{\partial \check{\tilde{q}} ^{\check{\textrm{I}}}} \, \, , \qquad \check{\omega} \left (\left \{\check{\textrm{q}}^{\check{\textrm{I}}} \right \}\right ) = \omega \left (\left \{\textrm{q}^{\textrm{I}} \right \}\right )
\end{equation}
The quantum Hamiltonian operator annihilates the states $\psi$, resulting in the WDW equation:
\begin{equation} \label{eq:10}
\begin{split}
    \frac{\widehat{\mathcal{H}}}{N} \psi &= -\frac{\hbar ^{2}}{2} \left [ \gamma ^{\textrm{IJ}}\,\tilde{\rho}_{\textrm{J}} \, \Xi ^{\textrm{K}}{}_{\textrm{I}} \, \frac{\partial}{\partial \textrm{q}^{\textrm{K}}} \left( \rho^{\textrm{L}} \frac{\partial \psi}{\partial \textrm{q}^{\textrm{L}}} \right ) + \omega \psi \right ] + \widetilde{V} \psi \\
    &= -\frac{\hbar ^{2}}{2} \left [ \gamma ^{\textrm{KL}}\frac{\partial ^{2} \psi}{\partial \textrm{q}^{\textrm{K}} \partial \textrm{q}^{\textrm{L}}} + \gamma ^{\textrm{IJ}} \, \tilde{\rho}_{\textrm{J}} \, \Xi ^{\textrm{K}}{}_{\textrm{I}} \left (\frac{\partial \rho ^{\textrm{L}}}{\partial \textrm{q}^{\textrm{K}}} \frac{\partial \psi}{\partial \textrm{q}^{\textrm{L}}} \right ) + \omega \psi \right ] + \widetilde{V} \psi \\
    &= -\frac{\hbar ^{2}}{2} \left [ \nabla^{2} \psi + \left (\gamma ^{\textrm{IJ}}\tilde{\rho}_{\textrm{J}} \, \Xi ^{\textrm{K}}{}_{\textrm{I}} \, \partial _{\textrm{K}} \rho ^{\textrm{L}} - \frac{\partial _{\textrm{J}} \left (\gamma ^{\textrm{JL}} \sqrt{-\gamma} \right )}{\sqrt{-\gamma}}\right )\nabla _{\textrm{L}} \psi + \omega \psi \right ] + \widetilde{V} \psi \\
    &\equiv -\frac{\hbar ^{2}}{2} \left [ \nabla^{2} \psi + \mathcal{P}^{\textrm{L}}\nabla _{\textrm{L}} \psi + \omega \psi \right ] + \widetilde{V} \psi = 0
\end{split}
\end{equation}
For $M=1$, we set $\Xi^{\textrm{q}}{}_{\textrm{q}} = \delta ^{\textrm{q}}{}_{\textrm{q}}$ while $v_{\textrm{q}} = \left(v^{\textrm{q}} \right )^{-1} \Rightarrow \tilde{v}^{\textrm{q}} = \gamma ^{\textrm{qq}} \left(v^{\textrm{q}} \right )^{-1}$ and $\tilde{\rho}_{\textrm{q}} = (\rho ^{\textrm{q}})^{-1}$, which satisfy all the conditions above. The states for the $M$-dimensional case are defined as a direct extension of (\ref{pathintegral2}) for $M$ fields $\tilde{q}^{\textrm{I}}$, with an initial state $\textrm{q}^{\textrm{I}}_{i}$ and final state $\textrm{q}^{\textrm{I}}$\cite{Reference64}: \pagebreak
\begin{equation} 
    \psi\left( \left \{\textrm{q}^{\textrm{I}} \right \} \right)=\int_0^{+\infty}\!\!d\ell \; \Delta_{\rm FP}(\ell) \prod _{\textrm{I} = 0}^{M}\int_{\tilde{q}^{\textrm{I}}(0)=\textrm{q}^{\textrm{I}}_i}^{\tilde{q}(1)=\textrm{q}^{\textrm{I}}}\mathcal{D} \tilde{q}^{\textrm{I}}   \; e^{{i\over \hbar}\int_{0}^{1}d x^0\, L \left (\ell, \left \{ \tilde{q}^{\textrm{I}} \right \}, \left \{\dot{\tilde{q}}^{\textrm{I}} \right \} \right)}
\end{equation}
It follows trivially then that the ambiguities in the path integral picture are once again emerging from the multiple ways we may choose a path integral measure $\prod_{\textrm{I}}\mathcal{D}\tilde{q}^{\textrm{I}}$ after a field redefinition. It is now straightforward to define a Hilbert space inner product for a specific choice of path integral measure $\mathcal{D}\tilde{q}$\cite{Reference64}:
\begin{equation} \label{eq:11}
    \langle \psi_{1}  ,  \psi_{2} \rangle _{\mathcal{D}\tilde{q}} = \int \limits_{\mathcal{T}} d^{M} \textrm{q} \, \sqrt{-\gamma } \, \mu \left (\left \{ \textrm{q}^{\textrm{I}}\right \} \right) \, \psi_{1} ^{*} \psi_{2}
\end{equation}
where we again define a positive measure function $\mu$. There's no a priori reason to assume that the inner products between two different choices of path integral measure will lead to the same result, which is the same issue of the non-uniqueness of the quantum theory we saw earlier. We will have to investigate this in a similar manner. We demand that the Hamiltonian operator is hermitian, so for two generic complex functions $f_{1}, \, f_{2}$ we have:
\begin{equation} \label{eq:12}
     \Bigg \langle f_{1} \, , \, \frac{\widehat{\mathcal{H}}}{N} f_{2} \Bigg \rangle = \Bigg \langle \frac{\widehat{\mathcal{H}}^{\dagger}}{N} f_{1} \, , \, f_{2} \Bigg \rangle + {\rm boundary\, terms}
\end{equation}
where the boundary terms have to vanish, and the Hamiltonian and its hermitian conjugate must be algebraically the same: 
\begin{equation} \label{eq:13}
    \frac{\widehat{\mathcal{H}}}{N} f = \frac{\widehat{\mathcal{H}}^{\dagger}}{N} f
\end{equation}
Using integration by parts, we obtain an expression for the hermitian conjugate:
\begin{equation} \label{eq:14}
    \frac{\widehat{\mathcal{H}}^{\dagger}}{N} f = -\frac{\hbar ^{2}}{2}\left \{\frac{1}{\mu \sqrt{-\gamma }} \frac{\partial}{\partial \textrm{q}^{\textrm{L}}} \left [\rho ^{\textrm{L}*} \frac{\partial}{\partial \textrm{q}^{\textrm{K}}}  \left (\gamma ^{\textrm{IJ}} \, \tilde{\rho} _{\textrm{J}}^{*} \,\Xi ^{\textrm{K}*}{}_{\textrm{I}}\,\mu \sqrt{-\gamma } \,f \right ) \right] + \omega ^{*} f \right \} + \widetilde{V} f 
\end{equation}
The boundary terms can once again be shown to vanish with the method outlined in Appendix A. For the application of (\ref{eq:13}), we shall analyze the ambiguity functions as in the 1-dimensional case. Namely, each ambiguity can be expressed as:
\begin{equation} \label{eq:14'}
\begin{split}
    \rho ^{\textrm{I}} = r^{\textrm{I}} \, e^{i\theta} \, , & \quad \tilde{\rho}_{\textrm{J}} = \tilde{r}_{\textrm{J}} \, e^{-i\theta} \, , \quad \Xi ^{\textrm{K}*}{}_{\textrm{I}} = \Xi ^{\textrm{K}}{}_{\textrm{I}}\\
    \text{where} \quad r^{\textrm{I}} , \, \tilde{r}_{\textrm{J}} \in \mathbb{R} \, \, \, &, \quad  \gamma^{\textrm{IJ}} \, \tilde{r}_{\textrm{J}} \, \Xi ^{\textrm{K}}{}_{\textrm{I}} \, r^{\textrm{L}}= \gamma ^{\textrm{KL}} \, \, , \quad \theta \in \left (-\frac{\pi}{2} , \frac{\pi}{2} \right ]
\end{split}
\end{equation}
$r^{\textrm{I}}, \, \tilde{r}_{\textrm{J}}$ must be real due to the diffeomorphisms of $\mathcal{T}$ being strictly real. By extension, the imaginary component of the ambiguities is given by a phase involving the scalar $\theta$. The fields of definition for these quantities are chosen to ensure invertibility in the mapping between the two frames, and $e^{\pm i\theta}$ must be a global phase in order to respect the transformation properties of $\rho ^{\textrm{I}} , \, \tilde{\rho}_{\textrm{J}}$. The quantity $\Xi ^{\textrm{K}}{}_{\textrm{I}}$ is taken to be real because, without loss of generality, we can absorb its complex phase into the $\rho$ and $\tilde{\rho}$ functions. Plugging these into (\ref{eq:14}) and applying (\ref{eq:13}):
\begin{equation} \label{eq:constraints1}
\begin{split}
    &\Bigg[\gamma^{\textrm{IJ}}\,\tilde{r}_\textrm{J} \, \Xi ^{\textrm{K}}{}_{\textrm{I}} \, \partial_{\textrm{K}}(r^{\textrm{L}}) -\partial_{\textrm{K}}(\gamma^{\textrm{IJ}}\,\tilde{r}_\textrm{J} \,\Xi ^{\textrm{K}}{}_{\textrm{I}}) \, r^{\textrm{L}} - \partial_{\textrm{K}}(\gamma^{\textrm{IJ}}\,\tilde{r}_\textrm{J} \,\Xi ^{\textrm{L}}{}_{\textrm{I}}) \, r^{\textrm{K}} - \gamma^{\textrm{IJ}}\,\tilde{r}_\textrm{J} \, \Xi ^{\textrm{L}}{}_{\textrm{I}}\, \partial _{\textrm{K}}(r^{\textrm{K}})\\
    &\qquad \qquad \qquad \qquad \qquad \qquad \qquad \qquad \qquad \qquad \qquad \quad-2\gamma ^{\textrm{KL}} \frac{\partial _{\textrm{K}}\Big (\sqrt{-\gamma}\,\mu \Big )}{\sqrt{-\gamma}\,\mu}\Bigg]\!\partial _{\textrm{L}}f \\
    = &\!\left[-2i \, \text{Im}\omega + {\rho^{\textrm{L}*}\partial _{\textrm{L}} \partial _{\textrm{K}}\displaystyle \Big({\gamma^{\textrm{IJ}} \tilde{\rho} _{\textrm{J}}^{*} \, \Xi ^{\textrm{K}}{}_{\textrm{I}}}\sqrt{-\gamma}\, \mu\Big) + {\partial _{\textrm{L}}(\rho^{\textrm{L} *})}\partial _{\textrm{K}}\Big({\gamma^{\textrm{IJ}} \tilde{\rho}_{\textrm{J}}^{*}} \, \Xi ^{\textrm{K}}{}_{\textrm{I}}\sqrt{-\gamma}\, \mu\Big)\over \displaystyle \sqrt{-\gamma}\, \mu}\right]\!f
\end{split}
\end{equation}
In order for this to hold for any function $f$, the brackets on either side must vanish. From the lhs we obtain a constraint for the measure $\mu$\cite{Reference64}:
\begin{equation} \label{eq:constraints2}
\begin{split}
    \frac{\nabla ^{\textrm{L}} \mu}{\mu} &= \frac{1}{2}\left [\gamma^{\textrm{IJ}}\tilde{r}_\textrm{J} \, \Xi ^{\textrm{K}}{}_{\textrm{I}} \, \partial_{\textrm{K}}(r^{\textrm{L}}) -\partial_{\textrm{K}}(\gamma^{\textrm{IJ}}\tilde{r}_\textrm{J} \,\Xi ^{\textrm{K}}{}_{\textrm{I}}) \, r^{\textrm{L}} - \partial_{\textrm{K}}(\gamma^{\textrm{IJ}}\tilde{r}_\textrm{J} \,\Xi ^{\textrm{L}}{}_{\textrm{I}}) \, r^{\textrm{K}} - \gamma^{\textrm{IJ}}\tilde{r}_\textrm{J} \, \Xi ^{\textrm{L}}{}_{\textrm{I}}\, \partial _{\textrm{K}}(r^{\textrm{K}}) \right ] \\
    &\qquad \qquad \qquad \qquad \qquad \qquad \qquad \qquad \qquad \qquad \qquad \qquad \qquad \qquad- \gamma ^{\textrm{KL}} \frac{\partial _{\textrm{K}}\Big (\sqrt{-\gamma} \Big )}{\sqrt{-\gamma}}\\
    &= \frac{1}{2} \left [2 \, \gamma^{\textrm{IJ}}\tilde{r}_\textrm{J} \, \Xi ^{\textrm{K}}{}_{\textrm{I}} \, \partial_{\textrm{K}}(r^{\textrm{L}}) -\partial_{\textrm{K}}(\gamma^{\textrm{IJ}}\tilde{r}_\textrm{J} \,\Xi ^{\textrm{K}}{}_{\textrm{I}} \, r^{\textrm{L}}) - \partial_{\textrm{K}}(\gamma^{\textrm{IJ}}\tilde{r}_\textrm{J} \,\Xi ^{\textrm{L}}{}_{\textrm{I}} \, r^{\textrm{K}}) \right ] - \gamma ^{\textrm{KL}} \frac{\partial _{\textrm{K}}\Big (\sqrt{-\gamma} \Big )}{\sqrt{-\gamma}} \\
    &= \gamma^{\textrm{IJ}}\,\tilde{r}_\textrm{J} \, \Xi ^{\textrm{K}}{}_{\textrm{I}} \, \partial_{\textrm{K}}(r^{\textrm{L}}) - \partial _{\textrm{K}}\gamma ^{\textrm{KL}}- \gamma ^{\textrm{KL}} \frac{\partial _{\textrm{K}}\Big (\sqrt{-\gamma} \Big )}{\sqrt{-\gamma}} \\
    &\equiv \Pi^{\textrm{L}} - \frac{\partial _{\textrm{K}} \Big (\gamma ^{\textrm{KL}} \sqrt{-\gamma} \Big )}{\sqrt{-\gamma}} \equiv \tilde{\Pi}^{\textrm{L}}
\end{split}
\end{equation}
$\tilde{\Pi}^{\textrm{K}}$ is none other than the real part of $\mathcal{P}^{\textrm{K}}$ defined back in (\ref{eq:10}). Solving for $\mu$ we get:
\begin{equation} \label{eq:15}
   \mu = \frac{\mathcal{C}}{\sqrt{-\gamma}} \exp{\left \{ \int d\textrm{q}^{\textrm{K}} \, \gamma _{\textrm{KL}} \Big ( \Pi ^{\textrm{L}} - \partial _{\textrm{M}} \gamma ^{\textrm{LM}} \Big )\right \}}
\end{equation}
where $\mathcal{C}$ an arbitrary positive integration constant which we can set equal to 1. Using this constraint in the rhs of equation (\ref{eq:constraints1}) and demanding that this bracket should also be 0, we find two additional constraints. One comes from the imaginary terms: \pagebreak
\begin{equation} \label{eq:constraints3}
\begin{split}
    2\,\text{Im}\omega &= \gamma^{\textrm{IJ}} \partial _{\textrm{I}} \partial _{\textrm{J}} \theta + \Pi ^{\textrm{I}} \, \partial _{\textrm{I}}\theta + \frac{r^{\textrm{L}} \partial _{\textrm{K}}\theta}{\sqrt{-\gamma} \, \mu} \partial _{\textrm{L}} \left (\gamma ^{\textrm{IJ}} \, \tilde{r}_{\textrm{J}} \, \Xi ^{\textrm{K}}{}_{\textrm{I}} \, \sqrt{-\gamma} \, \mu \right) \\
    &=\gamma^{\textrm{IJ}} \partial _{\textrm{I}} \partial _{\textrm{J}}  \theta + \Pi ^{\textrm{I}} \, \partial _{\textrm{I}}\theta +  \frac{\nabla ^{\textrm{K}} \mu}{\mu} \partial _{\textrm{K}}\theta - \Pi ^{\textrm{K}} \, \partial _{\textrm{K}}\theta + \frac{\partial _{\textrm{L}} \Big (\gamma ^{\textrm{KL}} \sqrt{-\gamma} \Big )}{\sqrt{-\gamma}} \partial _{\textrm{K}}\theta\\
    &= \nabla ^{2} \theta + \tilde{\Pi}^{\textrm{I}} \, \partial _{\textrm{I}} \theta 
\end{split}
\end{equation}
and another from the real terms:
\begin{equation}
    \frac{1}{\sqrt{-\gamma} \, \mu} \partial _{\textrm{L}} \left [ r^{\textrm{L}} \partial _{\textrm{K}} \left ( \gamma ^{\textrm{IJ}} \, \tilde{r}_{\textrm{J}} \, \Xi ^{\textrm{K}}{}_{\textrm{I}} \, \sqrt{-\gamma} \, \mu \right ) \right ] = 0
\end{equation}
The constraints (\ref{eq:15}) and (\ref{eq:constraints3}) are generalizations of the 1-dimensional analogues (\ref{mu}) and (\ref{omega}) respectively, and it is straightforward to reproduce them from the currently obtained relations. At this stage we again choose to define the new states $\Psi = \sqrt{\mu} \, \psi$ such that the Hilbert inner product can written as:
\begin{equation}
     \langle \psi_{1} \, , \, \psi_{2} \rangle _{\mathcal{D}\tilde{q}} = \int \limits_{\mathcal{T}} d^{M} \textrm{q} \, \sqrt{-\gamma }  \, \Psi_{1} ^{*}\, \Psi_{2}
\end{equation}
Therefore the uniqueness of the quantum theory depends on these states satisfying a WDW equation without ambiguities. Applying the Hamiltonian operator upon $\Psi$ we find:
\begin{equation} \label{Mdimwdw}
    \sqrt{\mu}\,{\widehat{\mathcal{H}}\over N}\,\psi\equiv-{\hbar^2\over 2}\big[\nabla^2\Psi+Y^\textrm{K} \nabla_{\!\textrm{K}}\Psi+W\Psi\big]\!+\widetilde{V}\Psi = 0
\end{equation}
where we have defined:
\begin{equation}
    Y^{\textrm{K}} = \gamma ^{\textrm{IJ}}\tilde{\rho}_{\textrm{J}} \, \Xi ^{\textrm{K}}{}_{\textrm{I}} \, \partial _{\textrm{K}} \rho ^{\textrm{L}} -{\partial _{\textrm{I}}(\gamma^{\textrm{IK}}\sqrt{-\gamma\mu})\over \sqrt{-\gamma\mu}}-\gamma^{\textrm{IK}}\, {\nabla_{\!\textrm{I}}\sqrt{\mu}\over \sqrt{\mu}}  
\end{equation}
\begin{equation}
    W = \omega - {\nabla^2\!\sqrt{\mu}\over \sqrt{\mu}}-Y^{\textrm{K}} \,{\nabla_{\!\textrm{K}}\sqrt\mu\over \sqrt{\mu}}
\end{equation}
We first use the constraint (\ref{eq:constraints2}) to simplify the vector function:
\begin{equation}
    Y^{\textrm{K}} = i \, \nabla ^{\textrm{K}}\theta
\end{equation}
and then we can use it in conjunction with (\ref{eq:constraints3}) to simplify the scalar:
\begin{equation}
    W = \text{Re}\omega - \frac{1}{2}\nabla _{\textrm{I}} \,\tilde{\Pi}^{\textrm{I}} - \frac{1}{4} \gamma _{\textrm{IJ}} \tilde{\Pi}^{\textrm{I}} \tilde{\Pi}^{\textrm{J}} + \frac{i}{2} \nabla ^{2} \theta
\end{equation}
where we have used: \pagebreak
\begin{equation}
    \frac{\nabla ^{2} \sqrt{\mu}}{\sqrt{\mu}} = \nabla _{\textrm{I}} \left ( \frac{\nabla ^{\textrm{I}} \sqrt{\mu}}{\sqrt{\mu}} \right ) + \gamma _{\textrm{IJ}} \left (\frac{\nabla ^{\textrm{I}} \sqrt{\mu}}{\sqrt{\mu}} \right ) \left (\frac{\nabla ^{\textrm{J}} \sqrt{\mu}}{\sqrt{\mu}} \right )
\end{equation}
The resulting WDW equation after these simplifications is\cite{Reference64}:
\begin{equation}
    -{\hbar^2\over 2}\left [\nabla^2\Psi+i\, \nabla^{\textrm{K}}\theta \,  \nabla_{\!\textrm{K}}\Psi+ \left (\text{Re}\omega - \frac{1}{2}\nabla _{\textrm{I}} \,\tilde{\Pi}^{\textrm{I}} - \frac{1}{4} \gamma _{\textrm{IJ}} \tilde{\Pi}^{\textrm{I}} \tilde{\Pi}^{\textrm{J}} \right )\Psi + \frac{i}{2} \nabla ^{2} \theta \, \Psi \right ]\!+\widetilde{V}\Psi = 0
\end{equation}
At this stage in the 1-dimensional model we moved to the path integral analysis to determine the ambiguity functions which helped us derive the values of $Y^{\textrm{q}}$ and $W$. This method, however, is not applicable for a general curved $M$-dimensional manifold. This has to do with the need to have a constant metric appearing in the exponent of (\ref{transamp}) which would allow the calculation of the Gaussian path integrals. Even though we can always find a coordinate transformation such that $\mathcal{T}$ is locally flat, we end up having to integrate an infinite series of terms with ever increasing powers of $\varepsilon ^{-1}$ as $\varepsilon \rightarrow 0$: 
\begin{equation}
    U^{(n)}\left (\vec{\textrm{Q}}_{i}, \vec{\textrm{Q}}, \ell \right ) = \frac{1}{\mathcal{N}} \int d^{M}\tilde{q} \, \exp{\left \{\frac{i}{2\varepsilon \hbar} \gamma _{\textrm{IJ}}(\vec{\textrm{Q}}) \left (\tilde{q}^{\textrm{I}} - \textrm{Q}^{\textrm{I}} \right )\left (\tilde{q}^{\textrm{J}} - \textrm{Q}^{\textrm{J}} \right ) \right \}} \, f
\end{equation}
where $\vec{\textrm{Q}}_{i} ,\, \vec{\textrm{Q}}$ the initial and final points respectively, expanding around the latter using a mid-point discretization:
\begin{equation}
\begin{split}
    \gamma _{\textrm{IJ}}\left (\frac{\vec{\tilde{q}} +\vec{\tilde{Q}}}{2} \right ) = \gamma _{\textrm{IJ}} (\vec{\textrm{Q}}) &+ \frac{1}{2} \left (\tilde{q}^{\textrm{K}} - \textrm{Q}^{\textrm{K}} \right ) \frac{\partial \gamma _{\textrm{IJ}}}{\partial \tilde{q}^{\textrm{K}}}\Bigg |_{\vec{\tilde{q}}=\vec{\textrm{Q}}} \\
    &+ \frac{1}{8} \left (\tilde{q}^{\textrm{K}} - \textrm{Q}^{\textrm{K}} \right )\left (\tilde{q}^{\textrm{L}} - \textrm{Q}^{\textrm{L}} \right ) \frac{\partial ^{2} \gamma _{\textrm{IJ}}}{\partial \tilde{q}^{\textrm{K}} \partial \tilde{q}^{\textrm{L}}}\Bigg |_{\vec{\tilde{q}}=\vec{\textrm{Q}}} + \, ...
\end{split}
\end{equation}
and $f$ contains the rest of the terms which complete the transition amplitude. Summing the resulting Gaussian integrations term by term is not mathematically rigorous, so we cannot follow with the rest of the analysis. We can make some reasonable assumptions, though. Namely, since the metric of $\mathcal{T}$ and any coordinate transformation are strictly real, it is expected that the ambiguities emerging from the path integral should be real. Therefore:
\begin{equation}
    Y^{\textrm{K}} = 0 \, \, \, , \qquad  W = \omega - \frac{1}{2}\nabla _{\textrm{I}} \,\tilde{\Pi}^{\textrm{I}} - \frac{1}{4} \gamma _{\textrm{IJ}} \tilde{\Pi}^{\textrm{I}} \tilde{\Pi}^{\textrm{J}}
\end{equation}
which brings the WDW equation to the form:
\begin{equation}
    -{\hbar^2\over 2}\left [\nabla^2\Psi + \left (\omega - \frac{1}{2}\nabla _{\textrm{I}} \,\tilde{\Pi}^{\textrm{I}} - \frac{1}{4} \gamma _{\textrm{IJ}} \tilde{\Pi}^{\textrm{I}} \tilde{\Pi}^{\textrm{J}} \right )\Psi \right ]\!+\widetilde{V}\Psi = 0
\end{equation}
Even though we still have no way to determine $W$ exactly and see if it is free of ambiguities, we notice that at the semi-classical limit $\hbar \rightarrow 0$ this term does not contribute. The semi-classical WDW equation then is just the $M$-dimensional equivalent of (\ref{finaleq}):
\begin{equation}
     \frac{\widehat{\mathcal{H}}}{N} \Psi \stackrel{\mathclap{\normalfont\mbox{$\hbar \rightarrow 0$}}}{=\joinrel=\joinrel=\joinrel=} -{\hbar^2\over 2}\nabla^2\Psi +\widetilde{V}\Psi = 0
\end{equation}
This means that, at least at the semi-classical level, the Hilbert space inner products for a given system are universal and therefore the quantum theory is unique. In the literature it has been suggested that $W \propto \mathcal{R}$, where $\mathcal{R}$ the Ricci scalar of the manifold $\mathcal{T}$\cite{Reference48}. While this is a natural choice, we cannot corroborate it using the methodology we have developed. We hence leave it as an open possibility which would be interesting to revisit in the future for an exact calculation of this higher order term.

\subsection{Universality in the special case of flat manifolds}

Despite the limitations explained above when tackling the more general curved $M$-dimensional models, we can still apply it in the case of flat manifolds of arbitrary dimensionality i.e. those with a vanishing Riemann curvature tensor. For such manifolds we can always find a coordinate transformation such that the metric becomes the Minkowski metric $\eta _{\textrm{AB}}$. In this frame, the problem of the infinite series of Gaussian integrals does not exist, hence we can carry on the calculations normally. \par
We define the transition amplitude for a generic metric $\gamma_{\textrm{IJ}}$\cite{Reference64}:
\begin{equation}
\begin{split}
    U(\vec{\textrm{q}}_{i},\vec{\textrm{q}},\ell)&=\prod_{\textrm{I}}\int_{\textrm{q}^{\textrm{I}}_i}^{\textrm{q}^{\textrm{I}}}\mathcal{D} q^{\textrm{I}}   \; e^{{i\over \hbar}\int_{0}^{1}d x^0\, L(\ell,\vec{q},\vec{\dot{q}})}\\
    &=\prod_{\textrm{I}}\int_{\textrm{q}^{\textrm{I}}_i}^{\textrm{q}^{\textrm{I}}}\mathcal{D} q^{\textrm{I}}   \; e^{{i\over \hbar}\int_{0}^
    \ell d t\,  \scriptsize \big(\gamma_{\textrm{IJ}}(\vec{q})({d q^{\textrm{I}}\over d t})({d q^{\textrm{J}}\over d t})-\widetilde{V}(\vec{q}) \scriptsize \big)} 
\end{split}
\end{equation}
where we again apply the gauge $N = \ell \, , \, \ell \in [0,1]$, and for the sake of brevity we skip the tilde notation for $q$. We perform the aforementioned field redefinition from coordinates $q^{\textrm{I}}$ to $q_{0}^{\textrm{A}}$ with mapping $\mathcal{Q}^{\textrm{I}}(\vec{q}_{0})$ such that:
\begin{equation}
    \gamma _{\textrm{IJ}}(\vec{q})  \rightarrow \eta _{\textrm{AB}} = \gamma _{\textrm{IJ}}\left(\vec{\mathcal{Q}}(\vec{q}_{0}) \right) \, \frac{\partial \mathcal{Q}^{\textrm{I}}}{\partial q^{\textrm{A}}_{0}} \frac{\partial \mathcal{Q}^{\textrm{J}}}{\partial q^{\textrm{B}}_{0}}
\end{equation}
and a field of definition $q^{\textrm{A}}_{0} \in [q^{\textrm{A}}_{0-} , q^{\textrm{A}}_{0+}]$. Note that the boundary values $q^{\textrm{A}}_{0\pm}$ in each direction have to be chosen so that the overall mapping is well-defined and by extension invertible. Said inverse mapping is in turn defined as:
\begin{equation}
    q^{\textrm{A}}_{0} =\Big(\mathcal{Q}^{-1} \Big)^{\textrm{A}}\left (\vec{\mathcal{Q}}(\vec{q}_{0}) \right) \equiv \textrm{Q}^{\textrm{A}}_{0}(\vec{q})
\end{equation}
The initial and final points in each direction are:
\begin{equation}
    \vec{\textrm{q}}_0 = \vec{\textrm{Q}}_0(\vec{\textrm{q}})\, , \qquad \vec{\textrm{q}}_{0i}= \vec{\textrm{Q}}_0(\vec{\textrm{q}}_i)
\end{equation}
and the scalar potential in this new coordinates becomes:
\begin{equation}
    \widetilde{V}_{0}(\vec{q}_{0}) = \widetilde{V}\left (\vec{\mathcal{Q}}(\vec{q}_{0}) \right )
\end{equation}
We finally get the new transition amplitude in flat coordinates:
\begin{equation} 
    U(\vec{\textrm{q}}_{i},\vec{\textrm{q}},\ell)= \prod_{\textrm{A}}\int_{\textrm{q}^{\textrm{A}}_{0i}}^{\textrm{q}^{\textrm{A}}_0}\mathcal{D} q^{\textrm{A}}_0 \;\mathcal{J}_0  \; e^{{i\over \hbar}\int_{0}^
    \ell d t\, \left(\displaystyle {1\over 2}\eta _{\textrm{AB}}\left ({d q^{\textrm{A}}_0\over d t} \right ) \left ({d q^{\textrm{B}}_0\over d t} \right )+ \widetilde{V}_0(\vec{q}_{0})\!\right)} \equiv U_0(\vec{\textrm{q}}_{0i},\vec{\textrm{q}}_0,\ell)
\end{equation}
The Jacobian determinant in this model is now not a single element like in 1 dimension: 
\begin{equation} \label{newjacobian}
    \mathcal{J}_0=\prod_{t\in[0,\ell]}J \big(\vec{q}_{0}(t) \big) = \prod_{t \in [0,\ell]} \left|\text{det} \,\textrm{J}^{\textrm{I}}{}_{\textrm{A}}\big (\vec{q}_{0}(t) \big ) \right | \, \,, \qquad \textrm{J}^{\textrm{I}}{}_{\textrm{A}} \big (\vec{q}_{0}(t) \big ) \equiv \left (\frac{\partial q^{\textrm{I}}(t)}{\partial q^{\textrm{A}}_{0}(t)} \right )
\end{equation}
We follow the same discretization scheme as before, dividing the time interval in $n$ segments of length $\varepsilon$ and performing mid-point discretization for the potential $\widetilde{V}_{0}$:
\begin{equation}
\begin{split}
    U_0^{(n)}(\vec{\textrm{q}}_{0i},\vec{\textrm{q}}_0,\ell)=&{1\over \mathcal{N}(\varepsilon,\vec{\textrm{q}}_{0i},\vec{\textrm{q}}_0)} \prod_{k'=1}^{n-1} \left(\prod _{\textrm{A}}\int_{\textrm{q}^{\textrm{A}}_{0-}}^{\textrm{q}^{\textrm{A}}_{0+}}{d q_{0k'}\over \mathcal{N}(\varepsilon,\vec{\textrm{q}}_{0i},\vec{\textrm{q}}_0)} \, J(\vec{q}_{0k'})\right)\\
    &\!\times \exp{\Bigg \{{{i\over \hbar}\sum_{k=0}^{n-1}\varepsilon \Big[{1\over 2}\eta _{\textrm{AB}}\left(\displaystyle{q^{\textrm{A}}_{0(k+1)}-q^{\textrm{A}}_{0k}\over \varepsilon}\right) \left (\displaystyle{q^{\textrm{B}}_{0(k+1)}-q^{\textrm{B}}_{0k}\over \varepsilon}\right ) - \widetilde{V}_0\textstyle({\vec{q}_{0k}+\vec{q}_{0(k+1)}\over 2})\Big]}\Bigg \}}
\end{split}
\end{equation}
Now we isolate the last slice of the segment:
\begin{equation} 
\begin{split}
    U_0^{(n)}(\vec{\textrm{q}}_{0i},\vec{\textrm{q}}_0,\ell)= \prod _{\textrm{A}}\int_{\textrm{q}^{\textrm{A}}_{0-}}^{\textrm{q}^{\textrm{A}}_{0+}}{d q^{\textrm{A}}_{0(n-1)}\over \mathcal{N}(\varepsilon,\textrm{q}_{0i},\textrm{q}_0)} \, J(\vec{q}_{0(n-1)})\, e^{-{i\over 2\hbar\varepsilon}(\textrm{q}^{0}_{0}-q^{0}_{0(n-1)})^2}  \,& e^{{i\over 2\hbar\varepsilon}\sum _{k=1}^{M-1}(\textrm{q}^{k}_{0}-q^{k}_{0(n-1)})^2} \\ 
    \times\, e^{-{i\over \hbar}\varepsilon V_0{\textstyle(}{\vec{q}_{0(n-1)}+\vec{\textrm{q}}_{0}\over 2}{\textstyle)}} \, &   U_{0}^{(n-1)}(\vec{\textrm{q}}_{0i},\vec{q}_{0(n-1)},\ell-\varepsilon)
\end{split}
\end{equation}
We may Taylor-expand the various quantities in the integrand in powers of $\varepsilon$ around the final point $\vec{\textrm{q}}_{0}$ for each direction, which will lead us to a number of Gaussian integrals. We again need to introduce the regularization $i\kappa \, , \, \kappa >0$ in the exponentials of each direction for them to converge, and we recall the generalized integral identities\cite{Reference64}:
\begin{equation}
    \int d^{M} q_{0}\, e^{\alpha \, \eta _{\textrm{AB}} \, q^{\textrm{A}}_{0} q^{\textrm{B}}_{0}}=\left (\pi\over \alpha \right )^{M/2}\, , \qquad \int d^{M} q_{0}\, \,q^{\textrm{A}}_{0} q^{\textrm{B}}_{0}\, e^{\alpha \, \eta _{\textrm{AB}} \, q^{\textrm{A}}_{0} q^{\textrm{B}}_{0}}=-{1\over 2\alpha} \left (\pi\over \alpha \right )^{M/2} \eta ^{\textrm{AB}}
\end{equation}
where we have $\alpha = \frac{i}{2\hbar \varepsilon}(1-i\kappa)$. Keeping terms up to order $\varepsilon$, we find:
\begin{equation} \label{Mdim}
    \begin{split}
U_0^{(n)}(\vec{\textrm{q}}_{0i},\vec{\textrm{q}}_0,\ell)={1\over \mathcal{N}(\varepsilon,\vec{\textrm{q}}_{0i},\vec{\textrm{q}}_0)} &\left (2\pi\hbar\varepsilon\over i \right )^{M/2}\,\Big(1-i\,{\varepsilon\over \hbar}\, \widetilde{V}_0(\vec{\textrm{q}}_0)-{\hbar\varepsilon\over 2i}\,\eta ^{\textrm{AB}}{\partial ^{2} \over \partial q_{0(n-1)}^{\textrm{A}} q_{0(n-1)}^{\textrm{B}}} +{\mathcal{O}}(\varepsilon^2)\Big)\\
&~~~~~~\times\big[J(\vec{q}_{0(n-1)})\, U_0^{(n-1)}(\vec{\textrm{q}}_{0i},\vec{q}_{0(n-1)},\ell-\varepsilon)\big]\Big|_{\vec{q}_{0(n-1)}=\vec{\textrm{q}}_0}
\end{split}
\end{equation}
If we examine the continuum limit by taking $n\rightarrow +\infty \Leftrightarrow \epsilon \rightarrow 0$ and demand consistency between the rhs and lhs, we get the value for the normalization constant of each slice:
\begin{equation}
    \mathcal{N}(\varepsilon,\vec{\textrm{q}}_{0i},\vec{\textrm{q}}_0)= \left (2\pi\hbar\varepsilon\over i \right )^{M/2} J(\vec{\textrm{q}}_0)
\end{equation}
In this same limit, the equation (\ref{Mdim}) becomes the differential equation:
\begin{equation}
    \begin{split}
&i\hbar\,{\partial \over \partial\ell}\, U_0(\vec{\textrm{q}}_{0i},\vec{\textrm{q}}_0,\ell)=\\
&\,~~~~\left\{-{\hbar^2\over 2}\eta ^{\textrm{AB}}\left[{\partial^2\over \partial\textrm{q}_0^{\textrm{A}} \partial \textrm{q}_{0}^{\textrm{B}}}+{2\over J(\vec{\textrm{q}}_0)}\, {\partial J\over \partial \textrm{q}_0^{\textrm{A}}}(\textrm{q}_0)\, {\partial\over \partial\textrm{q}^{\textrm{B}}_0}+{1\over J(\vec{\textrm{q}}_0)}\, {\partial ^2 J\over \partial\textrm{q}_0^{\textrm{A}} \partial \textrm{q}_{0}^{\textrm{B}}}(\vec{\textrm{q}}_0)\right]+ \widetilde{V}_0(\vec{\textrm{q}}_0)\right\} U_0(\vec{\textrm{q}}_{0i},\vec{\textrm{q}}_0,\ell)
\end{split}
\end{equation}
Before transforming this back to the original frame, it is useful to examine the transformation properties of this new Jacobian. The quantity $\textrm{J}^{\textrm{I}}{}_{\textrm{A}}$ defined in (\ref{newjacobian}), despite the two indices, transforms as a vector, since only one of the frame will have its coordinates transformed:
\begin{equation}
    \textrm{J}^{\textrm{I}}{}_{\textrm{A}} \rightarrow \check{\textrm{J}}^{\check{\textrm{I}}}{}_{\textrm{A}} = \frac{\partial \check{q}_{0}^{\check{\textrm{A}}}}{\partial q^{\textrm{K}}} =  \textrm{J}^{\textrm{K}}{}_{\textrm{A}} \left (\frac{\partial \check{q}^{\check{\textrm{I}}}}{\partial q^{\textrm{K}}} \right )
\end{equation}
For later ease of use, we define the inverse of this vector:
\begin{equation}
    (\textrm{J}^{-1})_{\textrm{J}}{}^{\textrm{B}} = \frac{\partial q_{0}^{\textrm{B}}}{\partial q^{\textrm{J}}} \, \, , \qquad \textrm{J}^{\textrm{I}}{}_{\textrm{A}} \, (\textrm{J}^{-1})_{\textrm{J}}{}^{\textrm{B}} = \delta ^{\textrm{I}}{}_{\textrm{J}} \, \delta ^{\textrm{A}}{}_{\textrm{B}} \, \Rightarrow \, \textrm{J}^{\textrm{I}}{}_{\textrm{A}} = \eta _{\textrm{AB}} \, (\textrm{J} ^{-1})_{\textrm{J}}{}^{\textrm{B}} \, \gamma ^{\textrm{IJ}}
\end{equation}
Lastly, it is notable to mention that the Jacobian will transform as a scalar density\cite{Reference64}:
\begin{equation}
    J \rightarrow \check{J} = J \, \left |\text{det}\left ( \frac{\partial \check{q}^{\check{\textrm{I}}}}{\partial q^{\textrm{K}}}\right ) \right | 
\end{equation}
or in terms of the way the covariant derivative acts upon it:
\begin{equation}
    \nabla _{\textrm{I}} J = \partial _{\textrm{I}} J + \Gamma ^{\textrm{K}}{}_{\textrm{IK}} J
\end{equation}
Using these and the known transformation properties of the metric, the differential equation in the original coordinates reads:
\begin{equation} \label{Mdimeq}
    i\hbar\,{\partial \over \partial\ell}\, U(\vec{\textrm{q}}_{i},\vec{\textrm{q}},\ell)= \left\{ -\frac{\hbar ^{2}}{2}\!\left[\gamma^{\textrm{IJ}}{\partial^2\over \partial\textrm{q}^{\textrm{I}} \partial \textrm{q}^{\textrm{J}}} + P^{\textrm{I}} {\partial\over \partial\textrm{q}^{\textrm{I}}}+\Omega \right]\!+ \widetilde{V}\right\}U(\vec{\textrm{q}}_{i},\vec{\textrm{q}},\ell)
\end{equation}
where we have defined the quantities:
\begin{equation}
    P^{\textrm{I}} = \gamma ^{\textrm{IJ}} \frac{2}{J} \frac{\partial J}{\partial \textrm{q}^{\textrm{J}}} + \gamma ^{\textrm{JK}} (\textrm{J}^{-1})_{K}{}^{\textrm{A}} \frac{\partial \textrm{J} ^{\textrm{I}}{}_{\textrm{A}}}{\partial \textrm{q} ^{\textrm{J}}}
\end{equation}
\begin{equation}
    \Omega = \gamma^{\textrm{IJ}} J^{-1} \left ( \frac{\partial ^{2} J}{\partial \textrm{q}^{\textrm{I}} \partial \textrm{q}^{\textrm{J}}} + (\textrm{J}^{-1})_{J}{}^{\textrm{A}} \frac{\partial \textrm{J} ^{\textrm{K}}{}_{\textrm{A}}}{\partial \textrm{q}^{\textrm{I}}} \frac{\partial J}{\partial \textrm{q}^{\textrm{K}}} \right )
\end{equation}
It is easy to verify that for the 1-dimensional case where we have $J^{-1} = \sqrt{-\gamma}$, these expressions reduce to (\ref{newamb}). We may additionally bring the equation in a covariant form:
\begin{equation}
    i\hbar\,{\partial \over \partial\ell}\, U(\vec{\textrm{q}}_{i},\vec{\textrm{q}},\ell)= \left\{ -\frac{\hbar ^{2}}{2}\!\left[\nabla ^{2} + \tilde{P}^{\textrm{I}} \nabla _{\textrm{I}}+\Omega \right]\!+ \widetilde{V}\right\}U(\vec{\textrm{q}}_{i},\vec{\textrm{q}},\ell)
\end{equation}
where for brevity we have redefined the vector quantity:
\begin{equation}
    \tilde{P}^{\textrm{I}} = P^{\textrm{I}} + \gamma ^{\textrm{JK}} \Gamma ^{\textrm{I}}{}_{\textrm{JK}}
\end{equation}
We define the quantum states $\psi$ just like before:
\begin{equation}
    \psi (\vec{\textrm{q}}) = \int _{0}^{+\infty} d\ell \, \,U(\vec{\textrm{q}}_{i},\vec{\textrm{q}},\ell)
\end{equation}
which leads to equation (\ref{Mdimeq}) yielding:
\begin{equation} 
    i\hbar\, \big[U(\vec{\textrm{q}}_{i},\vec{\textrm{q}},\ell)\big]^{\ell=+\infty}_{\ell=0}=  \widehat{\mathcal{H}}\,\psi\equiv  -\frac{\hbar ^{2}}{2}\!\left[\gamma^{\textrm{IJ}}{\partial^2 \psi\over \partial\textrm{q}^{\textrm{I}} \partial \textrm{q}^{\textrm{J}}} + P^{\textrm{I}} {\partial \psi\over \partial\textrm{q}^{\textrm{I}}}+\Omega \psi \right]\!+ \widetilde{V}\psi
\end{equation}
For this to match the WDW equation, we must ensure again that the values of the transition amplitude at the boundaries of the line segment vanish. We can employ the same method as in the 1-dimensional model which does lead to the lhs to become exactly 0, but we shall skip over this as it is trivial to generalize. Since the equation above is the WDW equation, the derived functions $P^{\textrm{I}}$ and $\Omega$ match the ambiguity functions derived in the previous section:
\begin{equation}
    P^{\textrm{I}} = \gamma ^{\textrm{JK}} \tilde{\rho}_{\textrm{J}} \frac{\partial \rho ^{\textrm{I}}}{\partial \textrm{q}^{\textrm{K}}} \, \, \,  , \qquad \Omega = \omega
\end{equation}
Notice how the implicit contraction of the $\alpha$ indices makes $P^{\textrm{I}}$ a scalar in $\mathbb{A}$. The dependence of the ambiguities on this space is hidden in the path integral picture. The matching of functions above also means that the Hermiticity criterion leads to the constraints (\ref{eq:constraints2}) and (\ref{eq:constraints3}) applying for $P^{\textrm{I}}$ and $\Omega$ as well. It is not difficult to find that based on the decomposition (\ref{eq:14'}) we have:
\begin{equation}
    \Pi ^{\textrm{I}} = \textrm{Re} P^{\textrm{I}} \Rightarrow \tilde{\Pi}^{\textrm{I}} = \text{Re}\tilde{P}^{\textrm{I}}
\end{equation}
since $\gamma ^{\textrm{JK}} \Gamma ^{\textrm{I}}{}_{\textrm{JK}} = - (\sqrt{-\gamma})^{-1}\partial _{\textrm{J}} \left (\gamma ^{\textrm{IJ}} \sqrt{-\gamma} \right )$. However, as seen above, both $P^{\textrm{I}}$ and $\Omega$ are real functions because any coordinate transformation - and therefore any Jacobian - is strictly real. If we define $\Psi = \sqrt{\mu} \, \psi$ and bring the WDW equation in the form (\ref{Mdimwdw}), we find:
\begin{equation}
    Y^{\textrm{K}} = 0
\end{equation}
\begin{equation}
\begin{split}
    W &= \Omega - \frac{1}{2}\nabla_{\textrm{I}} \tilde{P}^{\textrm{I}} - \frac{1}{4} \gamma _{\textrm{IJ}} \tilde{P}^{\textrm{I}} \tilde{P}^{\textrm{J}} \\
    &= - \frac{1}{4} \gamma _{\textrm{IJ}} \,(\textrm{J}^{-1})_{\textrm{K}}{}^{\textrm{A}}  (\textrm{J}^{-1})_{\textrm{L}}{}^{\textrm{B}} \, (\nabla ^{\textrm{K}} \textrm{J}^{\textrm{I}}{}_{\textrm{A}})(\nabla ^{\textrm{L}} \textrm{J}^{\textrm{J}}{}_{\textrm{B}}) - \frac{1}{2} \nabla _{\textrm{I}} \left [(\textrm{J}^{-1})_{\textrm{A}}{}^{\textrm{K}} (\nabla ^{\textrm{K}} \textrm{J}^{\textrm{I}}{}_{\textrm{A}}) \right]
\end{split}
\end{equation}
It is evident from the transformation properties of the various quantities above that $W$ is indeed a scalar. Additionally, the first result confirms that - at least for flat manifolds - our previous assertion that the path integral formulation would lead to a WDW that is universal at the semi-classical level, since only the correction from $W$ remains and that contributes only at order $\mathcal{O}(\hbar^{2})$. \par
Let us now assume that the original coordinate system was itself Minkowskian up to a positive multiplicative constant. That means $\gamma_{\textrm{IJ}} = \vartheta ^{2} \, \eta _{\textrm{IJ}}$ and consequently $\textrm{J}^{\textrm{I}}{}_{\textrm{A}} = \pm \vartheta \, \delta ^{\textrm{I}}{}_{\textrm{A}}$. In this particular case, we find that $W=0$ by using the result retrieved above. However, since $W$ is a scalar quantity, it doesn't change value after a change of coordinates, hence it remains 0 for any reference frame with a generic metric $\gamma _{\textrm{IJ}}$. As a result, the WDW equation is independent of ambiguities just like in 1 dimension:
\begin{equation}
    -{\hbar^2\over 2}\nabla^2\Psi +\widetilde{V}\Psi = 0
\end{equation}
and the resulting quantum theory is unique. This seems to hint at the fact $W$ is indeed something proportional to the Ricci scalar $\mathcal{R}$ as that becomes 0 for flat manifolds, but that remains inconclusive. Regardless, it is remarkable that even for arbitrary dimensions it is provable that the WDW equation is universal for these special cases. As we shall see, this seemingly overly specific category includes some rather interesting and physically important models.

\subsection{Explicit calculation for the Starobinsky model}

We recall from the previous chapter the form of the classical action for the Starobinsky model in the Einstein frame:
\begin{equation}
    S_{G} = \kappa ^{2}\int d^{4}x\, \sqrt{\tilde{g} }  \left [\widetilde{R} - \frac{1}{2} \partial _{\mu }\phi \partial ^{\mu }\phi - V(\phi) \right ]
\end{equation}
where $V(\phi) = \tilde{\mu}^{2} \,  (1 - e^{-\alpha \phi })^{2}$ the inflationary potential and $\alpha = \sqrt{\frac{2}{3}}$. If we use the FRLW ansatz for the metric $\tilde{g}_{\mu\nu}$, the Ricci scalar in terms of the degrees of freedom $a(t)$ and $N(t)$ becomes:
\begin{equation}
    \widetilde{R} = \frac{6}{a^{2}} + \frac{6}{Na^{2}} \frac{d}{dt} \left (\frac{\dot{a}a}{N} \right )
\end{equation}
The Lagrangian of the theory then takes the form:
\begin{equation}
    L = N v_{3} \left (-\frac{3a\dot{a}^{2}}{N^{2}} + \frac{a^{3}}{2} \frac{\dot{\phi}^{2}}{N^{2}} + 3a - a^{3}V(\phi) \right )
\end{equation}
We may perform a general field redefinition such that the old parameters are expressed via the mapping:
\begin{equation}
    a = \mathcal{A}(q^{0} ,q^{1}) \, \, \, , \qquad \phi = \Phi(q^{0}, q^{1})
\end{equation}
The Lagrangian expressed in this new set of parameters is then brought ito a non-linear $\sigma$-model form:
\begin{equation}
    L = N \left (\gamma _{\textrm{IJ}} \frac{\dot{q}^{\textrm{I}} \dot{q}^{\textrm{J}}}{2N^{2}} - \widetilde{V}(q^{0}, q^{1}) \right )
\end{equation}
where $\widetilde{V}(q^{0}, q^{1}) = -v_{3}\mathcal{A}(q^{0},q^{1}) \left [3 - \mathcal{A}^{2}(q^{0},q^{1}) \, V(\Phi(q^{0},q^{1}) \right ]$ the combined potential. The metric in this frame is easy to compute\cite{Reference64}:
\begin{equation}
    \gamma _{\textrm{IJ}} = v_{3} \mathcal{A}
    \begin{pmatrix}
        -6 (\partial _{0}\mathcal{A})^{2} + \mathcal{A}^{2} (\partial _{0} \Phi)^{2} & -6(\partial _{0} \mathcal{A})(\partial _{1} \mathcal{A}) + \mathcal{A}^{2}(\partial _{0} \Phi)(\partial _{1} \Phi) \\
        -6(\partial _{0} \mathcal{A})(\partial _{1} \mathcal{A}) + \mathcal{A}^{2}(\partial _{0} \Phi)(\partial _{1} \Phi) & -6 (\partial _{1}\mathcal{A})^{2} + \mathcal{A}^{2} (\partial _{1} \Phi)^{2}
    \end{pmatrix}
\end{equation}
Alternatively in the original frame:
\begin{equation} \label{oldframe}
    \tilde{\gamma}_{\textrm{AB}} = \frac{v_{3}}{2}
    \begin{pmatrix}
        -6a & 0 \\
        0 & a^{3}
    \end{pmatrix}
\end{equation}
It is made more clear here how the metric of the target space is in fact of Lorentzian signature, and it is also straightforward to verify that the Riemann tensor is 0:
\begin{equation}
    \mathcal{R}^{\textrm{A}}{}_{\textrm{BCD}} = 0 \Rightarrow \mathcal{R} = 0
\end{equation}
Given that the target space is flat, we can always find a change of coordinates such that the metric becomes Minkowskian. It is rather difficult to retrieve analytically hence we shall not provide the exact mapping, but it is sufficient to know that it exists. This allows us to use the results from the last section, namely that the resulting theory after quantization is unique and the WDW equation is universal. We choose the latter frame for the metric (\ref{oldframe}) and we shall make an explicit calculation from the WDW equation using the WKB approximation. Our result will have to be the same in any other frame of reference. \par
Rather than solving for $\psi$, we opt to calculate $\Psi$ directly. This removes the need to calculate the precise values for the ambiguity functions $(\rho^{\beta \gamma}_{\alpha}) ^{\textrm{I}} , \, (\tilde{\rho}^{\alpha}_{\beta \gamma})_{\textrm{J}}$. We use the equation:
\begin{equation}
    \nabla ^{2} \Psi - \frac{2\widetilde{V}}{\hbar ^{2}} \Psi = 0
\end{equation}
and plugging in all the relevant quantities we get the equation:
\begin{equation}
     -\partial _{a}^{2} \Psi + \frac{6}{a^{2}} \partial _{\phi}^{2} \Psi - \frac{1}{a} \partial _{a} \Psi - \frac{12 v_{3}a\widetilde{V}}{\hbar ^{2}} \Psi = 0
\end{equation}
We then use the WKB approximation:
\begin{equation}
     \Psi = \exp{\left [\frac{i}{\hbar} \left (\mathcal{F}_{0} + \frac{\hbar}{i} \mathcal{F}_{1} + \mathcal{O}(\hbar ^{2}) \right ) \right ]}
\end{equation}
Just like in the 1-dimensional calculation, we find terms in orders of $\hbar$ and by taking the semi-classical limit $\hbar \rightarrow 0$, we end up with two differential equations\cite{Reference64}:
\begin{equation} \label{first}
    \mathcal{O}(\hbar ^{-2}): \quad (\partial _{a} \mathcal{F}_{0})^{2} - \frac{6}{a^{2}} (\partial _{\phi} \mathcal{F}_{0})^{2} -12v_{3}a\widetilde{V} = 0
\end{equation}
\begin{equation} \label{second}
    \mathcal{O}(\hbar^{-1}): \quad -2(\partial_{a} \mathcal{F}_{1})(\partial _{\phi} \mathcal{F}_{0}) - \partial _{a}^{2} \mathcal{F}_{0} + \frac{12}{a^{2}} (\partial _{\phi} \mathcal{F}_{1})(\partial _{a} \mathcal{F}_{0}) + \frac{6}{a^{2}} \partial _{\phi}^{2} \mathcal{F}_{0} - \frac{1}{a} \partial _{a}\mathcal{F}_{0} = 0
\end{equation}
These equations are still difficult to solve analytically, but we may simplify the first (and by extension the second one) by taking the regime of interest at $\phi \rightarrow +\infty$. As explained in the preceding chapter, this corresponds to the inflationary regime where the potential is almost at a constant value. Let us now define $x = e^{-\alpha \phi}$. By Taylor-expanding around the value of $\phi$ for the inflationary regime, we are equivalently expanding around $x = 0$\cite{Reference64}:
\begin{equation}
    \mathcal{F}_{0} (a,x) = \sum _{k=0}^{+\infty} \frac{1}{k!} \mathcal{F}_{0}^{(k)} (a,0) \, x^{k}
\end{equation}
where the upper index in brackets signifies the degree of partial derivatives acting on the function with respect to $x$. This allows us to decompose equation (\ref{first}) in powers of $x$:
\begin{equation}
\begin{split}
    \left ((\partial _{a} \mathcal{F}_{0}^{(0)}) + x (\partial _{a} \mathcal{F}_{0}^{(1)}) + \frac{x^{2}}{2} (\partial _{a} \mathcal{F}_{0}^{(2)}) + \mathcal{O}(x^{n \geq 3}) \right )^{2}& - \\\frac{4x^{2}}{a^{2}}\left (\mathcal{F}_{0}^{(1)} + x \mathcal{F}_{0}^{(2)} + \frac{x^{2}}{2} \mathcal{F}_{0}^{(3)} + \mathcal{O}(x^{n \geq 3}) \right )^{2} &+
     36v_{3}^{2} a^{2} \left (1 - \frac{\tilde{\mu} ^{2}a^{2}}{3} (1 -x)^{2} \right ) = 0
\end{split}
\end{equation}
We can break this into several smaller differential equations by powers of $x$. The first term in the Taylor expansion comes from the equation of order $\mathcal{O}(x^{0})$:
\begin{equation}
    (\partial _{a} \mathcal{F}_{0}^{(0)})^{2} + 36v_{3}^{2} a^{2} \left (1 - \frac{\tilde{\mu} ^{2}a^{2}}{3} \right ) = 0
\end{equation}
This can be calculated exactly, yielding:
\begin{equation} \label{refsol3}
    \mathcal{F}_{0}^{(0)} = \mp i\frac{6v_{3}}{\tilde{\mu}^{2}} \left (1 - \frac{\tilde{\mu}^{2}a^{2}}{3} \right )^{3/2} + K_{0}
\end{equation}
where $K_{0}$ an arbitrary integration constant. Notice how this matches the 1-dimensional result in (\ref{1dresult}), with $\tilde{\mu}$ playing the role of the cosmological constant. This was expected, since in the inflationary regime the Starobinsky model may be approximated as $R^{2}$ gravity, and that gives a GR model with a matter field and a cosmological constant in the Einstein frame. The integration constant which is included accounts for the kinetic energy of $\phi$ which quantum mechanically can be arbitrary, even though classically it is very small\cite{Reference15}\cite{Reference64}. However, that this is simply the leading term. As we take into account higher order contributions from the dynamics of $\phi$, we shall find additional corrections that deviate from the 1-dimensional result. We may now plug the result above in the equation of order $\mathcal{O}(x)$:
\begin{equation}
    (\partial _{a} \mathcal{F}_{0}^{(0)})(\partial _{a} \mathcal{F}_{0}^{(1)}) + 12v_{3}^{2} \tilde{\mu} ^{2}a^{4} = 0
\end{equation}
which gives us the next term in the expansion:
\begin{equation}
    \mathcal{F}_{0}^{(1)} = \pm i \frac{6v_{3} (\tilde{\mu}^{2}a^{2} +6)}{\tilde{\mu}^{2}} \left (1 - \frac{\tilde{\mu}^{2} a^{2}}{3} \right )^{3/2} + K_{1}
\end{equation}
where $K_{1}$ another integration constant. We proceed to do the same for terms of order $\mathcal{O}(x^{2})$:
\begin{equation}
    (\partial _{a} \mathcal{F}_{0}^{(0)})(\partial _{a} \mathcal{F}_{0}^{(2)}) + 2(\partial _{a} \mathcal{F}_{0}^{(1)})^{2} + \frac{4}{a^{2}}(\mathcal{F}_{0}^{(1)})^{2} - 12v_{3}\tilde{\mu}^{2}a^{4} = 0
\end{equation}
We can continue this iterative process to get all terms in the Taylor expansion analytically if we like, but we shall restrict ourselves to these terms. Generally, each iteration for powers $n\geq 3$ in the expansion can be solved with\cite{Reference64}:
\begin{equation}
    \partial _{a} \mathcal{F}_{0}^{(n)} = (\partial _{a} \mathcal{F}_{0}^{(0)})^{-1} \sum _{k=1}^{n} \left [-\binom{n}{k} (\partial _{a} \mathcal{F}_{0}^{(k)})(\partial _{a} \mathcal{F}_{0}^{(n-k)}) + \frac{4}{a^{2}} \binom{n}{k-1} \mathcal{F}_{0}^{(k)} \mathcal{F}_{0}^{(n-k)} \right ]
\end{equation}
Similarly, since in theory we can get a full explicit expression for $\mathcal{F}_{0}$, we may then plug it into (\ref{second}) to obtain $\mathcal{F}_{1}$. It is not clear whether we would be able get an explicit expression for it from that, but overall we can calculate it analytically up to an arbitrary order of $x$. \par
In our work, we have also touched upon the special case of dS JT gravity, which is another 2d flat minisuperspace model where our conclusions from section 5.2.1 can be applied \cite{Reference64}. This model too is of great interest for quantum cosmology and high-energy physics, and has been the subject of extensive research in the literature \cite{Reference86, Reference87, Reference88, Reference89, Reference90, Reference91, Reference92, Reference93, Reference94, Reference95, Reference96, Reference97, Reference98}. However, since the subject of this thesis is more specifically focused on the case of the Starobinsky model and how it can be utilized in our later calculations, we shall refer the reader to our aforementioned paper for a showcase of this and relevant comments in relation to our research.

\pagebreak % Experiment 1

\fancyhead[RE,LO]{No boundary proposal and the Hartle-Hawking wave function}

\chapter{No boundary proposal and the Hartle-Hawking wave function}

Given a certain model such as the ones showcased in the previous chapter, we are able to calculate a collection of physical state solutions $\psi$ (or $\Psi$) which, as we have said, are meant to express the probability distribution to measure the Universe having a certain geometric configuration. However, this description is somewhat deceptive, as it overlooks certain additional issues that we encounter and which we need to address prior to drawing any meaningful conclusions\cite{Reference22}\cite{Reference47}. \par
The first major issue is that it is actually not so trivial to interpret the wave function solutions as probability amplitudes. A naively defined Hilbert space inner product:
\begin{equation}
     \langle \psi_1,  \psi_2 \rangle = \int \limits_{\mathcal T} d \textrm q \, \sqrt{-\gamma} \, \psi_1^* \psi_2
\end{equation}
is not strictly positive definite, meaning the measurable quantities may give negative probabilities. Equivalently, if we define the current\cite{Reference22}\cite{Reference47}:
\begin{equation}
    J_{\textrm{I}} = \frac{i}{2} \left (\psi \,\nabla _{\textrm{I}} \,\psi ^{*} - \psi ^{*} \,\nabla _{\textrm{I}} \,\psi \right )
\end{equation}
then this cannot possibly be a probability current, meaning we cannot establish a continuity equation for the Hilbert space over $\mathcal{T}$. There have been several suggested ways in which this weakness is amended, such as DeWitt's own suggestion in defining a Klein-Gordon inner product\cite{Reference22}. For our work, we have opted for the definition in (\ref{amp}) which ensures the measured probabilities take strictly positive-definite values. \par
Despite this, another issue that comes up has to do with the normalizability of the wave functions. The solutions of the WDW equation are not always able to be normalized, hence we once again cannot interpret the results of the Hilbert space inner products as real probabilities. The definition we introduced in general is also insufficient to amend this issue. At a first glance, this would suggest that our analysis thus far was futile because it does not yield true measurable results. Despite that, we do have one way in which we can make meaningful predictions. Namely, we can take the ratio between two non-normalized probability densities $|\psi_{1}|^{2}/|\psi_{2}|^{2}$. Even though the two states might not be normalizable, their behaviour with respect to their parameters is reliably derived with the methods we have shown. The result of the ratio then gives a meaningful behaviour for the relative probability of the two states up to a multiplicative constant. \par
To explain how this may prove useful, we first need to explain one of the major outstanding issues of inflationary models. To be more precise, for inflation itself to begin, as well as for the appropriate amount of e-folds for the resolution of the horizon and flatness problems of cosmology to occur, slow-roll inflation requires the Universe to have specific initial conditions. As we have shown, we need the Universe to begin in a state of large $V(\phi)$ (plateau region), which in the case of the Starobinsky model also means a very large value of the inflaton field $\phi$ itself. This is not trivial to achieve, and it has been argued that these constitute rather "unnatural" initial conditions for the Universe to be in\cite{Reference63}. \par
This is where we believe the solutions of the WDW equation may prove useful. By calculating the state solutions for both the inflationary and non-inflationary regimes in the context of e.g. the Starobinsky model, it could provide a way to quantify how this relative probability evolves as $\phi$ changes values. If the state of large $\phi$ which corresponds to the inflationary regime is dominant under certain conditions, then we could argue that the initial conditions ripe for inflation to occur may be explained probabilistically. In other words, inflationary initial conditions could be simply statistically more probable in the context of quantum mechanics. Inflation then becomes a stochastic process: the state of the Universe fluctuates until it "lands" on the correct regime, which would be inevitable with enough iterations. \par
This is our goal for the upcoming sections. We will introduce the no-boundary proposal for the wave function as per Hartle and Hawking \cite{Reference1}\cite{Reference9}, and explain its significance. We will then present our own results by applying those methods in order to obtain the corresponding result for pure $R^{2}$ theory as an inflationary regime approximation to the Starobinsky model \cite{Reference33}. Finally, we will examine various scenarios arising from those types of solutions.

\section{The Hartle-Hawking no boundary proposal}

The approach used by Hartle and Hawking to calculate a solution to the Wheeler-DeWitt equation in minisuperspace was by using a Euclidean path integral in order to find an effective "ground state" \cite{Reference1}\cite{Reference10}\cite{Reference16}. We say effective because there is no true ground state for a wave function solution, since the energy of the Universe is 0 without any excitations. Thus, a state of minimum excitation is defined instead, which could be seen as the new state of the lowest energy. By making a Wick rotation $t \rightarrow -i\tau$ which converts the path integral into a Euclidean one, we turn the exponents of the sum real and convergent. Higher excitation states that correspond to higher values of the exponent will therefore contribute less. By taking the limit $\tau' \rightarrow -\infty$ those contributions vanish, leaving us with the ground state state as the sole contributor. If the ground state has 0 energy and therefore a wave function does not exist, then the next highest contributor will be the state of minimum excitation whose energy we may normalize to 0. To illustrate this better, we start with the case of conventional quantum mechanics \cite{Reference1}\cite{Reference16}:

\begin{equation} \label{eq:4.17}
\begin{split}
    \langle x,0 | 0,t' \rangle &= \sum _{n} \psi _{n} (x) \psi^{*}_{n}(0) \, e^{iE_{n}t'} \\
    &= \mathcal{N}_{0} \int _{x(0) = 0} ^{x(t')=x} \mathcal{D}x(t) \, e^{iS[x(t)]} \xrightarrow{\text{$t\rightarrow -i\tau$}} \mathcal{N}_{0} \int _{x(0) = 0} ^{x(\tau ')=x} \mathcal{D}x(\tau) \, e^{-I[x(\tau)]}
\end{split}
\end{equation}
where $I[x(\tau)]$ the classical action in Euclidean space and $\mathcal{N}_{0}$ is a normalization constant (not to be confused with the one calculated in the last chapter). It becomes apparent how the aforementioned limit provides the state of minimum excitation:
\begin{equation} \label{eq:4.18}
    \lim _{\tau' \rightarrow -\infty} {\sum _{n} \psi _{n} (x) \psi^{*}_{n}(0) \, e^{E_{n}\tau '}} = \psi _{0} (x,0) = \mathcal{N}_{0} \int ^{x(0) = x} \mathcal{D}x(\tau) \, e^{-I[x(\tau)]}
\end{equation}
As we can see, this means that eventually we sum up all the Euclidean paths that are bounded only from above at $\tau = 0$ and for all negative values thereafter. This can easily be generalized to the case of summing over field configurations in the case of QFT\cite{Reference44}:
\begin{equation} \label{eq:4.19}
    \psi _{0} [\phi (\vec{x}), 0] = \mathcal{N}_{0} \int ^{\phi (x) = \phi (\vec{x},0)} \mathcal{D}\phi \, e^{-I[\phi(x)]}
\end{equation}
In our case the degrees of freedom are the induced metrics $h_{ij}$, thus we sum over all 4-geometries that have boundaries described by the induced metrics, just as we explained in the previous chapter. The lower boundary for the transition amplitude will be a hypersurface at the time slice $t=t'$ given by $h'_{ij}$, and the upper boundary some other hypersurface given by $h_{ij}$ at the time slice $t=t_{0}$. Hence \cite{Reference1}:
\begin{equation} \label{eq:4.20}
    \langle h_{ij}|h'_{ij} \rangle = \mathcal{N}_{0} \int ^{g_{i j} (t_{0}) = h_{ij}}_{g_{ij} (t') = h'_{ij}} \mathcal{D}g \, e^{iS[g(x)]} 
\end{equation}
Based on the minisuperspace approximation, these 4-geometries have to be closed, homogeneous and isotropic. We further demand that these geometries contain no singularities. "Closed" as we have seen normally implies no boundaries, or equivalently that we take FLRW metrics for $k=1$. In our case, we do admit time-like boundaries due to slicing, albeit the geometry of the bounded surface can still be seen as a piece of a closed manifold. An example would be the case of a sphere like in figure \ref{fig:2.2}. A constant time slice cleaves the sphere into two distinct parts that each have a boundary on that surface, but the bounded 4-geometries still have a compact geometry. \par 
All these greatly reduce the amount of configurations we sum over, but there is still an enormous class of such geometries that fulfill these requirements. That is because they do not have to satisfy the Einstein field equations or describe the Universe accurately in order to be summed, much like when summing particle paths we also include classically forbidden ones. However, we expect that such cases will be suppressed and their contribution will be very small the more they deviate from the classical solutions. \par
Furthermore, we see that there is no absolute measure of time to make the necessary Wick rotation, as expected. The rotation instead affects the time slices defining the hypersurfaces themselves, as the 4-geometries now become Euclidean. Given the upper boundary for the time slice taken $t_{0} \rightarrow \tau _{0}$ and taking the limit $\tau ' \rightarrow -\infty$ like above, the 4-geometries bounded by hypersurfaces of $\tau > \tau _{0}$ will not contribute. The lower bound which corresponds to $\tau '$ will match to a single point in Euclidean space. This eliminates the lower boundary, hence why this methodology is dubbed as the "no boundary proposal"\cite{Reference1}. So finally we have:
\begin{equation} \label{eq:4.21}
    \psi _{0} (h_{ij}) = \mathcal{N}_{0} \int ^{g_{ij} (\tau = \tau _{0}) = h_{ij}} \mathcal{D}g \, e^{-I[g(x)]}
\end{equation}
This addresses the difficulties usually associated with calculating path integrals involving gravity \cite{Reference23}\cite{Reference24}. The classical Euclidean action in the case of the Hartle-Hawking analysis will match the Einstein-Hilbert gravitational action in the presence of a cosmological constant, plus the corrective GYH boundary term. Therefore:
\begin{equation} \label{eq:4.22}
    I = -\kappa ^{2} \int d\tau \int d^{3}x \, \sqrt{g} \, (R-2\Lambda ) \, - 2 \kappa ^{2}\oint \limits_{\partial \mathcal{V}} d^{3}x \, \sqrt{h} \, K 
\end{equation}
We may further add a scalar field degree of freedom $\phi$, in which case the total action will also include the field's kinetic term plus its coupling to classical gravity \cite{Reference1}. In non-Euclidean space:
\begin{equation} \label{eq:4.23}
    S_{M} = -\frac{1}{2} \int \limits_{\mathcal{V}} d^{4}x \, \sqrt{g} \, \left (\partial _{\mu} \phi \partial ^{\mu} \phi + \xi R\phi ^{2} \right )
\end{equation}
where $\xi$ is the coupling constant. In the Hartle-Hawking analysis, the choice of field is arbitrarily taken to be a conformally coupled one i.e. $\xi = \frac{1}{6}$, meaning the field's equations of motion are conformally invariant \cite{Reference1}\cite{Reference37}\cite{Reference7}. By contrast, a minimal coupling as we have implicitly assumed in the previous chapter would be $\xi = 0$. The former choice allows the matter component of the path integral to be calculated separately. To specify the Euclidean action (\ref{eq:4.22}), we recall that the metric in question is the FLRW metric, thus the scalar curvature is given by (\ref{eq:1.2}) for $k=1$, while the boundary term is calculated like in (\ref{eq:3.74}). We shall once again take the FLRW metric with a generic lapse function:
\begin{equation} \label{eq:4.24}
    ds^{2} = \sigma ^{2}[-N^{2}(t) \, dt^{2} + a^{2}(t) \, d\Omega _{3}^{2}]
\end{equation}
where we have also multiplied by $\sigma ^{2} = \frac{\kappa ^{-2}}{12v_{3}} = \frac{2l_{P}^{2}}{3\pi }$ ($l_{P} = m_{P}^{-1}$ being the Planck length) to simplify the factors of $\kappa ^{2}$ and other future constants. The scalar field considered is also demanded to be homogeneous ($\phi = \phi(t)$) as part of the minisuperspace approximation. We shall redefine it and introduce some new constant parameters in order to further simplify the form of the action\cite{Reference1}:
\begin{equation} \label{eq:4.25}
    \phi = \frac{\chi }{\sqrt{v_{3} \sigma^{2}}a} \, \,  , \qquad \Lambda = \frac{3\ell ^{2} }{\sigma^{2}}
\end{equation}
The newly defined field $\chi$ is trivially also spatially homogeneous. The constant $\ell ^{2}$ is defined analogously to the previous chapter. Before we proceed to write the action, it is important to specify what the classical solutions for the 4-geometries are. Since we assume that they are closed in the sense that they have positive constant scalar curvature, the classical metric solutions will satisfy:
\begin{equation} \label{eq:4.26}
    R_{\mu \nu}^{cl} = \Lambda g_{\mu \nu}^{cl} 
\end{equation}
For $\Lambda >0$ ($\Lambda <0$), the classical solutions are dS (AdS) spaces of curvature $R_{cl} = 4\Lambda$. In addition, we know from the derived expression (\ref{eq:2.7}) that a dS space described by a FLRW metric will have curvature $R = 12H^{2}$. For the AdS case, the equivalent result will be $R=-12H^{2}$. Considering that now we have an extra factor of $\sigma ^{2}$ for physical lengths, classically we have:
\begin{equation} \label{eq:4.27}
    \Lambda = \pm \frac{3H^{2}}{\sigma ^{2}}
\end{equation}
Combining this with the redefinition of (\ref{eq:4.25}) we get:
\begin{equation} \label{eq:4.28}
    H^{2} = \pm \ell ^{2}
\end{equation}
It may seem odd that $H$ appears dimensionless, since conventionally it has units of mass. However, this is simply due to the manner in which we wrote down the metric (\ref{eq:4.24}). The presence of the factor $\sigma ^{2}$ supplies the coordinates with their dimensions, so the coordinates as well as the scale factor are dimensionless. \par
In the present analysis we are only interested in dS spaces because that corresponds to inflationary model solutions. We then intend on performing the path integral calculation via the semi-classical approximation. Recall from (\ref{eq:2.30}) that in this approximation we have very small quantum corrections which we effectively overlook. The leading term will be classical by inserting the instanton solutions $g_{cl}$\cite{Reference10}\cite{Reference16}:
\begin{equation} \label{eq:4.29}
    \int \mathcal{D}g \, e^{iS[g(x)]} \approx e^{iS_{cl}}
\end{equation}
where $S_{cl} = S[g_{cl}]$. Making a Wick rotation and using the method of steepest descents, we may approximate the path integral as $\sim e^{-I_{cl}}$. The instantons serve as saddle points and the real exponents of the fluctuations inside the path integral contribute negligibly as excited states. The calculated states can then be written as\cite{Reference1}:
\begin{equation} \label{eq:4.30}
    \psi _{0}(h_{ij}) \approx \mathcal{N}_{0} \, \Delta _{\textrm{FP}} \, e^{-I_{cl}(h_{ij})}
\end{equation}
$\Delta _{\textrm{FP}}$ is the Fadeev-Popov determinant which we have examined before. Hartle and Hawking do not explicitly calculate the term, but as we have proven in Appendix B, it is an irrelevant constant which we may set equal to 1. Now we are ready to write the action evaluated at the classical dS space solutions:
\begin{equation} \label{eq:4.31}
    S = \frac{1}{2} \int dt\, \left (\frac{N}{a} \right )\, \left [\left (\frac{a}{N} \, \dot{a} \right )^{2} + a^{2} - \ell ^{2} a^{4} + \left (\frac{a}{N} \frac{d\chi }{dt} \right )^{2} - \chi ^{2} \right ]
\end{equation}
We have the expected two true degrees of freedom: the scale factor $a$ and the scalar matter field (here $\chi$ instead of $\phi$). Taking the Wick rotation $t \rightarrow -i\tau$, the resulting Euclidean action is:
\begin{equation} \label{eq:4.32}
    I = \frac{1}{2} \int d\eta \, \left [-\left (\frac{da}{d\eta } \right )^{2} - a^{2} + \ell ^{2} a^{4} + \left (\frac{d\chi }{d\eta } \right )^{2} + \chi ^{2} \right ]
\end{equation}
where we have also introduced the conformal Euclidean time $d\eta = \frac{d\tau}{a}$ and gauge-fixed $N=1$. If the values of the degrees of freedom on the boundary are $a_{0}, \, \chi _{0}$, then:
\begin{equation} \label{eq:4.33}
    \psi_{0} (a_{0},\chi _{0}) \approx \mathcal{N}_{0} \, e^{-I_{cl}(a_{0}, \chi _{0})}
\end{equation}
As we mentioned earlier, the fact we have taken the scalar field to be conformally coupled has allowed us to decouple it from the gravitational part of the action. That is because the conformal invariance of the field equations ensures the conformal transformation undertaken $d\tau \rightarrow d\eta$ does not create any unwanted cross terms. The form of the matter component is just a one-dimensional harmonic oscillator kind of contribution, hence \cite{Reference1}:
\begin{equation} \label{eq:4.34}
    \psi_{0} (a_{0}, \chi _{0}) = e^{-\frac{\chi _{0} ^{2}}{2}} \, \bar{\psi} _{0} (a_{0})
\end{equation}
As for the gravitational component, we stated that the classical solutions correspond to dS spaces which in Lorentzian signature have spacelike boundaries on $t \rightarrow \pm \infty$. In Euclidean space dS spaces turn into 4-spheres, just like we showed back in chapter 3. The two boundaries no longer exist, as they now match the two poles of the 4-sphere. Despie that, since we are take time slices $\tau \leq \tau _{0}$, that means there is a 3-sphere acting as a boundary for that value. The lower bound which we have sent to $-\infty$ will indeed match to the south pole of the 4-sphere. The scalar curvature remains unchanged by the Wick rotation, meaning that this 4-sphere has radius $\sqrt{\frac{3}{\Lambda}}$, or equivalently $H^{-1}$. There are two distinct cases for the value of $a_{0}$. If $a_{0} H < 1$, then the boundary 3-sphere may serve as a boundary on the instanton 4-sphere. If $a_{0}H > 1$, then the radius of the boundary 3-sphere exceeds that of the 4-sphere and cannot serve as a real boundary solution. The first of these cases has two further subcases: one in which the part of the 4-sphere bounded by the spacelike hypersurface (3-sphere) covers less than its hemisphere, and one in which it covers more than its hemisphere. An illustration of this is shown below.
\begin{figure}[H]
\centering
\begin{tikzpicture} [scale=0.5]
\begin{scope}
  \clip (-6,0) circle (4cm);
  \fill[fill=magenta!30] (-10,-4) rectangle (-2,-2.1) ;
  \end{scope}
  \draw[thick] (-6,0) circle (4cm);
  \begin{scope}
  \clip (6,0) circle (4cm);
  \fill[fill=magenta!30] (2,-4) rectangle (10,2) ;
  \end{scope}
  \draw[thick] (6,0) circle (4cm);
  \draw[fill=magenta!20] (-9.3,-2.1) arc (180:360:3.3 and 1) ;
  \draw[dashed, fill=magenta!20] (-2.7,-2.1) arc (0:180:3.3 and 1) ;
  \draw[fill=magenta!20] (2.6,2) arc (180:360:3.4 and 1) ;
  \draw[dashed, fill=magenta!20] (9.4,2) arc (0:180:3.4 and 1) ;
  \draw[dashed, ->] (-6,-2.1) -- node[above]{$a_{0}$} (-2.7,-2.1) ;
  \draw[dashed, ->] (6,2) -- node[above]{$a_{0}$} (9.4,2) ;
  \draw[dashed, ->] (-6,-2.1) -- node[above]{$a_{0}$} (-2.7,-2.1) ;
  \draw[dashed, ->] (-6,0) -- node[above]{$H^{-1}$} (-10,0) ;
  \draw[dashed, ->] (6,0) -- node[above]{$H^{-1}$} (2,0) ;
  \draw [fill=black] (-6,0) circle (2pt) ;
  \draw [fill=black] (6,0) circle (2pt) ;
\end{tikzpicture}
\caption{The case of a boundary 3-sphere of radius $a_{0} <H^{-1}$ where $H^{-1}$ the radius of the Euclidean 4-sphere. The time slices ("rings") of radius $a_{0}$ form these 3-spheres, while the shaded parts represent the bounded 4-geometries. The left figure showcases the bounded 4-geometry covering less than a hemisphere, while the right one showcases it covering more than a hemisphere.}
\label{fig:4.1}
\end{figure}
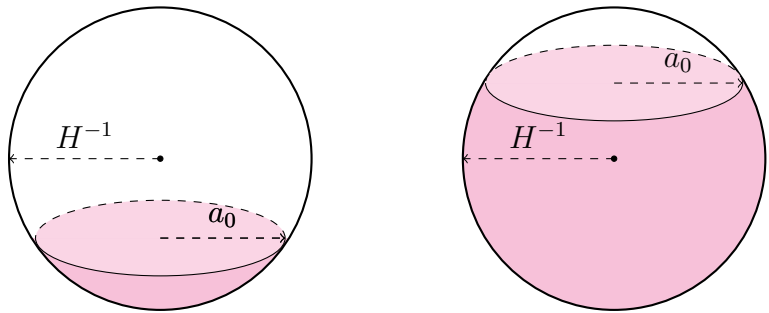
There are two methods for calculating the Euclidean action. The first is to directly calculate the instanton in (\ref{eq:4.32}) with the proper bounds and setting $a=H^{-1} \cosh{(Ht)}$, corresponding to the closed dS space we expect\cite{Reference16}\cite{Reference31}\cite{Reference33}. The second method used by Hartle and Hawking is to calculate an appropriate contour integral chosen so that it passes from the extremal points of the Euclidean action which are the same instantons. For the $a_{0}H <1$ case these turn out to be real, hence the Euclidean action and wave function are also found to be real. Specifically, the extrema are \cite{Reference1}:
\begin{equation} \label{eq:4.35}
    I_{\pm} = -\frac{1}{3H^{2}} \left [1 \pm (1 - H^{2}a_{0}^{2})^{\frac{3}{2}} \right ]
\end{equation}
where the $(+)$ value corresponds to the case of the bounded region covering more than a hemisphere, whereas the $(-)$ value for covering less than a hemisphere. It is then argued that the dominant contribution will be from the $I_{-}$ extremum. Even though the exponent is smaller, it ultimately comes down to the choice of contour. Some interesting observations can be made when approximating the probability for $Ha_{0} \rightarrow 1$. Both actions approach the value $-\frac{1}{3H^{2}}$, but for $I_{-}$ this is a minimum value, while for $I_{+}$ it is a maximum. Therefore when approaching this limit $I_{-}$  is decreasing until it reaches its minimum in contrast to $I_{+}$ that is increasing until it reaches its maximum. Since in the exponent we have $-I_{\pm}$, this leads to the inverse behaviours for the wave function. As we approach boundaries with radii close to that of the early Universe, the probability distribution is increasing in the case of $e^{-I_{-}}$, and decreasing for $e^{-I_{+}}$. At that limit, the boundary 3-sphere of radius $a_{0} = H^{-1}$ matches to the time slice $\tau = 0$ on the Euclidean 4-sphere. This serves as the smallest possible size of the Universe in Lorentzian space (see figures \ref{fig:2.1}, \ref{fig:2.3}). In other words, it is argued that it is more plausible for the probability distribution to predict a Universe that is small in its early stages, having a decreasing probability to occur for higher values than $H^{-1}$. When $a_{0}H >1$ there are no real extrema anymore. Consequently, the contour integral passes through complex values of the Euclidean action and the exponential becomes complex. We proceed to write the final solutions for both cases \cite{Reference1}:
\begin{equation} \label{eq:4.36}
    \bar{\psi} _{0} (a_{0}) \approx 
    \begin{cases}
    &\mathcal{N}_{0} \, \exp{ \left \{ \displaystyle\frac{1}{3H^{2}} \left [1-(H^{2}a_{0}^{2})^{\frac{3}{2}} \right ] \right \} } \, , \quad Ha_{0} < 1 \\
    &2 \cos{\left [\displaystyle\frac{(H^{2}a_{0}^{2} - 1)^{\frac{3}{2}}}{3H^{2}} - \frac{\pi}{4} \right ]} \, , \quad Ha_{0} > 1
    \end{cases}
\end{equation}
Notice how these solutions as well as the ones for an opposite sign in the exponent match the solutions (\ref{refsol1}) and (\ref{refsol2}) which we found in chapter 5. The inverse of the sphere radius $H$ plays the role of the constant $\tilde{\ell}$, while the ambiguity function takes the value:

\begin{equation}
    \rho ^{\textrm{q}} = \mathcal{A} ^{-\frac{3}{4}} |\mathcal{A}'|^{-\frac{3}{2}}
\end{equation}
which matches the form we know it takes in the path integral calculation\cite{Reference31}. Having calculated these answers, we choose the physical interpretation described earlier. The wave function describes the probability of the early Universe existing in the assumed configuration, which sometimes expressed as "emergence of a Universe out of nothing" \cite{Reference18}\cite{Reference20}\cite{Reference21}\cite{Reference17}. Given this interpretation, the limit $Ha_{0} \approx 1$ mentioned earlier provides us with the probability peak of the Universe emerging in that state. The wave function itself will be $\sim \exp{\left (\frac{1}{3H^{2}} \right )}$, meaning that it is far more likely for a Universe of smaller Hubble radius $H^{-1}$ to exist. From (\ref{eq:4.26}), this implies a very small cosmological constant. While this is valid for the present observable Universe, taking inflation into account invalidates the result. Therefore, Hartle-Hawking's result predicts a Universe that does not undergo inflation. \par
By contrast, Zeldovich and Starobinsky, Rubakov, and Linde and Vilenkin calculated this wave function for a reversed Wick rotation $t \rightarrow i\tau$ \cite{Reference17}\cite{Reference20}. Using the same method of steepest descents, the contour of integration had to be changed. This Linde-Vilenkin choice, as we shall call it, gives the same complex extrema that lead to the form of the wave function (\ref{eq:4.36}) when $Ha_{0} >1$, bar a small phase difference. The major change comes in the case of $Ha_{0}<1$. There the resulting wave function has the opposite sign in the exponent:
\begin{equation} \label{eq:4.37}
    \bar{\psi} _{0} (a_{0}) \approx \mathcal{N}_{0} \, \exp{ \left \{ -\frac{1}{3H^{2}} \left [1-(H^{2}a_{0}^{2})^{\frac{3}{2}} \right ] \right \} }
\end{equation}
The argument for the dominant contribution being that of $I_{-}$ is as follows. The action $I_{-}$ is smaller, therefore the exponential is less suppressed. The limit $Ha_{0} \approx 1$ now gives a wave function $\sim \exp{\left (-\frac{1}{3H^{2}} \right )}$, so the probability peak lies at a large value of $\Lambda$. This is consistent with inflation, as the early Universe still undergoing slow roll is at a roughly constant large value of $V(\phi)$. By extension, the Linde-Vilenkin sign choice favours a Universe emerging in an inflationary state - the complete opposite of Hartle-Hawking. This ambiguity in choosing a contour in such a way as to lead to completely different physical predictions shows that something is missing. Our own analysis from chapter 5 is did not prove adequate to resolve this, since the states for both sign choices are equally acceptable solutions of the WDW equation. We may then conclude that we are probably missing a crucial piece of information such as an initial condition that could determine the correct sign choice.\par
For the sake of completeness, another way to view this wave function would be the probability that an observer detects themselves to be in such a Universe, or possibly even a tunneling wave function through the barrier potential that appears in the $\sigma$-model Lagrangian \cite{Reference14}:
\begin{equation} \label{eq:4.38}
    \widetilde{V}(a, \phi ) = v_{3}a^{2} \left (6\kappa ^{2} - a^{2} V(\phi) \right )
\end{equation}
as it appears in (\ref{eq:4.15}). In the latter case, we see that the wave function for $Ha_{0}>1$ corresponds to the incident or transmitted waves that may exceed the barrier, whereas for $Ha_{0} <1$ it tunnels through the barrier as it cannot surpass it. With the Linde-Vilenkin sign choice, we do in fact get exponential damping for that regime.

\section{Scale factor path integral and modulus integral}

Hartle and Hawking's work has a significant drawback in that it doesn't make it clear how one could exactly perform the calculation themselves or apply the method to different systems. It is therefore worthwhile to approach the same calculation in greater detail and with the necessary mathematical technicalities addressed. We start from the definition (\ref{pathintegral2}) as a basis, but we make a Wick rotation to Euclidean space, we express it in terms of the scale factor parameter as it appears in the present chapter, and we use the gauge fixing of Appendix B\cite{Reference31}:
\begin{equation}
     \psi(a_0)=\Delta_{\rm FP}\int_0^{+\infty}\!\!d\ell \int_{a(0)=0}^{a(1)=a_{0}}\mathcal{D} a   \; e^{-{1\over \hbar}I[\ell ^{2},a]}
\end{equation}
where as we have proved in the same Appendix that we can freely set $\Delta_{\textrm{FP}}=1$ and the Euclidean action is given by:
\begin{equation}
    I[\ell ^{2},a] = 3sv_{3} \int _{0}^{1} d\tau \, \left [-\frac{a}{\ell} \left (\frac{da}{d\tau} \right )^{2} + \ell \, V(a) \right ]
\end{equation}
We will see that the factor $s=\pm1$ which we introduced in the WDW equation solutions of chapter 5 is none other than the choice between Hartle-Hawking and Linde-Vilenkin. Just like the calculation of the previous section, the validity of the following analysis using the method of steepest descents holds assuming we work in the semiclassical limit.

\subsection{Instanton solutions}

We begin by extremizing the Euclidean action with respect to $a(\tau) , \, \ell$ with the extremum point at $(\bar{\ell}^{2}, \bar{a})$:
\begin{equation} \label{extrem}
    \frac{dI}{d\ell} \Bigg |_{(\bar{\ell}^{2}, \bar{a})} = 3sv_{3}\int _{0}^{1} d\tau \, \left [-\frac{\bar{a}}{\bar{\ell}^{2}} \left (\frac{d\bar{a}}{d\tau} \right )^{2} + V(\bar{a}) \right ] = 0
\end{equation}
If we now express the scale factor as\cite{Reference31}:
\begin{equation} \label{bc}
\begin{split}
    &a(\tau) = \bar{a}(\tau) + \delta a(\tau) \\
\text{where} \qquad \bar{a}(0) = 0 \, , \quad &\bar{a}(1) = a_{0} \quad \text{and} \quad \delta a(0) = \delta a(1) = 0
\end{split}
\end{equation}
then the equation of motion for $a$ becomes:
\begin{equation} \label{extrem2}
    \frac{\delta I}{\delta a} \Bigg |_{(\bar{\ell}^{2}, \bar{a})} = 0 \quad \Leftrightarrow \quad -\frac{3}{\bar{a}^{2}} \left (\frac{d\bar{a}}{d\tau} \right )^{2} - 2 \frac{d}{d\tau} \left (\frac{1}{\bar{a}} \frac{d\bar{a}}{d\tau} \right ) + \frac{\bar{\ell}^{2}}{\bar{a}^{2}} V_{a}(\bar{a}) = 0
\end{equation}
with the subscript in the potential denoting partial derivative with respect to $a$. The only way for (\ref{extrem}) to hold is for the special case of the integrand vanishing, which leads to none other than the Friedmann equations. To see this differently, solving for the e.o.m. in (\ref{extrem2}) we get:
\begin{equation}
    -\frac{\bar{a}}{\bar{\ell}^{2}} \left (\frac{d\bar{a}}{d\tau} \right )^{2} + V(\bar{a}) = \frac{\mathcal{E}}{3v_{3}}
\end{equation}
where mathematically $\mathcal{E}$ is an arbitrary integration constant, but physically is related to the total energy of the Universe. The extremization (\ref{extrem}) is then tantamount to setting $s\mathcal{E} = 0$, hence the total energy of the Universe is 0 (as we expect). By extension, in order to extremize the action in terms of both parameters it suffices to solve the resulting Friedmann equation:
\begin{equation}
    \left (\frac{d(H\bar{a})}{d(H \bar{\ell} \tau)} \right )^{2} + (H\bar{a})^{2} = 1
\end{equation}
where we remind that we have set $H = \tilde{\ell} = \sqrt{\ell/3}$. This yields the solution:
\begin{equation}
    \bar{a}(\tau) = |\sin {(H\bar{\ell}\tau )} |
\end{equation}
Here we have implicitly imposed $\bar{a}(0)=0$, while the second boundary condition $\bar{a}(1)=a_{0}$ fixes the modulus $\bar{\ell}$. We encounter once again the different solutions outlined in the previous section, depending on the field of definition of $Ha_{0}$. We shall focus on $0 < Ha_{0} <1$ in this section. In this regime, $\bar{\ell}$ takes discrete real values of which only two gives us two smooth instanton solutions\cite{Reference31}:
\begin{equation} \label{instsol}
\begin{split}
    &H\bar{a}_{\epsilon}(\tau) = \sin {(H\bar{\ell}_{\epsilon}\tau)} \, , \quad \epsilon = \pm 1 \\
    \text{where} \qquad H\bar{\ell}_{+}(\tau) &= \arcsin{(Ha_{0})} \, , \quad H\bar{\ell}_{-}(\tau) = \pi - \arcsin{(Ha_{0})}
\end{split}
\end{equation}
Technically, there are other possible solutions:
\begin{equation}
    H\bar{a}_{n}^{\epsilon}(\tau) = \sin {(H\bar{\ell}_{n}^{\epsilon}\tau)} \, , \qquad \text{where} \qquad \bar{\ell}^{\epsilon}_{n} = \bar{\ell}_{\epsilon} + n\pi  \, , \quad n\in \mathbb{N}^{*} 
\end{equation}
These correspond to spacetimes that resembles "necklaces with $n$ beads", hence they are not smooth (as they are not differentiable)\cite{Reference67}. Given the requirement that the path integral sums over smooth manifolds, we omit such solutions. \par
The Euclidean space line segments are:
\begin{equation}
    (d\bar{s}_{\textrm{E}}^{\epsilon})^{2} = \frac{1}{H^{2}} \left [d(H\bar{\ell}_{\epsilon}\tau)^{2} + \sin ^{2}{(H\bar{\ell}_{\epsilon} \tau)} \,d\Omega _{3}^{2} \right ] \, , \quad \tau \in [0,1]
\end{equation}
These describe portions 4-spheres of radius $H^{-1}$. Indeed, these are none other than the two cases shown in (\ref{fig:4.1}), where $\epsilon =-1$ describes the one on the right (covering more than a hemisphere), while $\epsilon = +1$ the one on the left (covering less than a hemisphere). The Euclidean actions are therefore:
\begin{equation}
    \bar{I}_{\epsilon} = s\frac{2v_{3}}{H^{2}} \left [1 - \epsilon \left (1 - (Ha_{0})^{2} \right )^{3/2} \right]
\end{equation}
in full agreement with (\ref{eq:4.35}).
 
\subsection{Expansion to quadratic order}

Let us now expand the Euclidean action around the extremal solutions $(\bar{\ell}^{2}_{\epsilon} ,\bar{a}_{\epsilon})$. We first split the parameters as the extremal values plus a small perturbation\cite{Reference31}:
\begin{equation}
    \ell = \bar{\ell}_{\epsilon} = \delta \ell \, , \quad \bar{a}(\tau) = \bar{a}_{\epsilon}(\tau) + \delta a(\tau)
\end{equation}
with the same boundary conditions in (\ref{bc}). Keeping up to the second power of the perturbations, the action becomes:
\begin{equation} \label{eucact}
    I[\ell ^{2},a] = \bar{I}_{\epsilon} + 3sv_{3} \int _{0}^{1} d\tau \, \bar{\ell}_{\epsilon} \left [ \delta a \mathcal{Q}_{\epsilon} \delta a + 2 \delta a V_{a}(\bar{a}_{\epsilon}) \frac{\delta \ell}{\bar{\ell}_{\epsilon}} + \frac{\delta \ell}{\bar{\ell}_{\epsilon}} V(\bar{a}_{\epsilon} )\frac{\delta \ell}{\bar{\ell}_{\epsilon}} \right ] 
\end{equation}
where we have defined the quadratic differential operator:
\begin{equation} \label{operator2}
    \mathcal{Q}_{\epsilon} = -\frac{\bar{a}_{\epsilon}}{\bar{\ell}_{\epsilon}^{2}} \frac{d^{2}}{d\tau ^{2}} + \frac{1}{\bar{\ell}_{\epsilon}^{2}} \frac{d\bar{a}_{\epsilon}}{d\tau} \frac{d}{d\tau} + \frac{1}{2} V_{aa}(\bar{a}_{\epsilon})
\end{equation}
All linear terms with respect to the fluctuations go away due to the classical e.o.m. If we view the fluctuation $\delta a$ as an element of the Hilbert space of $L^{2}([0,1])$ functions, i.e. those functions that vanish at the boundaries $\tau = 0$ and $\tau = 1$, then we may define an inner product within this space\cite{Reference31}:
\begin{equation}
    (\delta a_{1}, \delta a_{2})_{\bar{\ell}_{\epsilon}} = \int _{0}^{1} d\tau \, \bar{\ell}_{\epsilon} \, \delta a_{1} \delta a_{2}
\end{equation}
The operator (\ref{operator2}) then satisfies:
\begin{equation}
     (\delta a_{1}, \mathcal{Q}_{\epsilon}\delta a_{2})_{\bar{\ell}_{\epsilon}} = (\mathcal{Q}_{\epsilon} ^{T}\delta a_{1}, \delta a_{2})_{\bar{\ell}_{\epsilon}}
\end{equation}
where the adjoint can be retrieved through integration by parts:
\begin{equation}
    \mathcal{Q}_{\epsilon} ^{T} = -\frac{\bar{a}_{\epsilon}}{\bar{\ell}_{\epsilon}^{2}} \frac{d^{2}}{d\tau ^{2}} - \frac{3}{\bar{\ell}_{\epsilon}^{2}} \frac{d\bar{a}_{\epsilon}}{d\tau} \frac{d}{d\tau} - \frac{2}{\bar{\ell}_{\epsilon}^{2}} \frac{d^{2}\bar{a}_{\epsilon}}{d\tau ^{2}} + \frac{1}{2} V_{aa}(\bar{a}_{\epsilon})
\end{equation}
Both the operator and its adjoint are endomorphisms i.e. their values upon acting on a fluctuation keeps it within the same space of functions. This isn't ensured immediately from the form of the function, as the derivatives don't necessarily vanish at the boundaries. We explicitly demand that this is true while expanding the Hilbert space vectors in the orthonormal Fourier basis $\left \{\sqrt{2/\bar{\ell}_{\epsilon}} \, \sin{(k\pi \tau )} \, , \, k\in \mathbb{N}^{*} \right \}$\cite{Reference31}. \par
We may construct a symmetric and antisymmetric operator from these two:
\begin{equation}
    \mathcal{S}_{\epsilon} = \frac{\mathcal{Q}_{\epsilon} + \mathcal{Q}_{\epsilon} ^{T}}{2} \, , \quad \mathcal{A}_{\epsilon} = \frac{\mathcal{Q}_{\epsilon} - \mathcal{Q}_{\epsilon} ^{T}}{2}
\end{equation}
with the latter not to be confused with the field redefinitions of the scale factor from the previous chapter. By construction, these new operators satisfy the relations:
\begin{equation}
    (\delta a_{1}, \mathcal{S}_{\epsilon}\delta a_{2})_{\bar{\ell}_{\epsilon}} = (\mathcal{S}_{\epsilon} \delta a_{1}, \delta a_{2})_{\bar{\ell}_{\epsilon}} \, , \qquad (\delta a_{1}, \mathcal{A}_{\epsilon}\delta a_{2})_{\bar{\ell}_{\epsilon}} = -(\mathcal{A}_{\epsilon} \delta a_{1}, \delta a_{2})_{\bar{\ell}_{\epsilon}}
\end{equation}
We immediately conclude that the inner product of $\mathcal{A}_{\epsilon}$ is 0, while $\mathcal{S}_{\epsilon}$ is self-dual, therefore:
\begin{equation}
    (\delta a_{1}, \mathcal{Q}_{\epsilon}\delta a_{2})_{\bar{\ell}_{\epsilon}} = (\delta a_{1}, \mathcal{S}_{\epsilon}\delta a_{2})_{\bar{\ell}_{\epsilon}}
\end{equation}
Using (\ref{instsol}), the symmetric operator can be written as:
\begin{equation}
     \mathcal{S}_{\epsilon} = -\frac{\bar{a}_{\epsilon}}{\bar{\ell}_{\epsilon}^{2}} \frac{d^{2}}{d\tau ^{2}} + \frac{1}{\bar{\ell}_{\epsilon}^{2}} \frac{d\bar{a}_{\epsilon}}{d\tau} \frac{d}{d\tau} + 2H^{2}\bar{a}_{\epsilon}
\end{equation}
If we want to diagonalize the integrand in (\ref{eucact}), then $\mathcal{S}_{\epsilon}$ must be invertible in the regime $0 < H a_{0} < 1$. We shall see how that comes about later in this section. For now let's assume it holds, in which case the diagonalization yields:
\begin{equation}
    \delta a \,\mathcal{S}_{\epsilon} \delta a + 2 \delta a V_{a}(\bar{a}_{\epsilon}) \frac{\delta \ell}{\bar{\ell}_{\epsilon}} + \frac{\delta \ell}{\bar{\ell}_{\epsilon}} V(\bar{a}_{\epsilon}) \frac{\delta \ell}{\bar{\ell}_{\epsilon}} = \delta \check{a} \,\mathcal{S}_{\epsilon} \delta \check{a} + \frac{\delta \ell}{\bar{\ell}_{\epsilon}} \left [V(\bar{a}_{\epsilon}) -  V_{a}(\bar{a}_{\epsilon}) \mathcal{S}_{\epsilon}^{-1} V_{a}(\bar{a}_{\epsilon}) \right ]\frac{\delta \ell}{\bar{\ell}_{\epsilon}}
\end{equation}
where we have defined:
\begin{equation}
    \delta \check{a} = \delta a + \frac{\delta \ell}{\bar{\ell}_{\epsilon}} \mathcal{S}_{\epsilon}^{-1} V_{a}(\bar{a}_{\epsilon})
\end{equation}
Again, if we take the derivative of the potential at face value it does not vanish at the boundaries:
\begin{equation}
V_a(\bar{a}_\epsilon) = 1-3(H \bar{a}_\epsilon)^2\quad \Longrightarrow\quad V_a(\bar{a}_\epsilon(0))=1\, ,~~ V_a(\bar{a}_\epsilon(1))=1-3(H a_0)^2
\label{pb}
\end{equation}
which means it doesn't belong to the space of functions we are working in. To amend this, we define it on $(0,1)$, and extend it such that it is odd on $(-1,0)\cup (0,1)$ and 2-periodic on $\mathbb{R} \setminus \mathbb{Z}$. Fourier expanding it will only yield modes $\left \{\sqrt{2/\bar{\ell}_{\epsilon}} \, \sin{(k\pi \tau )} \, , \, k\in \mathbb{N}^{*} \right \}$, vanishing for $\tau \in \mathbb{Z}$\cite{Reference31}. Taking $V_{a}$ to be this Fourier expansion ensures that it does vanish at the boundaries, belonging to the same space. \par
We move on with the steepest descent method, which yields\cite{Reference31}:
\begin{align} \label{fluctpsi}
\psi(a_0)=&\; \Delta_{\rm FP}\sum_{\epsilon=\pm1 } e^{-{1\over \hbar} \bar I_\epsilon} \int d\delta \ell \,\exp\Big\{\!-\!{3sv_3\over \hbar}\,   \mathcal{K}_\epsilon  \Big({\delta\ell\over \bar{\ell}_\epsilon}\Big)^2\Big\}\nonumber \\
&\; \int_{\textstyle \substack{\;\delta a(0)=0 \\\; \delta a(1)=0}}\mathcal{D} \delta a  \, \exp\Big\{\!-\!{3sv_3\over \hbar}\,( \delta \check a, \mathcal{S}_\epsilon \delta \check a)_{\bar{\ell}_\epsilon}\Big\}\,(1+\mathcal{O}(\hbar))
\end{align} 
where:
\begin{equation}
  \mathcal{K}_\epsilon = \int_0^1 d\tau \bar{\ell}_\epsilon\, \big[V(\bar{a}_\epsilon)-V_a(\bar{a}_\epsilon)\mathcal{S}_\epsilon^{-1} V_a(\bar{a}_\epsilon)\big]
 \label{barI}
\end{equation}
Given that as we have shown $\mathcal{S}_\epsilon^{-1}V_a(\bar{a}_\epsilon)$ belongs to the given Hilbert space, the same will be true of $\delta \check{a}$. This allows us to make a field redefinition in the path integral above $\delta a \rightarrow \delta \check{a}$ with the same vanishing boundary conditions. As a result we have:
\begin{equation}
\mathcal{D}\delta a= \mathcal{D}\delta \check a \, \left|{\mathcal{D}\delta a\over \mathcal{D}\delta \check a}\right|= \mathcal{D}\delta \check a
\end{equation}
which leads to the expression for the wave function:
\begin{align}
\psi(a_0)&=\Delta_{\rm FP}\sum_{\epsilon=\pm 1} e^{-{1\over \hbar} \bar{I}_{\epsilon}}\,Z_\epsilon(a_0) \int d\delta \ell \,\exp\Big\{\!-\!{3sv_3\over \hbar}\,  \mathcal{K}_\epsilon  \Big({\delta\ell\over \bar{\ell}_\epsilon}\Big)^2\Big\}\,(1+\mathcal{O}(\hbar))\nonumber \\
\text{where} \quad Z_\epsilon(a_0)&= \int_{\textstyle \substack{\;\delta \check a(0)=0 \\ \;\delta \check a(1)=0}}\mathcal{D} \delta \check a  \, \exp\Big\{\!-\!{3sv_3\over \hbar}\,( \delta \check a, \mathcal{S}_\epsilon \delta \check a)_{\bar{\ell}_\epsilon}\Big\}
\label{psii}
\end{align}
For the rest of this section we shall write $\delta \check{a}$ as $\delta a$ to simplify the expressions.

\subsection{Scale factor quadratic fluctuations}

We start by diagonalizing $\mathcal{S}_{\epsilon}$ in an orthonormal, which we can do since it's a self-dual endomorphism of the Hilbert space\cite{Reference31}:
\begin{equation}
    \mathcal{S}_\epsilon\phi^\epsilon_k = \nu^\epsilon_k\phi^\epsilon_k\, , ~~ k\in\mathbb{N}^*\, , \quad~~\text{where}~~\quad  (\phi^\epsilon_k,\phi^\epsilon_{k'})_{\bar{\ell}_\epsilon}=\delta_{kk'}\, ,~~ \nu^\epsilon_k\in\mathbb{R}
\end{equation}
where we have denoted $\phi^\epsilon_k$ the eigenfunctions with eigenvalues $\nu^\epsilon_k$. We can Fourier expand the fluctuation:
\begin{equation}
    \delta a(\tau)=\sum_{k\ge 1} \boldsymbol\delta \mathbf{a}_k\,  \phi^\epsilon_k(\tau)
\end{equation}
from which we can get the path integral measure:
\begin{equation}
    ||\delta a||_{\bar{\ell}_\epsilon}^2 =  (\delta a,\delta a)_{\bar{\ell}_\epsilon} = \sum_{k\ge 1} (\boldsymbol\delta \mathbf{a}_k)^2 \quad\Longrightarrow\quad \mathcal{D}\delta a = \bigwedge_{k\ge 1}d \boldsymbol\delta \mathbf{a}_k
\end{equation}
The expression for the quadratic-level function of scale factor fluctuations:
\begin{equation}
\begin{split}
    Z_\epsilon(a_0)&=\prod_{k\ge 1}\int d\boldsymbol\delta \mathbf{a}_k\; e^{-{3sv_3\over \hbar} \nu^\epsilon_k(\boldsymbol\delta \mathbf{a}_k)^2}= \prod_{k\ge 1} \sqrt{\hbar\, \pi\over 3sv_3\, \nu^\epsilon_k}\nonumber\\
&= \left({3sv_3\over \hbar \, \pi}\right)^{1\over 4}{1\over \sqrt{\det \mathcal{S}_\epsilon}} \, , \qquad \text{where} \quad \det\mathcal{S}_\epsilon = \prod_{k\ge 1}\nu^\epsilon_k
\end{split}
\end{equation}
where we made use of zeta-function regularization. This quantity is gauge-invariant in the sense that it is independent on the choice of metric. While we won't provide an explicit proof of this, it has been demonstrated in the literature\cite{Reference16}\cite{Reference68}. \par
An important point to raise with regards to the path integral is that we make sure that the mode $\boldsymbol\delta \mathbf{a}_k$ is summed over all real values in $(-\infty , +\infty)$ for $s \nu_k^\epsilon>0$, but over all imaginary values in $(-i \infty ,+i\infty)$ for $s \nu_k^\epsilon<0$. The value $\nu_k^\epsilon=0$ is skipped, since at that point $\mathcal{S}_{\epsilon}$ is no longer invertible. Because of this prescription, all path integrals are convergent and therefore $Z_{\epsilon}(a_{0})$ exists\cite{Reference31}. \par
In order to compute these determinants, we shall employ Coleman's method\cite{Reference16}. We start with the following differential equation and boundary conditions:
\begin{equation} \label{difsystem}
\begin{split}
    &\mathcal{S}_\epsilon \varphi^\epsilon_\nu = \nu \varphi^\epsilon_\nu \, , \\
    &\varphi^\epsilon_\nu(\tau_\epsilon)=0\, , \qquad {d\varphi^\epsilon_\nu\over d\tau(\tau_\epsilon)}=1
\end{split}
\end{equation}
We have eigenvalues $\nu \in \mathbb{R}$, while $\tau_{\epsilon} \in (0,1)$ is a regulator which will be set equal to 0 at the end\cite{Reference31}. The differential operator $\mathcal{S}_{\epsilon}$ as we have seen is of order 2, therefore with the two b.c.s above we can find a unique solution. When we have $\varphi^\epsilon_\nu(1)=0$, then there exists a $k\in \mathbb{N}^{*}$ such that $\nu = \nu ^{\epsilon}_{k}$. We may arbitrarily order these eigenvalues in ascending order($\nu _{1}^{\epsilon} < \nu _{2}^{\epsilon} < \dots$). In addition to that, we know that the dimension of the eigenspace of $\mathcal{S}_{\epsilon}$ is strictly equal to 1, otherwise we would get more than 1 independent solutions for the same $k$ given the same b.c.s. Let's generalize the form of the differential operator\cite{Reference31}:
\begin{equation}
    \mathcal{S}_\epsilon^{(i)} = -{\bar{a}_\epsilon\over \bar{\ell}_\epsilon^2}\, {d^2\over d\tau^2}-{1\over \bar{\ell}_\epsilon^2}\, {d\bar{a}_\epsilon\over d \tau}\, {d\over d\tau}+W^{(i)}_\epsilon\, ,\quad i\in\{1,2\}
\end{equation}
where $W^{(i)}_{\epsilon}(\tau)$ arbitrary functions defined on $[\tau_{\epsilon},1]$. The solutions shall then be:
\begin{equation}
\begin{split}
    \varphi_\nu^{\epsilon(i)}(1)=0\quad &\Longleftrightarrow \quad \text{$\exists \, \, \, k\in\mathbb{N}^*$ such that $\nu=\nu_k^{\epsilon(i)}$} \\
&\Longleftrightarrow \quad  \det(\mathcal{S}^{(i)}_\epsilon-\nu)\equiv \prod_{k\ge 1}(\nu_k^{\epsilon(i)}-\nu)=0
\end{split}
\end{equation}
for which we get\cite{Reference16}:
\begin{equation}
    {\det(\mathcal{S}^{(1)}_\epsilon-\nu)\over \det(\mathcal{S}^{(2)}_\epsilon-\nu)}={\varphi_\nu^{\epsilon(1)}(1)\over \varphi_\nu^{\epsilon(2)}(1)}
\end{equation}
Since the arbitrary functions $W^{(i)}_{\epsilon}(\tau)$ are bounded within their field of definition, the lhs and rhs both tend to 1 as $\nu \rightarrow \infty$ on the complex plane. This means the equality is true for any $\nu \in \mathbb{C}$, implying that there is a constant with respect to $\tau$ and independent of $W^{i}_{\epsilon}(\tau)$\cite{Reference31}:
\begin{equation}
    \mathcal{N}_\epsilon\equiv {\det(\mathcal{S}^{(i)}_\epsilon-\nu)\over\varphi_\nu^{\epsilon(i)}(1)}\, , \quad i\in\{1,2\}
\end{equation}
By construction, this constant is evidently dependent on the regulator $\tau_{\epsilon}$ and the scale factor boundary value $a_{0}$. If we take $\nu = 0$, we get an expression for the determinant of the differential operator we seek to calculate:
\begin{equation} \label{ndet}
    \det \mathcal{S}^{(i)}_\epsilon=\mathcal{N}_\epsilon\, \varphi^{\epsilon(i)}_{0}(1)
\end{equation}
In order to carry on with the calculation, we need to calculate $\varphi^{\epsilon(i)}_{0}(1)$ from (\ref{difsystem}) for $\nu = 0$. To simplify things, we first make the following redefinition\cite{Reference31}:
\begin{equation}
    \Phi_\epsilon(\theta)=H\varphi^\epsilon_{0}(\tau)\, , \quad \text{where} \quad \theta = H \bar{\ell}_\epsilon \tau \in[\theta_*,H\bar{\ell}_\epsilon]\, , \quad\theta_*=H\bar{\ell}_\epsilon\tau_\epsilon
\end{equation}
where $\theta _{*}$ is a regulator in the field of definition of $\theta$. Thus the system (\ref{difsystem}) turns into:
\begin{equation}
\begin{split}
    -&\sin\theta  \,{d^2\Phi_\epsilon\over d\theta^2}- \cos\theta \, {d\Phi_\epsilon\over d\theta}-2\sin\theta \,\Phi_\epsilon= 0\, , \\
    &\qquad\Phi_\epsilon(\theta_*)=0\, , \quad  {d \Phi_\epsilon\over d\theta}(\theta_*)={1\over \bar{\ell}_\epsilon}
\end{split}
\end{equation}
The general solution for the differential equation we find is:
\begin{equation}
    \Phi_\epsilon(\theta)= A_\epsilon \cos \theta + B_\epsilon \Big[\!-\!1+{\cos \theta \over 2}\log\!\Big({1+\cos \theta\over 1-\cos \theta}\Big)\Big]
\end{equation}
where $A_{\epsilon} ,B_{\epsilon}$ constants which we can calculate by applying the two b.c.s. The introduction of the regulator is now clear. Had we allowed $\tau_{\epsilon}=0$ and by extension $\theta _{*} = 0$ from the beginning, then we would find that $A_{\epsilon} = B_{\epsilon} = 0$\cite{Reference31}. We need the regulator to extract useful non-trivial results, and then applying the appropriate limit. The result in this case is:
\begin{equation}
    A_\epsilon\underset{\theta_*\to 0}{\sim}{\theta_*\over \bar{\ell}_\epsilon}\log{1\over \theta_*}\, , \quad~~B_\epsilon\underset{\theta_*\to 0}{\sim}-{\theta_*\over \bar{\ell}_\epsilon}
\end{equation}
Both of these do in fact tend to 0, but now that we have their approximate behaviour, we can see that $B_{\epsilon}$ tends to 0 faster than $A_{\epsilon}$, meaning the term involving the latter dominates at that limit. This allows us to obtain a form for the determinant of the operator:
\begin{equation} \label{detS}
\begin{split}
    \det \mathcal{S}_\epsilon &= {\mathcal{N}_\epsilon \over H}\, \Phi_\epsilon(H \bar{\ell}_\epsilon) \underset{\theta_*\to 0}{\sim} {\theta_*\over H}\log{1\over \theta_*}\,  {\mathcal{N}_\epsilon\over  \bar{\ell}_\epsilon} \, \cos(H \bar{\ell}_\epsilon)\nonumber \\
&\!\:\!\!\!\underset{\theta_*\to 0}{\sim} {\theta_*\over H}\log{1\over \theta_*}\, {\mathcal{N}_\epsilon\over \bar{\ell}_\epsilon} \, \epsilon\, \sqrt{1-(H a_0)^2}
\end{split}
\end{equation}
The final piece of the puzzle is to calculate the constant factor $\mathcal{N}_{\epsilon}$. To do so, we will compute the determinant of any operator $\mathcal{S}_{\epsilon}^{(1)}$ as the infinite products of its eigenvalues. This requires a suitable form for $W^{(1)}_{\epsilon}(\tau)$ which would make the computation possible. We start from\cite{Reference31}:
\begin{equation}
\begin{split}
    \mathcal{S}^{(1)}_\epsilon \psi^\epsilon_k &= \nu_k^{\epsilon(1)}\,  \psi^\epsilon_k  \, , \\
\psi^\epsilon_k(\tau_\epsilon)=0\, &, \quad  \psi^\epsilon_k(1)=0
\end{split}
\end{equation}
where the eigenstates $\psi_{k}^{\epsilon}$ shouldn't be confused with the quantum mechanical wave function. The b.c. on the left gives a linear relation between the two constants which we obtain from the solution of the differential equation. The b.c. on the right enforces that $\nu _{k}^{\epsilon}$ is an eigenvalue of the differential operator acting on the Hilbert space of the functions $L^{2}([\tau_{\epsilon},1])$. \par
We choose to introduce a new Euclidean time variable $\zeta$ and functions $\check{\psi}^{\epsilon}_{k}$ such that\cite{Reference31}:
\begin{equation} \label{varredef}
    \bar{\ell}_\epsilon\,  d\tau = \sqrt{\bar{a}_\epsilon(\tau)}\, d \zeta\, , \quad ~~\psi^\epsilon_k(\tau)=\bar{a}^{-{1/4}}_\epsilon(\tau)\, \check\psi^\epsilon_k(\zeta)
\end{equation}
The differential equation subsequently takes the canonical form:
\begin{equation} \label{redefsys}
\begin{split}
    -{d^2\check\psi^\epsilon_k\over d\zeta^2}+&\!\big(W^{(1)}_\epsilon-{1\over 16} \,{1\over \bar{a}_\epsilon} \, [1+3(H \bar{a}_\epsilon)^2]\big)\,\check \psi^\epsilon_k=\nu_k^{\epsilon(1)}\, \check \psi^\epsilon_k \, ,  \\
&\check \psi^\epsilon_k(\zeta(\tau_\epsilon))=0\, , \quad  \check \psi^\epsilon_k(\zeta(1))=0
\end{split}
\end{equation}
A viable choice we can make is\cite{Reference31}:
\begin{equation}
    W^{(1)}_\epsilon={1\over 16} \,{1\over \bar{a}_\epsilon} \, [1+3(H \bar{a}_\epsilon)^2]
\end{equation}
allowing us to easily calculate the solutions to the system:
\begin{equation}
    \check \psi_{k}^{{\epsilon}}(\zeta)= C_k\sin\!\left[\sqrt{\nu_k^{\epsilon(1)}}\, \big(\zeta-\zeta(\tau_\epsilon) \big)\right]\, ,\quad  \text{where}\quad \sqrt{\nu_k^{\epsilon(1)}} = {k\pi\over \zeta(1)-\zeta(\tau_\epsilon)}\, , \quad k\in \mathbb{N}^*
\end{equation}
where $C_{k}$ are arbitrary constants. The infinite product of the eigenvalues using zeta regularization gives us:
\begin{equation} \label{resdet}
    \det\mathcal{S}_\epsilon^{(1)} = \prod_{k\ge 1} k^2\prod_{k'\ge 1}\Big({\pi\over \zeta(1)-\zeta(\tau_\epsilon)}\Big)^2 = 2\big[\zeta(1)-\zeta(\tau_\epsilon)\big]
\end{equation}
Now we need to return to the expression (\ref{ndet}) and calculate the determinant once again, upon which we will equate it with the expression above. This time the system reads:
\begin{equation}
\begin{split}
    \mathcal{S}^{(1)}_\epsilon \varphi^{\epsilon(1)}_0 &= 0 \, , \\
\varphi^{\epsilon(1)}_0(\tau_\epsilon)=0\, &, \quad {d\varphi^{\epsilon(1)}_0\over d\tau}(\tau_\epsilon)=1
\end{split}
\end{equation}
and we introduce the same change of variable $\tau \rightarrow\zeta$ as (\ref{varredef}):
\begin{equation}
    \varphi^{\epsilon(1)}_0(\tau) = \bar{a}^{-{1/4}}_\epsilon(\tau)\,\check \varphi^{\epsilon(1)}_0(\zeta)
\end{equation}
The system takes an analogous form to (\ref{redefsys}):
\begin{equation}
\begin{split}
    -{d^2\check \varphi_0^{\epsilon(1)}\over d \zeta^2}= 0& \, , \\
\check \varphi^{\epsilon(1)}_0(\zeta(\tau_\epsilon))=0\, , \quad {d\check \varphi^{\epsilon(1)}_0\over d\zeta}&(\zeta(\tau_\epsilon))={\bar{a}_\epsilon(\tau_\epsilon)^{3\over 4}\over \bar{\ell}_\epsilon}
\end{split}
\end{equation}
yielding the solutions:
\begin{equation}
    \check \varphi^{\epsilon(1)}_0(\zeta)={\bar{a}^{3/4}_\epsilon(\tau_\epsilon)\over \bar{\ell}_\epsilon} \, \big[\zeta-\zeta(\tau_\epsilon)\big]\quad \Longrightarrow\quad \varphi^{\epsilon(1)}_0(\tau) ={\bar{a}^{3/4}_\epsilon(\tau_\epsilon)\over \bar{\ell}_\epsilon} \,\bar{a}^{-{1/4}}_\epsilon(\tau)\,  \big[\zeta(\tau)-\zeta(\tau_\epsilon)\big]
\end{equation}
The expression of the determinant becomes:
\begin{equation}
\begin{split}
     \det \mathcal{S}_\epsilon^{(1)}&=\mathcal{N}_\epsilon\, \varphi^{\epsilon(1)}_0(1) \\
     &= \mathcal{N}_\epsilon\, {\bar{a}^{3/4}_\epsilon(\tau_\epsilon)\over \bar{\ell}_\epsilon} \,a_0^{-{1/4}}\,  \big[\zeta(1)-\zeta(\tau_\epsilon)\big]
\end{split}
\end{equation}
Setting this equal to the result (\ref{resdet}):
\begin{equation}
    \mathcal{N}_\epsilon = 2\, \bar{a}^{-{3/4}}_\epsilon(\tau_\epsilon)\, \bar{\ell}_\epsilon \,a_0^{1/4} \underset{\theta_*\to 0}{\sim} 2\, \Big({H\over \theta_*}\Big)^{3/4} \bar{\ell}_\epsilon\, a_0^{1/4}
\end{equation}
Finally, having calculated this constant, we plug it back into (\ref{detS}):
\begin{equation}
    \det \mathcal{S}_\epsilon \underset{\theta_*\to 0}{\sim} 2\, \Big({\theta_*\over \lambda}\Big)^{1/4}\log{1\over \theta_*} \, \epsilon\, a_0^{1/4}\sqrt{1-(H a_0)^2}
\end{equation}
When we are in the classically allowed regime $0 < Ha_{0} <1$, then $\det \mathcal{S}_{+} > 0$ and $\det\mathcal{S}_{-} <0$, ensuring that the operators $\mathcal{S}_{\pm}$ are invertible.

\subsection{Quadratic fluctuations of the length}

We move on to the calculation of the path integral over the segment length fluctuations $\delta \ell$. In order to do that, we must first find the exact expression for the quantity (\ref{barI}). To determine the quantity $\mathcal{S}_{\epsilon}^{-1}V_{a}(\bar{a}_{\epsilon}$, we must find the unique solution to the differential equation\cite{Reference31}:
\begin{equation}
\begin{split}
    &\mathcal{S}_\epsilon f_\epsilon = V_a(\bar{a}_\epsilon) \, , \\
f_\epsilon(\tau_\epsilon)&=0\, , \quad f_\epsilon(1)=0
\end{split}
\end{equation}
where we once again introduce the regulator $\tau_{\epsilon}$. As we stated earlier, we have that $V_{a}(\bar{a}_{\epsilon}) = 1 - 3(H\bar{a}_{\epsilon})^{2}$ on the open set $(\tau_{\epsilon},1)$ with vanishing b.c.s where it is also discontinuous. To simplify calculations, let us once again perform a redefinition\cite{Reference31}:
\begin{equation}
     F_\epsilon(\theta)=H f_\epsilon(\tau)
\end{equation}
which brings the system above to the form:
\begin{equation}
\begin{split}
    -\sin\theta  \,{d^2F_\epsilon\over d\theta^2}- &\cos\theta \, {d F_\epsilon\over d\theta}-2\sin\theta \,F_\epsilon= 1-3\sin^2\theta\, ,\quad \theta\in(\theta_*,H \bar{\ell}_\epsilon)\, ,  \\
&F_\epsilon(\theta_*)=0\, , \quad F_\epsilon(H \bar{\ell}_\epsilon)=0
\end{split}
\end{equation}
We find that the general solution is:
\begin{equation}
    F_\epsilon(\theta)= -\theta \cos\theta+D_\epsilon \cos \theta + E_\epsilon \Big[\!-\!1+{\cos \theta \over 2}\log\!\Big({1+\cos \theta\over 1-\cos \theta}\Big)\Big]
\end{equation}
where $D_{\epsilon}, E_{\epsilon}$ are constants determined by the b.c.s. This time the role of the regulator is as follows: had we taken the lower bound to be at $\tau = 0$ and thus $\theta = 0$, then the solution would become $F_{\epsilon} (\theta) = -\theta \cos \theta$. Therefore, $F_{\epsilon} (H\bar{\ell}_{\epsilon}) = -H\bar{\ell}_{\epsilon} \, \epsilon \sqrt{1-(Ha_{0})^{2}}$. For the our regime of interest ($0<Ha_{0}<1$) this can never become 0, yielding no solution\cite{Reference31}. Now that we have the regulator $\theta_{*}$, we take the limit $\theta_{*}\rightarrow 0$ to find the expressions for the integration constants:
\begin{equation}
\begin{split}
    D_\epsilon &= H \bar{\ell}_\epsilon\left\{1+{\epsilon\over \log{1\over \theta_*}}\left[{-1\over \sqrt{1-(H a_0)^2}}+{1\over 2}\log\bigg({1+\sqrt{1-(H a_0)^2}\over 1-\sqrt{1-(H a_0)^2}}\bigg)\right] \!\Big(1+\mathcal{O}\Big({1\over \log{1\over \theta_*}}\Big)\Big)\right\} , \\
E_\epsilon &=-{H \bar{\ell}_\epsilon\over \log{1\over \theta_*}}\,\Big(1+\mathcal{O}\Big({1\over \ln{1\over \theta_*}}\Big)\Big)
\end{split}
\end{equation}
The form of our exact solution becomes:
\begin{equation}
\begin{split}
    \mathcal{S}_\epsilon^{-1}V_a(\bar{a}_\epsilon) (\tau&)= \bar{\ell}_\epsilon(1-\tau)\cos(H\bar{\ell}_\epsilon\tau)\\
&+{\bar{\ell}_\epsilon\over \log{1\over \theta_*}}\,\Bigg\{\epsilon\left[{-1\over \sqrt{1-(H a_0)^2}}+{1\over 2}\log\bigg({1+\sqrt{1-(H a_0)^2}\over 1-\sqrt{1-(H a_0)^2}}\bigg)\right] \!\Big(1+\mathcal{O}\Big({1\over \log{1\over \theta_*}}\Big)\Big)\cos(H\bar{\ell}_\epsilon\tau)\\
&~~~~~~~~~~~\,+ \Big(1+\mathcal{O}\Big({1\over \log{1\over \theta_*}}\Big)\Big)\Big[1-{\cos (H\bar{\ell}_\epsilon\tau) \over 2}\log\!\Big({1+\cos (H\bar{\ell}_\epsilon\tau)\over 1-\cos (H\bar{\ell}_\epsilon\tau)}\Big)\Big]\Bigg\}
\end{split}
\end{equation}
When plugging this into the expression for $\mathcal{K}_{\epsilon}$, it greatly simplifies as the integration over $\tau$ sends any term independent of $\theta _{*}$ to 0. This leads to the behaviour:
\begin{equation}
    \mathcal{K}_\epsilon \underset{\theta_*\to 0}{\sim} {\bar{\ell}_\epsilon^2\over \log{1\over \theta_*}}
\end{equation}
It is straightfroward to calculate the path integral over $\delta \ell$:
\begin{equation}
    \int d\delta \ell \,\exp\Big\{\!-\!{3sv_3\over \hbar}\,  \mathcal{K}_\epsilon  \Big({\delta\ell\over \bar{\ell}_\epsilon}\Big)^2\Big\}\underset{\theta_*\to 0}{\sim}\sqrt{{\pi\hbar\over 3sv_3}\, \log{1\over \theta_*}}
\end{equation}
The final step is to take all of the components we have calculated and put them back together using (\ref{fluctpsi}). This leads us to the expression\cite{Reference31}:
\begin{equation}
    \psi(a_0) = \widetilde{\mathcal{C}}_s(\theta_*)\sum_{\epsilon=\pm 1} {1\over \sqrt{\epsilon}}\, {\exp\!\Big[\epsilon s\, \displaystyle {2v_3\over \hbar H^2}\,\big(1-(H a_0)^2\big)^{3/2}\Big]\over a_0^{1/8}\,\big(1-(H a_0)^2\big)^{1/4}}\, (1+\mathcal{O}(\hbar))\, ,\quad 0<H a_0<1
\end{equation}
where we have the regulator-dependent coefficient:
\begin{equation}
    \widetilde{\mathcal{C}}_s(\theta_*)=\alpha\,\sqrt{i\pi} \, \Big({\pi \hbar\over 3sv_3}\Big)^{1/4}\exp\!\Big[\!-\!s\, {2v_3\over \hbar H^2}\Big]\Big({H \over \theta_*}\Big)^{1/8}
\end{equation}
which acts as a multiplicative factor that doesn't affect the behaviour of the wave function. Another difference in this calculation is that the factor $1/\sqrt{\epsilon}$ makes these solutions explicitly complex rather than real. The form of the solution above clearly agrees with both the WDW result we obtained in (\ref{refsol1}) (given the form of $\rho^{\textrm{q}}$ for the path integral), and each distinct part agrees in its form with the Hartle-Hawking ($s=-1$) and Linde-Vilenkin ($s=1$) solutions of the previous section. Despite that, each choice leads to a different form of the wave function with different physical predictions\cite{Reference66}.

\section{Hartle-Hawking wave function for $R^{2}$ gravity}

We shall make use of the approximation of chapter 4 in which we showed that for the earliest stages of the evolution of the Universe, it is the quadratic term of the Starobinsky model action that dominates. Therefore we will carry a similar calculation as Hartle-Hawking, only now the action will be that of $R^{2}$ theory. The appropriate boundary term for the upper bound of the 4-geometries has already been calculated - see (\ref{eq:3.71}). Assuming once again the minisuperspace approximation for a closed Universe and carrying out the integrations with respect to the spatial coordinates, the total action becomes\cite{Reference33}:
\begin{equation} \label{eq:4.39}
     S_{G} = 36v_{3} \lambda \left [\int dt\, \left (\frac{\dot{a} ^{4}}{a} + 2\dot{a} ^{2}\ddot{a} + a \ddot{a} ^{2} + 2\frac{\dot{a} ^{2}}{a} + 2\ddot{a} + \frac{1}{a} \right ) - 2(\dot{a} ^{3} + \ddot{a} \dot{a} a + \dot{a} )\Big |_{\partial \mathcal{V}} \right ]
\end{equation}
In order to use the semi-classical approximation and apply the method of steepest descents, first we perform the Wick rotation $t \rightarrow i\tau$ and then replace the scale factors with the known dS solutions $a = H^{-1} \cosh{(Ht)}$ the same way we did before. This time the sign choice for the Wick rotation is well-motivated. The arbitrariness of the sign of the Lorentzian action stems from the sign convention of the Riemann curvature tensor. However, in $R^{2}$ gravity we have a good argument to pick only one due to the Ricci scalar being squared. For $\lambda >0$ (as we assume here) the Lorentzian action is strictly positive-definite. The same, therefore, holds true for the Euclidean action. That does not mean a reverse Wick rotation is impossible, but using positive-definiteness for the Euclidean action is a plausible condition that helps determine the correct sign choice. \par
When calculating the instantons and inserting them in the action, we are careful to set Dirichlet boundary conditions where $R$ is fixed on the boundary. From that we simply get the same action solutions as (\ref{eq:4.31}) multiplied by a constant, minus the matter field components. One can think of this easily by considering that if $R$ is replaced by its classical dS solution $R_{dS}=4\beta$, the total $R^{2}$ action is just $4\beta$ times the Einstein-Hilbert action for the same solution. Therefore\cite{Reference33}:
\begin{equation} \label{eq:4.40}
    I_{G}^{cl} = -2\lambda \, \kappa ^{-2} R_{dS} \, I_{EH}^{cl} =  -8\lambda \, (36v_{3}H^{2}) I_{EH}^{cl}
\end{equation}
where we have replaced $\beta = 3H^{2} \sigma ^{-2}$ like in the first section, and the minus sign was included by considering that $I_{EH}^{cl}$ is calculated with the Hartle-Hawking Wick rotation sign choice. A crucial difference from the earlier results is the particular value of $\beta$, or equivalently $H$. In the Hartle-Hawking calculation, the curvature of dS space was given by the singular value of the cosmological constant $\Lambda$ which was fixed. For $R^{2}$ gravity, every dS space solution satisfies the classical Einstein equations, meaning that all constant positive values of the curvature are permitted. In other words, there are infinitely many classical spacetime solutions, each with its own probability distribution. $H$ is then effectively a second degree of freedom of the wave function. This could have been deduced just as well by considering that (\ref{eq:3.17}) are second-order differential equations with $\delta R \Big |_{\partial \mathcal{V}} = 0$ being a boundary condition. \par
Consequently, the result above also applies to the extrema instanton actions (\ref{eq:4.35}) for the $a_{0}H<1$ case, as well as their complex analogues in the case of $a_{0}H >1$. For the latter, once again both instanton actions contribute due to being complex conjugates. For the former, only one of the two will be the dominant contribution. Since the Euclidean action is positive-definite (because of which we have $0< e^{-I'_{cl}} < 1$), it is the smallest that contributes - just like in the Linde-Vilenkin result (\ref{eq:4.37}). Summarizing the results \cite{Reference33}:
\begin{equation} \label{eq:4.41}
    \psi _{0} (a_{0}) \approx 
    \begin{cases}
    &\mathcal{N}_{0} \, \exp{ \left \{ -96v_{3} \lambda \, \left [1-(H^{2}a_{0}^{2})^{\frac{3}{2}} \right ] \right \} } \, , \quad Ha_{0} < 1 \\
    &2 \cos{\left [96v_{3} \lambda \, (H^{2}a_{0}^{2} - 1)^{\frac{3}{2}} + \tilde{\theta} \right ]} \, , \quad Ha_{0} > 1
    \end{cases}
\end{equation}
where $\tilde{\theta}$ a new phase that is a function of $a_{0}$. Another way we can derive these solutions is by considering the transformation to the Einstein frame which was covered back in section 4.2. There we showed that the pure $R^{2}$ theory is fully equivalent with GR in the presence of a cosmological constant $\tilde{\Lambda} = \frac{M_{P}^{2}}{16\lambda}$ and a massless scalar field $\phi$. We can thereby directly find the Hartle-Hawking wave function utilizing the exact same methodology of section 6.1. There is some additional nuance once again. This time, we have a much clearer picture of how the state of least excitation differs from any excited states. In the case of a constant field $\phi$ whose kinetic term is by extension 0, the action (\ref{eq:3.23}) becomes:
\begin{equation} \label{eq:4.42}
    S_{G} = \int d^{4}x \, \sqrt{\tilde{g} } \, \frac{M_{P}^{2}}{2} \left (\widetilde{R} - 2 \tilde{\Lambda} \right )
\end{equation}
As expected, it is in the same form of the Einstein-Hilbert action that was used in the original Hartle-Hawking calculation. The state of least excitation is derived trivially:
\begin{equation} \label{eq:4.43}
    \psi _{0} (a_{0}) \approx 
    \begin{cases}
    &\mathcal{N}_{0} \, \exp{ \left \{ -\displaystyle\frac{1}{3\widetilde{H} ^{2}} \, \left [1-(\widetilde{H} ^{2}a_{0}^{2})^{\frac{3}{2}} \right ] \right \} } \, , \quad \widetilde{H}a_{0} < 1 \\
    &2 \cos{\left [\displaystyle\frac{(\widetilde{H} ^{2}a_{0}^{2} - 1)^{\frac{3}{2}} }{3\widetilde{H} ^{2}} + \tilde{\theta} \right ]} \, , \quad \widetilde{H} a_{0} > 1
    \end{cases}
\end{equation}
The only thing that changed compared to (\ref{eq:4.36}) are the signs of the arguments, matching those found by the same Wick rotation in the Linde-Vilenkin case, as well as the introduction of the quantity $\widetilde{H}$ as opposed to $H$. That is done in order to distinguish between the $H$ values of quadratic gravity as defined in the original Jordan frame, and the corresponding values when switching to the Einstein frame. Full equivalency between (\ref{eq:4.41}) and (\ref{eq:4.43}) can be established if we recall that by construction:
\begin{equation} \label{eq:4.44}
\begin{split}
    &\tilde{\Lambda} = \frac{3\widetilde{H} ^{2}}{\sigma ^{2}} \\
    \Rightarrow \, &\frac{1}{3\widetilde{H} ^{2}} = \frac{16\lambda}{\sigma ^{2} M_{P}^{2}} = 96v_{3} \lambda
\end{split}   
\end{equation}
In this picture, the existence of $\widetilde{H}$ as a modifiable parameter matches the fact that the near-constant value of the potential $V(\phi)$ in this regime in reality changes. As a result, there is a dynamic change of the behaviour of the wave function as the inflaton field rolls down in this region. What is also omitted is the kinetic contribution, in which case the second degree of freedom would be $\phi$ itself\cite{Reference33}:
\begin{equation}
    \psi _{0} (a_{0}, \phi) \approx 
    \begin{cases}
    &\mathcal{N}_{0} \, \exp{ \left \{ -\displaystyle\frac{1}{3\widetilde{H} ^{2}(\phi)} \, \left [1-(\widetilde{H}(\phi) ^{2}a_{0}^{2})^{\frac{3}{2}} + K_{0}(\phi) \right ] \right \}}  \, , \quad \widetilde{H}a_{0} < 1 \\
    &2 \cos{\left [\displaystyle\frac{(\widetilde{H} ^{2}(\phi)a_{0}^{2} - 1)^{\frac{3}{2}} }{3\widetilde{H} ^{2}(\phi)} + \tilde{\theta}  + K_{0}(\phi) \right ]} \, , \quad \widetilde{H} a_{0} > 1
    \end{cases}
\end{equation}
where $K_{0}(\phi)$ is the near-constant kinetic contribution. Just like $\widetilde{H}(\phi)$, this too can take any arbitrary value, albeit for this one we know that classically it should be very small due to the mechanics of the slow roll. This once again matches our obtained results for the Starobinsky model in this regime in the case we only keep the leading term (\ref{refsol3}). Since we concern ourselves with the state of least excitation, this term will suffice. The constant $\tilde{\mu}^{2}$ corresponds to $\widetilde{H}^{2}$ up to some multiplicative constant. The sign, as we have explained, is fixed  so that the Euclidean action is positive definite. Otherwise, the WDW equation solutions with either sign choice are equally mathematically valid. \par
If we now check the limit $\tilde{\Lambda} \gg M_{P}^{2}$ for which $\widetilde{H}^{2} \gg 1$, we find that the wave function for $a_{0}\widetilde{H} <1$ asymptotically tends to unity. The emergent Universe has a peak in probability for very small Hubble radii ($\widetilde{H} ^{-1} \ll 1$) and for large values of the cosmological constant in the Einstein frame, or equivalently large values of the Starobinsky potential $V(\phi)$. These are precisely the conditions necessary for a Universe in its infancy to lie within the slow roll regime. It is therefore statistically more likely for an nascent Universe to emerge in the inflationary phase.

\pagebreak

\section{Excited states for $R^{2}$ gravity wave function}

Let us consider a dynamic field with non-negligible kinetic energy. We may interpret these contributions essentially as excited states. We begin by finding the classical equation of motion for the inflaton in Euclidean space for the approximate form of the potential $V(\phi) \approx -v_{3}a(3 - \tilde{\mu}^{2}a^{2})$ (having set the GR constant equal to $1/2$):
\begin{equation} \label{eq:4.45}
    \frac{d\phi}{dx^{0}_{\textrm{E}}} (x^{0}_{\textrm{E}})= i\sqrt{6} \, c \frac{N}{a^{3}}
\end{equation}
where $c$ is an arbitrary real integration constant. Next we carefully define the lapse function in this case to be \cite{Reference33}:
\begin{equation} \label{eq:4.46}
    N = \frac{1}{\sqrt{1 - c^{2}a^{-4} - H^{2}a^{2}}} = \frac{H^{2}a^{2}}{\sqrt{-H^{3}a^{3} + H^{2}a^{2} - H^{4}c^{2}}} \equiv \frac{H^{2}a^{2}}{\sqrt{\mathcal{P}(H^{2}c, \, H^{2}a^{2})}}
\end{equation}
which helps solve the Friedmann equations for the classical $a(x^{0}_{\textrm{E}})$. Note that we will be symbolizing the Hubble parameter as $H$ without tilde despite working in the Einstein frame for the sake of simplicity. When calculating the solutions of the polynomial $\mathcal{P}$ of $u \equiv H^{2} a^{2}$ appearing within the square root, we come across a certain limit that determines the nature of said solutions. When $H^{4}c^{2} < \frac{4}{27}$ which we shall call the case of subcritical kinetic energy, all three polynomial roots are real. By contrast, when $H^{4}c^{2} > \frac{4}{27}$ - now called the case of supercritical kinetic energy -  only one root is real, while the other two are complex solutions that are conjugates of each other. We shall examine these more closely in the following subsections.

\subsection{Subcritical scalar kinetic energy}

Let us begin by examining the subcritical case. The polynomial roots we obtain are as follows\cite{Reference33}:
\begin{equation}
\begin{aligned} 
u_<&=-\kappa&& \quad \in(-\textstyle{1\over 3},0]\, ,\\
u_-&={1+\kappa\over 2}\bigg(1- \sqrt{1-3\kappa\over 1+\kappa}\bigg)&& \quad\in[0,\textstyle{2\over 3})\, ,\\
u_+&={1+\kappa\over 2}\bigg(1+ \sqrt{1-3\kappa\over 1+\kappa}\bigg)&& \quad\in(\textstyle{2\over 3},1]\, ,\\
\text{where}~~ \quad \kappa&={1\over 3}\bigg(2\cos\Big[{1\over 3}\cos^{-1}\!\Big(\!-1+{27\over 2}(H^2c)^2\Big)\Big]-1\bigg)&&\quad\in[0,\textstyle{1\over 3})\,.
\end{aligned}
\label{roots}
\end{equation}
For the range:
\begin{equation} \label{urange}
    H a_-\equiv \sqrt{u_-}~\le ~H a~\le~\sqrt{ u_+ }\equiv H a_+
\end{equation}
the lapse function is real, therefore the solutions of the Friedmann equation are Euclidean. It is also obvious that when there is no kinetic energy ($c=0$) the parameter $\kappa$ vanishes which means that the first two roots become $u_<=u_-=0$ and the last root $u_+=1$. This corresponds to the $[0,1]$ interval we encounter for the 1-dimensional minisuperspace in GR. If instead the kinetic energy approaches the critical value $4/27$, then we have $\kappa\underset{<}\to 1/3$, $u_<\underset{>}\to -1/3$, $u_-\underset{<}\to 2/3$, and $u_+\underset{>}\to 2/3$, so $Ha$ becomes constrained to the constant value $\sqrt{2/3}$. This is equivalent to perpetual but slow growth in Euclidean space. \par
A particularly interesting observation is how in $R^{2}$ theory the existence of the kinetic energy of $\phi$ inhibits $a$ from becoming 0, unless the former vanishes as well. Because of this, when $u$ is outside the range (\ref{urange}), the lapse function is instead purely imaginary. As a result, the cosmological evolution of the Universe happens in real (non-Euclidean) time, depending on whether the scale factor is in the $a\ge a_+$ or $0\le a\le a_-$ range. So when we have $a=a_{\pm}$, we get two scenarios of the Universe being created "out of nothing". These have been discussed in the literature\cite{Reference18}\cite{Reference20}\cite{Reference19}, but we shall showcase them briefly. If we choose to parametrize time with a newly-defined variable $\tilde{t}$ such that the lapse function is:
\begin{equation} \label{Nchoice}
    N_{\rm L}={1\over \sqrt{1+{\kappa\over H a^2}}}
\end{equation}
then we have:
\begin{equation} \label{possol}
    H a(\tilde t)=\left\{\begin{array}{l}\mathcal{B}\,\sqrt{\xi+\cosh ^2(H \tilde t)}\\
\mathcal{B}\,\sqrt{\xi-\sinh ^2(H \tilde t)}\end{array}\right.\equiv\sqrt{1+\kappa\over 2}\,\sqrt{1\pm\sqrt{1-3\kappa\over 1+\kappa}\, \cosh(2H\tilde t)}
\end{equation}
where we have defined the following quantities:
\begin{equation}
    \mathcal{B}= \sqrt{1+\kappa}\, \Big({1-3\kappa\over 1+\kappa}\Big)^{1/4}~,~~\quad \xi={1\over 2}\bigg(\sqrt{1+\kappa\over 1-3\kappa}-1\bigg)
\end{equation}
The first solution in (\ref{possol}) corresponds to the case $a\geq a_{+}$ while the second to $0 \leq a \leq a_{-}$. The former describes an inflationary phase where a deformation due to the kinetic energy of the inflaton occurs, causing a "Big Bounce" at $a=a_{+}$. Starting from the infinite past, the Universe undergoes a period of contraction, and upon the aforementioned bounce it enters a period of expansion. This mirrors the behaviour of the diagram (\ref{fig:2.1}). On the other hand, in the latter solution we have a Big Bang evolution. The Universe initially expands before reaching the terminal value $a=a_{-}$ where it starts contracting, leading to a "Big Crunch" at some point in the future. We present both scenarios in the following graph:

\begin{figure}[H]
\begin{center}
\includegraphics[scale=0.2]{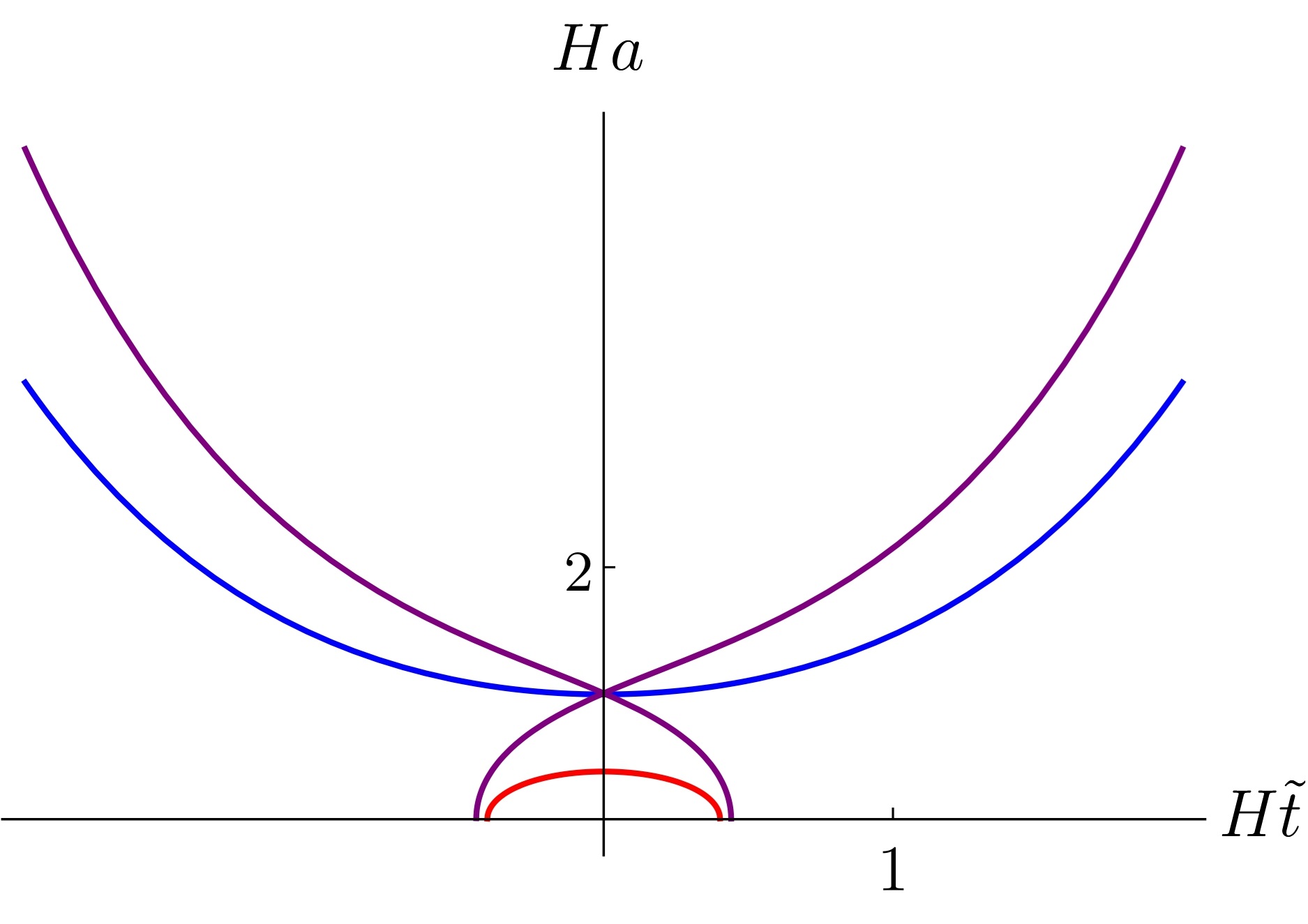}
\end{center}
\vspace{-0.6cm}
\caption{The two classical cosmological evolutions of the Big Bounce and Big Bang-Big Crunch which correspond to the subcritical case (setting $\kappa=1/8$) are depicted in blue and red respectively. Conversely, the two classical cosmological evolutions for the supercritical case (setting $\kappa=1$) which we will discuss more shortly are depicted in purple.}
\label{def_dS}
\end{figure}

If we restrict ourselves to the range between $a_{-}$ and $a_{+}$ where we have the evolution in Euclidean time, then the wave function oscillates between local maxima when $a=a_{+}$, and between local minima when $a=a_{-}$. The local maxima correspond to the hypersurface formed at the equator of the 4-sphere in (\ref{fig:4.1}), that is to say $a_{+} = H^{-1}$. As long as there exists kinetic energy for $\phi$ so that $a_{-}>0$, so the oscillation between the local minima at $a_{-}$ can be visualized as the Universe "bouncing" from the one to the other, causing a periodic evolution. To illustrate this more clearly, let's perform the Wick rotation:
\begin{equation}
    \tilde{t} = si\tilde{\tau} + \text{cst}
\end{equation}
where $s=\pm 1$ corresponding to the Linde-Vilenkin and Hartle-Hawking choice respectively. Both solutions (\ref{possol}) then take a sinusoidal form:
\begin{equation}
    H a(\tilde \tau)=\mathcal{B} \,\sqrt{\xi+\cos(H \tilde \tau)^2}\equiv\sqrt{1+\kappa\over 2}\,\sqrt{1+\sqrt{1-3\kappa\over 1+\kappa}\, \cos(2H \tilde \tau)}
\end{equation}
We summarize the results for the full range of values of $a$ in the graph that follows. The analytic continuation from Euclidean time $s\tilde{\tau}$ to real time $\tilde{t}$ connects the maxima of the oscillation with infinitely many Big Bounce solutions, and the minima with infinitely many Big Bang-Big Crunch solutions. The role of the kinetic energy in this scenario - as per Vilenkin - is allowing for switching between the two real-time evolution scenarios via the Euclidean regime. Therefore, rather than having a creation of a Universe "out of nothing" like with the Hartle-Hawking wave function in regular GR, we have phase transitions between a Universe undergoing a dS-type evolution and a Universe having started from a Big Bang. 

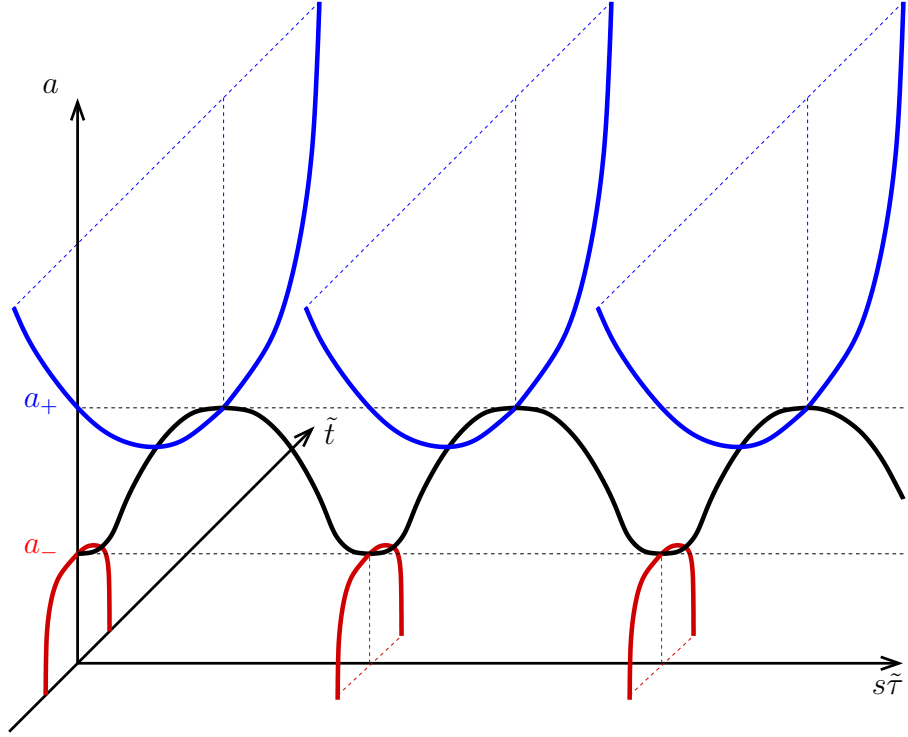
\begin{figure}[H]
\begin{center}
\includegraphics[scale=0.38]{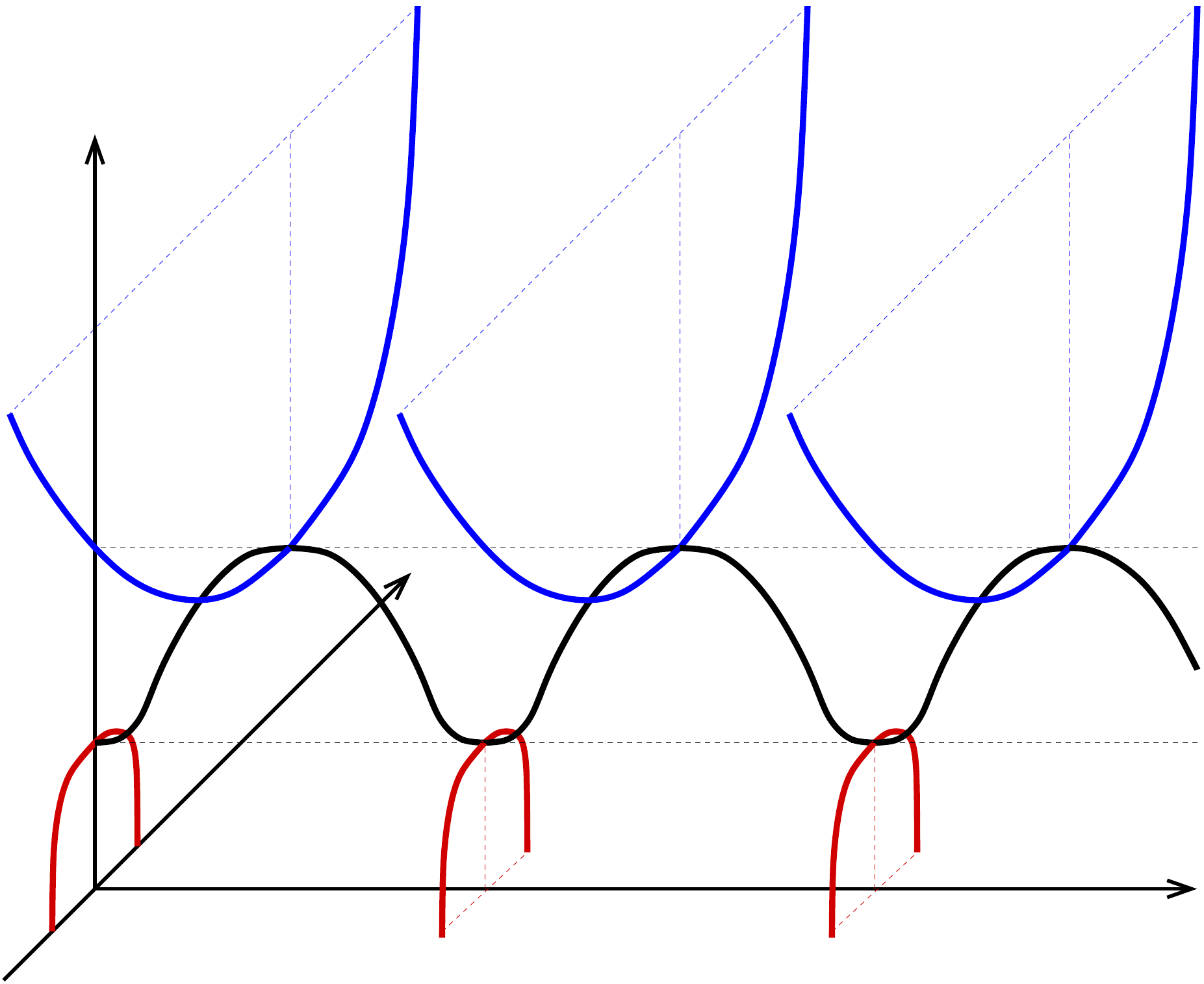}
\end{center}
\begin{picture}(0,0)
\put(390,34){$s \tilde \tau$}
\put(78,260){$a$}
\put(71,142){\color{blue} $ a_+$}
\put(71,87){\color{red} $a_-$}
\put(184,128){$\tilde t$}
\end{picture}
\vspace{-.8cm}
\caption{The Big Bounce and Big Bang-Big Crunch cosmological solutions are again depicted in blue and red respectively. The periodic solution in Euclidean time between $a_{-}$ and $a_{+}$ is marked in black. This trajectory can be analytically continued to an infinite number of evolutions for either of the aforementioned kinds.}
\label{osci}
\end{figure}

Given this novel feature of the instanton solutions in $R^{2}$ theory, we may then try to compute transition amplitudes between various excited states of the wave function for non-zero values of the kinetic energy of $\phi$. We will follow the outline of our relevant work on this closely\cite{Reference33}. \par
Since we have infinitely many such cases, we shall express the real time variable $x^{0}$ as such\cite{Reference33}:
\begin{equation} \label{euctime}
\begin{aligned}
x^0=s i (x^0_\textrm{E}-a_-)~,~ \text{where}~\left\{ \begin{array}{ll} x^0_{\textrm{E}}=a+2nd & \text{for}\quad a_- +2n \textrm{d} \le x^0_\textrm{E}\le a_+ +2n \textrm{d} \\
x^0_{\textrm{E}}=2a_ +-a+2n \textrm{d} & \text{for}\quad a_+ +2n \textrm{d} \le x^0_\textrm{E}\le 2a_+-a_-+2n \textrm{d} 
\end{array}\right.
\end{aligned}
\end{equation}
where $\textrm{d}=a_{+} - a_{-}$ and $n \in \mathbb{N}$ a natural integer. For the first set of values denoted as $(+,n)$, we have 4-geometries covering more than a hemisphere. These become:
\begin{equation}
\begin{aligned}
    I^{+}_{n}&=s v_3\,\Bigg(\int_{a_-}^{a_+}d x^0_\textrm{E}+\int_{a_+}^{a_++ \textrm{d} }d x^0_\textrm{E}+\cdots+\int_{a_++(2n-1) \textrm{d} }^{a_++2n \textrm{d} }d x^0_\textrm{E}+\int_{a_++2n \textrm{d} }^{a_0+2n \textrm{d} }d x^0_\textrm{E}\Bigg) \\
&\qquad \qquad\qquad\qquad\qquad\qquad \qquad \times Na^3\bigg[{3\over N^2 a^2}+{3\over a^2}-{1\over 2N^2}\left({d\phi\over d x^0_\textrm{E}}\right)^2-\Lambda\bigg]\\
&=6s v_3\,\Bigg(\int_{a_-}^{a_+}d a-\int_{a_+}^{a_-}d a+\cdots+\int_{a_-}^{a_+}d a+\int_{a_+}^{a_0}d x^0_\textrm{E}\Bigg)Na\,\Big[1-(H a)^2\Big]\,,
\end{aligned}
\label{sp}
\end{equation}
where we have applied the Friedmann equation solutions in the last step. $\Lambda$ corresponds to $\tilde{\Lambda}$ appearing in (\ref{eq:4.42}), omitting the tilde like we did with $H$ for the sake of consistency. \par
The second set of values which we denote as $(-,n)$, conversely correspond to 4-geometries covering less than a hemisphere. For them we get the same form of solutions, with the only difference being in the boundaries of the last integral. We can manipulate it to bring it to the same form as the one above:
\begin{equation}
    \int_{a_++2n \textrm{d} }^{a_0+2n \textrm{d} }d x^0_\textrm{E}\equiv \int_{a_+}^{a_0}d a\quad \longrightarrow  ~~\int_{a_++2n \textrm{d} }^{2a_+-a_0+2n \textrm{d} }d x^0_\textrm{E}\equiv -\int_{a_+}^{a_0}d a
\end{equation}
Therefore the two types of instanton solutions can be expressed together succinctly:
\begin{equation}
    I^\epsilon_{n}=(1+2n)\tilde{I}+\epsilon \Delta~,\quad~~ \epsilon=\pm1~,~~n\in\mathbb{N}
\end{equation}
where we define the quantities:
\begin{equation}
\begin{aligned}
    \tilde{I}(c)&= s\, {3 v_3\over H ^2}\int_{u_-}^{u_+} d u \, {u(1-u)\over \sqrt{(u+\kappa)(u-u_-)(u_+-u)}}~,\\
    \Delta(c,a_0)&= s\, {3 v_3\over H ^2}\int_{u_+}^{(H a_0)^2}\!\! d u \, {u(1-u)\over \sqrt{(u+\kappa)(u-u_-)(u_+-u)}}~, ~\,\text{for}~ ~ u_-\le (H a_0)^2\le u_+~.
    \end{aligned}
\label{ID}
\end{equation}
When $\phi$ is off-shell, we clearly see that the Lagrangian that appears in the Euclidean action can take either sign, but on-shell it has a definite sign due to the inequality $0\le \sqrt{u_-}\le H a\le \sqrt{u_+}\le 1$. This ensures $sI^{+}_{n} \geq 0$. This is consistent with what we discussed earlier, below (\ref{eq:4.39}): if we make the Linde-Vilenkin choice ($s=1$), that makes the Euclidean action positive-definite. \par
To calculate these Euclidean actions we apply the method of steepest descents as before. The probability amplitude for the Universe to transition from $a_{-}$ to $a_{0} \in [a_{-}, a_{+}]$ is then calculated as\cite{Reference33}:
\begin{equation}
    \widehat{A}(c,a_0)={1\over 2}\sum_{\epsilon}\sum_{n\ge 0} \mathcal{N} \mathcal{M}^{2n}\delta^{-\epsilon} \,  e^{-(1+2n)\tilde{I}}e^{-\epsilon \Delta}
\end{equation}
The factor of $1/2$ that appears is due to convention. The normalization factors that arise in addition to the main behaviour of the wave function ($e^{-(1+2n)I}e^{-\epsilon \Delta}$) are decomposed into distinct quantities for fluctuations of the fields around different backgrounds. $\mathcal{N}(c)\in \mathbb{R}_{+}$ appears when imposing the $a(x^{0}_{\textrm{E}} = a_{-}) \equiv a_{-}$ boundary condition for $x^{0}_{\textrm{E}} \in [a_{-},a_{+}]$. Likewise, $\mathcal{M}(c) \in \mathbb{R}_{+}$ appears for all ranges $x^0_{\textrm{E}} \in [a_++2k \textrm{d} ,2a_+-a_-+2k \textrm{d} ]$ or $x^0_{\textrm{E}} \in [a_-+2(k+1) \textrm{d} ,a_++2(k+1) \textrm{d} ]$, \, $k=0,\dots,n-1$. It is raised to the power of $2n$ because the Euclidean evolution is periodic over the span of $2\textrm{d}$, and due to time-reversal invariance at the extrema. The function $\delta(c,a_0)\in \mathbb{R}_+$ appears in this form (positive power) for the choice $\epsilon = -1$ in the Euclidean action. It comes from imposing the boundary condition $a(x^0_\textrm{E}=2a_+-a_0+2n \textrm{d}  )\equiv a_0$ in the range $x^0_\textrm{E}\in[a_++2n \textrm{d} ,2 a_+-a_0+2n \textrm{d} ]$. If we make the choice $\epsilon = +1$, then it appears as the inverse $1/\delta(c,a_0)$, while applying the boundary condition $a(x^0_\textrm{E}=a_0+2n \textrm{d}  )\equiv a_0$. It is noteworthy to mention that this function is independent of $n$. Taking these all into account and postulating that $\mathcal{M}(c) \,e^{-\tilde{I}(c)}<1$, we are able to evaluate the sums to calculate the probability amplitude:
\begin{equation}
    \widehat{A}(c,a_0)= {\mathcal{N} e^{-\tilde{I}}\over 1-\mathcal{M}^2e^{-2\tilde{I}}}\,\cosh\big(\Delta+\log \delta \big)~,\quad ~~\text{for}\quad  a_-\le a_0\le a_+
\end{equation}
The postulate above is potentially a good way to verify the correct value for $s$ if we calculated the Euclidean action from first principles. Depending on whether the inequality is true for $+1$ or $-1$, it would serve as an argument in favour of the one which satisfies it, since the one that doesn't satisfy it leads to a divergent probability amplitude. This doesn't seem to clear up based on our analysis of the WDW equation in the last chapter. Despite that, as we shall examine later, the postulate above will prove to be well-founded regardless. \par
At the particular value $a=a_{+}$ we see that $I_{n}^{+} = I_{n}^{-}$. This means that the two instanton actions contribute equally, which means $\delta (c, a_{+}) = 1$ and the extra "phase" in the hyperbolic cosine vanishes. $\Delta (c,a_{0})$ in this case evolves monotonically in the range $a_{0}\in [a_{-},a_{+}]$, vanishing at the upper boundary. If we now assume that $\log \delta$ acts as only a minor corrective term, we find that $\widehat{A}(c,a_{0})$ decreases as $a_{0}$ increases in the aforementioned range. \par
If we now want to examine the range $a_{0} \geq a_{+}$, the integrand of $\Delta$ becomes purely imaginary, therefore we must analytically continue into the complex plane. To that end, we define $\Delta = i\check{\Delta}$, where the latter is given by:
\begin{equation}
    \check\Delta(c,a_0)= s\, {3 v_3\over H ^2}\int_{u_+}^{(H a_0)^2}d u \, {u(1-u)\over \sqrt{(u+\kappa)(u-u_-)(u-u_+)}}~,~~\quad \text{for} \quad (H a_0)^2\ge u_+
\end{equation}
The Euclidean actions $I_{n}^{\epsilon}$ are the complex conjugates of each other. In turn, the hyperbolic function in the probability amplitude becomes sinusoidal:
\begin{equation} \label{cossol}
    \widehat{A}(c,a_0)= {\mathcal{N} e^{-\tilde{I}}\over 1-\mathcal{M}^2e^{-2\tilde{I}}}\,\cos\big(\check\Delta+\check\theta\big)~,\quad ~~\text{for} \quad  a_0\ge a_+
\end{equation}
where the quantity $\check{\theta}$ appears because $\delta$ in this case is an $a_{0}$-dependent phase $e^{i\check{\theta}} \, , \, \, \check{\theta}(a_{0}) \in \mathbb{R}$. At the limit $a_{0}=a_{+}$, we have $\check{\Delta} = 0$ and $e^{i\check{\theta}} = 1$ so that the amplitude is continuous between the two regimes. As $a_{0}$ becomes increasingly larger, $\check{\Delta}$ diverges as a polynomial power of the scale factor:
\begin{equation}
    \check\Delta(c,a_0)\sim - s{2v_3\over H ^2} (H a_0)^3~,~~\quad \text{when} \quad a_0\to +\infty
\end{equation}
If $\check{\theta}$ is much smaller than $\check{\Delta}$, then $\widehat{A}$ oscillates erratically around 0 with constant amplitude. \par
Lastly, we examine the $0 \leq a_{0} \leq a_{-}$ regime for which we have to also apply analytic continuation. This time we have more broadly complex integrands, hence:
\begin{equation} \label{ancodel}
\begin{split}
    &\Delta =-\tilde{I}+i\widehat \Delta \\
    \text{where}  \quad\widehat \Delta(c,a_0)&= s\, {3 v_3\over H ^2}\int_{u_-}^{(H a_0)^2} d u \, {u(1-u)\over \sqrt{(u+\kappa)(u_- -u)(u_+-u)}}~,~~ \text{for} \quad (H a_0)^2\le u_-
\end{split}
\end{equation}
For reasons that will be made clear shortly, we also define the following:
\begin{equation} \label{delred}
    \delta(c,a_0) = \mathcal{M}(c)\, \hat \delta(c,a_0)
\end{equation}
so the sum which gives the probability amplitude becomes:
\begin{equation} \label{sums}
\begin{aligned}
    {1\over 2}\sum_{\epsilon}\sum_{n\ge 0}\mathcal{N} \mathcal{M}^{2n-\epsilon}& \hat \delta^{-\epsilon}\, e^{-(1+2n-\epsilon)\tilde{I}}\,e^{-i\epsilon\widehat \Delta}\\
    &={1\over 2}\, \mathcal{N}\mathcal{M}^{-1}\hat\delta^{-1} \, e^{-i\widehat \Delta}+{1\over 2}\sum_{n\ge 1}\mathcal{N} \mathcal{M}^{-1}\mathcal{M}^{2n}e^{-2n\tilde{I}}\big(\hat\delta\,e^{i\widehat \Delta}+ \hat\delta^{-1} \,e^{-i\widehat \Delta}\big)
\end{aligned}
\end{equation}
When $n \geq 1$, the instanton actions for $(+,n)$ and $(-,n-1)$ are complex conjugates of each other. Since they are of the same magnitude, the newly defined quantity $\hat{\delta}$ will have to be a pure phase as $\delta$ was above: $\hat{\delta} = e^{i\hat \theta} \, , \, \, \hat \theta\in\mathbb{R}$. \par
Let us allow the values of the integers in the definition (\ref{euctime}) to take values $n \leq -1$. This allows Euclidean time to be defined wholly on $\mathbb{R}$. Instantons $(\epsilon,n)$ in this case would have appeared in our previous calculations had we taken the opposite contour choice $-s$, so the Euclidean actions $I_{n}^{\epsilon}$ evolve in the opposite direction compared to before. The only exception is for the case $\epsilon = -1 \, , \, \, n=-1$ when $a_{0}=a_{-}$, in which case the Euclidean time interval reduces to the singleton containing $x_{\textrm{E}}^{0} = a_{-}$. By extension, this amounts to an extra contribution not accounted for in (\ref{sums}). So for this extra instaton $(-,-1)$ we employ analytic continuation, obtaining $-\tilde{I}-(-\tilde{I}+i\widehat \Delta)=-i\widehat \Delta$, and the extra term in the sum must be ${1\over 2}\, \mathcal{N}\mathcal{M}^{-1}\hat \delta\, e^{i\widehat \Delta}$. With this in mind, the probability amplitude is calculated as:
\begin{equation}
    \widehat{A} (c,a_0)= \frac{\mathcal{N} \mathcal{M}^{-1}}{1-\mathcal{M}^2 e^{-2\tilde{I}}} \,\cos{\left (\widehat{\Delta} + \hat{\theta} \right )} \, ,\quad \text{for} \quad  0\le a_{0} \le a_{-}
\end{equation}
Checking for $a_{0}=a_{-}$, this result should match the one of (\ref{cossol}) for the same value. Since in this case we have $\widehat{\Delta} = 0 \, , \, \, \hat{\delta} = 1$ and consequently $\Delta = -\tilde{I} \, , \, \, \delta = \mathcal{N}$, we get the following relation from the argument from continuity:
\begin{equation} \label{valM}
    \mathcal{M} e^{-I}={1\over \cosh(-I+\log \mathcal{N})}
\end{equation}
This satisfies the inequality $\mathcal{M}(c) \, e^{\tilde{I}(c)} < 1$ from earlier, confirming that our postulate was correct. $\widehat{\Delta}$ in this regime is monotonic, decreasing until it vanishes at $a_{-}$. For the sake of ease, we can define:
\begin{equation}
    \widehat \Delta(c,0)=- s\, {3 v_3\over H ^2}\int^{u_-}_0 d u \, {u(1-u)\over \sqrt{(u+\kappa)(u_- -u)(u_+-u)}}\equiv -s\,{D(H^2 c)\over H^2}
\end{equation}
We again assume that the phase $\hat{\theta}$ only acts as a minor correction, therefore for small values $H \ll D(H^{2}c)$ the probability amplitude oscillates with a constant amplitude, while for large values $H \gg D(H^{2}c)$ it is nearly constant. To summarize the results for all regimes:
\begin{equation} \label{eq:4.47}
    \widehat{A}(c,a_0)=\mathcal{N} e^{-\tilde{I}}\coth ^2(-\tilde{I}+\log \mathcal{N})\times \left\{\begin{array}{ll}\cos \big(\check\Delta +\check \theta\big)~,&\text{for} \quad a_0\ge a_+ ~ \\
    \cosh \big(\Delta + \log \delta\big)~,&\text{for} \quad a_- \le a_0\le a_+ ~ \\
    \cosh(-\tilde{I}+\log \mathcal{N})\cos \big(\widehat\Delta + \hat\theta \big)~,&\text{for} \quad 0\le a_0 \le a_-~
    \end{array}\right.
\end{equation}
Notice how these are strictly real. Within $[a_{-},a_{+}]$ it is positive-definite, meaning that the transition amplitude connects the initial and final points purely in the Euclidean plane. If instead one of these points lies outside this regime, the behaviour is Euclidean within the first regime before transitioning to a cosmological time evolution with oscillatory behaviour. In the limit $c\rightarrow 0$ for which we get the state of minimum excitation, the amplitudes become:
\begin{align}
    \psi(a_0)&\equiv \widehat{A}(0,a_0) \nonumber\\
    &=\widetilde{\mathcal{N}} e^{-s {2v_3\over H ^2}}\times \left\{\begin{array}{ll}\cos  \left[-s\displaystyle {2v_3\over H^2}\!\left((H a_0)^2-1\right)^{3\over 2}+\check{\theta}\right],&\text{for}\quad a_0\ge a_+~\\
    \cosh\left[-s \displaystyle{2v_3\over H^2} \left(1-(H a_0)^2\right)^{3\over 2}+\log \delta\right],& \text{for}\quad 0\le a_0\le a_+~
    \end{array}\right. \nonumber\\
    \text{where} \quad \widetilde{\mathcal{N}} &\equiv \mathcal{N} \coth ^2\left[-s {2v_3\over H ^2}+\log \mathcal{N} \right]~
\end{align}
This matches the solutions found earlier in (\ref{eq:4.43}), with the exception that in the regime $0 \leq a_{0} \le a_{+}$ (or equivalently $0 \le Ha_{0} \le 1$) we now have the hyperbolic cosine rather than just a real exponential. That is because in the current calculation, we included both types of instanton action contributions $\epsilon = \pm1$ in the sum, whereas in the previous section we used the argument of Hartle and Hawking to omit the contribution from 4-geometries covering more than a hemisphere. Had we included this also, we would retrieve the same kind of function up to a phase $\log{\delta}$. In addition, we shall split the contributions $(\epsilon,0)$ from the rest:
\begin{equation}
    \coth ^2\left[-s {2v_3\over H ^2}+\log \mathcal{N}\right]=1+\sum_{n\ge 1} \mathcal{M}^{2n}\,e^{-s {2v_3\over H ^2}2n}
\end{equation}
since in the limit $c \rightarrow 0$ all other configurations are singular at $a_{0}=a_{-} = 0$. Such configurations form "necklaces of $n-1$ beads" which we have seen back in section 6.2 and which are irregular, thus violating the path integral axiom of only summing over smooth configurations\cite{Reference67}. Thus, when $c=0$, only the $(\epsilon,0)$ cases contribute, making $\widetilde{\mathcal{N}}=\mathcal{N}$. \par
Lastly, we must normalize using the Hilbert space inner product we defined in chapter 5\cite{Reference33}:
\begin{equation}
    \mathbf{A}(c,a_0)\equiv {\widehat{A}(c,a_0)\over \left (\displaystyle{\int_0^{+\infty} da_{0}' \, \sqrt{-\gamma} \,\mu(a_0')\, |\widehat{A}(c,a_0')|^2} \right )^{1/2}}
\end{equation}
where $\mu$ is the measure of integration and $\gamma$ the metric of the target space manifold in this frame. $\mu$ for this model is formally given by (\ref{eq:15}), since we have a 2-dimensional minisuperspace model. However, we only integrate over the scale factor $a_{0}'$ because here we take $c$ to be a fixed parameter. By extension, $\mu$ matches the 1-dimensional case of (\ref{mu}). This is ambiguity-dependent, but since we worked with the path integral, the ambiguity function of the 1-dimensional model is given by (\ref{pathamb}). It is (up to a constant factor):
\begin{equation}
    \mu (a_{0}) \sim (-\gamma)^{-1/4} \sim a_{0}^{-1/4}
\end{equation}

\subsection{Supercritical and critical scalar kinetic energy}

Let us move on to the case of the kinetic energy being supercritical i.e. when $(H^2c)^2> \frac{4}{27}$. The polynomial $\mathcal{P}$ now admits 3 solutions - one real and two complex which are conjugates of each other\cite{Reference33}:
\begin{equation}
    \begin{aligned}
u_<&=-\kappa&& \quad \in(-\infty, -\textstyle{1\over 3})\, ,\\
&\begin{array}{rl}
&\!\!\!\!\!\!\!\!\!\!\!\!\!\!\!\!\!\!\!\!\!\!\!\!\!\!\!\!\!u_-=\bar u_+=\displaystyle{1+\kappa\over 2}\bigg(1+i \,\sqrt{3\kappa-1\over 1+\kappa}\bigg)\\
&\!\!\!\!\!\!\!\!\!\!\!\!\!\!u_+=\displaystyle{1+\kappa\over 2}\bigg(1-i \,\sqrt{3\kappa-1\over 1+\kappa}\bigg)\phantom{\overset{\imath}{\overset{M}{M}}}\!\!\!\!\!\!\!\!
\end{array} \qquad \qquad  \qquad~ \Longrightarrow~~&& \left\{
\begin{array}{l}
 \text{Re} \, u_+\in( {2\over 3},+\infty)\\
  \text{Im} \, u_+\in (0,-\infty)
\end{array},\right.
\\
\text{where}~~ \quad \kappa&={1\over 3}\big(\alpha+\alpha^{-1}-1\big)&&\quad\in(\textstyle{1\over 3},+\infty)~,\phantom{\overset{\imath}{\overset{M}{M}}}\!\!\!\!\!\!\!\!
\\
\alpha&=\left[{27\over 2}\,(H ^2c)^2-1+3\sqrt{3}\, H^2|c|\sqrt{{27\over 4}\,(H ^2c)^2-1}\right]^{1/3}&&\quad \in(1,+\infty)~.
\end{aligned}
\label{roots2s}
\end{equation}
Since $H^{2}a^{2} > u_{<}$, the polynomial $\mathcal{P}$ is always negative, making the lapse function $N$ purely imaginary. The classical equations of motion are then cosmological and not passing through the Euclidean plane, hence there will be no instant solutions. Using the same time coordinate chosen in (\ref{Nchoice}), we get two classical equations of motion for $a$:
\begin{equation}
    H a(\tilde t)=\sqrt{1+\kappa\over 2}\,\sqrt{1\pm\sqrt{3\kappa-1\over 1+\kappa}\, \sinh(2H\tilde t)}
\end{equation}
These two evolutions are connected, since we can get one from the other with a time-reversal. The one with a positive sign describes a Universe beginning with a Big Bang before transitioning smoothly into a perpetual inflationary regime, with an inflection point at $H a=[\kappa(\kappa+\kappa)]^{1/4}$. This is similar to the scenario in the subcritical case where we have a phase transition from a Big Bang-Big Crunch evolution to a period of dS-like expansion. \par
Regardless, it is worthwhile to examine what we get by performing an analytic continuation and see where that leads us for the probability amplitude. From an abstract point of view, this implies a form of transition between a Universe with a complex scale factor $H^{-1}\sqrt{\bar u_+}$ and a Universe with a scale factor $a_{0} \in \mathbb{R}_{+}$. The quantities appearing in the instantons as seen back in (\ref{ID}) in this case become:
\begin{equation}
    \begin{aligned}
\tilde{I}(c)&= s  is\, {3 v_3\over H ^2}\int_{\bar u_+}^{u_+} d u \, {u(1-u)\over \sqrt{(u+\kappa)(u-\bar u_+)(u-u_+)}}~\\
\Delta(c,a_0)&= i \zeta s\, {3 v_3\over H ^2}\int_{u_+}^{(H a_0)^2}\!\! d u \, {u(1-u)\over \sqrt{(u+\kappa)(u-\bar u_+)(u-u_+)}}~, ~~\quad \text{for}~ ~ a_0\ge 0
\end{aligned}
\label{ID2}
\end{equation}
where $\zeta = \pm1$ is a factor of ambiguity due to having the ability to flip the sign of the polynomial $\mathcal{P} \rightarrow -\mathcal{P}$. $I(c)$ in this case is always real, while $\Delta(c)$ is generally complex. We can express these more succinctly by defining:
\begin{equation}
    F(c,u)=\int^u_{u_*} d u \,f(c,u)~,~~\quad \text{where}~~\quad f(c,u)={u(1-u)\over \sqrt{u^3-u^2+(H^2c)^2}}
\end{equation}
where $u_{*} \in \mathbb{R}$ an arbitrary constant. We may then express (\ref{ID2}) as:
\begin{equation}
    \begin{aligned}
\tilde{I}(c)&=-2\zeta s\, {3 v_3\over H ^2}\, \text{Im} F(c,u_+)~, \\
 \Delta(c,a_0)&=-{I(c)\over 2}+i\widetilde{\Delta}(c,a_0)~,\\
\text{where} \quad ~~\widetilde \Delta(c,a_0)&=\zeta s\, {3v_3\over H^2}\, \Big[F\big(c,(H a_0)^2\big)-\text{Re} F(c,u_+)\Big]\;.
\end{aligned}
\label{ID3}
\end{equation}
At the exact critical value $(H^{2}c)^{2} = 4/27$, for which we have $\kappa = 1/3$ and $u_{-}=u_{+} = 2/3$, the function $f$ is singular:
\begin{equation}
    f\Big( {\pm2\over 3\sqrt{3}},u\Big)\sim {2\over 9\Big(\displaystyle u-{2\over 3}\Big)}~,\quad ~~\text{when}\quad ~~ u\to {2\over 3}
\end{equation}
Despite that, the function $I(c)$ is being suppressed due to the boundaries of integration tending towards the same value. These two behaviours end up counteracting each other. As $(H^{2}c)^{2}$ goes from subcritical to supercritical values, we find that $I(c)$ is non-singular. Moreover, by matching the limits approaching the critical value from both the subcritical and supercritical regime:
\begin{equation}
    \lim _{(H^{2}c)^{2} \rightarrow (\frac{4}{27})^{+}} I(c) = \lim _{(H^{2}c)^{2} \rightarrow (\frac{4}{27})^{-}} I(c)
\end{equation}
we find that only $\zeta = +1$ is permitted, and that $I(c)$ increases monotonically as the value of $(H^{2}c)^{2}$ increases. To further illustrate this, we calculate the extreme cases\cite{Reference33}:
\begin{equation}
    \tilde{I}(c)=s\, {2v_3\over H^2}\times \left\{\begin{array}{lll}1&\quad \text{for $(H^2 c)^2=0$} &({subcritical})~\\
\displaystyle {\pi\over 3}&\quad \text{for $(H^2 c)^2=\displaystyle{4\over 27}$} &({critical})~\\
\displaystyle{{3\over 2}\,{\Gamma ({5\over 3})\over \Gamma ({1\over 6})}}\, \sqrt{\pi \alpha}\, \big(1+\mathcal{O}(\alpha^{-1})\big)&\quad \text{for $(H^2 c)^2\to +\infty$} &({supercritical})~
 \end{array}\right.
\end{equation}
$\Delta(c,a_{0})$ is also singular at $(H^{2}c)^{2}$:
\begin{equation}
    \Delta(c,a_0)\sim is \, {v_3\over H^2} {1\over 3} \log\Big[{4\over 27}-(H^2c)^2\Big]\; ,~~\quad \text{when} \quad ~~ (H^2 c)^2\to {4\over 27}
\end{equation}
but this time we cannot go from subcritical to supercritical values on the real number line, as no continuity can be established. Rather, we need to perform a counterclockwise turn of angle $-\pi$ on the complex plane around the critical point. Using that $I(c)$ increases as we have shown, we may then state a similar postulate to the subcritical case:
\begin{equation} \label{postul}
    |\mathcal{M}(c)|\, e^{-\tilde{I}(c)}<1
\end{equation}
At this stage we have to be careful about the way in which we take the critical limit. Taking it naively such that $\kappa \rightarrow 1/3$, the classical equations of motion for the scale factor approaching from both the subcritical and the supercritical regime lead to a constant $Ha \rightarrow \sqrt{2/3}$. That is because at this limit the time for which the Big Bang and Big Crunch occur will tend to infinity, as opposed to the extrema and inflection points. This effectively leads to a Universe that experiences no growth and is in a constant "plateau". To amend this, we have to fix the times for which the Big Bang and Big Crunch occur and allow the extrema and inflection points to appear at infinity instead. Doing so, we obtain four equations of motion\cite{Reference33}:
\begin{equation}
    H a(\tilde t)=\sqrt{2\over 3}\,\sqrt{1\pm e^{\pm{2H \tilde t}}}
\end{equation}
The only solution that exists on the Euclidean plane is the case of $\tilde t = 0$ for which indeed we have a constant $Ha = \sqrt{2/3}$. Coming back to $\Delta(c,a_{0})$, the fact it is now complex means it is advantageous to redefine $\delta (c,a_{0})$ analogously to (\ref{delred}):
\begin{equation}
    \delta (c,a_{0}) = \mathcal{M}^{1/2}(c) \tilde{\delta}(c,a_{0})
\end{equation}
which in conjunction with the postulate (\ref{postul}) leads to the probability amplitude:
\begin{equation}
    \widehat{A}(c,a_0)= {\mathcal{N} \mathcal{M}^{-{1\over 2}}e^{-{\tilde{I}\over 2}}\over 1-\mathcal{M}^2e^{-2\tilde{I}}}\,{1\over2}\,\Big[|\tilde \delta |^{-1}\, e^{-i(\tilde \Delta+\tilde \theta)}+\mathcal{M} e^{-\tilde{I}}\, |\tilde \delta|\, e^{i(\tilde \Delta+\tilde \theta)}\Big]\;,\quad ~~\text{for}\quad  a_0\ge 0
\end{equation}
where we analyzed $\tilde \delta\equiv |\tilde \delta|e^{i\tilde \theta}$. An interesting possibility to examine is whether the result above is obtainable by analytic continuation in the subcritical case in the same regime $a_{0} > 0$, performing the rotation in the complex plane mentioned earlier. We notice that the expression for $\Delta(c,a_{0})$ in (\ref{ancodel}) matches the form of (\ref{ID3}), so we should get the same transition amplitudes. Performing the analytic continuation outlined above, the Euclidean actions from the subcritical case take the form:
\begin{equation}
    I_{n}^\epsilon=\left(1+2n-{\epsilon\over 2}\right)\tilde{I}+\epsilon \, i\widetilde \Delta
\end{equation}
which remain positive as long as $n \ge 0$. That means that for this case we do not include the contribution from the instanton $(-,-1)$ like we did in the subcritical case. Since the two solutions are now connected, that means we can replace $\mathcal{M}(c)$ with the analytically continued form of the value we found earlier in (\ref{valM}). The amplitude finally reads:
\begin{equation}
    \begin{aligned}
\widehat{A}(c,a_0)= \mathcal{N} e^{-\tilde{I}}\cosh ^{1\over2} &(-\tilde{I}+\log \mathcal{N})\coth ^2(-\tilde{I}+\log \mathcal{N})\\
\times \,&{1\over2}\,\Big[|\tilde \delta |^{-1}\, e^{-i(\tilde \Delta+\tilde \theta)}+\cosh (-\tilde{I}+\log \mathcal{N})\, |\tilde \delta|\, e^{i(\tilde \Delta+\tilde \theta)}\Big]\;,\quad ~~\text{for} \quad  a_0\ge 0
\end{aligned}
\label{Asucri2}
\end{equation}
Using the same assumption that the measure of the added phase $|\tilde{\delta}|$ is very small, the first term in the brackets dominates, leading to $\widehat{A}$ being complex. For large values of the scale factor, we note that $\widetilde{\Delta}$ mirrors the behaviour of $\check{\Delta}$ in the subcritical case. Consequently, the probability $|\mathbf{A}(c,a_{0})|^{2}$ for the supercritical case is oscillatory. The difference is that here we do not have nodes where the probability becomes 0, contrary to the subcritical case\cite{Reference33}.

%\section{Probability ratios for the inflationary regime}

\pagebreak % Experiment 2

\fancyhead[RE,LO]{Conclusion}

\chapter{Conclusion}

In this thesis, we have outlined a thorough summary of inflationary cosmology and its merits, as well as the specific models of Starobinsky and $R^{2}$ gravity. We have also introduced the fundamentals of quantum cosmology in the context of the Wheeler-DeWitt equation, as well as Hartle and Hawking's no boundary proposal. The premise was to use these as a starting point to make meaningful quantum calculations for plausible cosmological models using the WDW equation, and to apply Hartle-Hawking's methods to an inflationary model of modified gravity. The end goal is to potentially determine a quantum probabilistic argument for the emergence of the Universe with initial conditions which favour inflation. \par
Our first contribution was to develop a rigorous mathematical formalism from first principles to handle operator ordering ambiguities in the WDW equation for minisuperspace models. We did so by introducing ambiguity functions in the canonical quantization, while noting that they appear in the path integral due to the ambiguity of choosing a measure after field redefinitions. Using an argument from Hamiltonian hermiticity, we obtained some constraints for those functions. In turn, we used them to prove the uniqueness of the quantum theory to all orders for 1d and flat arbitrary-dimensional target space manifolds. Despite that this could not be generalized to any arbitrary curved manifold, we found that universality for such cases holds at the semi-classical level. The mathematical description corroborates past known results, but also serves as a correction to errors found in the literature. \par
For the second contribution, we used Hartle-Hawking's method to calculate an analogous state of minimum excitation in the case of $R^{2}$ gravity. We did so in two ways: one directly using the Jordan frame of the modified gravity action, and one using the Einstein frame transformation into a GR model with a cosmological constant and a massless scalar matter field $\phi$. We emphasized that the choice of contour of integration plays a significant role. The effective sign choice that comes from Hartle and Hawking's choice inhibits the emergence of an inflationary Universe. The opposite choice by Linde-Vilenkin instead favours it. We made an argument for the better suitability of the latter in our model, but it is not conclusive. \par
We expanded upon the aforementioned result by examining excited states due to non-zero kinetic energy for the field $\phi$. We saw that these scenarios do not correspond to the emergence of a Universe out of nothing in some geometrical configuration. Rather, they describe various transitions from a Big Bang-Big Crunch type evolution to a deSitter-like one, or the inverse. We specifically made computations for both subcritical and critical/supercritical values of the kinetic energy. By making use of the formalism from our first work, we arrived at a formula with which we can calculate transition probability amplitudes. \par
In future works, a natural extension would be to to see if we can find a method to prove the uniqueness of quantum theory for any type of target space manifold beyond the semi-classical level. Furthermore, it is important to construct a more robust argument from first principles such that a choice between Hartle-Hawking and Linde-Vilenkin is made. If that's impossible and both are acceptable, then we have to prove that explicitly and explain how the Universe would most likely emerge. Finally, it would be interesting to calculate the path integral for the state solutions in the case of $R^{2}$ gravity exactly. That potentially involves finding the path integral-specific values for the WDW ambiguity functions in that model.

\pagebreak % Results and Discussion

%\input{Chapters/Chapter7} % Conclusion

%% ----------------------------------------------------------------
% Now begin the Appendices, including them as separate files

\addtocontents{toc}{\vspace{2em}} % Add a gap in the Contents, for aesthetics

\appendix % Cue to tell LaTeX that the following 'chapters' are Appendices

\fancyhead[RE,LO]{Hermiticity of the Wheeler-DeWitt Hamiltonian}

\chapter{Hermiticity of the WDW Hamiltonian}

We start by defining the following integral\cite{Reference31}:
\begin{equation} \label{eq:16}
    \mathcal{J} = \int \limits_{\mathcal{T}} d^{M} \textrm{q} \, \sqrt{-\gamma } \, \mu \, \widetilde{\Xi}  \,  \left [\left (\frac{\widehat{\mathcal{H}}}{N} f _{1}  \right )^{*} f _{2} - f _{1} ^{*} \frac{\widehat{\mathcal{H}}}{N} f _{2} \right ]
\end{equation}
$\widetilde{\Xi}$ is an arbitrarily chosen test function such that it vanishes at the boundaries of the target space manifold $\mathcal{T}$. We manipulate the first term of the integral above and rewrite it as such:
\begin{equation} \label{eq:a1}
    \int \limits_{\mathcal{T}} d^{M} \textrm{q} \, \sqrt{-\gamma } \, \mu \, \widetilde{\Xi}  \left (\frac{\widehat{\mathcal{{H}}}}{N} f _{1}  \right )^{*} f _{2} = \Bigg \langle \frac{\widehat{\mathcal{H}}}{N} f _{1} \, , \, \widetilde{\Xi}  f _{2} \Bigg \rangle = \Bigg \langle f _{1} \, , \, \frac{\widehat{\mathcal{H}}^{\dagger}}{N} \left ( \widetilde{\Xi}  f _{2} \right ) \Bigg \rangle
\end{equation}
Hence the integral is now calculated as follows\cite{Reference31}\cite{Reference64}: 
\begin{equation} \label{eq:a2}
\begin{split}
    &\mathcal{J}  =\int \limits_{\mathcal{T}} d^{M} \textrm{q} \, \sqrt{-\gamma } \, \mu \, f _{1} ^{*} \left \{ \frac{\widehat{\mathcal{H}}^{\dagger}}{N} \left ( \widetilde{\Xi} f _{2} \right ) - \widetilde{\Xi} \frac{\widehat{\mathcal{H}}}{N} f _{2} \right \} \\
    &= -\frac{\hbar ^{2}}{2} \int \limits_{\mathcal{T}} d^{M} \textrm{q} \, f _{1}^{*} \Bigg \{ \frac{\partial}{\partial \textrm{q}^{\textrm{L}}} \left( \rho ^{\textrm{L}*}\frac{\partial (\gamma^{\textrm{IJ}} \tilde{\rho}_{\textrm{J}}^{*} \, \Xi^{\textrm{K}}{}_{\textrm{I}} \, \mu \, \sqrt{-\gamma} \,\widetilde{\Xi})}{\partial \textrm{q}^{\textrm{K}}} f _{2} \right ) + \rho ^{\textrm{L}*}\frac{\partial (\gamma^{\textrm{IJ}} \,  \tilde{\rho}_{\textrm{J}} ^{*} \,\Xi^{\textrm{K}}{}_{\textrm{I}} \, \mu \, \sqrt{-\gamma} \,\widetilde{\Xi})}{\partial \textrm{q}^{\textrm{L}}} \frac{\partial f_{2}}{\partial \textrm{q}^{\textrm{K}}} \\
    &\qquad \qquad \qquad \qquad \qquad -\widetilde{\Xi} \, \mu \, \sqrt{-\gamma}\, \left (\gamma ^{\textrm{IJ}} \tilde{\rho}_{\textrm{J}} \,\Xi^{\textrm{K}}{}_{\textrm{I}} \,\frac{\partial \rho ^{\textrm{L}} }{\partial \textrm{q}^{\textrm{K}}} \frac{\partial f _{2}}{\partial \textrm{q}^{\textrm{L}}} + 2i\, \text{Im} \omega \, f_{2}\right ) \Bigg \} \\
    &= \frac{\hbar ^{2}}{2} \int \limits_{\mathcal{T}} d^{M} \textrm{q} \, \mu \, \sqrt{-\gamma} \, \widetilde{\Xi} \, \Bigg \{ f_{1}^{*} \left (\gamma ^{\textrm{IJ}} \tilde{\rho}_{\textrm{J}} \,\Xi^{\textrm{K}}{}_{\textrm{I}} \,\frac{\partial \rho ^{\textrm{L}} }{\partial \textrm{q}^{\textrm{K}}} \frac{\partial f _{2}}{\partial \textrm{q}^{\textrm{L}}} + 2 i\, \text{Im} \omega \, f_{2} \right ) \\
    & \qquad \qquad + \gamma ^{\textrm{IJ}}  \, \tilde{\rho}_{\textrm{J}} ^{*}\,\Xi^{\textrm{K}}{}_{\textrm{I}} \left [\frac{\partial}{\partial \textrm{q}^{\textrm{L}}} \left (\rho ^{\textrm{L}*}f_{1}^{*} \frac{\partial f_{2}}{\partial \textrm{q}^{\textrm{K}}} \right )  + \frac{\partial}{\partial \textrm{q}^{\textrm{K}}} \left (f_{1}^{*} \frac{\partial (\rho ^{\textrm{L}*}f_{2})}{\partial \textrm{q}^{\textrm{L}}} \right )-  \frac{\partial ^{2}}{\partial \textrm{q}^{\textrm{K}} \partial\textrm{q}^{\textrm{L}}} \left (\rho ^{\textrm{L}*} f _{1} ^{*} f_{2}\right ) \right ] \Bigg \}  
\end{split}
\end{equation}
where we performed integration by parts in the last step, and we have used the vanishing property of the test function to get rid of any boundary terms. Notice how the last line of equation (\ref{eq:a2}) has $\widetilde{\Xi}$ as a common factor in the integrand, just like in the definition of $\mathcal{J}$ in (\ref{eq:16}). The test function is arbitrary, which means the two integrands are the same:
\begin{equation} \label{eq:a3}
\begin{split}
    \left (\frac{\widehat{\mathcal{H}}}{N} f _{1}  \right )^{*}& f _{2} - f _{1} ^{*} \frac{\widehat{\mathcal{H}}}{N} f _{2} = \frac{\hbar ^{2}}{2} \Bigg \{ f_{1}^{*} \left (\gamma ^{\textrm{IJ}} \tilde{\rho}_{\textrm{J}} \,\Xi^{\textrm{K}}{}_{\textrm{I}} \,\frac{\partial \rho ^{\textrm{L}} }{\partial \textrm{q}^{\textrm{K}}} \frac{\partial f _{2}}{\partial \textrm{q}^{\textrm{L}}} + 2 i\, \text{Im} \omega \, f_{2} \right ) \\
    & + \gamma ^{\textrm{IJ}}  \, \tilde{\rho}_{\textrm{J}} ^{*}\,\Xi^{\textrm{K}}{}_{\textrm{I}} \left [\frac{\partial}{\partial \textrm{q}^{\textrm{L}}} \left (\rho ^{\textrm{L}*}f_{1}^{*} \frac{\partial f_{2}}{\partial \textrm{q}^{\textrm{K}}} \right )  + \frac{\partial}{\partial \textrm{q}^{\textrm{K}}} \left (f_{1}^{*} \frac{\partial (\rho ^{\textrm{L}*}f_{2})}{\partial \textrm{q}^{\textrm{L}}} \right )-  \frac{\partial ^{2}}{\partial \textrm{q}^{\textrm{K}} \partial\textrm{q}^{\textrm{L}}} \left (\rho ^{\textrm{L}*} f _{1} ^{*} f_{2}\right ) \right ]  \Bigg \} 
\end{split}
\end{equation}
Since the functions $f_{1}, \, f_{2}$ are arbitrary as well, we may choose $f _{1} = f_{2} =\psi$ where $\psi =\psi ^{*}$ a real function that constitutes a physical state of the system. We can use the fact that $\psi$ is a solution of the WDW equation: 
\begin{equation} \label{eq:a4}
    \frac{\widehat{\mathcal{H}}}{N} \psi = 0
\end{equation}
to set the lhs as zero. Our equation then reads:
\begin{equation} \label{eq:a5}
    \left (\gamma ^{\textrm{IJ}} \tilde{\rho}_{\textrm{J}} \,\Xi^{\textrm{K}}{}_{\textrm{I}} \,\frac{\partial \rho ^{\textrm{L}} }{\partial \textrm{q}^{\textrm{K}}} \frac{\partial \psi}{\partial \textrm{q}^{\textrm{L}}} + 2 i\, \text{Im} \omega \, \psi \right ) - \gamma ^{\textrm{IJ}} \, \Xi^{\textrm{K}}{}_{\textrm{I}} \,\tilde{\rho}_{\textrm{J}} ^{*}\frac{\partial \rho ^{\textrm{L}*}}{\partial \textrm{q}^{\textrm{K}}} \frac{\partial \psi}{\partial \textrm{q}^{\textrm{L}}}  = 0
\end{equation}
This is none other than equation (\ref{eq:constraints1}) before decomposing $\rho ^{\textrm{L}} , \, \tilde{\rho} _{\textrm{J}}, \, \rho ^{\textrm{L}*} , \, \tilde{\rho} _{\textrm{J}}^{*}$ and having $\psi$ rather than a generic $f$ function\cite{Reference42}\cite{Reference64}. As a result, we are led to the same constraints (\ref{eq:constraints2}) and (\ref{eq:constraints3}) which correspond to the other criterion:
\begin{equation}
    \frac{\widehat{\mathcal{H}}}{N}f = \frac{\widehat{\mathcal{H}}^{\dagger}}{N}f
\end{equation}
If this criterion is satisfied, then the boundary terms from (\ref{eq:12}) vanish and the Hamiltonian is hermitian. To conclude, since we have now proved that this holds for arbitrary dimensions $M$ of the manifold $\mathcal{T}$, it trivially follows for the 1-dimensional model as well.

\pagebreak	% Appendix Title

\fancyhead[RE,LO]{Faddeev-Popov Jacobian calculation in minisuperspace}

\chapter{Faddeev-Popov Jacobian calculation in minisuperspace}

We recall the form of the path integral in the 1-dimensional case after employing the Faddeev-Popov trick, but we also perform a Wick rotation to Euclidean time $x^{0} \rightarrow x^{0}_{\textrm{E}} = ix^{0}$:
\begin{equation} 
    \psi(\textrm{q})=\int_0^{+\infty}\!\!d\ell \; \Delta_{\rm FP}(\ell) \int_{q(x^{0}_{\textrm{E}i})=\textrm{q}_i}^{q(x^{0}_{f})=\textrm{q}}\mathcal{D} q   \; e^{-{1\over \hbar}\int_{x^{0}_{i}}^{x^{0}_{\textrm{E}f}}d x^0_{\textrm{E}}\, L(\ell,q,\dot q)}
\end{equation}
The form in which the Faddeev-Popov determinant appears is\cite{Reference31}:
\begin{equation} \label{a:FP}
    \frac{1}{\Delta _{\textrm{FP}}[g_{00}[\ell]]} = \int _{0}^{+\infty} d\ell' \int _{\text{Diff}[g_{00}[\ell ']]} \mathcal{D} \xi \, \delta \left [g_{00}[\ell] - g^{\xi}_{00}[\ell '] \right ]
\end{equation}
where we have defined a Dirac delta functional of the metric components, and $\xi (x^{0}_{\textrm{E}}) = x^{\xi 0}_{\textrm{E}}$ is an infinitesimal diffeomorphism in the $x^{0}_{\textrm{E}}$ direction:
\begin{equation}
    \xi = \text{Id} +\delta \xi \quad  \Rightarrow \quad x^{\xi 0}_{\textrm{E}} = x^{0}_{\textrm{E}} + \delta x^{0}_{\textrm{E}}
\end{equation}
The spacetime metric components transform as such:
\begin{equation}
     g^{\xi}_{00}(x^{\xi 0}_{\textrm{E}}) = \left (\frac{dx^{0}_{\textrm{E}}}{dx^{\xi 0}_{\textrm{E}}} \right )^{2} g_{00}(x^{0}_{\textrm{E}}) \, , \qquad a^{\xi }(x^{\xi 0}_{\textrm{E}}) = a(x^{0}_{\textrm{E}})
\end{equation}
By integrating over all diffeomorphisms (i.e. time reparametrizations) in the presence of the Dirac delta introduced with the Faddeev-Popov trick, we cancel out redundant geometries that stem from this particular symmetry $\text{Diff}[g_{00}[\ell]]$\cite{Reference31}. We also recall that we have gauge-fixed the timelike component of the metric. We started from the fiducial metric $g_{00}[1]$ over a line segment $\left [x^{0}_{\textrm{E}i}, \, x^{0}_{\textrm{E}f} \right ]$ as we saw back in chapter 4, and then chose:
\begin{equation}
    g_{00}[\ell] = \ell ^{2} g_{00}[1]
\end{equation}
It is important to note that it's obvious from the definition of the length $\ell$ in (\ref{line}) that it is a diffeomorphism-invariant quantity. Combining this with the fact the Dirac delta is also invariant, we find that it is true for the determinant too\cite{Reference31}:
\begin{equation}
\begin{split}
    \frac{1}{\Delta _{\textrm{FP}}[g^{\xi}_{00}[\ell]]} &= \int _{0}^{+\infty} d\ell' \int _{\text{Diff}[g_{00}[\ell ']]} \mathcal{D} \xi' \, \delta \left [g^{\xi}_{00}[\ell] - g^{\xi '}_{00}[\ell '] \right ] \\
    &= \int _{0}^{+\infty} d\ell' \int _{\text{Diff}[g_{00}[\ell ']]} \mathcal{D} \xi' \, \delta \left [g_{00}[\ell] - g^{\xi ^{-1} \circ \xi}_{00}[\ell '] \right ] \\
    &= \int _{0}^{+\infty} d\ell' \int _{\text{Diff}[g_{00}[\ell ']]} \mathcal{D} \xi'' \, \delta \left [g_{00}[\ell] - g^{\xi ''}_{00}[\ell '] \right ] = \frac{1}{\Delta _{\textrm{FP}}[g_{00}[\ell]]}
\end{split}
\end{equation}
where we defined $\xi '' = \xi ^{-1} \circ \xi$ and we used $\mathcal{D} \xi '' = \mathcal{D} \xi '$. We further note that for a 1-dimensional line segment, there is an orientation reversal isometry, which means the path integral over diffeomorphisms becomes:
\begin{equation}
    \int _{\text{Diff}[g_{00}[\ell ']]} \mathcal{D} \xi \rightarrow 2 \int _{\text{Diff}[g_{00}[\ell ']]_{\text{Id}}} \mathcal{D} \xi
\end{equation}
The Dirac delta now only enforces $\ell ' =\ell$ and $\xi = \text{Id}$. The total variation of the fiducial metric is:
\begin{equation}
    \delta g_{00}[\ell] = g_{00}^{\text{Id}+\delta \xi}[\ell +\delta \ell] - g_{00}[\ell] = -2\nabla _{0} (\delta x_{0\textrm{E}}) + 2g_{00}[\ell] \frac{\delta \ell}{\ell} + \, ...
\end{equation}
where $\nabla _{0}$ the covariant derivative defined on $g_{00}$ and we have omitted terms of order $\mathcal{O}\left ((\delta x_{0\textrm{E}})^{2} \right )$ and above. The Jacobian determinant now reads:
\begin{equation}
    \frac{1}{\Delta _{\textrm{FP}}[g_{00}[\ell]]} = 2\int  d(\delta \ell) \int _{\delta x^{0}_{\textrm{E}}(x^{0}_{\textrm{E}i}) = 0}^{\delta x^{0}_{\textrm{E}}(x^{0}_{\textrm{E}f})=0} \mathcal{D} (\delta x^{0}_{\textrm{E}}) \, \delta \left [2\nabla _{0} (\delta x_{0\textrm{E}}) - 2g_{00}[\ell] \frac{\delta \ell}{\ell} \right ]
\end{equation}
It is useful to analyze the Dirac delta with a Fourier transform with a new parameter $\beta ^{00}$:
\begin{equation}
\begin{split}
    \frac{1}{\Delta _{\textrm{FP}}[g_{00}[\ell]]} = 2\int  d(\delta \ell) &\int _{\delta x^{0}_{\textrm{E}}(x^{0}_{\textrm{E}i}) = 0}^{\delta x^{0}_{\textrm{E}}(x^{0}_{\textrm{E}f})=0} \mathcal{D} (\delta x^{0}_{\textrm{E}}) \int \mathcal{D} (\beta^{00}) \\
    \times \, &\exp{\left \{ 2i\pi \int _{x^{0}_{i}}^{x^{0}_{f}} dx^{0}_{\textrm{E}} \, \sqrt{-g_{00}[\ell]} \, \beta ^{00} \left (2\nabla _{0} (\delta x_{0\textrm{E}}) - 2g_{00}[\ell] \frac{\delta \ell}{\ell} \right ) \right \}}
\end{split}
\end{equation}
where $\mathcal{D} \beta ^{00}$ is assumed to be invariant under diffeomorphisms. The positioning of the indices for the quantities in the exponential can be performed arbitrarily, hence there is no ambiguity involved in the way that it is written. This will be made clear shortly. Furthermore, for the sake of brevity, we shall introduce the following notation\cite{Reference31}:
\begin{equation} \label{a:inner}
    (f, \, h)_{\ell} = \int _{x^{0}_{\textrm{E}i}}^{x^{0}_{\textrm{E}f}} dx^{0}_{\textrm{E}} \, \sqrt{-g_{00}[\ell]} \, f^{0...0} h^{0...0} (g_{00}[\ell])^{m}
\end{equation}
where $f^{0...0} , \, h^{0...0}$ contravariant vectors with $m$ indices each. If we allow $m \in \mathbb{Z}$, then the negative values correspond to covariant indices, while $(g_{00}[\ell])^{m}$ becomes a positive power of the inverse metric component. Let us switch all the integration variables to Grassmann numbers:
\begin{equation}
    \delta \ell \rightarrow \lambda \, , \qquad \delta x^{0}_{\textrm{E}} (x_{0\textrm{E}}) \rightarrow c_{0} (x^{0}_{\textrm{E}}) \, , \qquad \beta ^{00} (x^{0}_{\textrm{E}}) \rightarrow b^{00} (x^{0}_{\textrm{E}})
\end{equation}
The variables $c^{0} , \, b^{00}$ correspond to the symmetry we gauge away, therefore they are ghost fields. Given that all integration variables are Grassmann numbers, the operation is now reverted up to an irrelevant constant factor $\alpha$, so the determinant takes the form:
\begin{equation}
\begin{split}
     2\Delta _{\textrm{FP}}[g_{00}[\ell]] &= \alpha \int _{c_{0}(x^{0}_{i})}^{c_{0}(x^{0}_{f})} \mathcal{D}c \int \mathcal{D}b \int d\lambda \, \exp{\left \{ 2i\pi \left (b, \, 2\nabla c - 2g[\ell] \frac{\lambda}{\ell} \right )_{\ell} \right \}} \\
     &= 4i\alpha \pi \int _{c_{0}(x^{0}_{i})}^{c_{0}(x^{0}_{f})} \mathcal{D}c \int \mathcal{D}b \, \left (b, \frac{g[\ell]}{\ell} \right )_{\ell} \exp{\left \{ 4i\pi \,  (b, \,\nabla c )_{\ell} \right \}}
\end{split}
\end{equation}
We now specify the fiducial metric to be the one chosen back in chapter 4 in order to simplify our calculations, setting $g_{00}[1] = 1$ over the segment $[0,1]$. By extension we have $g_{00}[\ell] = \ell ^{2}$ and $\nabla _{0} = \partial _{0}$. The timelike coordinate $x^{0}_{\textrm{E}}$ is simply the Euclidean cosmological time $t_{\textrm{E}}$. For simplicity, we may further define $t_{\textrm{E}} = \ell \,\tau$, with $\tau$ being a dimensionless parameter\cite{Reference31}. \par
Both $\beta^{00}$ and $\delta x_{0\textrm{E}}$ are defined in the vectorial space of functions that are square integrable in the domain $[0,1]$, which we symbolize as $L^{2}([0,1])$. We can therefore expand them in the basis $\left \{\frac{1}{\sqrt{2}} , \, \cos{(k\pi \tau )} \, , \, k\in \mathbb{N} \right \}$. In addition, since $\delta x_{0\textrm{E}}$ has vanishing boundary conditions, we can define the latter in $\left \{\sin{(k\pi \tau )} \, , \, k\in \mathbb{N} \right \}$ instead. The notation (\ref{a:inner}) can be seen as an inner product over the respective Hilbert space of the two functions\cite{Reference31}:
\begin{equation}
    (f, \, h)_{\ell} = \int _{0}^{1} d\tau \, \ell \, f^{00}(\tau) h^{00}(\tau) (\ell ^{2})^{2} \, \, , \qquad (f, \, h)_{\ell} = \int _{0}^{1} d\tau \, \ell \, f_{0}(\tau) h_{0}(\tau) (\ell ^{2})^{-1}
\end{equation}
where the former corresponds to $\beta^{00}$ and the latter to $\delta x_{0\textrm{E}}$. Here it is made clear how there is no ambiguity in choosing index positions for $\beta^{00}$, $\delta x_{0\textrm{E}}$, $b^{00}$, and $c_{0}$. Even if we had defined the new variables as, for example, $b^{0}{}_{0}$ and $c^{0}$, we can always raise or lower indices with $g^{00}$ and $g_{00}$ respectively. This corresponds to simply dividing or multiplying by a power of $\ell ^{2}$ in the inner products above, leaving them completely unchanged. \par
The variables can now be expanded in an orthonormal basis\cite{Reference31}:
\begin{equation}
\begin{split}
    &\beta ^{00}(\tau) = \sum _{k \geq 0} \boldsymbol{\beta}_{k} \, \chi_{k}^{00}(\tau) \, \,  , \qquad \delta x_{0\textrm{E}} = \sum _{k \geq 0} \boldsymbol{\gamma}_{k} \, \sigma_{0,k}(\tau) \\
    \Rightarrow \, &b ^{00}(\tau) = \sum _{k \geq 0} \boldsymbol{b}_{k} \,\chi_{k}^{00}(\tau) \, \,  , \qquad c_{0} = \sum _{k \geq 0} \boldsymbol{c}_{k} \, \sigma_{0,k}(\tau)
\end{split}
\end{equation}
where $\boldsymbol{\beta}_{k} , \, \boldsymbol{\gamma}_{k}$ (not to be confused with the target space metric seen elsewhere in the text) and $\boldsymbol{b}_{k} , \, \boldsymbol{c}_{k}$ regular and anticommuting constants respectively. The modes in turn are:
\begin{equation}
\begin{split}
    \chi _{0}^{00}(\tau) = \frac{\ell ^{-2}}{\sqrt{\ell}} \, \, , \qquad &\chi _{k}^{00}(\tau) = \ell ^{-2} \sqrt{\frac{2}{\ell}} \, \cos{(k\pi \tau)} \, , \, \, \, k\in \mathbb{N} \\
    &\sigma _{0,k}(\tau) = \ell \sqrt{\frac{2}{\ell}} \, \sin{(k\pi \tau)} \, , \, \, \, k\in \mathbb{N}
\end{split}
\end{equation}
which then satisfy $(\chi_{k}, \, \chi_{k'})_{\ell} = \delta _{kk'}$ and $(\sigma_{k}, \, \sigma _{k'})_{\ell} = \delta _{kk'}$. Using these, we obtain the inner products of the quantities appearing in the expression for the determinant:
\begin{equation}
    \left (b, \, \frac{g[\ell]}{\ell} \right )_{\ell} = \frac{\boldsymbol{b}_{0}}{\sqrt{\ell}} \, \, , \qquad 4i\pi \, (b , \, \nabla c)_{\ell} = \frac{4i\pi ^{2}}{\ell} \sum _{k\geq 1} k \, \boldsymbol{b}_{k} \, \boldsymbol{c}_{k}
\end{equation}
Moreover, we must correctly define the path integral measures in this new orthonormal basis. For the regular (non-Grassmann) parameters, we have:
\begin{equation}
\begin{split}
    \big ||\beta | \big |^{2}_{\ell} = (\beta, \, \beta)_{\ell} = \sum _{k\geq 0} \boldsymbol{\beta}_{k}^{2} \qquad  &\Rightarrow \qquad \mathcal{D}\beta = \bigwedge _{k\geq0} d\boldsymbol{\beta}_{k} \\
     \big ||\delta x_{\textrm{E}} | \big |^{2}_{\ell} = (\delta x_{\textrm{E}}, \, \delta x_{\textrm{E}})_{\ell} = \sum _{k\geq 1} \boldsymbol{\gamma}_{k}^{2} \qquad  &\Rightarrow \qquad \mathcal{D}\delta x_{\textrm{E}}= \bigwedge _{k\geq1} d\boldsymbol{\gamma}_{k}
\end{split}
\end{equation}
Equivalently for the anticommuting variables:
\begin{equation}
    \mathcal{D}b = \prod _{k\geq0} d\boldsymbol{b}_{k} \, \, \qquad \mathcal{D}c = \prod _{k\geq1} d\boldsymbol{c}_{k}
\end{equation}
Combining all of these, the expression for the determinant now becomes\cite{Reference31}:
\begin{equation} \label{a:result}
\begin{split}
    \Delta _{\textrm{FP}}[\ell ^{2}] &= 2i\pi \alpha \int d\boldsymbol{b}_{0} \, \frac{\boldsymbol{b}_{0}}{\sqrt{\ell}} \prod _{k\geq1} \int d\boldsymbol{c}_{k} \, d\boldsymbol{b}_{k} \, \exp{\left \{ \frac{4i\pi ^{2}}{\ell} k \, \boldsymbol{b}_{k} \, \boldsymbol{c}_{k} \right \}} \\
    &= \frac{2i\pi \alpha}{\sqrt{\ell}} \prod_{k\geq1} \frac{4i\pi ^{2}}{\ell} \, k \\
    &= \alpha \sqrt{2i\pi} 
\end{split}
\end{equation}
We got the third line by utilizing the zeta regularization identities\cite{Reference65}:
\begin{equation}
    \prod _{k\geq1} z = \frac{1}{\sqrt{z}} \, \, , \qquad \prod _{k\geq1}k = \sqrt{2\pi}
\end{equation}
$\alpha$, despite being arbitrary, is restricted to values such that $\Delta _{\textrm{FP}}$ comes out strictly real and positive-definite\cite{Reference31}\cite{Reference65}. The result (\ref{a:result}) confirms that the Jacobian determinant is a trivial constant which we could then set equal to 1, as we have done in chapter 4. \par
In conclusion, we should mention that even though this calculation was done in the 1-dimensional model, it trivially holds in the $M$-dimensional case. We still have a FRLW metric with a component $g_{00}$ that is defined on the line segment of length $\ell$  with the same diffeomorphism invariance, so the Faddeev-Popov determinant is still defined as (\ref{a:FP}). All other steps shown in this appendix follow with the same logic applied, eventually leading to the same result.

\pagebreak % Appendix Title

\addtocontents{toc}{\vspace{2em}}  % Add a gap in the Contents, for aesthetics
\backmatter

%% ----------------------------------------------------------------
\label{Bibliography}
\fancyhead[LO,RE]{\emph{Bibliography}}  % Change the left side page header to "Bibliography"
\bibliographystyle{unsrtnat}  % Use the "unsrtnat" BibTeX style for formatting the Bibliography
\bibliography{Bibliography}  % The references (bibliography) information are stored in the file named "Bibliography.bib"

@article{Reference1,
	Author = "J.B. Hartle and S.W. Hawking",
	Journal = "Physical Review D",
	Number = "12",
	Title = "{Wave function of the Universe}",
	Volume = "28",
	Year = "1983"}

@article{Reference2,
	Author = "L. Álvarez-Gaumé and A. Kehagias and C. Kounnas and D. Lüst and A. Riotto",
	Title = "{Aspects of Quadratic Gravity}",
	Year = "2015",
	note = "[arXiv:1505.07657v2 [hep-th]]" }

@article{Reference3,
	Author = "C. Kounnas and D. Lüst and N. Toumbas",
	Title = "{$R^{2}$ inflation from scale invariant supergravity and anomaly free superstrings with fluxes}",
	Year = "2014",
	note = "[arXiv:1409.7076v2 [hep-th]]" }

@article{Reference4,
	Author = "A. Guarnizo and L. Castañeda and J. M. Tejeiro",
	Title = "{Boundary Term in Metric \(f(R)\) Gravity: Field Equations in the Metric Formalism}",
	Year = "2010",
	note = "[arXiv:1002.0617v4 [gr-qc]]" }

@book{Reference5,
    author    = "S. Weinberg",
    title     = "{Gravitation and Cosmology: Principles and Applications of the General Theory of Relativity}",
    year      = "1972",
    publisher = "John Wiley \& Sons Inc.",
    address   = "Massachusetts Institute of Technology"}

@book{Reference6,
    author    = "M. Spradlin and A. Strominger and A. Volovich",
    title     = "{Les Houches Lectures on de Sitter Space}",
    year      = "2001",
    note      = "[arXiv:hep-th/0110007v2"}

@book{Reference7,
    author    = "F. Mandl and G. Shaw",
    title     = "{Quantum Field Theory}",
    year      = "1984",
    publisher = "John Wiley \& Sons Inc."}

@article{Reference8,
	Author = "S. Zerbini",
	Title = "{One-loop Effective Action for \(f(R)\) Gravity with Applications to Cosmology}",
	Year = "2015",
	note = "CP3 Lecture, Odense"}

@article{Reference9,
	Author = "J.B. Hartle",
	Title = "{The State of the No-Boundary Wave Function}",
	Year = "2012",
	note = "Hawking70, Cambridge"}

@book{Reference10,
    author    = "R.P. Feynman and A.R. Hibbs",
    title     = "{Quantum Mechanics and Path Integrals}",
    year      = "1965",
    publisher = "Dover Publications Inc.",
    address   = "New York"}

@article{Reference11,
	Author = "T. Hartman",
	Title = "{Lecture Notes on Classical de Sitter Space}",
	note = "email: hartman@cornell.edu"}

@article{Reference12,
	Author = "A.D. Linde",
	Journal = "Reports on Progress in Physics",
	Title = "{The inflationary universe}",
	Volume = "47",
	Year = "1984"}

@article{Reference13,
    author    = "R.A. McNees IV",
    title     = "{Conventions, Definitions, Identities, and Other Useful Formulae}",
    year      = "2010",
    Journal   = "Physics: Faculty Publications and Other Works, Faculty Publications",
    url       = "http://jacobi.luc.edu/notes.html",
    note      = "Loyola University Chicago"}

@book{Reference14,
    author    = "A.D. Linde",
    title     = "{Particle Physics and Inflationary Cosmology}",
    year      = "1990",
    publisher = "CRC Press"}

@book{Reference15,
    author    = "E.W. Kolb and M.S. Turner",
    title     = "{The Early Universe}",
    year      = "1990",
    publisher = "Addison-Wesley Publishing Company"}

@book{Reference16,
    author    = "S.R. Coleman",
    title     = "{Aspects of Symmetry: Selected Erice Lectures}",
    year      = "1985",
    publisher = "Cambridge University Press"}

@article{Reference17,
	Author = "A.D. Linde",
	Journal = "Physical Review D",
	Title = "{Quantum creation of an open inflationary Universe}",
	Volume = "58",
	Year = "1998"}

@article{Reference18,
	Author = "A. Vilenkin",
	Journal = "Physical Review D",
	Title = "{Birth of inflationary universes}",
	Volume = "27(12)",
	Year = "1983"}

@article{Reference19,
	Author = "A. Vilenkin",
	Journal = "Physical Review D",
	Title = "{Boundary conditions in quantum cosmology}",
	Volume = "33(12)",
	Year = "1986"}

@article{Reference20,
	Author = "A. Vilenkin",
	Journal = "Physics Letters",
	Title = "{Creation of universes from nothing}",
	Volume = "117B(1,2)",
	Year = "1982"}

@article{Reference21,
	Author = "A. Vilenkin",
	Journal = "Physical Review D",
	Title = "{Quantum creation of universes}",
	Volume = "30(2)",
	Year = "1984"}

@article{Reference22,
	Author = "B.S. DeWitt",
	Journal = "Physical Review",
	Title = "{Quantum Theory of Gravity. I. The Canonical Theory}",
	Volume = "160(5)",
	Year = "1967"}

@article{Reference23,
	Author = "G.W. Gibbons and S.W. Hawking",
	Journal = "Physical Review D",
	Title = "{Action integrals and partition functions in quantum gravity}",
	Volume = "15(10)",
	Year = "1977"}

@article{Reference24,
	Author = "G.W. Gibbons and S.W. Hawking and M.J. Perry",
	Journal = "Nuclear Physics",
	Title = "{Path integrals and the indefiniteness of the gravitational action}",
	Volume = "B138",
	Year = "1978"}

@article{Reference25,
	Author = "C. Kounnas and H. Partouche",
	Title = "{Instanton Transition in Thermal and Moduli deformed de Sitter Cosmology}",
	Year = "2007",
	note = "[arXiv:0705.3206v3 [hep-th]]" }

@article{Reference26,
	Author = "C. Kounnas and N. Toumbas and J. Troost",
	Title = "{A Wave-function for Stringy Universes}",
	Year = "2007",
	note = "[arXiv:0704.1996v3 [hep-th]]" }

@article{Reference27,
	Author = "R.M. Wald",
	Journal = "Physical Review D",
	Title = "{Asymptotic behavior of homogeneous cosmological models in the presence of a positive cosmological constant}",
	Volume = "28(8)",
	Year = "1983"}

@article{Reference28,
	Author = "A.A. Starobinsky",
	Journal = "Pisma Astron. Zh.",
	Title = "{The perturbation spectrum evolving from a nonsingular, initially de Sitter cosmology, and the microwave background anisotropy}",
	Volume = "9",
	Year = "1983"}

@article{Reference29,
	Author = "A.A. Starobinsky",
	Journal = "Physics Letters",
	Title = "{A new type of isotropic cosmological models without singularity}",
	Volume = "91B(1)",
	Year = "1980"}

@article{Reference30,
	Author = "V.F. Mukhanov  and G.V. Chibisov",
	Journal = "Pisma Zh. Eksp. Teor. Fiz.",
	Title = "{Quantum Fluctuation and Nonsingular Universe}",
	Volume = "33",
	Year = "1981"}

@article{Reference31,
	Author = "H. Partouche and N. Toumbas and B. de Vaulchier",
	Title = "{Wavefunction of the universe: Reparametrization invariance and field redefinitions of the minisuperspace path integral}",
	Year = "2021",
	note = "[arXiv:2103.15168v1 [hep-th]]" }

@article{Reference32,
	Author = "H. Partouche and N. Toumbas and B. de Vaulchier",
	Title = "{Gauge fixing and field redefinitions of the Hartle–Hawking wavefunction path integral}",
	Year = "2021",
	note = "[arXiv:2105.04818v1 [hep-th]]" }

@article{Reference33,
	Author = "E. Kaimakkamis and H. Partouche and N. Toumbas",
	Title = "{Wave function of the universe in $R^{2}$ theory}",
	Year = "TBD",
	note = "{Work in progress}" }

@book{Reference34,
    author    = "S. Weinberg",
    title     = "{Cosmology}",
    year      = "2008",
    publisher = "Oxford University Press"}

@article{Reference35,
	Author = "Planck Collaboration",
	Journal = "A\&A",
	Title = "{Planck 2018 results I-XII}",
	Volume = "641(A1-A12)",
	Year = "2020"}

@book{Reference36,
    author    = "M.P. Hobson and G.P. Efstathiou and A.N. Lasenby",
    title     = "{General Relativity: An Introduction for Physicists}",
    year      = "2006",
    publisher = "Cambridge University Press"}

@book{Reference37,
    author    = "N.D. Birrell and P. C.W. Davies",
    title     = "{Quantum Fields in Curved Space}",
    year      = "1982",
    publisher = "Cambridge University Press"}

@article{Reference38,
	Author = "W. Israel",
	Journal = "Physical Review",
	Title = "{Event Horizons in Static Vacuum Space-Times}",
	Volume = "164(5)",
	Year = "1967"}

@book{Reference39,
    author    = "C.W. Misner and K.S. Thorne and J.A. Wheeler",
    title     = "{Gravitation}",
    year      = "2017",
    publisher = "Princeton University Press"}

@book{Reference40,
    author    = "C. Kiefer",
    title     = "{Quantum Gravity, Third Edition}",
    year      = "2012",
    publisher = "Oxford University Press"}

@book{Reference41,
    author    = "V.F. Mukhanov and S. Winitzki",
    title     = "{Introduction to Quantum Effects in Gravity}",
    year      = "2007",
    publisher = "Cambridge University Press"}

@article{Reference42,
	Author = "E. Kaimakkamis and H. Partouche and K. Sil and N. Toumbas",
	Journal = "Class. Quant. Grav.",
	Title = "{The exact Wheeler-DeWitt equation for the scale-factor minisuperspace model}",
	Volume = "42",
	Year = "2025",
    note = "[arXiv:2412.18532 [hep-th]]"}

@article{Reference43,
	Author = "E. Kaimakkamis and K. Sil",
	Journal = "PoS \textbf{CORFU2023}",
	Title = "{A formalism for the ambiguities of the Wheeler-DeWitt equation}",
	Volume = "227",
	Year = "2024",
    note = "[arXiv:2407.03077 [gr-qc]]"}

@book{Reference44,
    author    = "M.E. Peskin and D.V. Schroeder",
    title     = "{An Introduction to Quantum Field Theory}",
    year      = "1995",
    publisher = "Perseus Book Publishing"}

@article{Reference45,
	Author = "S.W. Hawking",
	Journal = "Phys. Lett. B",
	Title = "{The cosmological constant is probably zero}",
	Volume = "134",
	Year = "1984"}

@article{Reference46,
	Author = "J.J. Halliwell and S.W. Hawking",
	Journal = "Phys. Rev. D",
	Title = "{The origin of structure in the Universe}",
	Volume = "31",
	Year = "1985"}

@article{Reference47,
	Author = "S.W. Hawking and D.N. Page",
	Journal = "Nucl. Phys. B",
	Title = "{Operator ordering and the flatness of the Universe}",
	Volume = "264",
	Year = "1986"}

@article{Reference48,
	Author = "J.J. Halliwell",
	Journal = "Phys. Rev. D",
	Title = "{Derivation of the Wheeler-De Witt equation from a path integral for minisuperspace models}",
	Volume = "38",
	Year = "1988"}

@article{Reference49,
	Author = "J.J. Halliwell and J. Louko",
	Journal = "Phys. Rev. D",
	Title = "{Steepest descent contours in the path integral approach to quantum cosmology. 1. The de Sitter minisuperspace model}",
	Volume = "39",
	Year = "1989"}

@article{Reference50,
	Author = "J.B. Hartle and S.W. Hawking and T. Hertog",
	Journal = "Phys. Rev. D",
	Title = "{The classical Universes of the no-boundary quantum state}",
	Volume = "77",
	Year = "2008",
    note = "[arXiv:0803.1663 [hep-th]]"}

@article{Reference51,
	Author = "J. Feldbrugge and J.L. Lehners and N. Turok",
	Journal = "Phys. Rev. D",
	Title = "{Lorentzian Quantum Cosmology}",
	Volume = "95",
	Year = "2017",
    note = "[arXiv:1703.02076 [hep-th]]"}

@article{Reference52,
	Author = "J. Diaz Dorronsoro and J.J. Halliwell and J.B. Hartle and T. Hertog and O. Janssen",
	Journal = "Phys. Rev. D",
	Title = "{The real no-boundary wave function in Lorentzian quantum cosmology}",
	Volume = "96",
	Year = "2017",
    note = "[arXiv:1705.05340 [gr-qc]]"}

@article{Reference53,
	Author = "W. Nelson and M. Sakellariadou",
	Journal = "Phys. Rev. D",
	Title = "{Unique factor ordering in the continuum limit of LQC}",
	Volume = "78",
	Year = "2008",
    note = "[arXiv:0806.0595 [gr-qc]]"}

@article{Reference54,
	Author = "D. He and D. Gao and Q. Y. Cai",
	Journal = "Phys. Lett. B",
	Title = "{Dynamical interpretation of the wavefunction of the Universe}",
	Volume = "748",
	Year = "2015",
    note = "[arXiv:1507.06727 [gr-qc]]"}

@article{Reference55,
	Author = "H. Partouche and N. Toumbas and B. de Vaulchier",
	Journal = "PoS \textbf{CORFU2021}",
	Title = "{Wavefunction of the Universe: Diffeomeorphism invariance and field redefinitions}",
	Volume = "159",
	Year = "2022"}

@article{Reference56,
	Author = "A. Kehagias and H. Partouche and N. Toumbas",
	Journal = "JHEP",
	Title = "{Probability distribution for the quantum Universe}",
	Volume = "12",
	Year = "2021",
    note = "[arXiv:2110.03050 [hep-th]]"}

@article{Reference57,
	Author = "E. Anderson",
	Journal = "Class. Quant. Grav.",
	Title = "{Relational motivation for conformal operator ordering in quantum cosmology}",
	Volume = "27",
	Year = "2010",
    note = "[arXiv:0905.3357 [gr-qc]]"}

@book{Reference58,
    author    = "M. Henneaux and C. Teitelboim",
    title     = "{Quantization of gauge systems}",
    year      = "1992",
    publisher = "Princeton University Press"}

@article{Reference59,
	Author = "C.J. Isham",
	Journal = "NATO Sci. Ser. C",
	Title = "{Canonical quantum gravity and the problem of time}",
	Volume = "409",
	Year = "1993",
    note = "[arXiv:gr-qc/9210011v1 [gr-qc]]"}

@article{Reference60,
	Author = "D.N. Page",
	Journal = "Class. Quant. Grav.",
	Title = "{A fractal set of perpetually bouncing Universes?}",
	Volume = "1",
	Year = "1984"}

@article{Reference61,
	Author = "M. Demianski and R. de Ritis and C. Rubano and P. Scudellaro",
	Journal = "Phys. Rev. D",
	Title = "{Scalar fields and anisotropy in cosmological models}",
	Volume = "46",
	Year = "1992"}

@article{Reference62,
	Author = "G.W. Gibbons and S.W. Hawking and J.M. Stewart",
	Journal = "Nucl. Phys. B",
	Title = "{A Natural Measure on the Set of All Universes}",
	Volume = "281",
	Year = "1987"}

@article{Reference63,
	Author = "S.M. Carroll and Jennifer Chen",
	Journal = "International Journal of Modern Physics D",
	Title = "{Does Inflation Provide Natural Initial Conditions for the Universe?}",
	Volume = "14(12)",
	Year = "2005",
    note = "[arXiv:gr-qc/0505037v1 [gr-qc]]"}

@article{Reference64,
	Author = "V. Franken and E. Kaimakkamis and H. Partouche and N. Toumbas",
	Title = "{Ordering-Independent Wheeler-DeWitt Equation for Flat Minisuperspace Models}",
	Year = "2025",
	note = "[arXiv:2512.23656 [hep-th]]" }

@book{Reference65,
    author    = "J. Polchinski",
    title     = "{String theory. Vol. 1: An introduction to the bosonic string}",
    year      = "1998",
    publisher = "Cambridge University Press"}

@article{Reference66,
	Author = "D. He and Q.Y. Cai",
	Journal = "Phys. Lett. B",
	Title = "{Wheeler-DeWitt equation rejects quantum effects of grown-up universes as a candidate for dark energy}",
	Volume = "809",
	Year = "2020",
    note = "[arXiv:2009.05187 [gr-qc]]"}

@article{Reference67,
	Author = "J.J. Halliwell and R.C. Myers",
	Journal = "Phys. Rev. D",
	Title = "{Multiple sphere configurations in the path integral representation of the wave function of the universe}",
	Volume = "40",
	Year = "1989"}

@article{Reference68,
	Author = "C.G. Callan and S.R. Coleman",
	Journal = "Phys. Rev. D",
	Title = "{The fate of the false vacuum. II. First quantum corrections}",
	Volume = "16",
	Year = "1977"}

@article{Reference69,
	Author = "T. Christodoulakis and J. Zanelli",
	Journal = "Nuovo Cim. B",
	Title = "{Operator Ordering in Quantum Mechanics and Quantum Gravity}",
	Volume = "93",
	Year = "1986"}

@article{Reference70,
	Author = "H. Singh and M.K. Nandy",
	Journal = "Int. J. Theor. Phys.",
	Title = "{Imprint of Operator Ordering on the Quantum Nature of DeWitt-regular Black Holes}",
	Volume = "64",
	Year = "2025"}

@article{Reference71,
	Author = "V. Mondal and H.S. Sahota and K. Lochan",
	Journal = "JHEP",
	Title = "{Exact path integrals on half-line in quantum cosmology with a fluid clock and aspects of operator ordering ambiguity}",
	Volume = "128",
	Year = "2025",
    note = "[arXiv:2501.11680 [gr-qc]]"}

@article{Reference72,
	Author = "R.L. Arnowitt and S. Deser and C.W. Misner",
	Journal = "Phys. Rev.",
	Title = "{Dynamical Structure and Definition of Energy in General Relativity}",
	Volume = "116",
	Year = "1959"}

@article{Reference73,
	Author = "E. Anderson",
	Journal = "Class. Quant. Grav.",
	Title = "{Relational motivation for conformal operator ordering in quantum cosmology}",
	Volume = "27",
	Year = "2009",
    note = "[arXiv:0905.3357 [gr-qc]]"}

@article{Reference74,
	Author = "A.H. Guth",
	Journal = "Phys. Rev. D",
	Title = "{The Inflationary Universe: A Possible Solution to the Horizon and Flatness Problems}",
	Volume = "23",
	Year = "1981"}

@article{Reference75,
	Author = "K.S. Stelle",
	Journal = "Phys. Rev. D",
	Title = "{Renormalization of Higher Derivative Quantum Gravity}",
	Volume = "16",
	Year = "1977"}

@article{Reference76,
	Author = "K.S. Stelle",
	Journal = "Gen. Rel. Grav.",
	Title = "{Classical Gravity with Higher Derivatives}",
	Volume = "9",
	Year = "1978"}

@article{Reference77,
	Author = "P. Teyssandier and P. Tourenc",
	Journal = "J. Math. Phys.",
	Title = "{The Cauchy problem for the R+R**2 theories of gravity without torsion}",
	Volume = "24",
	Year = "1983"}

@article{Reference78,
	Author = "B. Whitt",
	Journal = "Phys. Lett. B",
	Title = "{Fourth Order Gravity as General Relativity Plus Matter}",
	Volume = "145",
	Year = "1984"}

@article{Reference79,
	Author = "J.D. Barrow and S. Cotsakis",
	Journal = "Phys. Lett. B",
	Title = "{Inflation and the Conformal Structure of Higher Order Gravity Theories}",
	Volume = "214",
	Year = "1988"}

@article{Reference80,
	Author = "J.D. Barrow",
	Journal = "Nucl. Phys. B",
	Title = "{The Premature Recollapse Problem in Closed Inflationary Universes}",
	Volume = "296",
	Year = "1988"}

@article{Reference81,
	Author = "T. Chiba",
	Journal = "Phys. Lett. B",
	Title = "{1/R gravity and scalar - tensor gravity}",
	Volume = "575",
	Year = "2003",
    note = "[arXiv:astro-ph/0307338 [astro-ph]]"}

@article{Reference82,
	Author = "N.H. Barth",
	Journal = "Class. Quant. Grav.",
	Title = "{The Fourth Order Gravitational Action for Manifolds With Boundaries}",
	Volume = "2",
	Year = "1985"}

@article{Reference83,
	Author = "R. Cassadio and A. Gruppuso",
	Journal = "Int. J. Mod. Phys. D",
	Title = "{On boundary terms and conformal transformations in curved space-times}",
	Volume = "11",
	Year = "2002",
    note = "[arXiv:gr-qc/0107077 [gr-qc]]"}

@article{Reference84,
	Author = "E. Dyer and K. Hinterbichler",
	Journal = "Phys. Rev. D",
	Title = "{Boundary Terms, Variational Principles and Higher Derivative Modified Gravity}",
	Volume = "79",
	Year = "2009",
    note = "[arXiv:0809.4033 [gr-qc]]"}

@article{Reference85,
	Author = "A. Balczerak and M.P. Dabrowski",
	Journal = "JCAP",
	Title = "{Gibbons-Hawking Boundary Terms and Junction Conditions for Higher-Order Brane Gravity Models}",
	Volume = "1",
	Year = "2009",
    note = "[arXiv:0804.0855 [hep-th]]"}

@article{Reference86,
	Author = "R. Jackiw",
	Journal = "Nucl. Phys. B",
	Title = "{Lower Dimensional Gravity}",
	Volume = "252",
	Year = "1985"}

@article{Reference87,
	Author = "C. Teitelboim",
	Journal = "Phys. Lett. B",
	Title = "{Gravitation and Hamiltonian Structure in Two Space-Time Dimensions}",
	Volume = "126",
	Year = "1983"}

@article{Reference88,
	Author = "J. Maldacena and G.J. Turiaci and Z. Yang",
	Journal = "JHEP",
	Title = "{Two dimensional Nearly de Sitter gravity}",
	Volume = "01",
	Year = "2021",
    note = "[arXiv:1904.01911 [hep-th]]"}

@article{Reference89,
	Author = "J. Cotler and K. Jensen",
	Journal = "JHEP",
	Title = "{Emergent unitarity in de Sitter from matrix integrals}",
	Volume = "12",
	Year = "2021",
    note = "[arXiv:1911.12358 [hep-th]]"}

@article{Reference90,
	Author = "J. Cotler and K. Jensen and A. Maloney",
	Journal = "JHEP",
	Title = "{Low-dimensional de Sitter quantum gravity}",
	Volume = "06",
	Year = "2020",
    note = "[arXiv:1905.03780 [hep-th]]"}

@article{Reference91,
	Author = "K.K. Nanda and S.K. Sake and S.P. Trivedi",
	Journal = "JHEP",
	Title = "{JT gravity in de Sitter space and the problem of time}",
	Volume = "02",
	Year = "2024",
    note = "[arXiv:2307.15900 [hep-th]]"}

@article{Reference92,
	Author = "J. Cotler and K. Jensen",
	Journal = "JHEP",
	Title = "{Non-perturbative de Sitter Jackiw-Teitelboim gravity}",
	Volume = "12",
	Year = "2024",
    note = "[arXiv:2401.01925 [hep-th]]"}

@article{Reference93,
	Author = "E. Alonso-Monsalve and D. Harlow and P. Jefferson",
	Journal = "",
	Title = "{Phase space of Jackiw-Teitelboim gravity with positive cosmological constant}",
	Volume = "",
	Year = "2024",
    note = "[arXiv:2409.12943 [hep-th]]"}

@article{Reference94,
	Author = "W. Buchmuller and A. Hebecker and A. Westphal",
	Journal = "JHEP",
	Title = "{DeWitt wave functions for de Sitter JT gravity}",
	Volume = "06",
	Year = "2025",
    note = "[arXiv:2412.09211 [hep-th]]"}

@article{Reference95,
	Author = "I. Dey and K.K. Nanda and A. Roy and S.K. Sake and S.P. Trivedi",
	Journal = "",
	Title = "{JT Gravity in de Sitter Space and Its Extensions}",
	Volume = "",
	Year = "2025",
    note = "[arXiv:2501.03148 [hep-th]]"}

@article{Reference96,
	Author = "J. Held and H. Maxfield",
	Journal = "",
	Title = "{The Hilbert space of de Sitter JT: a case study for canonical methods in quantum gravity}",
	Volume = "",
	Year = "2024",
    note = "[arXiv:2410.14824 [hep-th]]"}

@article{Reference97,
	Author = "D. Louis-Martinez and J. Gegenberg and G. Kunstatter",
	Journal = "Phys. Lett. B",
	Title = "{Exact Dirac quantization of all 2-D dilaton gravity theories}",
	Volume = "321",
	Year = "1994",
    note = "[arXiv:gr-qc/9309018 [gr-qc]]"}

@article{Reference98,
	Author = "B. Alexandre and A. Etkin and F.S. Rassouli",
	Journal = "Phys. Rev. D",
	Title = "{Unimodular Jackiw-Teitelboim gravity and de Sitter quantum cosmology}",
	Volume = "111",
	Year = "2025",
    note = "[arXiv:2501.17213 [gr-qc]]"}

@book{Reference99,
    author    = "M. Bojowald",
    title     = "{Foundations of Quantum Cosmology}",
    year      = "2020",
    publisher = "IOP Publishing"}

@book{Reference100,
    author    = "G. Calcagni",
    title     = "{Classical and Quantum Cosmology}",
    year      = "2017",
    publisher = "Springer"}
\nocite{*}
\end{document}